\documentstyle[epsf,epsfig,12pt]{article}
\def\markphotonatomur{\begin{picture}(2,2)(0,0)
                             \put(2,1){\oval(2,2)[tl]}
                             \put(0,1){\oval(2,2)[br]}
                     \end{picture}
                    }
\def\markphotonatomdr{\begin{picture}(2,2)(0,0)
                             \put(1,0){\oval(2,2)[bl]}
                             \put(1,-2){\oval(2,2)[tr]}
                     \end{picture}
                    }

\def\photonatomright{\begin{picture}(3,1.5)(0,0)
                                \put(0,-0.75){\tencircw \symbol{2}}
                                \put(1.5,-0.75){\tencircw \symbol{1}}
                                \put(1.5,0.75){\tencircw \symbol{3}}
                                \put(3,0.75){\tencircw \symbol{0}}
                      \end{picture}
                     }

\def\photonatomup{\begin{picture}(1.5,3)(0,0)
                             \put(-0.75,3){\tencircw \symbol{3}}
                             \put(-0.75,1.5){\tencircw \symbol{2}}
                             \put(0.75,1.5){\tencircw \symbol{0}}
                             \put(0.75,0){\tencircw \symbol{1}}
                   \end{picture}
                  }

\def\photonright{\begin{picture}(30,1.5)(0,0)
                     \multiput(0,0)(3,0){10}{\photonatomright}
                  \end{picture}
                 }

\def\markphotonurh{\begin{picture}(16,16)(0,0)
                     \multiput(0,0)(2,2){8}{\markphotonatomur}
                  \end{picture}
                 }
\def\markphotondrh{\begin{picture}(16,16)(0,0)
                     \multiput(0,0)(2,-2){8}{\markphotonatomdr}
                  \end{picture}
                 }
\def\photonrighthalf{\begin{picture}(30,1.5)(0,0)
                     \multiput(0,0)(3,0){5}{\photonatomright}
                  \end{picture}
                 }

\def\photonup{\begin{picture}(1.5,30)(0,0)
                  \multiput(0,0)(0,3){10}{\photonatomup}
               \end{picture}
              }

\def\fermionup{\begin{picture}(1,30)(0,0)
                     \put(0,0){\vector(0,1){15}}
                     \put(0,15){\line(0,1){15}}
               \end{picture}
              }

\def\fermiondown{\begin{picture}(1,30)(0,-30)
                       \put(0,0){\vector(0,-1){15}}
                       \put(0,-15){\line(0,-1){15}}
                 \end{picture}
                }

\def\fermionlefthalf{\begin{picture}(15,1)(0,0)
                           \put(15,0){\vector(-1,0){7.5}}
                           \put(7.5,0){\line(-1,0){7.5}}
                     \end{picture}
                    }

\def\fermionrighthalf{\begin{picture}(15,1)(0,0)
                            \put(0,0){\vector(1,0){7.5}}
                            \put(7.5,0){\line(1,0){7.5}}
                      \end{picture}
                     }
\def\fermionul{\begin{picture}(15,15)(0,0)
                        \put(0,0){\vector(-1,1){7.5}}
                        \put(-7.5,7.5){\line(-1,1){7.5}}
                  \end{picture}
                 }

\def\fermionur{\begin{picture}(15,15)(0,0)
                        \put(-15,-15){\vector(1,1){7.5}}
                        \put(-7.5,-7.5){\line(1,1){7.5}}
                  \end{picture}
                 }

\def\fermiondl{\begin{picture}(15,15)(0,0)
                        \put(15,15){\vector(-1,-1){7.5}}
                        \put(7.5,7.5){\line(-1,-1){7.5}}
                  \end{picture}
                 }

\def\fermiondr{\begin{picture}(15,15)(0,0)
                        \put(0,0){\vector(1,-1){7.5}}
                        \put(7.5,-7.5){\line(1,-1){7.5}}
                  \end{picture}
                 }

\newenvironment{Feynman}[3]{\begin{center}
                            \setlength{\unitlength}{#3 mm}
                            \begin{picture}(#1)(#2)
                            \thicklines
                           }{\end{picture} \end{center}}
\def\fermionloopright{\begin{picture}(15,15)(0,0)
                             \put(7.5,0){\oval(15,15)}
                  \end{picture}
                 }
\def\blob{\begin{picture}(5,5)(0,0)
                             \put(0,0){\circle*{5}}
                  \end{picture}
                 }
\def\mcross{\begin{picture}(5,5)(0,0)
                             \put(-2.5,-2.5){\line(1,1){5}}
                             \put(-2.5,2.5){\line(1,-1){5}}
                  \end{picture}
                 }
\mark{{}{}}
\newcommand {\rW} {\mbox{$r_{_W} $}}
\newcommand {\rZ} {\mbox{$r_{_Z} $}}

\def\nll{ \nonumber \\}
\def\z0{Z}
\def\gf{G_{\mu}}
\def\zm{M_{_Z}}
\def\bm{m_b}
\def\cm{m_c}
\def\gev{{\hbox{GeV}}}
\def\tev{{\hbox{TeV}}}
\def\nb{{\hbox{nb}}}
\def\msb{{\overline{MS}}}
\def\als{\alpha_{_S}}
\def\tm{m_{t}}
\def\hm{M_{_H}}
\def\wm{M_{_W}}
\def\gn{\Gamma_{\nu}}
\def\ge{\Gamma_{e}}
\def\gmu{\Gamma_{\mu}}
\def\gt{\Gamma_{\tau}}
\def\gl{\Gamma_{l}}
\def\gu{\Gamma_{u}}
\def\gd{\Gamma_{d}}
\def\gc{\Gamma_{c}}
\def\gs{\Gamma_{s}}
\def\gb{\Gamma_{b}}
\def\gz{\Gamma_{_Z}}
\def\gh{\Gamma_{h}}
\def\gi{\Gamma_{\rm {inv}}}
\def\afb{A_{_{\rm {FB}}}}
\def\alr{A_{_{\rm {LR}}}}
\def\gv{g_{_V}}
\def\ga{g_{_A}}
\def\barf{\overline f}

\def\barb{\overline b}
\def\bart{\overline t}

\def\gvf{g^f_{_{V}}}
\def\gaf{g^f_{_{A}}}
\def\gvl{g^l_{_{V}}}
\def\gal{g^l_{_{A}}}

\def\gsvb{g^b_v}
\def\gsab{g^b_a}
\def\ord {\cal O}
\def\ical{\cal I}
\def\shat{\hat s}
\def\chat{\hat c}
\def\vhat{\hat v}

\def\fvf{F_{_V}^f}
\def\faf{F_{_A}^f}

\def\acal{\cal A}

\def\stes{\sin^2\theta_{\rm{eff}}}

\def\piv{\Pi_{_V}}
\def\pia{\Pi_{_A}}
\def\dr{\Delta r}
\def\drl{\Delta r_{_L}}
\def\dgvf{\delta g_{_V}^f}
\def\dgaf{\delta g_{_A}^f}
\def\dgvl{\delta g_{_V}^l}
\def\dgal{\delta g_{_A}^l}
\def\i3f{I^{(3)}_f}

\def\osp2{16\,\pi^2}
\def\ap2{\left(p^2\right)}
\def\stw{s_{\theta}}
\def\ctw{c_{\theta}}
\def\stws{s_{\theta}^2}
\def\stwf{s_{\theta}^4}
\def\ctws{c_{\theta}^2}
\def\Szg{\Sigma_{_{Z\gamma}}}
\def\Szz{\Sigma_{_{ZZ}}}
\def\Sww{\Sigma_{_{WW}}}
\def\Swwg{\Sigma_{_{WW}}^{^G}}
\def\Stg{\Sigma_{_{3Q}}}
\def\Stt{\Sigma_{_{33}}}
\def\Pgg{\Pi_{\gamma\gamma}}
\def\Pf{\Pi_{_F}}

\def\rhoz{\rho_{_Z}}
\def\rhozr{\rho^{\scriptscriptstyle R}_{_Z}}
\def\gfd{\gamma_5}
\def\gau{\gamma^{\alpha}}
\def\gad{\gamma_{\alpha}}
\def\mev{{\hbox{MeV}}}
\def\gev{{\hbox{GeV}}}
\def\tmo{\times 10^{-1}}
\def\tmt{\times 10^{-2}}
\def\tmth{\times 10^{-3}}
\def\tmf{\times 10^{-4}}
\def\tmfv{\times 10^{-5}}
\def\srt{\sqrt{2}}
\def\xsf{\sigma_{_F}}
\def\xsb{\sigma_{_B}}
\def\chig{\chi_{\gamma}}
\def\chiz{\chi_{_Z}}
\def\s0h{\sigma^h_0}
\def\ba{\begin{eqnarray}}
\def\ea{\end{eqnarray}}
\def\nl{\nonumber \\}
\def\baral{\bar{\alpha}}
\def\beq{\begin{equation}}
\def\eeq{\end{equation}}
\def\bea{\begin{eqnarray}}
\def\eea{\end{eqnarray}}
\def\barr{\begin{array}}
\def\earr{\end{array}}
\def\bc{\begin{center}}
\def\ec{\end{center}}
\def\btab{\begin{tabular}}
\def\etab{\end{tabular}}

\def\nn{\nonumber}
\def\ra{\rightarrow}

\def\sg{\Sigma^{^{\gamma\gamma}}}
\def\sz{\Sigma^{^{ZZ}}}
\def\sw{\Sigma^{^{WW}}}
\def\sgz{\Sigma^{^{\gamma Z}}}
\def\szr{\hat{\Sigma}^{^Z}}
\def\szzr{\hat{\Sigma}^{^{ZZ}}}
\def\swwr{\hat{\Sigma}^{^{WW}}}
\def\sgzr{\hat{\Sigma}^{^{\gamma Z}}}
\def\sggr{\hat{\Sigma}^{^{\gamma\gamma}}}
\def\dmz{\frac{\delta M_{_Z}^2}{M_{_Z}^2}}
\def\dmw{\frac{\delta M_{_W}^2}{M_{_W}^2}}

\def\dmmw{\delta M_{_W}^2}

\def\dro{\Delta\rho}

\def\al{\alpha}
\def\alpi{\frac{\alpha}{4\pi}}

\def\m{\mu}
\def\g{\gamma}
\def\G{\Gamma}
\def\Gmu{G_{\mu}}
\def\gamu{\gamma_{\mu}}
\def\gimu{\gamma^{\mu}}

\def\gafi{\gamma_5}

\def\Gr{\hat{\Gamma}_{\mu}}
\def\Pig{\Pi^{^{\gamma}}}
\def\Pigr{\hat{\Pi}^{^{\gamma}}}
\def\Pizr{\hat{\Pi}^{^Z}}

\def\Pgzrz{\hat{\Pi}^{^{\gamma Z}}(M_{_Z}^2)}

\def\noi{\noindent}
\def\epmf{{\rm e}^+{\rm e}^- \rightarrow {\rm f}\bar{\rm f}}

\def\irg{s+\hat{\Sigma}^{^{\gamma\gamma}}(s)}
\def\irz{s-M_{_Z}^2+\hat{\Sigma}^{^{ZZ}}(s)}

\def\mixs{\left[ \hat{\Sigma}^{^{\gamma Z}}(s) \right] ^2}

\def\mz{M_{_Z}^2}

\def\real{{\cal R}e}

\def\Dr{\Delta r}

\def\veps{\varepsilon}

\def\dlog{\mbox{Li}_2}

\def\dz{\delta Z}

\def\fvge{F_V^{\g e}}
\def\fage{F_A^{\g e}}

\def\fvgf{F_V^{\g f}}
\def\fagf{F_A^{\g f}}
\def\fvzf{F_V^{Z  f}}
\def\fazf{F_A^{Z  f}}

\begin{document}
 
\title{\bf Electroweak Working Group Report}
\author{
D.~Bardin$^{ab}$,
W.~Beenakker$^{c}$,
M.~Bilenky$^{ad}$,
W.~Hollik$^{e}$,
M.~Martinez$^{f}$, \\
G.~Montagna$^{gl}$,
O.~Nicrosini$^{h*}$,
V.~Novikov$^{i}$,
L.~Okun$^{i}$,
A.~Olshevsky$^{d}$,  \\
G.~Passarino$^{jk}$,
F.~Piccinini$^{gl}$,
S.~Riemann$^{a}$,
T.~Riemann$^{a}$,
A.~Rozanov$^{im}$,    \\
F.~Teubert$^{f}$,
M.~Vysotsky$^{i}$
}
\date{}
 
\maketitle
 
\begin{itemize}
\item[$^a$]
             Deutsches Elektronen-Synchrotron DESY  \\
             Institut f\"ur Hochenergiephysik IfH, Zeuthen, Germany
\item[$^b$]
             Bogoliubov Laboratory of Theoretical Physics \\
             Joint Institute for Nuclear Research, Dubna, Moscow Region,
             Russia
\item[$^c$]
             Deutsches Elektronen-Synchrotron DESY, Hamburg, Germany
\item[$^d$]
             Laboratory for Nuclear Problems \\
             Joint Institute for Nuclear Research, Dubna, Moscow Region,
             Russia
\item[$^e$]
             Institut f\"ur Theoretische Physik \\
             Universit\"at Karlsruhe, Karlsruhe, Germany
\item[$^f$]
             Institut de Fisica d'Altes Energies \\
             Universitat Autonoma de Barcelona, Barcelona, Spain
\item[$^g$]
             Dipartimento di Fisica Nucleare e Teorica,
             Universit\`a di Pavia, Pavia, Italy
\item[$^h$]
             CERN, TH Division, Geneva, Switzerland
\item[$^i$]
             ITEP, Moscow, Russia
\item[$^j$]
             Dipartimento di Fisica Teorica,
             Universit\`a di Torino, Torino, Italy
\item[$^k$]
             INFN, Sezione di Torino, Torino, Italy
\item[$^l$]
             INFN, Sezione di Pavia, Pavia, Italy
\item[$^m$]
             CPPM, IN2P3-CNRS, Marseille, France
 
\end{itemize}
 
\noindent
emails:
\\
bardindy@cernvm.cern.ch, t00bee@dsyibm.desy.de, bilenky@ifh.de,
woh@dmumpiwh.mppmu. \\
mpg.de, martinez@ifae.es, montagna@pv.infn.it,
nicrosini@pv.infn.it, novikov@vxitep.itep.ru, \\
okun@vxitep.itep.ru,
olshevsk@vxcern.cern.ch, giampiero@to.infn.it, piccinini@pv.infn.it,\\
riemanns@cernvm.cern.ch, riemann@cernvm.cern.ch, rozanov@cernvm.cern.ch, \\
teubert@ifae.es, vysotsky@vxitep.itep.ru
 
\vspace{.2cm}
 
\footnoterule
\noindent
$^{*}${\footnotesize On leave from INFN, Sezione di Pavia, Italy.}
 
\newpage
\tableofcontents

 
\newpage
 
\section{Theoretical basics}
 
\subsection{Introduction}
 
The radiative corrections (RC) for observables related to the
$Z$ resonance have been described in great detail in the 1989 CERN Yellow
Report `$Z$ Physics at LEP~1'~\cite{yr89}.
About $1.5 \times 10^7 \;Z$ decays have been recorded and analysed
during the years of operation of the four LEP experiments ---
from autumn of 1989 to the end of 1994. In order to match the actual
experimental precision, a completely revised analysis of radiative corrections
at the $Z$ resonance is needed.
The aim of this contribution to the present Report is threefold.
We will:
 
\begin{itemize}
\item[(i)]
introduce a common language for presentation of the results emerging from
different approaches;
\item[(ii)]
update the predictions of $Z$ resonance observables within the Minimal
Standard Model (MSM);
\item[(iii)]
estimate the intrinsic theoretical uncertainties of these predictions,
which are mainly caused by the neglect of higher order contributions.
\end{itemize}
 
The results, which are presented in this Report, are based on several
different approaches and on a comparison of their numerical predictions.
\\
The findings of the Report are based on the following computer codes:
 
{\tt BHM}~\cite{bhm}, {\tt LEPTOP}~\cite{leptop}, {\tt TOPAZ0}~\cite{topaz0},
{\tt WOH}~\cite{woh}, and {\tt ZFITTER}~\cite{zfitter}.
 
\noindent
The design of some of these codes can be traced back to the 1989
Workshop on `$Z$-Physics at LEP~1'; others have been developed later.
All of them contain a so-called electroweak library, which allows the
calculation of virtual higher order electroweak and QCD corrections
to selected quantities for example, a weak mixing angle or a partial or
total $Z$ width.
Strictly speaking, these are {\it `pseudo-observables'} and not directly
measurable in an experiment. Some packages (codes) contain, in addition,
virtual and real photonic corrections (bremsstrahlung) with simple kinematical
and geometrical cuts similar to those used in experiments, thus allowing
for the
calculation of (idealized) measurable cross-sections and asymmetries,
which we will call {\it `realistic observables'}.
 
Apart from a comparison of the numerical results of different codes, we
attempted a simulation of the theoretical uncertainties of each code.
To do so we had to reach an agreement on the principles on which these
simulations are based. We hope that as a result of this project
the prospects of the quantitative tests of the MSM at LEP and SLC,
and their sensitivity to potential New Physics beyond the MSM,
will become clearer.
 
The Report is organized as follows.
In the first section we define the exact
framework for the analysis of the electroweak data in the MSM, beginning with
`Input parameters' and ending with `Basic notions', which are
needed for discussing the numerical results collected
in the subsequent sections.
These deal with pseudo-observables and realistic observables in the sense
introduced above. The first three sections represent
a homogenous presentation of the material related to or obtained with the
above mentioned codes.
In the fourth and fifth sections we have collected items particular
to the different codes, such as
explicit analytical formulae and descriptions
of the design and of the use of the packages.
All the authors of the report agree on the following
conclusions from this study:
 
\begin{itemize}
 
\item The differences between results of different codes are small compared to
existing experimental uncertainties. Thus improvement of experimental accuracy
at LEP~1 and SLC is welcome even at the present level of theoretical accuracy.
 
\item At present the most promising are measurements of $g_{_V}/g_{_A}$
in various $P$- and $C$- violating asymmetries and polarizations.
 
\item The real bottleneck for improved theoretical accuracy in $g_{_V}/g_{_A}$
is presented by the uncertainty of the input parameter
$\baral\equiv\alpha(M_{_Z})$. The improved accuracy of this important
parameter calls for new accurate measurements of the cross-section
${\rm e}^+{\rm e}^-\to$ hadrons at low energies (Novosibirsk and Bejing
accelerators,
etc.)
 
\item The estimates of theoretical uncertainties are highly subjective and
their values partly reflect the internal philosophy of the actual
implementation of radiative corrections in a given code.
 
\item In many cases the one-loop approximation in the electroweak gauge
coupling is adequate enough at the present level of experimental accuracy.
At the same time, however,
it should be stressed that a complete evaluation of
the sub-leading corrections, ${\ord}(\gf^2\zm^2 m_t^2)$ would greatly reduce
the uncertainty that we observe, one way or the other, for all observables.
 
\item In case the next generation of experiments at LEP~1 and SLC
improves accuracy considerably (a problem
not only of statistics but mainly of systematics) the full program of
two-loop electroweak calculations should be carried out.
 
\end{itemize}
 
\subsection{Input parameters}
 
Within the MSM, any measured quantity can be calculated in terms of
a small set of input parameters. Once all possible relations between the
parameters of the Lagrangian in the MSM are exploited, we may choose a set
of them which, in the electroweak part, consists of
{\it one interaction constant} ---
say, the fine structure constant $\alpha$ --- of the {\it masses} of all
particles and of the CKM fermionic {\it mixing angles}.
The latter ones are of little importance for $Z$ resonance observables, which
are dominated by flavour-diagonal neutral current interactions.
 
In practical calculations one prefers to make use of the most precisely
measured parameters. Three of them characterize the gauge sector of the MSM:
 
\begin{eqnarray}
\alpha &\equiv& \alpha(0) = 1/137.0359895(61),  \nll
\gf &=& 1.16639(2)\times 10^{-5}\,\gev^{-2},  \nll
\zm &=& 91.1887 \pm 0.0044\,\gev,
\label{inputset1}
\end{eqnarray}
 
\noindent
where $\gf$, the four-fermion $\mu$ decay coupling constant,
is defined through the muon lifetime $\tau_{\mu}$ by
 
\begin{equation}
\frac{1}{\tau_{\mu}}  =
\frac{G^2_{\mu}m^5_{\mu}}{192\pi^3}\left(1-\frac{8m^2_e}{m^2_{\mu}}\right)
\Biggl[1+\frac{\alpha}{2\pi}
\left(1+\frac{2\alpha}{3\pi}\ln\frac{m_{\mu}}{m_e}\right)
\left(\frac{25}{4}-\pi^2\right)\Biggr]\;.
\label{taumu}
\end{equation}
 
\noindent
Light fermion masses contribute via the electromagnetic coupling constant
$\alpha$, running from very small momenta up to $\zm$, the typical scale of
the $Z$ resonance, and yielding
 
\begin{equation}
\alpha(M_{_Z}) \equiv {\bar \alpha}
= \frac{\alpha}{1-\Delta \alpha} \;,
\label{alphabar}
\end{equation}
 
\noindent
with $\Delta \alpha$ consisting of {\it leptonic} and {\it hadronic}
contributions
 
\begin{equation}
\Delta \alpha = \Delta \alpha_l + \Delta \alpha_h\;.
\end{equation}
 
\noindent
The leptonic contribution is known explicitly:
 
\begin{equation}
 \Delta \alpha_l = \frac{\alpha}{3\pi}
\sum_l\Biggl[-\frac{5}{3}- 4\frac{m^2_l}{\zm^2}
 +\beta_l \Biggl(1 + 2 \frac{m^2_l}{\zm^2} \Biggr)
 \ln \frac{\beta_l+1}{\beta_l-1}\Biggr]\;,
\label{lepexact}
\end{equation}
 
\noindent
where
 
\begin{equation}
\beta_l = \sqrt{1 - 4 \frac{m^2_l}{\zm^2} } \; .
\end{equation}
 
\noindent
With lepton masses taken from the last edition of the Review of Particle
properties~\cite{rpp94}, from~(\ref{lepexact}) we get
 
\begin{equation}
 \Delta \alpha_l = 0.0314129\;.
\label{7}
\end{equation}
 
\noindent
In the small mass approximation,~(\ref{lepexact}) reduces to the well-known
expression
 
\begin{equation}
 \Delta \alpha_l = \frac{\alpha}{3\pi}
\sum_l\Biggl(\ln\frac{M^2_{_Z}}{m^2_l}-\frac{5}{3}\Biggr) = 0.0314177\;.
\label{lepapprox}
\end{equation}
 
\noindent
The hadronic contribution is being calculated by a dispersion relation,
 
\begin{equation}
 \Delta \alpha_h = \frac{\alpha M^2_{_Z}}{3\pi} {\cal R}e
 \int^{\infty}_{4m^2_{\pi}}\frac{ds}{s(M^2_{_Z} - i\epsilon
 - s)} R(s)\;,
\label{dispersion}
\end{equation}
 
\noindent
from experimental data on the cross-section ratio 
$R(s)=\sigma_{{\rm e}^+{\rm e}^-}(s)/\sigma_0(s)$, 
where $\sigma_{{\rm e}^+{\rm e}^-}(s)= \sigma({\rm e}^+{\rm e}^- \to \gamma^*
\to\,$hadrons) and $\sigma_0=4/3\pi\alpha^2/s$.
In 1990, the value $\Delta \alpha_h$ has been updated to
 
\begin{equation}
 \Delta \alpha_h = 0.0282(9)\;,
\end{equation}
 
\noindent
leading to $\alpha^{-1}(\zm) = 128.87 \pm 0.12$~\cite{jalpha}
{\footnote{
Very recently, three new analyses have been published.
The first,~\cite{morrisls},
leads to $\Delta \alpha_h = 0.02666 \pm 0.00075$ and
$\alpha^{-1}(\zm) = 129.08 \pm 0.10$. The second,~\cite{martzepp}, leads to
$\Delta \alpha_h = 0.02732 \pm 0.00042$ and $\alpha^{-1}(\zm) = 128.99 \pm
0.06$. The third,~\cite{eideljeg},
leads to $\Delta \alpha_h = 0.0280 \pm 0.0007$ and
$\alpha^{-1}(\zm) = 128.899 \pm 0.090$.}}.
 
By using ${\bar \alpha}$ instead of ${\alpha}$, the electromagnetic coupling
constant at the correct scale for $Z$ physics is chosen and one automatically
avoids the problem of the light quark mass singularities.
So the relevant input parameter is $\baral$,
and not $\alpha$ in ---
spite of the extremely high accuracy of the latter.
It should be stressed that the quantity $\baral$ used by us
is not only numerically, but also in principle, different from
the quantity $\hat{\alpha}(M_Z) = 1/127.9(1)$
(see, for example Ref. \cite{rpp94}, page 1304).
The latter is defined in the modified minimal subtraction scheme
($\overline{MS}$) in terms of a bare electric charge, while the former is a
physical quantity expressed in terms of charged fermion masses.
 
The treatment of the light fermion masses (all but the top quark mass)
in the electroweak corrections is not the same in different codes.
First we should mention that the exact expression~(\ref{lepexact})
for the photonic vacuum polarization is not the only place where
fermionic power corrections, $m^2_f/\zm^2,$ may appear
{\footnote {One should emphasize that the dispersion integral~(\ref{dispersion})
automatically includes all these power corrections for the hadronic part of
the photonic vacuum polarization.}}.
There are fermionic contributions to the self-energy insertions of the heavy
gauge bosons, which also contain power mass corrections.
Moreover additional power corrections appear, due to the presence of axial
couplings. The latter may be explicitly calculated for leptons; for quarks,
they are not taken into account by the dispersion integral~(\ref{dispersion}).
In principle, there are also finite mass corrections in
three- and four-point functions.
In some codes these finite mass terms are neglected completely,
some other codes retain them in the two- and three-point functions.
In the latter case, for the quarks some {\it effective} masses are used,
which are selected in order to reproduce with a sufficiently high accuracy
the hadronic vacuum polarization contribution $\Delta \alpha_h$
{\footnote {Details on {\it effective quark masses} be found in~\cite{effjeg}
or~\cite{effhol}.}}.
We emphasize that these finite mass terms have no numerical relevance.
They are being retained just in order to investigate the sensitivity of
corrections to finite light fermion masses.
They also help sometimes to compare results of independent calculations.
Such a comparison has been done once for $\Delta r$, and an agreement
of up to 12 digits (computer precision) was found~\cite{baknst}.
 
The $\tau$ lepton mass has been retained in all codes in the phase-space
corrections and in the Born-level matrix element.
The masses of the heavy quarks $b$ and $c$ also give rise to phase-space
corrections, but are treated together with QCD final state interaction
corrections (see the QCD part of this Report).
We take the following input values for $m_{\tau}$, $\bm$ and $\cm$~\cite{rpp94}:
 
\begin{eqnarray}
m_{\tau} &=& 1.7771 \pm 0.0005\,\gev , \nll
\bm &=& 4.7 \pm 0.3\,\gev , \nll
\cm &=& 1.55 \pm 0.35\,\gev ,
\label{inputset2}
\end{eqnarray}
 
\noindent
where $\bm,\cm$ are the pole masses converted in the actual
calculation of radiative corrections to $\msb$ masses using QCD
perturbation theory.
The meaning of these $\msb$ mass definitions and their relation
to the {\it pole} masses are explained in the QCD Part of this Report
by Chetyrkin, K\"uhn and Kwiatkowski.
The QCD part of the MSM is characterized by these running quark masses,
and by the strong coupling constant $\als(\zm)$, which will be assumed to
be~\cite{lepwg3} and~\cite{schaile}:
 
\begin{equation}
 \als(M_{_Z}) \equiv {\hat \alpha}_{_S} = 0.125 \pm 0.007\;.
\label{alphahat}
\end{equation}
\noindent
Note, that from pure QCD observables at the $Z$ peak one
obtains~\cite{bethke}
 
\begin{equation}
 \als(M_{_Z}) \equiv {\hat \alpha}_{_S} = 0.123 \pm 0.006\;.
\label{alphahate}
\end{equation}
 
\noindent
Finally, there are two yet unknown input parameters left: the top quark mass
and the Higgs boson mass,
 
\begin{equation}
m_t\;,\;\;\;\hm\;,
\label{inputset3}
\end{equation}
 
\noindent
although the situation has considerably improved with the recent CDF
indication~\cite{cdf} of evidence for the $t$ quark with $m_t = 174 \pm 17\,$
GeV. In the following, by $m_t$ we will always imply
the $t$-quark {\it pole} mass.
 
\subsection{Codes to calculate electroweak observables}
 
The radiative corrections for the measured physical observables
must be included into the theoretical predictions for them,
and the consistency of the theory is verified by comparison
with the data.
This particular step may assume different aspects related
to the actual implementation of the comparisons,
but in the end it usually
takes the form of some constraint on $m_t$ and ${\hat \alpha_{_S}}$
and to a lesser extent on $\hm$ and ${\bar \alpha}$.
It is our main goal to present and discuss the most accurate
theoretical predictions for the measured quantities and to introduce
a reliable estimate of the associated theoretical uncertainties.
 
The results of this Report have been obtained by a critical comparison and
examination of:
 
\begin{itemize}
 
\item[-] the variations in the results following from the different,
although not antithetic, formulations of the various groups;
 
\item[-] the {\it internal} estimates of the uncertainties induced by the
still missing higher order corrections.
 
\end{itemize}
 
The groups participating in this project can be identified by
the names of the codes that they have assembled, namely:
 
\begin{itemize}
 
\item[] {\tt BHM}~\cite{bhm}
\\ {\tt Burgers, Hollik, Martinez, Teubert};
\item[] {\tt LEPTOP}~\cite{leptop} -- ITEP Moscow group
\\ {\tt Novikov, Okun, Rozanov, Vysotsky};
\item[] {\tt TOPAZ0}~\cite{topaz0} -- Torino-Pavia group
\\ {\tt Montagna, Nicrosini, Passarino, Piccinini, Pittau};
\item[] {\tt WOH}~\cite{woh}
\\ {\tt Beenakker, Burgers, Hollik};
\item[] {\tt ZFITTER}~\cite{zfitter} -- Dubna-Zeuthen group
\\ {\tt Bardin, Bilenky, Chizhov, Olchevsky, S.Riemann, T.Riemann,
Sachwitz, \\ Sazonov, Sedykh, Sheer}.
\end{itemize}
 
\noindent
All these codes may be used to calculate pseudo-observables.
Three of them,~{\tt BHM}, {\tt TOPAZ0}, and {\tt ZFITTER},
may calculate also realistic observables.
One code,~{\tt TOPAZ0}, is additionally able
to calculate the full Bhabha cross-section,
including complete $s$ and $t$ channel exchange contributions.
Concerning the electroweak corrections, four codes are based on completely
independent theoretical computational schemes, while {\tt BHM/WOH} use
basically the same framework.
The treatment of QED corrections in the three QED dressers is based
on completely independent theoretical methods.
As a consequence, it is extremely difficult to describe them in a common
language. On the other hand, the treatment of QCD corrections is to a
large extent common in all five codes.
It is based on papers included in Part~II of this volume and on
references quoted therein.
 
In this Report, we will attempt to present the collected material
homogeneously as for long as possible. Because of this, all
complicated description of what is entering in the various,
alternative theoretical formulations, has been
shifted to the last two sections of the Report.
 
\subsection{The language of effective couplings}
 
A natural and familiar but approximate language for the basic ingredients
of the physical observables of the $Z$ resonance is
that of {\it effective couplings}.
How it will translate into the particular realizations and schemes of the
various codes can be studied in References
\cite{bhm}-\cite{zfitter} and in the references quoted in these.
The more theoretically oriented reader should, however, be aware that
basic differences do indeed exist and that the language of
effective couplings is not universally realized in all the approaches.
The basic notions of effective couplings, which are common to all
approaches, can be easily introduced. Nevertheless, this deserves to be
seen in several logical steps.
 
\subsubsection{The Born Approximation for 
${\rm e}^+{\rm e}^- \to {\rm f}\bar{\rm f}$}
 
The matrix element of the process
 
\begin{equation}
{\rm e}^+ {\rm e}^- \rightarrow (\gamma,Z)
\rightarrow {\rm f} \bar{\rm f},\;\;\;\;f \neq {\rm e}\;,
\label{eeann}
\end{equation}
 
\noindent
depicted in Fig.~\ref{figborn} may be written as follows:
 
\begin{figure*}[tbhp]
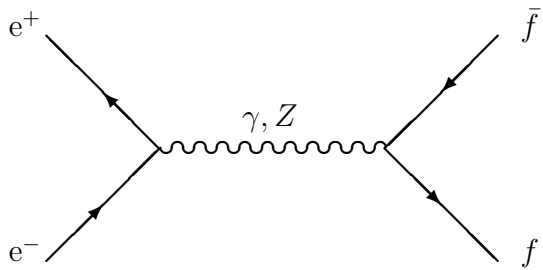

\begin{Feynman}{60,30}{0,0}{1.0}
\put(15,15){\fermionur}
\put(15,15){\fermionul}
\put(15,15){\photonright}
\put(26,18){$\gamma,Z$}
\put(45,15){\fermiondl}
\put(45,15){\fermiondr}
\put(-5,30){${\rm e}^+$}
\put(-5,0){${\rm e}^-$}
\put(63,0){$f$}
\put(63,30){$\bar{f}$}
\end{Feynman}
\caption[ ]
{Feynman graph for the reaction~(\ref{eeann}) in the Born approximation}
\label{figborn}
\end{figure*}
 
\begin{eqnarray}
{\cal M}^{Born}
&\sim& \frac{1}{s}
\Biggl[Q_eQ_f\gad \otimes \gau
+ \chi \gad \left(g^e_v - g^e_a\gfd \right)
\otimes
\gau \left(g^f_v - g^f_a\gfd \right) \Biggr]
\label{mborn}
\\
& = & \frac{1}{s}
\Biggl[Q_eQ_f\gad \otimes \gau
+ \chi \Biggl( g^e_v g^f_v \gad \otimes \gau
              -g^e_v g^f_a \gad \otimes \gau\gfd
\nonumber \\
& & -g^e_a g^f_v \gad \gfd \otimes \gau
    +g^e_a g^f_a \gad \gfd \otimes \gau \gfd
    \Biggr)\Biggr]\;,
\label{mborndef}
\end{eqnarray}
 
\noindent
with $\chi$ being the propagator ratio
 
\begin{equation}
\chi = \frac{s}{s-\zm^2+i s\Gamma_{_Z} / \zm}\;.
\label{proprat}
\end{equation}
 
\noindent
And in the Minimal Standard Model (MSM)
 
\begin{equation}
g_a^f =I_f^{(3)} \;,\;\;\;\;\; g_v^f = I_f^{(3)} -2Q^f \sin^2\theta \;,
\end{equation}
 
\noindent
where $g_a^f$, $g_v^f$ are vector and axial vector couplings of the $Z$-boson,
$Q^f$ are the electric charges of fermions in units of position charge,
$I_f^{(3)}$ the projection of weak isospin, and $\theta$ the weak mixing
angle in the Born
approximation. Its value depends on the choice of computational scheme.
\noindent
In~(\ref{mborn}) and (\ref{mborndef})
a short notation for bilinear combinations of spinors $u$ and $v$ is used:
 
\begin{equation}
A_{\beta} \otimes B^{\beta} = \left[ \bar v_e A_{\beta} u_e \right]
                        \times \left[ \bar u_f B^{\beta} v_f \right]\;.
\end{equation}
 
\noindent
In the matrix element~(\ref{mborn}) and (\ref{mborndef})
the contributions from
$\gamma$ and $Z$ exchange diagrams are unambiguously separated,
and the $Z$ exchange contribution is presented in a {\it factorized}
form~(\ref{mborn}).
 
\subsubsection{Electroweak non-photonic corrections}
 
\noindent
The {\it higher order} electroweak non-photonic corrections are
indicated symbolically by the blobs in Fig.~\ref{figmeff}.
They include all possible self-energy and  $Z{\bar f}f$ vertex insertions,
together with {\it non-photonic boxes} insertions (the $WW$ and $ZZ$ boxes in
the one-loop approximation), and lead to a slightly more complicated
structure in the matrix element, which is valid as long as we neglect the
external fermion masses:
 
\begin{figure*}[tbhp]
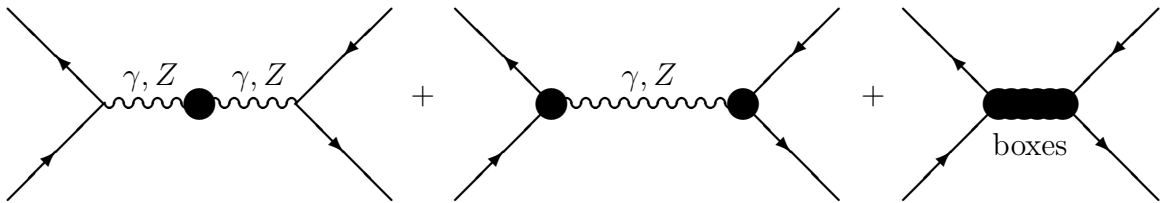

\begin{Feynman}{185,30}{0,0}{0.85}
 
\put(15,15){\fermionur}
\put(15,15){\fermionul}
\put(30,15){\circle*{5}}
\put(15,15){\photonright}
\put(18,18){$\gamma,Z$}
\put(35,18){$\gamma,Z$}
\put(45,15){\fermiondl}
\put(45,15){\fermiondr}
\put(63,15){+}
 
\put( 85,15){\fermionur}
\put( 85,15){\fermionul}
\put( 85,15){\circle*{5}}
\put( 85,15){\photonright}
\put( 96,18){$\gamma,Z$}
\put(115,15){\circle*{5}}
\put(115,15){\fermiondl}
\put(115,15){\fermiondr}
 
\put(133.5,15){+}
\put(155,15){\fermionur}
\put(155,15){\fermionul}
\put(155,15){\circle*{5}}
\put(157,15){\circle*{5}}
\put(159,15){\circle*{5}}
\put(161,15){\circle*{5}}
\put(163,15){\circle*{5}}
\put(165,15){\circle*{5}}
\put(165,15){\fermiondl}
\put(165,15){\fermiondr}
\put(154,7){boxes}
 
\end{Feynman}
\caption[ ]
{Feynman graphs for higher-order corrections to the reaction~(\ref{eeann})}
\label{figmeff}
\end{figure*}
 
\begin{eqnarray}
{\cal M}^{eff} &\sim& \frac{1}{s}
\Biggl\{\alpha(s) \gad \otimes \gau
+ \chi \Biggl[
 {\cal F}^{ef}_{vv}(s,t) \gad \otimes \gau
-{\cal F}^{ef}_{va}(s,t) \gad \otimes \gau \gfd
\nonumber \\
& &-{\cal F}^{ef}_{av}(s,t) \gad \gfd \otimes \gau
   +{\cal F}^{ef}_{aa}(s,t) \gad \gfd \otimes \gau
\gfd
\Biggr]\Biggr\}.
\label{meff}
\end{eqnarray}
 
The products of interaction constants in~(\ref{mborndef}) are replaced
by the {\it running} $\alpha(s)$ --- this time including the
imaginary part --- and by four {\it running electroweak form factors},
${\cal F}^{ef}_{ij}$.
In the MSM, the explicit expressions for the form factors result from an
order-by-order calculation and are certain explicit functions of the input
parameters~(\ref{inputset1}),~(\ref{alphabar}),~(\ref{inputset2}),
(\ref{alphahat}) and~(\ref{inputset3}).
 
Several comments should be given on the structure of the matrix
element~(\ref{meff}) after inclusion of higher order electroweak corrections:
 
\begin{itemize}
 
\item This structure is unique, but implies the introduction of complex valued
form factors which depend on the two Mandelstam variables $s,t$; the
dependence on $t$ is due to the weak box diagrams.
 
\item The separation into insertions for the $\gamma$ and $Z$
exchanges is lost. In the proposed realization, the first term
in~(\ref{meff}) contains the running QED coupling where only fermionic
insertions are retained. This ensures a gauge invariant separation.
 
\item The weak boxes are present in~(\ref{meff}) as non-resonating
($\sim s-\zm^2$) insertions to the electroweak form factors
${\cal F}^{ef}_{ij}$. At the $Z$ resonance, the one-loop weak
$WW$ and $ZZ$ box terms are small, with relative contribution $\leq 10^{-4}$.
If we neglect them, the $t$ dependence is turned off. The $t$-dependence would
also spoil factorization of the form factors into products of effective vector
and axial vector couplings.
 
\item
Full factorization is re-established by neglecting, in addition,
the other
non-resonating loop contributions,
such as the bosonic insertions to the photon
propagator, and photon-fermion vertex corrections.
All the neglected terms are of the order ${\ord}(\alpha\frac{\gz}{\zm})$.
The resulting effective vector and axial -vector couplings are complex
valued and dependent on $s$. The factorization is the result of a variety of
approximations, valid at the $Z$ resonance to the accuracy needed,
and indispensible in order to relate the pseudo-observables
to actually measured quantities.
 
\end{itemize}
 
In the complete codes --- {\tt BHM, TOPAZ0, ZFITTER} --- it is possible
to control the numerical influence of all of these approximations.
 
\subsubsection{$Z$ pole approximation}
 
\begin{figure*}[tbhp]
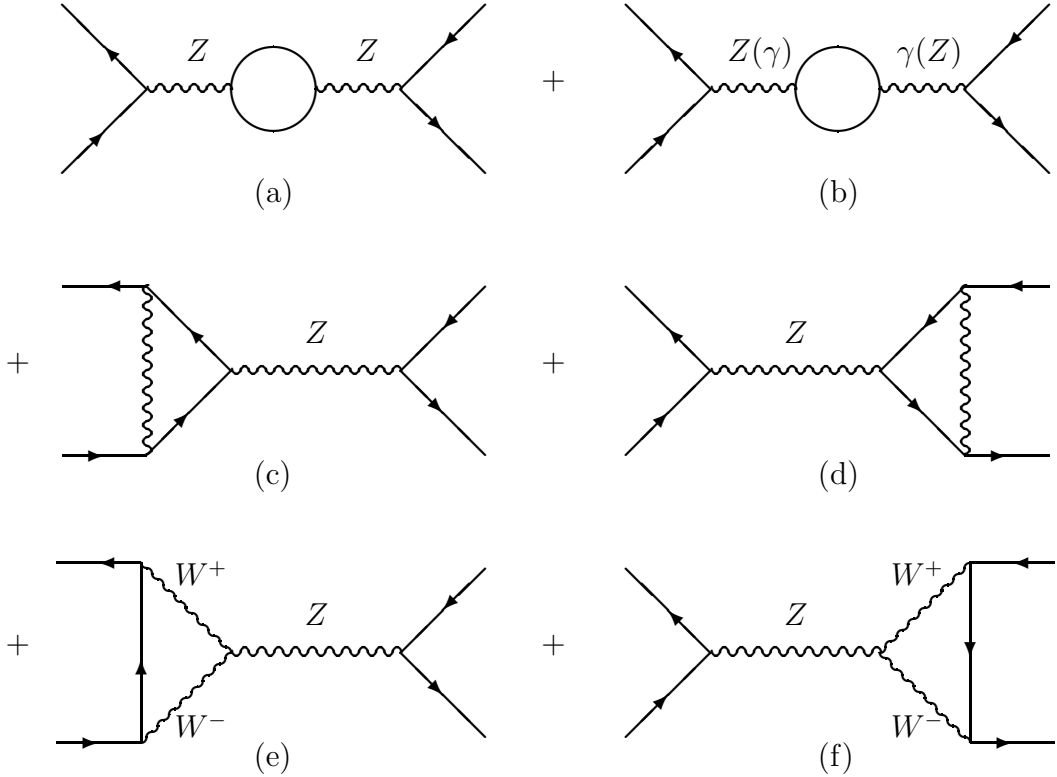

\begin{Feynman}{160,125}{0,0}{.75}
 
\put(15,115){\fermionul}
\put(15,115){\fermionur}
\put(15,115){\photonrighthalf}
\put(22,120){$Z$}
\put(30,115){\fermionloopright}
\put(45,115){\photonrighthalf}
\put(52,120){$Z$}
\put(60,115){\fermiondl}
\put(60,115){\fermiondr}
\put(34,95){(a)}
 
\put(85,115){+}
\put(115,115){\fermionul}
\put(115,115){\fermionur}
\put(115,115){\photonrighthalf}
\put(118,120){$Z$($\gamma$)}
\put(130,115){\fermionloopright}
\put(145,115){\photonrighthalf}
\put(148,120){$\gamma$($Z$)}
\put(160,115){\fermiondl}
\put(160,115){\fermiondr}
\put(134,95){(b)}
 
\put(-10,65){+}
\put(30,65){\fermionul}
\put(30,65){\fermionur}
\put(15,50){\photonup}
\put(0,80){\fermionlefthalf}
\put(0,50){\fermionrighthalf}
\put(30,65){\photonright}
\put(43,70){$Z$}
\put(60,65){\fermiondl}
\put(60,65){\fermiondr}
\put(34,45){(c)}
 
\put(85,65){+}
\put(115,65){\fermionul}
\put(115,65){\fermionur}
\put(115,65){\photonright}
\put(128,70){$Z$}
\put(145,65){\fermiondl}
\put(145,65){\fermiondr}
\put(160,50){\photonup}
\put(160,80){\fermionlefthalf}
\put(160,50){\fermionrighthalf}
\put(134,45){(d)}
 
\put(-10,15){+}
\put(14,31){\markphotondrh}
\put(20,0){$W^-$}
\put(-1,-1){\fermionrighthalf}
\put(14,29){\line(0,1){2}}
\put(14,-1){\fermionup}
\put(-1,31){\fermionlefthalf}
\put(14,-1){\markphotonurh}
\put(20,27){$W^+$}
\put(30,15){\photonright}
\put(43,20){$Z$}
\put(60,15){\fermiondl}
\put(60,15){\fermiondr}
\put(34,-5){(e)}
 
\put(85,15){+}
\put(115,15){\fermionul}
\put(115,15){\fermionur}
\put(115,15){\photonright}
\put(128,20){$Z$}
\put(145,15){\markphotondrh}
\put(147,0){$W^-$}
\put(161,-1){\fermionrighthalf}
\put(161,29){\line(0,1){2}}
\put(161,-1){\fermiondown}
\put(161,31){\fermionlefthalf}
\put(145,15){\markphotonurh}
\put(147,27){$W^+$}
\put(134,-5){(f)}
 
\end{Feynman}
\caption[]
{The one-loop Feynman graphs at the $Z$ boson pole}
\label{figlb1}
\end{figure*}

After the above-mentioned series of approximations, we arrive at the
so called $Z$ boson pole approximation, which is actually equivalent to the
setting $s=M_{_Z}^2$ in the form factors.
In the $Z$ pole approximation, the one-loop diagrams contributing to
the blobs in Fig.~\ref{figmeff} can be visualized in the diagrams of
Fig.~\ref{figlb1}.
 
\noindent
In Fig.~\ref{figlb1}~(a) all particles ($Z$, $H$, $W$, $q$, $l$, $\nu$)
contribute through their weak neutral currents, while in the $Z\gamma$
mixing (b) only charged ones contribute.
This, of course, is strictly true only in the unitary gauge, while in the
renormalizable gauge, Faddeev-Popov and Higgs-Kibble ghosts will
also appear.
The vertex diagrams (c) and (d) contain $Z$, $W$, as vertical lines.
The vertex diagrams (e) and (f) contain trilinear gauge boson interactions.
The vertex diagrams contain, in principle, also Higgs boson exchanges ---
as vertical lines in diagrams (c) and (d), and as $Z,H$ virtual states
in diagrams (e) and (d). They give, however, a negligible contribution
because of small $Hf\bar{f}$ Yukava couplings
{\footnote {In some renormalization schemes one should add also the self-energy
insertions for external fermion lines. In some other schemes, they vanish, when
the field renormalization counterterms contributions are added.}}.
 
\begin{figure}[tbh]
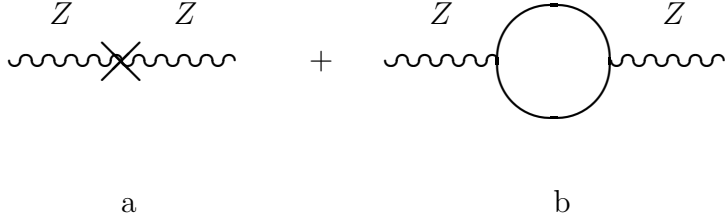

\begin{center}
\begin{Feynman}{100,30}{0,0}{1.0}
\put( 0,15){\photonrighthalf}
\put(5.5,20){$Z$}
\put(15,-5){a}
\put(15,15){\mcross}
\put(15,15){\photonrighthalf}
\put(22,20){$Z$}
\put(40,14){+}
\put(50,15){\photonrighthalf}
\put(56,20){$Z$}
\put(72.5,-5){b}
\put(65,15){\fermionloopright}
\put(80,15){\photonrighthalf}
\put(87,20){$Z$}
\end{Feynman}
\end{center}
\caption[]
{The $Z$-boson self-energy}
\label{figlb2}
\end{figure}

The set of the diagrams of Fig.~\ref{figlb1}
has to be complemented by the
contributions from the bare `unphysical' parameters (masses and couplings)
in order to get the physical amplitude of Fig.~\ref{figmeff}.
This is depicted in Fig.~\ref{figlb2} as an example,
the cross representing the contribution from the bare mass $\zm^0$.
 
\begin{figure}[tbh]
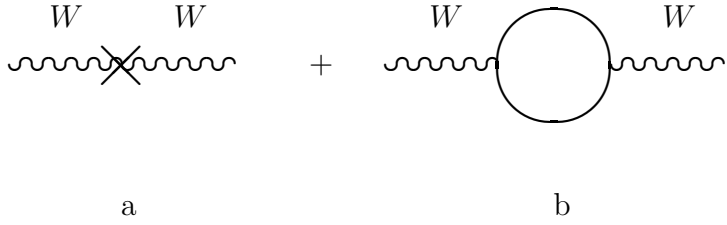

\begin{center}
\begin{Feynman}{100,30}{0,0}{1.0}
\put( 0,15){\photonrighthalf}
\put(5.5,20){$W$}
\put(15,-5){a}
\put(15,15){\mcross}
\put(15,15){\photonrighthalf}
\put(22,20){$W$}
\put(40,14){+}
\put(50,15){\photonrighthalf}
\put(56,20){$W$}
\put(72.5,-5){b}
\put(65,15){\fermionloopright}
\put(80,15){\photonrighthalf}
\put(87,20){$W$}
\end{Feynman}
\end{center}
\caption[]
{The $W$ boson self-energy}
\label{figlb3}
\end{figure}

The $W$ boson self-energy enters the analysis only through the $W$ mass (see
Fig.~\ref{figlb3}).
The $W$ self-energy contains all types of particles entering
 through their charged weak currents.
Explicit expressions for the decay width of the $W$ are not discussed
in this Report, (see the existing literature~\cite{wdec}).
 
If the mass of the $W$ boson were known with the same accuracy as that of
the $Z$, the conceptual picture of the electroweak corrections would be
as simple as that of QED. Every physical observable would be expressed
in terms of $\alpha$, $\zm$, $\wm$, the masses of all fermions and the mass of
the Higgs boson. Unfortunately, $\wm$ is not known with an accuracy comparable
to that of $\zm$.
Therefore, instead of $\wm$ we have to choose $G_{\mu}$ as
the most accurately measured dimensional observable.
 
\subsubsection{Effective couplings for $Z$ decay.}
 
Formulae~(\ref{mborndef}) and~(\ref{meff}) are very similar in their
structure, a property which makes the language of effective couplings so
convenient.
These formulae refer to a general $f {\bar f} \rightarrow f' {\bar f'}$
annihilation process. After the simplifications described in
1.4.2--1.4.3, the factorization allows us  to write for the form factors
 
\begin{equation}
{\cal F}^{ef}_{ij}(s=\zm^2,t=0)
={\cal G}^{e}_{i}(\zm^2){\cal G}^{f}_{j}(\zm^2)\;.
\end{equation}
 
\noindent
This implies that the $Z$ part of the amplitude~(\ref{meff}) is obtained from
that of~(\ref{mborn}) by substituting
 
\begin{equation}
g^{e,f}_{v,a} \to {\cal G}^{e,f}_{v,a}(\zm^2)\;.
\end{equation}
 
\begin{figure}[tbhp]
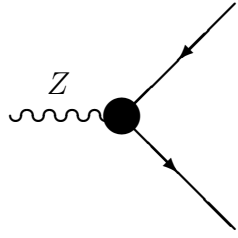

\begin{Feynman}{60,30}{0,0}{1.0}
\put(20,15){\photonrighthalf}
\put(25,18){$Z$}
\put(35,15){\circle*{5}}
\put(35,15){\fermiondl}
\put(35,15){\fermiondr}
\end{Feynman}
\caption[ ]
{Feynman graph for higher-order corections to the $Z \rightarrow f \bar f$
decay}
\label{figdec}
\end{figure}

\noindent
Simultaneously, we obtain for the matrix element of the decay process
$Z \rightarrow f {\bar f}$, Fig.~\ref{figdec}, as follows:
 
\begin{eqnarray}
{\cal M}^{eff}_{Z{\bar f} f } &=&
{\bar u}_f \gad \left[{\cal G}^f_v(\zm^2) - {\cal G}^f_a(\zm^2)\gfd
   \right] v_f \epsilon^{\alpha}_{_Z}\;,
\label{mdec}
\end{eqnarray}
 
\noindent
where $\bar{u}_f$, $v_f$, $\epsilon_{_Z}^{\alpha}$ are wave-functions
of the fermion $f$, antifermion $\bar{f}$ and $Z$ boson.
 
 
\begin{figure}[htbp]
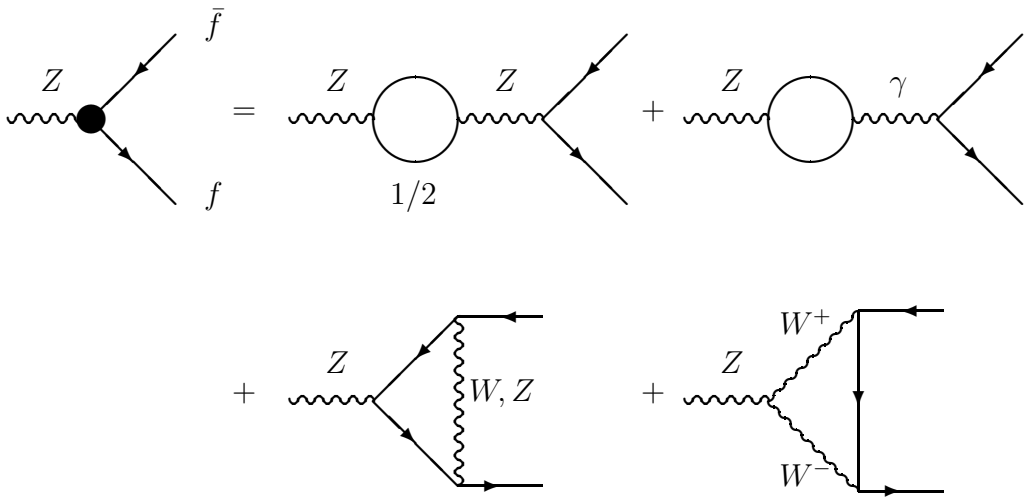

\begin{center}
\begin{Feynman}{180,90}{0,0}{0.75}
\put(0,65){\photonrighthalf}
\put(6.,70){$Z$}
\put(15,65){\blob}
\put(15,65){\fermiondl}
\put(15,65){\fermiondr}
\put(35,80){$\bar{f}$}
\put(35,50){${f}$}
\put(40,65){=}
 
\put(50,65){\photonrighthalf}
\put(56.5,70){$Z$}
\put(65,65){\fermionloopright}
\put(68,50){1/2}
\put(80,65){\photonrighthalf}
\put(86.5,70){$Z$}
\put(95,65){\fermiondl}
\put(95,65){\fermiondr}
 
\put(112.5,65){+}
\put(120,65){\photonrighthalf}
\put(126.5,70){$Z$}
\put(135,65){\fermionloopright}
\put(150,65){\photonrighthalf}
\put(156.5,70){$\gamma$}
\put(165,65){\fermiondl}
\put(165,65){\fermiondr}
 
\put(40,15){+}
\put(50,15){\photonrighthalf}
\put(56.5,20){$Z$}
\put(65,15){\fermiondr}
\put(80,0){\fermionrighthalf}
\put(80,0){\photonup}
\put(82,15){$W,Z$}
\put(80,30){\fermionlefthalf}
\put(65,15){\fermiondl}
 
\put(112.5,15){+}
\put(120,15){\photonrighthalf}
\put(126.5,20){$Z$}
\put(135,15){\markphotondrh}
\put(137,0){$W^-$}
\put(151,-1){\fermionrighthalf}
\put(151,29){\line(0,1){2}}
\put(151,-1){\fermiondown}
\put(151,31){\fermionlefthalf}
\put(135,15){\markphotonurh}
\put(137,27){$W^+$}
 
\end{Feynman}
\end{center}
\caption[ ]
{Feynman graphs of the $Z$ decay in the one-loop approximation}
\label{figlb4}
\end{figure}

\noindent
In the one-loop approximation, the blob in Fig.~\ref{figdec} can be
visualized as the diagrams of Fig.~\ref{figlb4}. Note that Fig.~\ref{figlb1}
(a) contributes both to initial and final state virtual corrections.
 
\noindent
The effective decay couplings ${\cal G}^f_{v,a}$, are nothing else but the
reduced electroweak form factors, which are now constant because of the
kinematical constraint $s=\zm^2$, which is equivalent to the $Z$ pole
approximation. They will be the basic objects in the following presentation.
We will use for them the simplified notations,
 
\begin{equation}
g^f_{_{V,A}} = {\cal R}e {\cal G}^f_{v,a}(\zm^2) \; ,
\label{24}
\end{equation}
 
\noindent
where the capital letters $V,A$ are introduced in order to distinguish
effective couplings $g_{_{V,A}}$, dressed by higher order interactions, from
their Born-like analogs $g_{v,a}$.
 
In the next two subsections we will define the quantities to be computed
and compared: the pseudo-observables, and the realistic observables.
 
\subsection{Pseudo-observables}
 
The {\it pseudo-variables} are related to measured cross-sections and
asymmetries by some de-convolution or {\it unfolding} procedure.
The concept itself of pseudo-observability is rather difficult to define.
One way to introduce it is to say that the experiments {\it measure} some
primordial (basically cross-sections and thereby asymmetries also)
quantities which are then reduced to secondary quantities under
some set of specific assumptions. Within these assumptions,
the secondary quantities, the pseudo-observables, also deserve the label of
{\it observability}.
Just to give an example, we quote the de-convoluted forward--backward
asymmetry,
where, typically, only the $Z$ exchange is included and initial and
final state QED corrections, plus eventual final state QCD corrections,
are assumed to be subtracted from the experimental data.
We have analyzed 25 of such pseudo-observables in details, namely:
 
\begin{eqnarray*}
{}\hbox{mass of the W} && \wm \nll
{\hbox{ hadronic peak cross-section}} &&
\sigma_h \nll
{\hbox{partial leptonic and hadronic widths}}&&
 \gn, \ge, \gmu, \gt, \gu, \gd, \gc, \gs , \gb  \nll
{}\hbox{the total width} &&     \gz  \nll
{\hbox{the total hadronic width}} && \gh  \nll
\hbox{the total invisible width} && \gi \nll
{\hbox{ratios}}&& R_l, R_b, R_c \\
{\hbox{asymmetries and polarization}}&&
 \afb^{\mu}, \alr^e, \afb^b, \afb^c, P^{\tau}, P^{b} \\
{\hbox{{effective sine}}}&& \stes^{\rm {lept}}, \stes^b \\
\end{eqnarray*}
 
\noindent
The {\it effective sine} are defined by
{\footnote {With lepton universality there exists only one
{\it leptonic effective sine}, $\stes^{\rm {lept}} =
\stes^l$.}}
 
\begin{equation}
4\,|Q_f|\stes^f = 1 - \frac{\gvf}{\gaf} \; ,
\end{equation}
 
\noindent
with $Q_f$ being the electric charge of the fermion $f$ in units of the
positron charge. By definition, the total and partial widths of the $Z$ boson
include final state QED and QCD radiation. The explicit formulae for partial
widths will be presented in subsection 1.9. Moreover, we have defined
 
\begin{eqnarray}
\gh &=& \gu + \gd + \gc + \gs + \gb \; ,
\label{gamhad}
\\
\gi &=& \gz - \ge -\gmu - \gt - \gh \; ,
\label{gaminv}
\\
R_l &=& \frac{\gh}{\ge} \; ,
\label{ratl}
\\
R_{b,c} &=& \frac{\Gamma_{b,c}}{\gh} \; ,
\label{ratq}
\\
\sigma_h&=& 12\pi\,\frac{\ge\gh}{\zm^2\gz^2} \; .
\label{sigh0}
\end{eqnarray}
 
\noindent
The quantity $\sigma_h$ is the de-convoluted hadronic peak cross-section,
which by definition includes only the $Z$ exchange. To this end we would like
to emphasize that in our calculations we indeed assumed that
 
\begin{equation}
\gi = 3\gn \; ,
\end{equation}
 
\noindent
then the total $Z$ width becomes
 
\begin{equation}
\gz = 3\gn + \ge + \gmu + \gt + \gh\;.
\end{equation}
 
\noindent
Unlike the widths, asymmetries and polarizations do not contain,
by definition, QED and QCD corrections; furthermore, they will only
refer to pure $Z$ exchange: they are nothing but simple combination     s
of the effective $Z$ couplings, introduced above,
 
\begin{eqnarray}
\afb^{f} & = & \frac{3}{4} {\cal A}^e {\cal A}^f,  \nll
\alr^e   & = & {\cal A}^e,                         \nll
P^{f}    & = & - {\cal A}^f,                       \nll
P_{_{\rm {FB}}}(\tau) &=& - \frac{3}{4}{\acal}^e,
\label{defasym}
\end{eqnarray}
 
\noindent
where we define
 
\begin{equation}
{\cal A}^f = \frac{2\gvf\gaf}{(\gvf)^2+(\gaf)^2}\;,
\end{equation}
 
\noindent
and $\gvf,\gaf$ defined by Eq.~(\ref{24}) are the effective neutral
current
vector and axial-vector couplings of the $\z0$ to a fermion pair $f \bar f$.
$P_{_{\rm {FB}}}(\tau)$ is the $\tau$ polarization
forward--backward asymmetry.
One should realize that as a consequence of the adopted definition of
pseudo-observables, there exist even more relations among them
than are indicated above; it is, for example, $\alr^e=-P^{\tau}$,
if in addition lepton universality is assumed, which is granted
automatically in the MSM. For this reason we will present numerical and
graphical results only for a selected subset of them.
 
\subsection{Realistic observables}
 
Realistic observables are the cross-sections $\sigma^f(s)$
and asymmetries $A^f_{_{\rm {FB}}}(s)$ of the reactions
 
\begin{eqnarray}
{\rm e}^+ {\rm e}^- \rightarrow
(\gamma, Z) \rightarrow f {\bar f} (n\gamma) \; ,
\label{basicffng}
\end{eqnarray}
 
\noindent
for a given interval of $s=4E^2$ around the $Z$ resonance,
including real and virtual photonic corrections (`dressed by QED').
We will present results for $f=\mu$ and $f=had$, both in a fully extrapolated
set-up or with simple kinematical cuts. For $f=e$ both the $s$ channel
exchange cross-section and the complete Bhabha reaction will be treated.
In total, we considered 14 cross-sections and asymmetries for
different combinations of flavour $f$ and {\it cuts}; they will be defined
in section 3. The kinematical cuts that will be imposed are of two types:
 
\begin{itemize}
 
\item $\theta_{\rm {min}} < \theta_- < \pi - \theta_{\rm {min}}$
 
\item $s' > s'_{\rm {min}}$
 
\end{itemize}
 
\noindent
and:
 
\begin{itemize}
 
\item $\theta_{\rm {min}} < \theta_- < \pi - \theta_{\rm {min}}$
 
\item $\theta_{\rm {acoll}} < \xi$
 
\item $E_{\pm} > E_{\rm {th}}  \;.$
 
\end{itemize}
 
\noindent
Where $s'$ is the invariant mass of two final-state fermions,
$\theta_-$ refers to the outgoing scattering angle of the outgoing
fermion, $\theta_{\rm {acoll}}$
is the acollinearity angle between the outgoing
fermions and $E_{\pm}$ the outgoing fermion (anti-fermion) energies. By
$s$ channel Bhabha scattering we mean that the $s-t$ and $t-t$ exchange
interferences have been subtracted from the observable under consideration.
 
We would like to mention in passing that the QED dressers have options which
allow a model independent interpretation of realistic observables. Since this
is beyond our present scope, we refer the reader for details
to the existing literature~\cite{miza}-\cite{smatr}.
 
\subsection{Calculational schemes}
 
Before entering into a detailed study of the numerical results it is important
to underline how an estimate of the theoretical uncertainty emerges from
the many sets of numbers obtained with the five codes.
First of all, one may distinguish between {\it intrinsic}
and {\it parametric}
uncertainties.
The latter are normally associated with a variation of the input parameters
according to the precision with which they are known. Typically, we have
$|\Delta\alpha^{-1}(\zm^2)| = 0.12$, $|\Delta m_b| = 0.3\,$GeV, $|\Delta m_c|
= 0.35\,$GeV etc. These uncertainties will eventually shrink when more
accurate measurements become available.
In this Report we are mainly devoted to a discussion of the intrinsic
uncertainties associated with missing non-leading higher-order
corrections, although some results on parametric uncertainties will be also
given. An essential ingredient of all calculations for radiative corrections
to pseudo-observable is the choice of the renormalization scheme.
There are many renormalization schemes in the literature:
 
\begin{itemize}
 
\item the on-shell schemes in various realizations~\cite{ossch1}-\cite{ossch7},
~\cite{leptop}
 
\item the $\gf$ scheme~\cite{gsch1}-\cite{gsch3}
 
\item the * scheme~\cite{ssch}
 
\item the $\msb$ scheme~\cite{msbsch1}-\cite{msbsch2}.
 
\end{itemize}
 
One cannot simulate the shift of a given quantity due to
a change in the renormalization scheme with one code alone.
Thus the corresponding theoretical
band in that quantity will be obtained from the differences in the prediction
of the codes, which use different renormalization schemes.
On top of that we should also take into account the possibility of having
different implementations of the full radiative corrections within one code,
--- within one well specified renormalization scheme.
Typically we are dealing here with the practical implementation of resummation
procedures, of the exact definition of leading versus non-leading higher order
corrections and so on.
Many of these implementations are equally plausible and differ by non-leading
higher order contributions,
which, however, may become relevant in view of the
achieved or projected experimental precision. This sort of intrinsic
theoretical uncertainty can very well be estimated by staying within each
single code. However, since there are no reasons to expect that these will be
the same in different codes, only the full collection of different sources
will, in the end, give a reliable information on how accurate an observable
may be considered from a theoretical point of view.
 
Of course, only a complete two-loop electroweak calculation, combined with
some re-summations of the leading corrections (renormalization group
improvements), would ultimately solve the problem of the
theoretical accuracy.
 
\subsection{Main features of different approaches}
 
In this subsection we will discuss common features of and main differences
between the electroweak libraries of the five codes.
 
\begin{itemize}
 
\item[] Common features:
 
\item All five codes use as input parameters the most accurately known
electroweak parameters $\gf, \zm$ and  $\baral$,
(the codes, with the exception of {\tt LEPTOP},
also use $\alpha$), in order to calculate the less precisely measured
pseudo-observables.
 
\item They use the same expressions for final state QED and QCD corrections
(radiation factors).
 
\item All codes include essentially the same internal gluon corrections of
the order of $\alpha\alpha_s$ in the $W$ and $Z$ self-energy quark loops.
 
\item All codes include leading {\it two-loop} corrections
of the order of $\gf^2 m_t^4$; all gluonic corrections of the order of
$\alpha_s\gf m_t^2$; leading gluonic corrections $\alpha\alpha_s^2$
in the vector boson self-energies.
 
\item[] Main differences:
 
\item Each code uses a different renormazation framework.
 
\item Some codes define an electroweak Born approximation, others give
no physical emphasis to a Born approximation and employ only the notion of an
{\it Improved Born Approximation (IBA)} which includes the leading electroweak
loop corrections.
 
\item Certain  codes include higher-order electroweak corrections,
which are documented in the literature but are numerically irrelevant,
such as the irreducible two-loop photon vacuum polarization and
the Higgs corrections proportional to $\alpha^2 m_{_H}^2$.
Some codes use, on top of these higher order corrections, resummation
of one-loop terms.
 
\item They differ by the choice of the definition of the weak mixing angle.
 
\end{itemize}

\subsubsection{\tt BHM/WOH, ZFITTER}
 
The three codes {\tt BHM}~\cite{bhm}, {\tt WOH}~\cite{woh} and
{\tt ZFITTER}~\cite{zfitter}
rely on different realizations of the on-mass-shell renormalization scheme.
They perform the renormalization procedure with systematic use of the
counterterm method for the basic parameters $\alpha, \wm, \zm$.
The {\tt BHM} and {\tt WOH} codes lead back to one approach and are thus not
completely independent of each other.
 
The description of the renormalization schemes may be found in~\cite{ossch7}
for {\tt BHM/WOH} and in~\cite{ossch3} for {\tt ZFITTER}.
The weak mixing angle formally appears as
 
\begin{eqnarray}
c^2_{_W} &=& \cos^2\theta_{_W} = (M_{_W}/M_{_Z})^2 , \nll
s^2_{_W} &=& \sin^2\theta_{_W} = 1 - c^2_{_W}.
\label{sirlin}
\end{eqnarray}
 
\noindent
This corresponds to the definition proposed by Sirlin~\cite{sirlinang}.
The masses in~(\ref{sirlin}) are the physical $W$ and $Z$ boson masses.
Thus, $s^2_{_W}$ is not an independent quantity and is used mainly for
internal bookkeeping.
For fixed values of $\tm$ and $\hm$ each observable, including the Fermi
constant $\gf$, is expressed in terms of $\alpha$, $\zm$ and $\wm$ with the
corresponding quantum corrections being taken into account. Because of lack
of precision in the $W$ mass, $\wm$, it is not taken as an experimentally
measured input quantity. It is instead replaced by the more accurate value of
$\gf$.
 
The quantum corrections for $\mu$ decay~(\ref{taumu}) after removing the
Fermi-model like QED corrections are contained in the parameter $\rho_{c}$
related to $\gf$ by the equation{\footnote{The naive Born approximation for
muon decay of Eq.~(\ref{gmu}), $\rho_c=1$, was used in the 70's for
predictions of the masses of the $W$ and $Z$ bosons with a crude accuracy by
using the value of $s_{_W}$ from the ratio of neutral and charged currents in
neutrino interactions.}}
 
\begin{eqnarray}
G_{\mu} &=& \frac{\pi}{\srt}\;\frac{\alpha}{s^2_{_W}c^2_{_W}\zm^2}
            \; \rho_{c}\;.
\label{gmu}
\end{eqnarray}
 
\noindent
Since this equation contains $\alpha$ instead of $\alpha(\zm)$
a large fraction of $\rho_{c}$ is of purely electromagnetic origin via
$\Delta \alpha$.
The parameter $\rho_{c}$ in~(\ref{gmu}) takes into account the $W$ propagator,
vertex and box corrections due to one-loop diagrams and presently available
higher order terms. The one-loop corrections were first calculated
in Refs.~\cite{sirlinang}-\cite{sirlinmu}.
The correction due to the photon self-energy is QED renormalization group
improved: a geometrical progression brings $\Delta \alpha$ into the denominator
of $\rho_c$. Hence, $\rho_{c}$ is usually written in the form
 
\begin{equation}
\rho_{c} = \frac{1}{1-\Delta r}\;,
\label{rhoc}
\end{equation}
 
\noindent
where $\Delta r$ contains all the one-loop corrections to the muon-decay with
the inclusion and the proper arrangement of the higher order terms.
(For more
details see the explanations presented in subsection 1.10.2.)
 
For given values of $\tm$ and $\hm$,
Eq.~(\ref{gmu}) fixes $\theta_{_W}$ and,
hence, $\wm$ by the experimental value of $\gf$.
In practice, Eq.~(\ref{gmu}) is solved with respect to $\wm$ iteratively,
since $\Delta r$ is a complicated function of $\wm$. This equation for
iteration of $M_{_W}$ reads
 
\begin{equation}
M_{_W}=M_{_Z}\sqrt{1
-\sqrt{1-\frac{4\pi\alpha}{\sqrt{2}G_{\mu}\zm^2[1-\Delta r(M_{_W})]}}}
\;.
\label{41}
\end{equation}
 
\noindent
Therefore, $\wm$ appears as an $\tm$, $\hm$, ${\hat \alpha}_s$ dependent
prediction, which can be compared with the experimental values from UA2
and CDF~\cite{wm}.
 
After fixing $\wm$ in this way all the other observables are expressed in
terms of $\gf, \alpha, \zm, \tm, \hm$ and ${\hat \alpha}_s$\footnote{
In leptonic processes the dependence on ${\hat \alpha}_s$ is of higher order;
it comes from gluonic exchanges between virtual quarks in $W$, $Z$, $Z\gamma$
and $\gamma$ self-energy diagrams with quark loops.}.
As a consequence of this procedure, purely leptonic Feynman graphs (e.g. a
vertex correction in the $Z \rightarrow \bar{\mu}\mu$ decay) also
turn out to be implicitly $\tm$ dependent through the $\wm$.
 
In conclusion, Eq.~(\ref{gmu}) uses the amplitude of $\mu$ decay with the
inclusion of weak corrections in order to establish the interdependence
between $\wm$, $\tm$ and $\hm$ and the best measured parameters $\alpha,
G_{\mu}, \zm$. Thus, $\wm$ appears as a prediction as well as an intermediate
parameter for the calculation of $Z$-boson observables.
 
The flowchart of the {\tt BHM/WOH}, {\tt ZFITTER} approach is shown in
Fig.~8. One should have in mind that in spite of a common presentation of the
basics of these codes there are certain differences in the realization of the
on-mass-shell renormalization schemes between {\tt BHM/WOH} on one side and
{\tt ZFITTER} on the other --- for example,
in the use of different gauges and in the
different treatment of field renormalization.
\clearpage

\begin{center}
{\large {\bf FLOWCHART OF ZFITTER/BHM/WOH}}\\[5mm]
 
\setlength{\unitlength}{1mm}
\begin{picture}(160,210)
\put(0,195){\framebox(160,20)}
\put(80,211){\makebox(0,0){\large
Select \underline{minimal set} of parameters in the MSM Lagranian:}}
\put(80,205){\makebox(0,0){\large
$\alpha_0, M_{_{W0}}, M_{_{Z0}}, M_{_{H0}}, m_{f0}$
(including $m_{to}$); note that $\alpha_{_{W0}}$,}}
\put(80,199){\makebox(0,0){\large
$\alpha_{_{Z0}}$ and VEV $\eta$ are \underline{not} among these.}}
\put(80,190){\line(0,1){5}}

\put(0,170){\framebox(160,20)}
\put(80,186){\makebox(0,0){\large
Define renormalization Z-factors for each bare parameter and each field}}
\put(80,180){\makebox(0,0){\large
(Z-matrices for Z$-\gamma$ and fermion  mixing --- for {\tt ZFITTER} only).}}
\put(80,174){\makebox(0,0){\large
Fix Z-factors on mass shell.
Use dimensional regularization ($1/\epsilon, \mu)$.}}
\put(80,165){\line(0,1){5}}

\put(0,150){\framebox(160,15)}
\put(80,160){\makebox(0,0){\large
Lagrangian now depends only on physical fields, couplings and}}
\put(80,154){\makebox(0,0){\large
masses, and on counterterms (Z-factors).}}
\put(80,145){\line(0,1){5}}

\put(0,130){\framebox(160,15)}
\put(80,141){\makebox(0,0){\large
Expand Z-factors; $Z_i$= 1+$\alpha f_i$, where $\alpha=\alpha(0)$
and $f_i$'s are functions}}
\put(80,135){\makebox(0,0){\large
of physical input $M_{_W}$, $M_{_Z}$, $M_{_H}$, $m_f$
and 1/$\epsilon$ and $\mu$.}}
\put(80,125){\line(0,1){5}}

\put(0,110){\framebox(160,15)}
\put(80,121){\makebox(0,0){\large
Calculate one--loop electroweak amplitudes with graphs, including}}
\put(80,115){\makebox(0,0){\large
loops and counterterms; 1/$\epsilon$ and $\mu$ drop out.}}
\put(80,105){\line(0,1){5}}

\put(0,90){\framebox(160,15)}
\put(80,101){\makebox(0,0){\large
Improve one-loop results by RG-techniques and by proper resummation}}
\put(80,95){\makebox(0,0){\large
of the higher-order e.w. terms. Define improved Born approximation.}}
\put(80,85){\line(0,1){5}}

\put(0,75){\framebox(160,10)}
\put(80,80){\makebox(0,0){\large
Select experimental inputs: $\alpha(0)$, $M_{_Z}$, $G_{\mu}$
($\tau_{\mu}$).}}
\put(80,70){\line(0,1){5}}

\put(0,55){\framebox(160,15)}
\put(80,66){\makebox(0,0){\large
Get $M_{_W}$ from $G_{\mu}$ = ($\pi$/ $\sqrt{2}$)~($\alpha$/$s^2_{_W}$
$c_{_W}^2$ $M_{_Z}^2$)~$\rho_c$, where $\rho_c$ depends}}
\put(80,60){\makebox(0,0){\large
on $m_t$, $M_{_H}$, $\alpha$(0),
$M_{_W}$, $M_{_Z}$ and $s_{_W}^2$ = 1 - $M_{_W}^2$/$M_{_Z}^2$.}}
\put(80,50){\line(0,1){5}}

\put(0,35){\framebox(160,15)}
\put(80,46){\makebox(0,0){\large
Calculate $Z^0$ decay observables, with $m_t$ and $M_{_H}$ free,}}
\put(80,40){\makebox(0,0){\large
in terms of $G_{\mu}$, $\alpha(0)$, $M_{_Z}$.}}
\put(80,30){\line(0,1){5}}

\put(0,15){\framebox(160,15)}
\put(80,26){\makebox(0,0){\large
Introduce gluonic corrections into quark loops and QED + QCD}}
\put(80,20){\makebox(0,0){\large
final state interactions
in terms of $\bar{\alpha}$, $\hat{\alpha}_s(M_{_Z})$, $m_b$
$(M_{_Z})$, $m_t$.}}
\put(80,10){\line(0,1){5}}

\put(0,-5){\framebox(160,15)}
\put(80, 6){\makebox(0,0){\large
Compare the results with electroweak experimental data,}}
\put(80, 0){\makebox(0,0){\large
exhibit $M_{_Z}$, $m_t$, $M_{_H}$, and $\hat{\alpha}_s(M_{_Z})$ dependence.}}

\end{picture}
 
\vspace{4mm}
 
Figure 8: {\tt BHM/WOH ZFITTER} flowchart
\end{center}
\clearpage
 
\subsubsection{\tt LEPTOP}
 
The authors of {\tt LEPTOP} avoid the use of counterterms in the formulation
of the theoretical framework.
They do not use resummation of one-loop electroweak corrections,
thus limiting themselves
to a simple one-loop approximation, completed by selected
leading electroweak two-loop corrections. In contrast to all the other codes,
{\tt LEPTOP} uses a Born approximation.
According to {\tt LEPTOP}, the weak mixing angle
$\theta$ is defined by ($s\equiv \sin\theta$, $c\equiv \cos \theta$):
 
\begin{eqnarray}
\gf &=& {{\pi{\bar \alpha}}\over {\srt s^2c^2 \zm^2}}\; ,
\label{sleptop}
\\
\sin^2 2\theta &=& {{4\pi{\bar \alpha}}\over {\srt\gf\zm^2}}\;.
\label{sleptop1}
\end{eqnarray}
 
\noindent
This gives $s^2$=0.23117(33), $c$=0.87683(19).
Such a $\theta$, which by definition does not depend on $\tm$
and $\hm$, is used to determine the so-called $\baral$-Born
approximations for electroweak observables.
 
The bare gauge couplings $\alpha^0$, $\alpha^0_{_{W}}$ and the bare mass
$\zm^0$ are expressed in terms of $\bar \alpha$, $G_{\mu}$ and $\zm$,
$1/\epsilon$ and $\mu$, where $\mu$ is the 't Hooft's scale parameter and
$2\epsilon = 4-D$ in dimensional regularization scheme.
The $W$ mass, $\wm$, is treated on an equal footing with
the other observables.
 
The $\baral$-Born approximation automatically includes purely
electromagnetic corrections. In terms of $s$ and $c$, the expressions for the
hadron-free pseudo-observables are very simple in that approximation
 
\begin{equation}
(\wm/\zm)^B = c\;,
\label{28}
\end{equation}
\begin{equation}
(g_{_A}^f)^B = I^{(3)}_f\;,
\label{29}
\end{equation}
\begin{equation}
(g_{_V}^f/g_{_A}^f)^B = 1-4|Q_f|s^2\;.
\label{30}
\end{equation}
 
\noindent
The $\baral$-Born approximation with the due account of the
final state QED and QCD radiation factors gives simple expressions for the
observables in the {\it hadronic} decays of $Z$ boson as well.
 
The electroweak corrections for all observables are calculated in the
framework of {\tt LEPTOP} in terms of $\alpha^0$, $\alpha_{_W}^0$, $M_{_Z}^0$,
$m_{t}^0$, $m_{_{H}}^0$, and ($1/\epsilon$, $\mu$) in one-loop approximation.
In this approximation $m_{t}^0$ and $M_{_H}^0$ can be replaced by
the physical masses $m_t$ and $M_{_H}$.
 
By expressing $\alpha^0$, $\alpha_{_W}^0$, $M_{_Z}^0$ in terms of $\baral$,
$G_{\mu}$, $M_{_Z}$ and ($1/\epsilon$, $\mu$) one derives formulae in which
terms ($1/\epsilon$, $\mu$) cancel out.
Thus introduction of counterterms is avoided.
 
The explicit analytical expressions for the {\tt LEPTOP} electroweak
corrections in terms of $G_{\mu}$, $\baral$, $c$, $s$, $t = (m_t/M_{_Z})^2$ and
$h=(M_{_H}/M_{_Z})^2$ are given in section 4.1.
 
The flowchart for the {\tt LEPTOP} approach is shown in Fig. 9.
\newpage
 
\begin{center}
{\large {\bf FLOWCHART OF LEPTOP}}\\[5mm]
 
\setlength{\unitlength}{1mm}
\begin{picture}(160,210)
 
\put(0,195){\framebox(160,15)}
\put(80,205){\makebox(0,0){\large
Select the three most accurate observables:}}
\put(80,199){\makebox(0,0){\large
$ \gf, \zm, \alpha(\zm) \equiv \bar{\alpha}$
}}
\put(80,190){\line(0,1){5}}
 
\put(0,170){\framebox(160,20)}
\put(80,185){\makebox(0,0){\large
Define angle $\theta$ ( $s \equiv \sin \theta, c \equiv \cos \theta$)}}
\put(80,180){\makebox(0,0){\large
in terms of $\gf, \zm$ and $\bar \alpha$:}}
\put(80,174){\makebox(0,0){\large
$\gf = (\pi / \sqrt{2}) \bar{\alpha} / s^2 c^2 \zm^2$}}
\put(80,165){\line(0,1){5}}
 
\put(0,150){\framebox(160,15)}
\put(80,160){\makebox(0,0){\large
Define the Born approximation for other electroweak}}
\put(80,154){\makebox(0,0){\large
observables in terms of $\gf, \zm$ and $\theta$.}}
\put(80,145){\line(0,1){5}}
 
\put(0,130){\framebox(160,15)}
\put(80,140){\makebox(0,0){\large
Introduce bare couplings ($\alpha^0, \alpha^0_{_Z}, \alpha^0_{_W}$),
masses ($M^0_{_Z}, M^0_{_W},$}}
\put(80,134){\makebox(0,0){\large
$ m^0_{q}$ (including $m^0_{t}$)) and VEV $\eta$
in the framework of MSM.}}
\put(80,125){\line(0,1){5}}
 
\put(0,110){\framebox(160,15)}
\put(80,120){\makebox(0,0){\large
Express $\alpha^0, \alpha^0_{_Z},M^0_{_Z}$
in terms of $\gf, \zm$ and $\bar \alpha$
in the one-loop}}
\put(80,114){\makebox(0,0){\large
approximation, using dimensional regularization ($1/ \epsilon, \mu$)
}}
\put(80,105){\line(0,1){5}}
 
\put(0,85){\framebox(160,20)}
\put(80,100){\makebox(0,0){\large
Express one-loop corrections to all other electroweak observables}}
\put(80,95){\makebox(0,0){\large
in terms of $\alpha^0, \alpha^0_{_Z},M^0_{_Z}, \tm , \hm$
and hence in terms of $\gf$,
}}
\put(80,89){\makebox(0,0){\large
$\zm,\bar \alpha, \tm, \hm$. Check cancellation of $1/ \epsilon$ and $\mu$.
}}
\put(80,80){\line(0,1){5}}
 
\put(0,60){\framebox(160,20)}
\put(80,75){\makebox(0,0){\large
Introduce gluonic corrections into quark loops and QED and QCD
}}
\put(80,70){\makebox(0,0){\large
                final state interactions for hadronic decays}}
\put(80,64){\makebox(0,0){\large
(in terms of $\bar \alpha, \hat{\alpha_{s}}(\zm), m_b(\zm), \tm$).
}}
\put(80,55){\line(0,1){5}}
 
\put(0,40){\framebox(160,15)}
\put(80,50){\makebox(0,0){\large
Compare the Born results and Born + one loop results with
}}
\put(80,44){\makebox(0,0){\large
            experimental data on $Z$-decays and $\wm$.
}}
\put(80,35){\line(0,1){5}}
 
\put(0,25){\framebox(160,10)}
\put(80,29){\makebox(0,0){\large
Make a global fit for three parameters  $\tm, \hm, \hat{\alpha_s}(\zm)$.
}}
\put(80,20){\line(0,1){5}}
 
\put(0,5){\framebox(160,15)}
\put(80,15){\makebox(0,0){\large
Predict  the central values of all electroweak observables
}}
\put(80,9){\makebox(0,0){\large
                          and their uncertainties.}}
\put(80,0){\makebox(0,0){\large
}}
\label{tel1}
\end{picture}
 
\vspace{5mm}
 
Figure 9: {\tt LEPTOP} flowchart
\end{center}
 
\clearpage
 
\subsubsection{\tt TOPAZ0}
 
The {\tt TOPAZ0} code~\cite{topaz0} uses the $\msb$ (modified minimal
subtraction) scheme for all types of interactions, including the electroweak
ones, as introduced in~\cite{msbsch2}. Its approach is quite different both
from those of {\tt BHM/WOH}, {\tt ZFITTER}, and {\tt LEPTOP}.
Its main steps are presented in the following items:
 
\begin{itemize}
 
\item The MSM Lagrangian is assumed.
Excluding fermion masses and mixing angles
the MSM is a three-parameter theory:
i.e. $g^0$, $M^0_{_W}$ and $\sin\theta_0$.
At the level of defining a renormalization scheme,
two essentially different
approaches may be distinguished: to prescribe precisely what a parameter
of the Lagrangian is or to prescribe precisely what a counterterm is.
It is a matter of convention, since only their combined effect appears in the
confrontation with the data,
and once the latter is chosen it would then seem
natural to follow the consensus with respect to QCD.
The bare parameters are fixed by considering three data points ---
$\alpha(0) $, $\gf$, and $\zm$.
These quantities are computed up to one-loop diagrams and
{\it fitting equations} are written:
 
\begin{equation}
p^0_i = f_i(\alpha,\gf,\zm,\Delta)\;, \qquad \qquad i=1,2,3\;,
\end{equation}
 
where $p^0_i$ are the bare parameters and $\Delta = -2/(n-4) + \gamma_E
- \ln\pi$. Everything is worked out explicitly in the
't Hooft--Feynman gauge.
 
\item The {\it fitting equations} are solved perturbatively (order-by-order
renormalization),
but gauge invariant (fermionic) higher order leading terms
are always re-summed.
 
\noindent
Strictly speaking, one usually solves an implicit equation
$f(a,\lambda) = 0$, where $a$ is some parameter and $\lambda$ is a coupling
constant, by expanding $a$ around $a_0$, the solution of $f(a_0,0) = 0$.
However, one could just as well expand $a$ around ${\bar a}$,
the solution of some other equation
$h({\bar a},\lambda) = 0$, as long as ${\bar a}$ is gauge invariant.
Thus in our implementation $\alpha(0)$ will evolve into $\alpha(\zm)$ and the
lowest-order approximation for $s_{\theta}^2 = \sin^2\theta_0$ is
 
\begin{eqnarray}
\shat^2 &=& \frac{1}{2}\,\left[\, 1 - \sqrt{1 - \frac{4\,\pi\alpha(\zm)}
{\srt\,\gf\zm^2\rhozr}}\;\;\right]\;,
\label{topazsin}
\\
(\rhozr)^{-1} &=& 1 + {{\gf\zm^2}\over {2\srt\pi^2}}
\times \{ {\hbox{UV finite combination of two-point functions}} \},
\label{topazroz}
\end{eqnarray}
 
\noindent
where $(\rho_{_Z}^{_R})^{-1}-1$ is determined by an ultraviolet finite
combination of two-point functions. The detailed definition of $\rhozr$ is
left for the section 4.2 --- Eqs.~(\ref{rhodef})
and~(\ref{nores} -- \ref{myres}).
 
Eq.(\ref{topazsin}) is an algebraic solution of the following
relation{\footnote {For analogous relations
for {\tt BHM/WOH}, {\tt ZFITTER} see
Eq.~(\ref{gmu}) and for {\tt LEPTOP} - Eq.(\ref{sleptop}).}}
 
\begin{equation}
G_{\mu}=\frac{\pi\baral}
        {\sqrt{2}\zm^2\shat^2\chat^2\rhozr}\; .
\label{topazroz1}
\end{equation}
 
\noindent
If no resummation at all is performed in {\tt TOPAZ0} then $\rhozr = 1$ and,
in this limit, one recovers the weak mixing angle as defined by {\tt LEPTOP}.
The $\rhozr$ parameter is strictly connected to the $\z0$ wave function
renormalization and usually all bosonic contributions are expanded up to first
order, in this case the $\rho$ parameter being instead
denoted by $\rhoz$, Eq.(\ref{rhodef}), but the possibility of a resummation
for a certain gauge invariant subset in the $\msb$ framework is built in.
It should be noticed that $\rhoz$ or its variant $\rhozr$ represent the
natural extension of the Veltman's $\rho$-parameter, as defined at low
energy~\cite{msbsch2}, to the scale $\zm$.
Its asymptotic behavior, for large $m_t$, is exactly the same as dictated by
$\Delta\rho$ --- see Eq.~(\ref{rholead}).
That is why all the higher-order
leading ($m_t$) corrections will simply be added to $\rhozr(\rhoz)$.
 
\item By a proper redefinition of the bare coupling $g^0$~\cite{gsch3},
corresponding to a gauge field re-diagonalization in the quadratic part of
the Lagrangian, in the $\xi = 1$ (t'Hooft--Feynman) gauge
the following properties are fulfilled: the sum of all $\z0\barf f$ vertices,
$\sum_v\{{\hbox{vertex}}\}$, is ultraviolet finite and the $\z0-\gamma$
transition satisfies $\Szg(0) = 0$.
In the renormalization procedure of {\tt TOPAZ0} it is observed that,
unlike QED, no one-to-one correspondence exists between the bare
parameters and experimental data points. Therefore no attempt is made to give
an all-orders relation as for $s_{_W}^2$, but rather it is observed that the
mixing angle defines the distribution of the vector current between $Z$ and
$\gamma$.
It is an accident of the minimal Higgs system (or more generally of
representations where the Higgs scalars only occur in doublets) that the
vector boson masses are not both free and beyond lowest order different
definitions of the weak mixing angle receive different radiative corrections.
Also, the $W$ mass has no special role in {\tt TOPAZ0},
being computed like any
other quantity, once the bare parameters are substituted through their
explicit expressions. Typically,
 
\begin{eqnarray}
\sin^2\theta_0 &=& \shat^2 + \left\{ {\ord}(\alpha)
{\hbox{UV finite bosonic corrections}}\right\} \nll
&&+ \left\{ {\ord}(\alpha)\;
{\hbox{$\Delta$-dependent terms}} \right\} \; ,
\end{eqnarray}
 
\noindent
and in computing for instance $Z \to f \bar f$ the latter cancel their
$\Delta$ dependence against the $Z-\gamma$ transition, leading to an ultraviolet
finite result. According to the chosen strategy, all bosonic terms are
expanded or, alternatively, a gauge invariant sub-set of the bosonic
corrections is re-summed in $\shat^2$ (in the ${\overline {MS}}$ environment).
In the second case a (usually small) remainder is left, which compensates
against vertex corrections.
The corrected $Z$ propagator, leading to mass renormalization and to a $Z$
wave-function renormalization factor, is ultraviolet finite by inspection.
 
\item The processes 
$Z \to {\rm f} \bar{\rm f} $, ${\rm e}^+{\rm e}^- \to {\rm f} \bar{\rm f}$
are computed up
to one-loop diagrams with lowest-order (partially re-summed) parameters plus
tree diagrams with one-loop corrected parameters. Ultraviolet finiteness is
checked both analytically and numerically (independence of $\Delta$). Roughly
speaking, the central objects in {\tt TOPAZ0}
are the $S$-matrix elements for
a given process, so particular care is devoted to the proper treatment of the
wave-function renormalization factors (see Ref.~\cite{veltd}).
 
\end{itemize}
 
\subsection{Phenomenology of $Z$ boson decays}
 
The decay width of $Z\to f\bar{f}$ is described by the following
equation used by all five codes:
 
\begin{equation}
\Gamma_{f}\equiv \Gamma(Z\to f\bar{f}) =4N_c^f\Gamma_0 [(g_{_V}^f)^2
R_{_V}^f +(g_{_A}^f)^2 R_{_A}^f] \;,
\label{A}
\end{equation}
 
\noindent
where $g_{_V}^f$ and $g_{_A}^f$ are effective electroweak couplings defined
in section 1.4.4, $N_c^f =1$ or $3$ for leptons or quarks ($f=l, q$). The
factors $R_{_V}^f$ and $R_{_A}^f$ describe the final state QED and QCD
interactions and take into account the mass $m_f$. The radiation factors
are universal, to a large extent, and will be described below in this
subsection.
 
The standard width $\Gamma_0$ in Eq.(\ref{A}) is known with very high
accuracy:
 
\begin{equation}
\Gamma_0 =\frac{G_{\mu}M_{_Z}^3}{24\sqrt{2}\pi} = 82.945(12)\;\;
{\rm {MeV}}\;.
\label{B}
\end{equation}
 
For the decay into leptons $l\bar{l}$ the radiation factors $R_{_{V,A}}^l$
are especially simple. For charged leptons:
 
\begin{equation}
\Gamma_l = 4\Gamma_0\Biggl[(g^l_{_V})^2\left(1+\frac{3}{4\pi}\baral\right)
+(g^l_{_A})^2\left(1-6\frac{m_l^2}{M_{_Z}^2}+\frac{3}{4\pi}\baral\right)
\Biggr]\; .
\label{C}
\end{equation}
 
\noindent
The mass term in Eq.~(\ref{C}) is negligible for $l = e, \mu$ and is
barely visible only for $l =\tau~~(m_{\tau}^2/M_{_Z}^2 = 3.8\times 10^{-4}$).
 
For the neutrino decay:
 
\begin{eqnarray}
\Gamma_{\nu} &=& 8(g^{\nu})^2\Gamma_0 \; , \nonumber \\
g^{\nu} &=& g^{\nu}_{_V} = g^{\nu}_{_A} \; .
\label{D}
\end{eqnarray}
 
For the decays into quarks $q\bar{q}$ the radiation factors are given
by~\cite{asku}-\cite{qcd}
 
\begin{eqnarray}
R_{_V}^q(s) &=& 1 + \delta_e^q + \delta_v a_s^2  - 12.76706 a_s^3
    +12 \frac{\bar m_q^2(s)}{s} a_s \delta_{vm}\;,
\label{78}
\\
R_{_A}^q(s) &=& 1+ \delta_e^q  + \left[\delta_v - 2I^{(3)}_q{\ical}^{(2)}\left(
              \frac{s}{m_t^2}\right)
           \right] a_s^2         \nll
          & &+\left[-12.76706 - 2I^{(3)}_q {\ical}^{(3)}\left(\frac{s}{m_t^2}
             \right)\right] \; a_s^3  \nll
          & &-6 \frac{\bar m_q^2(s)}{s}\delta^1_{am}
          -10\frac{\bar m_q^2(s)}{m^2_t} a_s^2 \delta^2_{am}\;,
\label{79}
\end{eqnarray}
 
\noindent
where
 
\begin{eqnarray}
\delta_e^q &=& \frac{3}{4} Q_q^2 a + a_s - \frac{1}{4} Q_q^2 a a_s \;,
\nll
\delta_v &=& 1.40923 +\left(\frac{44}{675} - \frac{2}{135}
\ln\frac{s}{m^2_t}
             \right)\frac{s}{m^2_t} \;, \nll
\delta_{vm} &=& 1+8.7 \; a_s + 45.15 \; a_s^2 \;, \nll
\delta^1_{am} &=& 1+\frac{11}{3} a_s + \left(11.286
            +\ln\frac{s}{m^2_t} \right) a_s^2 \;, \nll
\delta^2_{am} &=& \frac{8}{81}-\frac{1}{54}\ln\frac{s}{m^2_t} \;,
\label{80}
\end{eqnarray}
 
\noindent
and
 
\begin{eqnarray}
      a_s &=& \frac{\alpha_s(s)}{\pi}\;,
      \qquad a = \frac{\alpha(s)}{\pi} \;,\nll \nll
      {\ical}^{(2)}(x) &=&  - \frac{37}{12} + l + \frac{7}{81}x
                      +0.0132 \; x^2 \;,  \nll   \nll
      {\ical}^{(3)}(x) &=& - \frac{5651}{216} + \frac{8}{3}
  + \frac {23}{36}\,\pi^2 + \zeta(3)     +\frac{67}{18}l
                      + \frac{23}{12} l^2 \;,
\end{eqnarray}
 
\noindent
where $l = \ln x$, $m_t= m_{t(pole)}$ (and $s=\zm^2$ for $Z$ decay).
 
Note that we have automatically absorbed the finite mass terms
into the QCD correction factors and that in $R_{_A}$ the large logarithms
$\ln(m_f^2/s)$ have been absorbed through the use of running parameters
 
\begin{eqnarray}
{\bar m}(s) &=& {\bar m}(m^2)\,\exp\left\{
-\,\int_{a_s(m^2)}^{a_s(s)}\,dx \frac{\gamma_m(x)}{\beta(x)}\right\} \;,
\nll
m &=& {\bar m}(m^2)\,\left[ 1 + \frac{4}{3} a_s(m) + K a_s^2(m)\right]\;,
\label{82}
\end{eqnarray}
 
\noindent
where $m = m_{pole}$ and $K_b \approx 12.4, K_c \approx 13.3$.
The various codes may differ in the construction of $\gvf$ and $\gaf$
but, a general consensus has been reached for final state radiation.
In this way we obtain the hadronic and total $\z0$ width as
 
\begin{eqnarray}
\gh &=& \sum_{f=udcsb}\,\Gamma(\z0 \to f \bar f )
+ 4 \Gamma_0 N^f_c R^h_{_V} \;,
\nll
R_{_V}^h &=& -0.41318 \left(\sum_q v_q\right)^2 a_s^3  \;, \nll
\gz &=& \gh + \sum_{f=e\mu\tau} \Gamma_f + \gi \;.
\end{eqnarray}
 
\noindent
Actually, the hadronic width and the partial $q \bar q$ widths deserve
some
partial comment, while the complete theoretical framework is described
in~\cite{qcd}. The singlet QCD contribution which is simple and unambiguous
for the hadronic width becomes ambiguous, starting at ${\ord}(\als^2)$,
for individual $q \bar q$ channels.
In fact, some sort of agreement has recently been
reached on these matters but we want to summarize the roots of
the problem. From a pragmatic point of view there is a hierarchical
description where
 
\begin{equation}
\Gamma_{q} = \Gamma\left[\z0 \to q {\bar q}(g) + q' {\bar q}'\right]
\end{equation}
 
\noindent
for all $q'$, such that $m_{q'} < m_q$. However, we could have a
democratic description where the final states $q \bar q + q' \bar q'$ are
assigned for $\frac{1}{2}$ to $\Gamma_{q}$ and for the other $\frac{1}{2}$
to $\Gamma_{q'}$. The two descriptions agree --- fortunately for the
leading terms.
In general, one could also decide that such final states should
not be assigned to any specific channels in such a way that
 
\begin{equation}
\sum_q\,\Gamma_{q} \neq \gh
= \sum_q\,\Gamma(\z0 \to q \bar q)
 +  \sum_{q,q'}\,\Gamma(\z0 \to q \bar q q' \bar q') + \dots
\end{equation}
 
\noindent
In particular the ${\ord}(\als^3)$ contribution to $\Gamma_V^S$ cannot be
assigned to any specific flavour. In our approach the ${\ical}^{(2,3)}$
corrections terms are included according to isospin, i.e. proportional
to $I^{(3)}_q$, and therefore cancel in the hadronic width while summing
over the first four flavours $u,d,c,s$.
However, it must be pointed out that
no general consensus has been reached on this particular point, so that the
prediction for $\gu,\gd,\gc$ may differ from one code to the other,
although the total hadronic width will remain the same.
 
\subsection{Electroweak corrections: Basic notions }
 
Above, we introduced a universal language describing the general
features of radiative corrections, the language of {\it effective} couplings.
At a secondary level, each computational scheme, being a particular
{\it representation} of the same general concept, may use different building
blocks, which, however,
are so deeply related that one could even attempt the
construction of some sort of `Rosetta stone'. Since one of our main
motivations was to build up a theoretical framework that includes some
estimate of its own uncertainty, we must necessarily spend some time in
discussing some of the specific building blocks.
 
Indeed, even though there is a high degree of universality in
the realization of the theoretical uncertainty, in practice the way in
which this realization becomes effective is strictly related to the actual
implementation of higher-order radiative corrections into the various codes.
In order to discuss the numerous effects involved in the calculations
we have to introduce additional notions.
 
\subsubsection{Comparison of notations of different codes}
 
As explained earlier, all electroweak loop effects in $Z$ boson
decays are concealed in the effective couplings $g_{_V}^f$ and $g_{_A}^f$.
The treatment of these couplings, unlike that of the radiation factors
$R_{_{V,A}}^f$, differs from one code to another
(having several different realizations).
In order to concentrate on these realizations let us for a
moment forget about radiative factors $R_{_{V,A}}^f$. Then
 
\begin{equation}
\Gamma_{f}= 4N_c^f\Gamma_0 [(g_{_V}^f)^2 +(g_{_A}^f)^2] \; .
\label{a}
\end{equation}
 
\noindent
The five codes deal with three main realizations.
 
In {\tt BHM/WOH,ZFITTER} the following notations are introduced:
 
\begin{equation}
\rho_f =\frac{1}{1-\delta\rho_f} = 4(g_{_A}^f)^2 \; ,
\label{b}
\end{equation}
 
\begin{equation}
\frac{g_{_V}^f}{g_{_A}^f} =1-4|Q_f|s_{_W}^2\kappa_f \; ,
\label{c}
\end{equation}
 
\noindent
where
 
\begin{equation}
\kappa_f = 1+\delta\kappa_f \; ,
\label{d}
\end{equation}
 
\noindent
so that Eq. (\ref{a}) may be rewritten in the form
 
\begin{equation}
\Gamma_{f} =\Gamma_0 N_c^f\rho_f[4(I^3_f -2Q_f s_{_W}^2\kappa_f)^2 +1]\;.
\label{e}
\end{equation}
 
\noindent
Here the electroweak corrections affect $\delta\rho_f$, $\delta\kappa_f$ and
$s_{_W}$. They will be discussed in the next subsection.
 
In {\tt TOPAZ0},
 
\begin{equation}
\Gamma_{f} = 4\,\Gamma_0\,N_c^f\rhoz\,
\left[\left(I^{(3)}_f - 2\,Q_f\,\shat^2 + \dgvf\right)^2 +
\left(I^{(3)}_f + \dgaf\right)^2\right]\;,
\label{g}
\end{equation}
 
\noindent
where $\shat^2$ is defined in~(\ref{topazsin}), $\rhoz$ in (\ref{rhodef}) and
$\dgvf,\dgaf$ --- see~(\ref{defcorr}) ---
contain (part of) the bosonic corrections
as well as the vertex corrections.
 
In {\tt LEPTOP},
 
\begin{equation}
g_{_A}^f = I_f^{(3)}\left[1+\frac{3}{32\pi s^2c^2}\baral V_A^f\right] \; ,
\label{h}
\end{equation}
 
\begin{equation}
\frac{g_{_V}^f}{g_{_A}^f} = 1-4|Q_f|s^2 +\frac{3|Q_f|}{4\pi(c^2 -s^2)}
\baral V_R^f \; ,
\label{i}
\end{equation}
 
\noindent
and $V_{A,R}$ are simple functions of $(m_t/M_{_Z})^2$ and $(M_{_H}/M_{_Z})^2$,
as described in section 4.1.
 
\subsubsection{Basic notions of different codes}
 
Here we will concentrate on a specific {\it realization} of the language of
effective couplings in different codes.
In particular it is important to clarify the notion of {\it leading},
{\it non-leading} and eventually of {\it remainder} terms
in {\tt BHM/WOH}, {\tt TOPAZ0} and {\tt ZFITTER}.
 
We start our presentation by considering two well known objects, $\Delta r$
(which is used by {\tt BHM, TOPAZ0, ZFITTER}) and the partial $Z$ width.
There are many notions which are common to both these objects.
 
The quantity $\Delta r$ is nothing but the effective coupling of the
$\mu$-decay, when it is being implemented with higher order corrections, see
the couple of Eqs.~(\ref{gmu}) and~(\ref{rhoc}). In this case there is only
one effective coupling, as the $\mu$ decay is a purely weak process mediated
by a charged current.
For the partial $Z$ width, the situation is more involved, since we have a
decay which is mediated by the neutral current.
That is why it is described by
two effective electroweak couplings. The introduction of $\Delta r$ will be
very useful in clarifying the notion of leading and of remainder contributions
to radiative corrections.
However, from a general point of view this is only one
of the many specific realizations of the effective coupling in the       $\mu$
decay.
 
As an example of the leading--remainder splitting we consider $\rho_c$
and $\Delta r$ (\ref{rhoc},~\ref{41}), $\rho_f$ and $\delta\rho_f$
(\ref{b}), and $\kappa_f$ and $\delta\kappa_f$ (\ref{c},
\ref{d}) --- the electroweak corrections to $\Delta r$ and to $\Gamma_{f}$.
Next we subdivide $\rho_{c}$ as introduced
in Eq.~(\ref{rhoc}) and $\rho_f$ and
$\kappa_f$ of~(\ref{b},~\ref{c}) into a {\it leading} term,
$\Delta_{_L}$, and {\it remainder}, $\Delta_{\mathrm{rem}}$,
terms as follows:
 
\begin{eqnarray}
\rho_{c} &=&
\frac{1}{1-\Delta r} = \frac{1}{1- \Delta r_{_L}- \Delta r_{\mathrm{rem}}}
= \frac{1} {\left(1 - \Delta \alpha \right)
\left( 1 +{\displaystyle \frac {c^2_{_W}}{s^2_{_W}}} \Delta \rho_{_X} \right)
- \Delta r_{\mathrm{rem}} }\;,
\label{romus}
\\
\rho_f &=& \frac{1}{1-\delta \rho_f} = \frac{1} {
 1 - \Delta \rho  - \Delta \rho_{f,\mathrm{rem}} }\;,
\label{rofs}
\nll \\
\kappa_f &=& 1 + \delta \kappa_f =
\left(1 + \Delta \kappa_{f,\mathrm{rem}} \right)
\left(1 + \frac {c^{2}_{_W}}{s^{2}_{_W}} \Delta \rho_{_X} \right).
\label{kappafs}
\end{eqnarray}
 
\noindent
From Eqs.~(\ref{romus}--\ref{kappafs}) one may notice that each coupling
contains some {\it universal}, flavour-independent piece
($\Delta \alpha$, $\Delta r_{_L}$, $\Delta \rho$, $\Delta \rho_{_X}$),
and some flavour-dependent remainder.
The former comprise the {\it leading}
terms ($\Delta \alpha$, $\Delta \rho$)
they are {\it re-summed} to all orders
in accordance with the renormalization group equation for $\Delta \alpha$ and
with the proper inclusion of leading irreducible terms for
$\Delta \rho$~\cite{resum}.
The latter are normally small, since all potentially large contributions
have already been subtracted and shifted to the leading terms.
For this reason, they are placed freely either into the
denominators,~(\ref{romus},~(\ref{rofs}),
or as a factor of the leading contribution,~(\ref{kappafs});
in doing so we follow Ref.~\cite{bhr}.
In one case however, namely Eq.~(\ref{romus}),
there exist arguments in favour of
keeping the remainder in the denominator.
In Ref.~\cite{sirlin84} it was proved
that the right-hand side of Eq.~(\ref{romus})
actually means that, on the two-loop level at least, all fermionic mass
singularities are located exclusively in $\Delta \alpha$. This may be
argued by a partial expansion of a simplified case ($\Delta \rho=0$)
up to the two-loop order terms,
 
\begin{equation}
\rho_{c} \approx
 \frac{1}{1- \Delta \alpha - \Delta r_{\mathrm{rem}}}
\approx \frac{1} {\left(1 - \Delta \alpha \right)}
\left( 1 + \frac{\Delta r_{\mathrm{rem}}}{1- \Delta \alpha}\right),
\label{massing}
\end{equation}
 
\noindent
from which one can see that the scale of $\Delta r_{\mathrm{rem}}$ is
actually $\alpha(M_{_Z})$.
 
One should emphasize that since there is no $\Delta \alpha$
in Eqs.~(\ref{rofs}),~\ref{kappafs})
and since the non-leading terms of the order $\Delta_{_L}
\Delta_{\mathrm{rem}}$ are unknown, no arguments in favour of putting
$\Delta_{\mathrm{rem}}$ as it is done in~(\ref{rofs},~\ref{kappafs}) can be
presented. The size of these uncontrolled terms should be treated as an
intrinsic theoretical error.
 
All the remainder terms have a similar structure:
 
\begin{eqnarray}
\Delta r_{\mathrm{rem}} &=&
\Delta r^{\mathrm{1loop},\alpha}
+ \Delta r^{\mathrm{2loop},\alpha \alpha_s}
+ \frac{c^2_{_W}}{s^2_{_W}} \Delta {\bar \rho}_{_X}
- \Delta \alpha\;,
\label{drrem}
\\
\Delta \rho_{f,\mathrm{rem}} &=&
\Delta \rho_f^{\mathrm{1loop},\alpha}
+ \Delta \rho^{\mathrm{2loop},\alpha \alpha_s}
- \Delta {\bar \rho}\;,
\label{rhorem}
\\
\Delta \kappa_{f,\mathrm{rem}} &=&
\Delta \kappa_f^{\mathrm{1loop},\alpha}
+ \Delta \kappa^{\mathrm{2loop},\alpha \alpha_s}
- \frac {c^2_{_W}} {s^2_{_W}} \Delta {\bar \rho_{_X}}\;.
\label{kaprem}
\end{eqnarray}
 
\noindent
They contain all the terms known at present: the complete one-loop
${\cal O}(\alpha )$ corrections
(two-, three-, four-point functions) and complete
two-loop ${\cal O} (\alpha\alpha_s)$ insertions to two-point functions,
from which the leading ${\cal O}(\alpha )$ and ${\cal O}(\alpha\alpha_s)$
terms are subtracted:
 
\begin{eqnarray}
\Delta {\bar \rho}_{_{(X)}} &=&
\Delta {\bar \rho}^{\alpha} +
\Delta {\bar \rho}^{\alpha \alpha_s} + ({\bar X})
 \nonumber \\
 &=&  \frac{3 \alpha}{16 \pi s^2_{_W} c^2_{_W}} \;
\frac{m_t^2}{\zm^2} \left[ 1 - \frac{2}{3} \left(1 + \frac{\pi^2}{3}\right)
\frac {\alpha_s(m^2_t)}{\pi}\right] + ({\bar X})\;,
\label{drobar}
\end{eqnarray}
 
\noindent
where braces in~(\ref{drobar}) and below mean that this expression
describes simultaneously both quantities $\Delta {\bar \rho}$ and
$\Delta {\bar \rho}_{_{X}}$,
which appeared in Eqs.~(\ref{drrem})-(\ref{kaprem}).
 
The term ${\bar X}$ in~(\ref{drobar}) is a next-to-leading order term, whose
proper treatment is rather important:
 
\begin{equation}
{\bar X} = {\cal R}e
\left[ \frac{ \Pi_Z(\zm^2)}{\zm^2} - \frac{ \Pi_W(\wm^2)}{\wm^2}
      \right]^{\mathrm{1loop}}_{\overline {\mathrm{MS}}}
      - \Delta \rho^{\alpha}.
\label{xbar}
\end{equation}
 
\noindent
In Ref.~\cite{fansir} it was argued that in the on-mass-shell
renormalization
scheme this term should be re-summed together with $\Delta {\bar {\rho}}^{\alpha
}$, since in the ${\overline{\mathrm{MS}}}$ scheme it appears to be
automatically incorporated. Similar arguments in favour of such a resummation
were presented in Ref.~\cite{rwoh}.
In ${\bar X}$ the UV divergences are removed according to the
${\overline{\mathrm{MS}}}$ renormalization scheme with $\mu = \zm$.
The separation of ${\bar X}$ is not unique; it makes the resummation
dependent on the renormalization procedure.
 
The leading contribution $\Delta \rho$ is built out of the same terms
as~(\ref{drobar}). But they are normalized by $G_{\mu}$ rather than by
$\alpha/s^2_{_W}\wm^2$, as is required by the resummation
proposed in Ref.~\cite{resum}:
 
\begin{eqnarray}
\Delta \rho_{_{(X)}} &=& \Delta \rho^{\alpha}
+ \Delta \rho^{\alpha^2} + \Delta \rho^{\alpha \alpha_s}
+ \Delta \rho^{\alpha \alpha^2_s}
+ (X)    \nonumber \\
 &=& N_c x_t \left[ 1 + x_t \Delta \rho^{(2)} \left(\frac{m^2_t}{M^2_{_H}}
\right)
+ c_1 \frac{\alpha_s(m^2_t)}{\pi}
+ c_2 \left(\frac{\alpha_s(m^2_t)}{\pi}\right)^2 \right] + (X)\;,
\label{rholead}
\end{eqnarray}
 
\noindent
where
 
\begin{eqnarray}
x_t &=& \frac{G_{\mu}}{\srt} \frac{m_t^2}{8 \pi^2}\;,
\label{calt} \\
X &=& 2 s^2_{_W} c^2_{_W} \frac{G_{\mu} \zm^2}
{\srt \pi \alpha} {\bar X}\;,
\label{defx}
\end{eqnarray}
 
\noindent
and $x_t$ is the Veltman heavy top factor~\cite{velt}. The coefficients $c_1$
and $c_2$ describe
the first- and second-order QCD corrections for the leading
$x_t$ contribution to $\Delta \rho$, calculated
in Refs.~\cite{aasse} and~\cite{afmt}. Correspondingly:
 
\begin{eqnarray}
c_1 &=& - \frac{2}{3} \left(1+ \frac{\pi^2}{3}\right)\;,
\label{c1qcd} \\
c_2 &=& - \pi^2 \left( 2.155165 - 0.180981\;n_f\right)\;.
\label{c2qcd}
\end{eqnarray}
 
\noindent
The function $\Delta \rho^{(2)}(m^2_t/M^2_{_H})$ describes the leading
second order $x_t$ (two-loop electroweak) correction to $\Delta \rho$,
calculated first in the $M_{_H}=0$ approximation in Ref.~\cite{vanhoog}
and later in Ref.~\cite{barb} for an arbitrary relation between $M_{_H}$
and $m_t$.
 
The partial decay width $Z \to b {\bar b}$ contains an additional $m_t$
dependence due to vertex diagrams (see Fig.~10).
 
\begin{figure}[htbp]
\begin{Feynman}{120,30}{0,0}{1.0}
\put(5,15){\photonrighthalf}
\put(0,15){$Z$}
\put(20,15){\fermiondr}
\put(25,4){$t$}
\put(35,0){\fermionrighthalf}
\put(51,-1){$b$}
\put(35,0){\photonup}
\put(37,15){$W$}
\put(35,30){\fermionlefthalf}
\put(51,29){$\bar{b}$}
\put(20,15){\fermiondl}
\put(25,26){$\bar{t}$}
\put(70,15){\photonrighthalf}
\put(65,15){$Z$}
\put(85,15){\markphotondrh}
\put(88,2){$W^-$}
\put(101,-1){\fermionrighthalf}
\put(117,-2){$b$}
\put(101,29){\line(0,1){2}}
\put(101,-1){\fermiondown}
\put(102,15){$t$}
\put(101,31){\fermionlefthalf}
\put(117,30){$\bar{b}$}
\put(85,15){\markphotonurh}
\put(88,28){$W^+$}
\end{Feynman}
\end{figure}
\begin{center}
Figure~10: {Top quark exchange diagrams which contribute to $\Gamma_b$
(unitary gauge)}
\end{center}
 
As a consequence, the effective couplings $\rho_b$ and $\kappa_b$
contain additional leading terms of the order ${\cal O}(G_{\mu}m^2_t)$.
The complete one-loop approximation for the $Z \to b \bar b$ partial width
was calculated in Ref.~\cite{zbb}.
We first redefine remainder terms by an additional subtraction of the leading
one-loop term originating from these diagrams:
 
\begin{eqnarray}
\Delta \rho_{b,\mathrm{rem}} & \to & \Delta \rho_{b,\mathrm{rem}}
-2\Delta {\bar {\rho}}_b\;, \nll
\Delta \kappa_{b,\mathrm{rem}} & \to & \Delta \kappa_{b,\mathrm{rem}}
+\Delta {\bar {\rho}}_b\;,
\label{rembb}
\end{eqnarray}
where
\begin{equation}
\Delta {\bar {\rho}}_b = \frac{\alpha}{8\pi s^2_{_W}} \;
\frac{\tm^2}{\wm^2}\;.
\label{drhob}
\end{equation}
 
\noindent
Following papers~\cite{barb}
and~\cite{ftjr}, the two-loop order QCD and
electroweak leading terms in the $Zb{\bar b}$ vertex are implemented
by an additional re-definition of effective couplings $\rho_b$ and $\kappa_b$:
 
\begin{eqnarray}
\rho_b & \to & \rho_b(1+\tau_b)^2 \; , \\
\kappa_b & \to & \frac{\kappa_b}{1+\tau_b} \; ,
\end{eqnarray}
where $\tau_b$ is given by
 
\begin{equation}
\tau_b= - 2x_t\Biggl[ 1-\frac{\pi}{3}\als(\tm)
+x_t\tau^{(2)}\left(\frac{\tm^2}{\hm^2}\right)\Biggr] \; .
\label{taubb}
\end{equation}
 
\noindent
A compact analytic representation for the two-loop functions $\rho^{(2)}$ and
$\tau^{(2)}$ was also given in Ref~\cite{ftj}.
 
We have just discussed what is known in the literature about the treatment
of the corrections of the order ${\cal O}(G_{\mu}m^2_t)$. The
$Zb{\bar b}$ vertex also contains a logarithmically enhanced term,
${\cal O}[\alpha\ln(\tm^2/\wm^2)]$, whose contribution is comparable to
the leading one. Recently, QCD corrections were also calculated for this term,
(see contribution by Kwiatkowski and Steinhauser in the QCD Part of this
Report). This correction, however, can be nearly completely absorbed
into the final-state QCD corrections. What remains is approximately one
hundred times less than the QCD correction for the leading term
as given in~(\ref{taubb}).
 
What has been presented so far
in this subsection, is to an extent inherent
in all five codes. However, in order to reach a better understanding of our
set of theoretical predictions and of the strategy adopted to extract their
related {\it intrinsic} errors we have to devote more time to discussing
other realizations of the effective coupling language.
In {\tt TOPAZ0} --- see, Eq.(\ref{g}) --- the partial $Z$-widths reads:
 
\begin{equation}
\Gamma_f = 4\,\Gamma_0\,N_c^f\rhoz\,
\left[\left(I^{(3)}_f - 2\,Q_f\,\shat^2 + \dgvf\right)^2 +
\left(I^{(3)}_f + \dgaf\right)^2\right]\;,
\label{zfftopaz}
\end{equation}
 
\noindent
where $\shat^2$, defined by~(\ref{topazsin})
and~(\ref{cdshat}) (see also Ref.~\cite{gsch3}), and
is to be put in partial correspondence with the $s_{_W}^2\kappa_f$ of
Eq.~(\ref{c}).
Moreover, $\dgvf,\dgaf$ contain bosonic corrections as well
as vertex corrections. Again, the basic point under examination is the
leading--remainder splitting.
The main idea beyond this realization is to write a system of equations
that connect the $\msb$ parameters of the theory in terms of $\alpha, \gf$ and
$\zm$. These equations contain the effects of radiative corrections up to a
certain order in perturbation theory and must be solved consistently and
respecting gauge invariance. As a result of this procedure we end up in
Eq.~(\ref{zfftopaz}) with a $\rhoz$ defined according to Eq.~(\ref{rhodef}).
This parameter, which properly re-sums the gauge invariant fermionic
corrections, contains all isospin breaking terms and includes, beyond ${\ord}
(\alpha)$, all the presently known higher-order terms --- the ${\ord}
(\alpha^2)~\cite{barb},~\cite{ftj}, {\ord}(\alpha\als)$~\cite{aasse} and
${\ord} (\alpha\als ^2)$~\cite{afmt} corrections.
 
The bare weak mixing angle, according to the previous strategy, will always
be expanded around $\shat$, Eq.(~\ref{cdshat}), where $\rhozr$ from
Eq.~({\ref{nores}) and (\ref{myres}) is used.
These two quantities ($\rhozr$ and
$\rhoz$) may differ whenever a resummation of bosonic corrections is performed
for the weak mixing angle. Once the divergent terms are treated in the $\msb$
framework, $\shat^2$ (the leading term) will receive a correction (the
remainder) $\Delta s^2$.
At this point, when the $\z0$ wave function renormalization factor $\Pi_{_Z}$
is included, we usually expand all the remaining corrections up to first order
--- we decompose $\Pi_{_Z}$ into its leading, fermionic part and a
remainder $\Delta\Pi_{_Z}$ (see Eq.~(\ref{cdpiz})). This $\Delta\Pi_{_Z}$ is
always expanded. After inclusion of vertices and fermion wave-function
renormalization factors we end up with
 
\begin{eqnarray}
\dgvf &=& \frac{\alpha}{4\pi}\left[ {{2\fvf - \frac{1}{2} v_f \Delta\Pi_Z}\over
 {c^2s^2}} - 2 Q_f\Delta s^2 \right]\;, \nll
\dgaf &=& \frac{\alpha}{4\pi}\left[ {{2\faf - \frac{1}{2} \i3f \Delta\Pi_Z}\over
 {c^2s^2}} \right]\;,
\label{defcorr}
\end{eqnarray}
 
\noindent
where $s^2$ is again defined as~(\ref{sleptop}) and $F^f_{_{V,A}}$ are the
flavour-dependent vertex corrections.
Additional versions of this realization will be discussed in the
framework of the theoretical options. Here we need only mention some of
the problems connected to the resummation of bosonic contributions
in this realization (as well as in others). It is strictly related to the
question of gauge invariance. As it is well known, the vertices and the bosonic
parts of the vector boson self-energies are not separately gauge invariant.
On the one hand there is no rigorous procedure for re-summing the vertices,
and on the other hand any attempt to isolate the gauge variant parts and
to throw them away is not unambiguously defined.
Even when we identify the universal contributions from vertices and boxes
(for instance by working in the $R_{\xi}$ gauge), to be combined with the
vector boson self-energies in order to get the one-loop gauge invariant
dressed propagators, we still have the freedom to shift from one part
(to be re-summed) to the other some process-independent, $\xi$-independent,
UV finite function of $q^2$ with the proper asymptotic behavior.
Strict enforcement of gauge invariance is one of the roots of
Eq.~(\ref{defcorr}), in that $\Delta\Pi_{_Z}$ is not absorbed
into the $\rhoz$ parameter and that particular care must be devoted to the
proper definition of the remainder, $\Delta s^2$, when resummation of
irreducible terms is performed. This fact accounts for one of the many
structural differences with other realizations.
 
In {\tt LEPTOP}, due to the proper choice of the $\baral$-Born approximation,
the electroweak radiative corrections turned out to be small ---
for $m_t$ around $175$ GeV.
The smallness of corrections is a result of the mutual cancellation
of different equally important terms. Therefore, functions $V_{A,R}^f$
representing loop corrections are not subdivided into leading and remainder
terms and no resummation is performed.
 
\clearpage
 
\section{Options, theoretical uncertainties}
 
During the completion of this work it has become increasingly evident
that there is a need
to quantify the effect of our partial lack of knowledge of
the missing higher-order terms in radiative corrections. Thus we have
introduced the notion of {\it option}, which refers to a set of possible and
plausible alternative implementations of the full machinery of radiative
corrections within a given renormalization scheme.
In order to explain the treatment of higher-order terms and the interplay
between pure electroweak and QCD corrections we must once more remember
that the effective coupling description of the $\z0$ width has several
different realizations.
 
Quite independent of specific details, all the realizations single out
two main components in each observable:
 
\begin{equation}
 O = O_B + \Delta O \;.
\label{obulk}
\end{equation}
 
\noindent
The term $O_B$, giving the bulk of the answer, is often called the Born
approximation (the $\baral$-Born approximation in {\tt LEPTOP}, or
improved Born approximation in {\tt BHM/WOH}, {\tt TOPAZ0} and {\tt ZFITTER}),
or the leading contribution to $O$. The term $\Delta O$ represents a small
perturbative correction, often called remainder or non-leading contribution.
Different realizations usually have different ways of performing this
splitting so that, while they agree at the ${\ord}(\alpha)$, there are
differences which start at ${\ord}(\alpha^2)$.
 
Independent of the particular realization of $\gv,\ga$, these effective
couplings are complex valued functions, due to the imaginary parts of the
diagrams. This, however, will have some relevance only for realistic
distributions while for pseudo-observables they are taken to be strictly
real.
Given that the universal language has in practice many alternative
realizations and that the actual implementation of the options is
very much code-dependent, we have not created a common set-up for the
electroweak options. From this simple fact follows the need to discuss
at length the physical motivations behind our options. Although we have
tried, as much as possible, to illustrate them from a general perspective,
it is also evident that some space has been left to analyze certain
specific issues.
A description of the implementation of the options in the various codes,
with the relative flags, will be given in section~\ref{pecul}.
 
\subsection{Factorization of QCD Corrections}
 
No matter which realization we are using, one problem naturally emerges
when final-state QCD corrections are switched on. This problem is connected
with the folding of the non-corrected widths using the QCD factors
$R^f_{_{V,A}}$.
One can take the point of view that non-universal and flavour-dependent
couplings should also be folded or, to the contrary,
that only the universal
effective couplings should multiply the QCD radiation terms. This option
merely reflects our ignorance of the mixed ${\ord}(\alpha\als)$ corrections,
with the noticeable exception of the $b \bar b$ partial width where the leading
$m_t$ part of these corrections is actually known~\cite{ftjr}.
If we consider the $b \bar b$ partial width we can write
 
\begin{eqnarray}
\Gamma^b_0 &=& \frac{\gf\zm^3}{2\srt\,\pi}\,
\left[ (\gsvb)^2 + (\gsab)^2 \right] \;, \nll
\Gamma_{_{EW}}^b &=& \Gamma_{_{EW}}
(\z0 \to \barb b) \approx \Gamma^b_0\,\left[ 1 - 4\,x_t\,\frac
{\left(\gsvb+\gsab\right)\gsab}{(\gsvb)^2+(\gsab)^2}\right] \; .
\end{eqnarray}
 
\noindent
At this point naive factorization would imply
 
\begin{equation}
\Gamma_b = \Gamma_{_{EW}}^b \,\left( 1 + \frac{\als}{\pi} \right) \;,
\end{equation}
 
\noindent
while the computed FTJR term~\cite{ftjr} gives
 
\begin{equation}
\Gamma_b = \Gamma^b_0\,\left[1 + \frac{\als}{\pi}
- 4\,x_t\,\frac{\left(\gsvb+\gsab\right)\gsab}{(\gsvb)^2+(\gsab)^2}\left(
1 + \frac{3 - \pi^2}{3} \; \frac{\als}{\pi}\right)\right] \;,
\end{equation}
 
\noindent
so that the correct QCD coefficient in front of the heavy top factor turns
out to be $-2.290\;\als/\pi$ instead of $\als/\pi$.
Of course, an approximate
but factorized expression is still possible and can be written as
 
\begin{equation}
\Gamma_b = \Gamma^b_0\,\left[ 1 - 4\,x_t\,\frac{\left(
\gsvb+\gsab\right)\gsab}{(\gsvb)^2+(\gsab)^2}\,\left(1 - \frac{\pi^2}{3}
\;\frac{\als}{\pi}\right)\right]\,
\left( 1 + \frac{\als}{\pi}\right) \;,
\end{equation}
 
\noindent
where the uncertainty has now moved to order $\als^2\gf m_t^2$.
Actually, we
have not yet specified the scale of $\als$ in the previous equations but these
corrections have been implemented such that the universal QCD factor is
computed as $1 + \als(\zm)/\pi$, while the specific FTJR term
is computed with $\als(m_t)$.
As a consequence the $1 + \als(\zm)/\pi$ factor is not included for the
$b-$quark asymmetries and for the effective $b-$quark mixing angle.
In general, however, the ${\ord}(\alpha\als)$ corrections for the
$\z0 f \bar f$
vertex is presently not known and we must accept an intrinsic uncertainty
associated with the two procedures, i.e. factorization or non-factorization of
electroweak and QCD corrections. Once again, if $\Gamma^q_0$
is the (improved) Born $q \bar q$ partial width and
 
\begin{equation}
\Gamma^q_{_{\rm {EW}}} =
\Gamma^q_0\,\left(1 + \delta^{\rm {univ}}_{_{\rm {EW}}}\right)
+ \Delta^q_{_{\rm {EW}}}
\end{equation}
 
\noindent
then
 
\begin{equation}
{{\Delta^q_{_{\rm {EW}}}}\over {\Gamma^q_0}}\,{{\als(\zm)}\over {\pi}}
\end{equation}
 
\noindent
is roughly assumed as the corresponding uncertainty. This type of uncertainty
can be illustrated very well by asking how correct it is to shrink
an electroweak blob to a point before allowing for QCD radiation.
 
\subsection{Genuinely Weak Uncertainties}
 
In this section we briefly discuss the main ingredients which enter the
pure weak corrections to the pseudo-observables --- the resummation
of the one-particle irreducible vector boson self-energies, the scale in
vertex corrections and the linearization of the corresponding S-matrix
elements.
We have avoided in this chapter any intensive usage of the lengthy
formulae introduced in the first part of this Report and summarized in
some detail in section~\ref{pecul}.
Indeed, there is a simple set of problems in the implementation of
radiative corrections which is inherent to perturbation theory and does
not depend on any specific approach.
Only formulae belonging to different realizations
represent the technical transcription of these implementations.
Real progress is usually achieved whenever a new term becomes available,
otherwise we are left with heuristic arguments or with ingenious attempts to
improve upon perturbation theory. Different paths along this road represent
our present degree of inaccuracy and to understand these differences already
gives a hint on how to proceed. The structural and logical essence of these
differences can be explained without a massive use of equations.
 
\subsubsection{Leading--Remainder splitting}
 
Before we move to discuss the next source of uncertainty we must recall again
that, generally speaking, the effective couplings $\gvf,\gaf$
contain a leading and usually re-summed part and a non-leading (remainder)
one, quite independent of the specific realization. For instance,
in one of the realizations the building blocks are $\rho_f$ and $\kappa_f$,
while in another they are $\rhoz$,  $\shat^2$ and $\dgvf, \dgaf$, and in a
third they are $s^2,V_A$ and $V_R$.
The way in which the non-leading terms can be treated and the exact form of
the {\it leading--remainder} splitting give rise to several possible options
in the actual implementation of radiative corrections that in turn become
another source of theoretical uncertainty.
For instance, for all the objects $\dr$, $\Delta\rho_f$ and $\Delta\kappa_f$,
we can introduce the decomposition into leading and remainder. Since we
know how to proceed with all objects in the leading approximation the only
ambiguity is due to the treatment of the remainders. Clearly, after
the splitting $\dr = \drl + \dr_{\mathrm{rem}}$ there are in principle
several possible ways of handling the remainder{\footnote{In case of
$\dr$, the last two expansions are not valid indeed, see a discussion around
Eq.~(\ref{massing}). We present them here nevertheless for the sake of
illustration.}}:
 
\[\frac{1}{1 - \dr} = \frac{1}{1- \drl- \Delta r_{\mathrm{rem}}}
= \left\{
\begin{array}{ll}
{}\\
{1\over {1 - \drl - \dr_{\mathrm{rem}}}} \\
{}\\
{1\over {1 - \drl}}\,\left( 1 + {{\dr_{\mathrm{rem}}}
\over {1 - \drl}}\right) \\
{}\\
{{1 + \dr_{\mathrm{rem}}}\over {1 - \drl}} \\
{}\\
{{1} \over {1 - \drl}}+ \dr_{\mathrm{rem}} \; .\\
{}
\end{array}
\right. \]
\begin{equation}
\label{ifacr}
\end{equation}
 
\noindent
Actually, these options differ among themselves but the difference can be
related to the choice of the scale in the remainder term.
A complete evaluation of the sub-leading ${\ord}(\gf^2\zm^2 m_t^2)$
would greatly reduce the associated uncertainty. At the moment these
sub-leading terms are simulated just by varying the scale in the remainder.
Typical choices are described in section~\ref{pecul} and may vary from
realization to realization. A different approach to the problem~\cite{leptop}
will be illustrated in section~\ref{missord}.
 
\subsubsection{Scale in vertex corrections}
 
Another possible option, which is realized by some of the codes (but not all),
has to do with the scale of $\alpha$ in the non-leading corrections, in
particular the vertex corrections. To make this point clear we use
one of the {\it realizations} of the effective couplings and analyse
the $\z0 \to \barb b$ decay in some details. The essential ingredient will
be $\Delta\rho_b$ which we write as
 
\begin{eqnarray}
\Delta\rho_b &=& \Delta\rho_d + \delta\rho_b(m_t)  \nll
\delta\rho_b(m_t) &=& -\,\frac{\gf m_t^2}{2\srt\pi^2}\,\left[
1 + {\ord}(\gf m_t^2) + {\ord}(\als) \right] +
{\ord}\left(\alpha\ln\,m_t^2\right) + \delta\rho_b^{^{NL}} \; ,
\end{eqnarray}
 
\noindent
where $\delta\rho_b$ is a correction specific to the $b \bar b$ channel.
The question naturally arises as to what to use for the scale in
$\delta^{^{NL}}$
and in the sub-leading logarithmic term --- $\gf$, $\alpha(\zm)$ or
$\alpha(0)$. Obviously the same kind of option will be present in the
vertex corrections for light fermions where the expansion parameter is formally
$\alpha/(4\pi\sin^2\theta\cos^2\theta)$. Different ways to interpret $\alpha$
in these expressions give rise to different results, from $\alpha(0)/
(4\pi\sin^2\theta\cos^2\theta)$ to $\gf\zm^2/(2\sqrt{2}\pi^2)$.
The difference between possible identifications of coupling constants in the
${\ord}(\alpha)$ corrections represents, of course, effects of ${\ord}(\alpha^2
)$. The fact that the neutral current amplitude is automatically expressed in
terms of $\gf\zm^2$ is a possible heuristic argument to adopt the same strategy
in the evaluation of the presently known ${\ord}(\alpha)$ corrections,
but in order to be on the safe side, the differences should be considered as
a theoretical uncertainty, at least according to many authors.
For instance, we know from a specific calculation~\cite{kat} that for inclusive
quantities the final state QED radiation is controlled by $\alpha(\zm)$ and
not by $\alpha(0)$.
The option of variation of the scale of non-leading corrections has been
implemented by the majority of the codes.
 
\subsubsection{Linearization}
 
Another example to be discussed is the following. In almost any
realization we have the possibility of using an {\it expanded} versus a
{\it non-expanded} option --- to linearize our expressions. To give a partial
illustration of this possibility we use the realization of Ref.~\cite{topaz0}
where we have
 
\[\stes^l = \frac{1}{4}\left(1 - {\gvl\over \gal}\right) = \left\{
\begin{array}{ll}
\shat^2 + \frac{1}{2}\,\dgvl + \frac{1}{2}\,\left(4\,\shat^2 - 1\right)\dgal
& \mbox{expanded} \\
{} &\mbox{} \\
{{\shat^2 + \frac{1}{2}\,\left(\dgvl - \dgal\right)}\over {1 - 2\dgal}}
& \mbox{non-expanded} \;.  \\
{} &\mbox{}
\end{array}
\right. \]
 
\noindent
On the same footing, the $\dgvf,\dgaf$ terms in each partial width will
be expanded up to first order in the {\it expanded} option.
More generally, the difference between the two options in the evaluation
of $O^2$, where $O$ is given by Eq.~(\ref{obulk}), is equal to
$(\Delta O)^2$, a two-loop reducible but non-leading contribution,
and from this point of view we clearly do not need a specific exemplification.
By comparing the two options we obtain a rough estimate of the importance of
the missing non-leading two-loop effects. It goes almost without saying that
when this option is implemented on top of the leading--remainder
splitting and of the choice $\gf $ versus $\alpha(0)$ we can have rather
different behaviours in the global theoretical error according to the size of
the remainder.  Thus whenever we need to quantify the effect then the
remainder must be considered in detail and the effect becomes
realization-dependent.
This has some relevance for the de-convoluted leptonic forward--backward
asymmetry, where the leading term is small by itself due to accidental
cancellations. To summarize we can say that the theoretical error on $\afb^l$
is very sensitive to the sum of various factors:
 
\begin{itemize}
 
\item {\it expansion} versus {\it non-expansion}
 
\item definition of the {\it leading} part of $\stes^l$
 
\item scale of $\alpha$ for the non-leading terms.
 
\end{itemize}
 
A further exemplification of what we call option for radiative corrections,
in part connected to a selection of the scale in the coupling but generalizable
to all parameters appearing in the remainder, can be illustrated as follows.
Suppose that a given quantity $A$, function of the parameter $a$, is given in
perturbation theory by the following expansion:
 
\begin{eqnarray}
A &=& a + g\left[a^2 + f_1(a)\right] + g^2\left[ a^3 + f_2(a)\right] +
{\ord}(g^3) \nll
{} &=& {\bar a} + g f_1(a) + {\ord}(g^2) \nll
{\bar a} &=& a/(1 - g a) \; ,
\end{eqnarray}
 
\noindent
and that only the $f_1$ term is actually known. It could be decided that
${\bar a}$ is the effective expansion parameter (or that in the full expression
we change variable $a \to {\bar a}$), which of course in the truncated
expansion introduces the option,
 
\begin{eqnarray}
A &=& {\bar a} + g f_1(a) \nll
{}&=& {\bar a} + g f_1({\bar a}) \; ,
\end{eqnarray}
 
\noindent
so that $\Delta A = g^2 a^2 f'_1(a)$ would be our estimate of the associated
theoretical uncertainty. Sometimes it is only a rough estimate, since there is
no guarantee that the irreducible terms $g^2 f_2$ are essentially small in size.
 
\subsubsection{Resummation}
 
Another source of theoretical uncertainty is connected with the treatment
of the physical Higgs contribution. As we know for large Higgs masses
($\hm \gg \zm$) there is a correction term in $\Delta\rho$ which is only
logarithmic, in contrast to the heavy fermion case, the so-called Veltman's
screening effect~\cite{velth}. With respect to this correction
$\Delta\rho_{_H}$ we can:
 
\begin{itemize}
 
\item expand it to first order in $\alpha$ as is sometimes done for
all bosonic corrections;
 
\item re-sum the leading part of it, $\Delta\rho_{_H}^{_L}$,
for relatively large values of $\hm$, e.g. $\hm > \wm\exp{5/12}$;
 
\item re-sum in $\rho$ the whole physical Higgs contribution. This requires
a further comment, since this term is not UV finite by itself
and therefore the resummation procedure must be understood strictly
in the $\msb$ scheme with a scale $\mu = \zm$.
 
\end{itemize}
 
\noindent
As an additional and rather general comment, which somehow collects many of the
previous considerations, we stress once more that there are different ways of
implementing the resummation of the vector boson self-energies. These
choices, which in turn are deeply related to the proper definition of
remainder, differ from code to code --- at least in their default settings.
We have already illustrated the splitting $\drl - \dr_{\mathrm{rem}}$ and
simply add a few additional considerations. Resummation is very often the main
recipe for separating a small remainder from the bulk of the effect.
 
\begin{itemize}
 
\item One choice consists in a resummation which includes the
square of the $\z0-\gamma$ mixing-term~\cite{bhm,woh} with the option of
strictly keeping in the resummation only the one-loop irreducible terms.
The default thus corresponds to an additional mass counter-term which enters
the field and coupling renormalization constants and modifies the quantity
$\dr$. The resummation of the modified $\dr$ leads automatically to the
factorization property which takes into account the proper summation of all
the leading higher-order reducible terms.
 
\item Even more generally we can distinguish among complete expansion
of the one-loop self-energies, partial resummation of fermionic self-energies
or partial inclusion of bosonic self-energies in the resummation. There
are two considerations to be made at this point. Sometimes accidental
cancellations occur among the fermionic and the bosonic sectors, which would
suggest a similar treatment for both; however, the bosonic sector is
not gauge invariant by itself. Thus any resummation of bosonic parts must
properly identify some numerically relevant but gauge invariant sub-set.
 
\end{itemize}
 
\noindent
For completeness we recall that one of these identifications gives
in the $\xi = 1$ gauge~\cite{pinch}
 
\begin{eqnarray}
{\cal S}_{_{\gamma\gamma}}(p^2) &=&
S_{_{\gamma\gamma}}(p^2)|_{\xi=1} -4e^2p^2I(p^2) \;, \nll
{\cal S}_{_{Z \gamma}}(p^2) &=&
S_{_{Z \gamma}}(p^2)|_{\xi=1} - 2e^2{{\cos\theta}\over
{\sin\theta}}\left(2p^2 - \zm^2\right)I(p^2) \;,  \nll
{\cal S}_{_{ZZ}}(p^2) &=& S_{_{ZZ}}(p^2)|_{\xi=1} -4e^2{{\cos^2\theta}\over
 {\sin^2\theta}}
\left(p^2 - \zm^2\right)I(p^2) \;,
\end{eqnarray}
 
\noindent
where $I(p^2) = - B_0(p^2,\wm^2,\wm^2)/(16\pi^2)$, $B_0$ is a scalar
formfactor~\cite{ossch1} and ${\cal S}$ denotes a possible identification
of the gauge invariant part of $S$.
 
To spend an additional word on electroweak uncertainties and to understand the
implications of some of the procedures used as a possible estimator of the
theoretical error we consider a fictitious quantity $X$ defined as $X = 1 -
4\,\stes^l$.
Each code will define some effective Born approximation to $\stes^l$ and here
we distinguish between three basic possibilities: complete expansion of the
corrections (E), a resummation which only includes the fermionic self-energies
(FR), or a global resummation (R). We refer here to the complete set of
formulae given in section~\ref{basics} and simply quote the adopted strategies:
 
\begin{itemize}
 
\item {\tt BHM/WOH} adopted as the default of the resummation of the
entire set of self-energies, including the $\z0-\gamma$ mixing term;
 
\item {\tt LEPTOP}, which expands all those contributions not re-absorbed in
the running of $\alpha$ into ${\bar\alpha}$;
 
\item {\tt TOPAZ0}'s default, which re-sums in the $\msb$ framework the
$\Sigma_R$ term of Eq.~(\ref{myres}), while an option is left
in which only its fermionic content is re-summed;
 
\item {\tt ZFITTER}'s default, which also re-sums in the $\msb$ framework
the $X$ term of Eq.~(\ref{defx}), allowing asr options
the resummations of the leading terms only.
 
\end{itemize}
 
Let us define $X_0$ as the effective Born approximation of $X$ and define
a non-leading (remainder) part, $X = X_0 + [\alpha(\zm)/\pi]X_1 +
{\ord}(\alpha^2)$. The three different procedures, E or FR or R,
will find $X_0 = 0.07528, 0.08672, 0.07184$ (for $m_t = 175\,$GeV,
$\hm = 300\,$GeV and $\als = 0.125$). In fact, there is no unique way for
global resummation (R), so that the last number can vary a little (say, from
$0.07184$ to $0.07416$).
Using the corresponding value for $\stes^l$ we get $X_1 = -1.344, -5.976,
-0.016$. Whenever we compute $X^2$ and use the square of the remainder to
estimate part of the uncertainty (there are other options around) we are
bound to see rather different behaviors, depending on the adopted
leading--remainder splitting.
It should be clear that the remainder itself
is subject to several possible options, from the scale of the coupling to
the choice of those terms which are to be considered small and perturbative.
These options themselves contribute, sometimes sizably, to the final
uncertainty, quite independently from linearization (expansion) and in
the end every adopted procedure is somehow equivalent to a proper choice of
the scale. Thus a preferred set of options is equivalent, in some sense, to
an ideal optimization of the perturbative expansion.
 
\subsubsection{Estimate of the missing terms in higher orders \label{missord}}
 
While some of the realizations of the effective coupling description make use
of a (partial) resummation of higher-order terms, thus trying to improve upon
ordinary perturbation theory, we have other realizations where the mixing angle
is defined only in terms of $\gf, \zm$ and of the running constant
$\alpha(\zm)$ while everything else is strictly expanded.
Thus, a Born approximation is defined in terms of an $s^2$, the definition
of which was
given in Eq.~(\ref{sleptop1}) and the realization is constructed in terms of
the one-loop approximation with respect to the genuinely electroweak
interactions. As has became clear from the previous discussion, there is
no common set of implemented procedures for describing the theoretical
uncertainties but only some rather general guidelines. Certainly, when we move
to a concrete implementation, it happens that every code has its own internal
ways of estimating the missing higher-order effects. Codes which do not
foresee, for theoretical reasons, the possibility of expanding the remainders
with respect to the leading terms or of playing with the choice of the
scale in the remainders, introduce another set of options based on
a different estimate of the not-yet-calculated diagrams or terms
in a given diagram. The basic and alternative idea is that each of the Born
relations~(\ref{28}--\ref{30}) will receive a genuinely electroweak
correction proportional to $\baral V_i$, as given by Eq.~(\ref{31}).
The main point is related to the recent observation~\cite{nl} that the
sub-leading two-loop corrections to the vector boson self-energies, of order
${\ord}(\gf^2 m^2_t \zm^2 $),
could be numerically close to the leading ones, of
${\ord}(\gf^2 m_t^4)$. Thus the latter can be considered as an estimate
of the uncertainties in the $V_i$. At the same level the
${\ord}(\alpha\als^2 m_t^2)$ corrections, to be discussed in the next section,
can also be used as an estimate of the uncertainty. In order to have the
correct asymptotic behavior of the uncertainties for $m_t \gg \zm$, it is
assumed that these universal corrections are multiplied by a factor $2/t =
2\,\zm^2/m_t^2$. Indeed, the ${\ord}(t^2)$ is completely under control,
although the sub-leading ${\ord}(t)$ is not and thereby $1/t$ follows with
the usual safety factor of $2$.
Thus if the leading higher-loop corrections calculated in
Refs.~\cite{barb} and~\cite{afmt} are denoted respectively by
 
\begin{equation}
\delta V_i^{t^2} \; , \qquad \delta V_i^{\als^2}\;,
\end{equation}
 
\noindent
then the corresponding estimates of the missing terms are assumed to be
 
\begin{equation}
\Delta V = (2/t)\,\delta V \; ,
\quad {\hbox{with}} \quad t = {{m_t^2}\over {\zm^2}} \;.
\end{equation}
 
\noindent
There is also a specific correction to the $\z0 \to b \bar b$
vertex dependent on $M^2_{_H}/m^2_t$, which is presently computed up tp
${\ord}(\gf^2 m_t^4)$ term~\cite{barb}, while the sub-leading terms are
still unknown. If we denote it by $\delta \phi^{t^2}$ --- see the
second term in Eq.~(\ref{403})) --- then an upper bound on the
`Higgs theoretical uncertainty' can be estimated as
 
\begin{equation}
\Delta\Gamma_b = -{{\alpha(\zm)}\over \pi}\,\Gamma_0\delta\phi^{t^2} \;.
\end{equation}
 
\noindent
For $m_t = 175\,$GeV and $\hm = 300\,$GeV this amounts to $0.02\,$MeV, which
is much smaller than other uncertainties and could therefore be neglected.
 
\subsection{QCD Corrections on Electroweak Loops}
 
The effect of QCD corrections is not confined to the final-state radiation
or the ${\ord}(\alpha\alpha_s)$ vertex corrections to $\z0 \to b \bar b$
but it will influence all vector boson self-energies through virtual
gluon
exchanges within quark-loop insertions. All codes include these two-loop
diagrams~\cite{aasse} by first decomposing the $WW,ZZ,Z\gamma,$ and
$\gamma\gamma$ self-energies into two basic building blocks,
$\piv(m_i,m_j)$ and $\pia(m_i,m_j)$, which are given by the expansion
 
\begin{equation}
\Pi_{_{V,A}} = N_c\,\left(\Pi_{_{V,A}}^{(1)} + \frac{\als}{\pi}\,
\frac{C_F}{4}\,\Pi_{_{V,A}}^{(2)} + \dots \right)  \;,
\end{equation}
 
\noindent
with $C_F = \left(N_c^2 - 1\right)/\left(2\,N_c\right)$. For instance
for each isodoublet,
 
\begin{eqnarray}
\gamma - \gamma &\to& e^2\sum_{i=1}^2\,Q_i^2\Pi_{_V}\left(p^2,m_i,m_i\right)
\;, \nll
\z0 - \z0  &\to& e^2\sum_{i=1}^2\left[v_i^2\Pi_{_V}\left(p^2,m_i,m_i\right)
+a_i^2\Pi_{_A}\left(p^2,m_i,m_i\right)\right]  \;.
\end{eqnarray}
 
\noindent
In the limit where we neglect light quark masses, four different cases
have to be considered:
 
\begin{eqnarray}
\piv(m_t,m_t)\;, \qquad \pia(m_t,m_t)\;, \qquad
\Pi_{_{V,A}}(m_t,0)\;, \qquad \Pi_{_{V,A}}(0,0) \;.
\end{eqnarray}
 
\noindent
An important issue is related to the renormalization scale to choose for
$\als$. The default that we have adopted is to select $\mu = m_t$ for
contributions from the $t-b$ doublet while $\mu = \zm$ is assumed for
light quark contributions.
A practical difference emerges in the various implementations of $\Pi^{(2)}$,
where sometimes the full expression is used, while in other cases a
Taylor expanded (in $q^2/m_t^2$) version has been used.
A step forward has been made with the evaluation of the
${\ord}(\alpha\als^2)$ corrections to $\Delta\rho$, the AFMT term~\cite{afmt}.
Given the usual definition of $\Delta\rho$ as
 
\begin{eqnarray}
\rho &=& {1\over {1-\Delta\rho}}  \;, \nll
\Delta\rho &=&
N_c x_t \left( 1 + \delta^{^{\rm {EW}}} + \delta^{^{\rm {QCD}}}\right) \; ,
\end{eqnarray}
 
\noindent
we have the QCD contribution to $\Delta\rho$, up to three loops and in
the heavy top limit, as
 
\begin{eqnarray}
\Delta\rho^{^{\rm {QCD}}} &=& N_c x_t \delta^{^{\rm {QCD}}}  \nll
{} &=& N_c x_t a_s \left( \delta^{^{\rm {QCD}}}_2
+ a_s\delta^{^{\rm {QCD}}}_3 \right) \;,
\end{eqnarray}
 
\noindent
with $a_s = \als/\pi$. As stated, this correction has been computed in the
heavy top limit and therefore only the leading part of
$\delta^{^{\rm {QCD}}}_3$ is available.
Actually, this new calculation makes the QCD corrections to $\Delta\rho$ much
more stable with respect to the renormalization scale, since we now have
 
\begin{equation}
\delta^{^{\rm {QCD}}}(\mu) \approx -0.910\;\als(m_t)-1.069\;\als^2(m_t)
+ 2.609\;\als^3(m_t)
\ln\left(\frac{\mu^2}{m_t^2}\right) + {\ord}(\als^4) \;,
\label{afmtdef}
\end{equation}
 
\noindent
where $\als(m_t)$ is evaluated with five flavours and the $n_f$ that
appears in the final AFMT result is interpreted as the total number of
flavours contributing in the Feynman diagrams ---  $n_f = 6$.
Recently the same result has been cast~\cite{sir} into a slightly different
form by use of the notion that the corrections to $\Delta\rho$ in terms of
$m_t$ are almost entirely contained in ${\hat m}^2(m_t)/m_t^2$; ${\hat m}(m_t)$
being the running mass evaluated at the pole mass. The corresponding
differences in $\delta^{^{\rm {QCD}}}$ amount to $\approx5\%$ of
the total QCD correction.
Concerning the treatment of the AFMT term it should be noted that there is,
at present, some disagreement on the value for $n_f$ ---
for instance, {\tt LEPTOP} uses $n_f = 5$ on the basis of decoupling of
heavy fermions in vector theories, leading to a 2\% change in
$\delta^{^{\rm {QCD}}}$
{\footnote{A recent and independent evaluation of the QCD corrections to the
$\rho$-parameter has been presented by K.G.~Chetyrkin, J.H.~K\"uhn and
M.~Steinhauser~\cite{nafmt}, showing disagreement with the original AFMT
result. Meanwhile, a revised version of the AFMT calculation has appeared
in hep-ph 16.02.1995 showing agreement with the CKS result.
For understanding the effect we have added a {\bf Note in proof}.}}.
 
A strictly related topic concerns the inclusion and the magnitude of the
$t \bar t$ threshold effects on $\Delta\rho$ and therefore on all the
electroweak parameters.
These effects have recently been estimated by both dispersion
relation and perturbative methods~\cite{ttbth} however an uncertainty remains
as to their magnitude (see Ref.~\cite{kyr} for a detailed discussion).
Most of the QCD corrections to $\Delta\rho$ beyond the leading-order
QCD terms can be discussed and evaluated by absorbing them into the
${\ord}(\als)$ term computed with an adjusted scale $\mu = \xi m_t$:
 
\begin{equation}
\Delta\rho^{^{\rm {QCD}}} = N_c x_t\,\left[ 1 - \frac{2}{3}
\left(1+\frac{\pi^2}{3}\right)\,
\frac{\als(\xi m_t)}{\pi}\right]
+ \Delta\rho^{^{\rm {QCD}}}_{_{\rm {NL}}} \;,
\label{xi}
\end{equation}
 
\noindent
where $\Delta\rho^{^{\rm {QCD}}}_{_{\rm {NL}}}$
takes into account the non-leading top effect with the scale set to
$\mu = m_t$, as well as the light quark effects with $\mu = \zm$.
For instance the result of Ref.~\cite{sir} can be summarized
by saying that it corresponds to a very high accuracy, to a scale,
 
\begin{equation}
\xi = 0.321^{+0.110}_{-0.073} \;.
\label{xisir}
\end{equation}
 
\noindent
Incidentally, the original AFMT formulation corresponds in this language to
a scale $\xi= 0.444$. We note that the absorption of the non-leading
QCD terms into a rescaling of $\als$ has only been performed, so far, at the
level of the leading $\delta^{^{\rm {QCD}}}_2$ term, even if we have
at our disposal the full ${\ord}(\alpha\als)$ correction factor.
 
A comment is in order here. Our original idea was to incorporate the
$\bart t$ threshold effects through an opportune rescaling factor $\xi$.
However, the intrinsic theoretical error on $\xi$ deriving from the
threshold analysis is very difficult to assess, since non-relativistic
approximations also play a certain role. In view of the present situation,
the majority of the codes has agreed with a specific strategy:
 
\begin{itemize}
 
\item The default is represented by computing the pseudo-observables
to include the complete three-loop AFMT term at a scale of $\xi = 1$,
Eq.~(\ref{afmtdef}), which is equivalent to the use of Eq.~(\ref{xi})
with $\xi = 0.444$.
 
\item In order to estimate the size of the non-leading QCD effects, the
$\delta^{^{\rm {QCD}}}$ correction factor has been implemented according
to the formulation of Ref.~\cite{sir}, with a scale which gives
the maximum variation with respect to the AFMT term --- $\xi = 0.248$ ---
and the difference between this and the AFMT
calculation is used as an estimate of the corresponding uncertainty.
It is certainly not the ideal solution, but we cannot rely on other
analyses, not incorporating the ${\ord}(\alpha\als^2)$ term.
In this context a further subtlety arises, since the effective scale for the
AFMT term ($\xi = 0.444$) gives a correction outside the range
required by the present Green function analysis~\cite{kyr}.
In fact, the result of Ref.~\cite{sir} and the corresponding error estimate,
Eq.~(\ref{xisir}), also leave the AFMT expansion at the edge or even a
bit outside the error range.
 
\end{itemize}
 
\subsection{Parametric Uncertainties}
 
The parametric uncertainties are those related to the input parameters,
$\alpha(\zm)$, $\gf$ and the masses, $\zm, m_b$ etc.
Among them the largest uncertainty comes from $\alpha(\zm)$: $\alpha^{-1}(\zm)
= 128.87 \pm 0.12$ (see, however, the new
results~\cite{morrisls}-\cite{eideljeg}),
while the relative uncertainty in $\zm$ is an order of magnitude smaller and
that in $\gf$ is $50$ times smaller. For the $b$ quark mass we used $4.7 \pm
0.3\,$GeV as a sort of conservative average.
We have tried to compare the effects of variations in the input parameters
in a way which, hopefully, will also give useful informations in the future,
providing some sort of evolution of the uncertainties as a function of
the errors in the input parameters. Assuming independence from the actual
central values of $\alpha(\zm)$ and of $m_b$ we computed the derivatives
of the pseudo-observables with respect to $\alpha(\zm), m_b$. Assuming,
also, that the errors are small enough and that the dependence is therefore
approximately linear, this result will allow us to give the uncertainties,
even when the input parameters or the errors on them change with time.
Actually, a linear approximation is good enough for the derivative with respect
to $\bar{\alpha}^{-1}$  but not with respect to $m_b$. In order not
to have problems with the latter near a local extremum we have defined a
maximum derivative given by
 
\begin{eqnarray}
{\cal D}f &=& {\hbox{sign}}(f')\,{{\max\left(\delta_h,\delta_l\right)}\over
{\Delta x}}  \;, \nll
\delta_h &=& |\max(f) - {\bar f}| \;, \nll
\delta_l &=& |{\bar f} - \min(f)| \;,
\end{eqnarray}
 
\noindent
where ${\bar f}$ is the corresponding central value.
In order to avoid lengthy tables we have computed this set of derivatives
for our pseudo-observables at the standard reference point,
where $m_t = 175\,$GeV, $\hm = 300\,$GeV and $\alpha_s(\zm) = 0.125$.
The results are shown in Tables~\ref{ta1} and~\ref{ta2} for {\tt BHM},
{\tt LEPTOP}, {\tt TOPAZ0} and {\tt ZFITTER}.
 
\subsection{Structure of the Comparisons}
 
Having introduced the main ingredients of the calculations we are now
ready to explain in more detail the structure of the comparison. As already
pointed out, we focus on $25$ pseudo-observables and vary $m_t, \hm$
and $\als(\zm)$ in a given range of values. It is worth mentioning that
we could introduce at this point the notion of an {\it adapted} set-up, in
the sense that prior to the introduction of options we have made
an effort to show that different codes running under as similar
as possible conditions give very close answers.
 
In fact, each code will produce a set of results according to some well
specified {\it preferred} set-up relative to the various options
briefly discussed. As far as the external building blocks are concerned ---
QCD and QED correction factors ---
some effort has been made in order to reach a common default.
 
\begin{itemize}
 
\item The QED final state radiation for the partial widths is computed
at the scale $\zm$~\cite{kat}.
 
\item The FTJR correction~\cite{ftjr} has been split into a universal QCD
factor computed at a scale $\mu = \zm$ and an internal correction
factor computed at the scale $\mu = m_t$. The former is not included for
the $b$-quark asymmetries (effective mixing angle $\stes^b$).
 
\item The AFMT three-loop effect~\cite{afmt} with $n_f = 6$ is included in the
default with a scale $\mu = m_t$.
 
\item The ${\ord}(\alpha\alpha_s)$ vector boson self-energies ($\Pi^{(2)}$)
are included (almost always the full expression) with two scales:
$\mu = m_t$ for the $t-b$ doublet and $\mu = \zm$ for the light quarks.
 
\item The introduction of an effective scale $\xi m_t$ is not part of the
{\it preferred} set-up but rather is included in the uncertainty.
 
\end{itemize}
 
\noindent
This procedure for estimating the size of the non-leading QCD effects
has been implemented in all codes but {\tt LEPTOP}, which uses the
$(2/t)\times$AFMT term as an estimator of the uncertainty.
 
In this way the {\it preferred} set-up of a code refers to a specific choice
of the electroweak options, everything of course embedded in a given choice of
the renormalization scheme. The result of this procedure is given by five sets
of predictions for the $25$ pseudo-observables as functions of $m_t,~\hm$
and $\als(\zm)$. On top of our predictions, each group has adapted the
various codes to run under all its options (typically up to $2^5\div 2^6$).
It must be clearly realized that no common set of options has been created
and that the options of one code have been designed independent of the
options of all other codes. In the end, the $2^n$ electroweak options
are folded with 2 QCD options, the inclusion of AFMT being the default
versus a rescaling $m_t \to 0.248\, m_t$ in the ${\ord}(\alpha\als)$ term
representing the uncertainty.
{\tt LEPTOP}, however, is using a different procedure (see above).
Note, that electroweak--QCD factorization simulates ${\ord}(\alpha\als)
$ non-controlled terms, therefore it can also be considered a QCD option.
For each pseudo-observable $O$ we have collected
$O_{\rm {adapt}}$ and $O_+, O_-$, i.e.
 
\begin{eqnarray}
O_+ &=&  \max_i O_i  \nll
O_- &=&  \min_i O_i ~~~~
\end{eqnarray}
 
\noindent
where the index $i$ is running over the options. The differences $O_+ -
O_{\rm {adapt}}$ and $O_{\rm {adapt}} - O_-$ calculated by a given code
are our {\it internal} estimates of the theoretical uncertainty associated
with $O$, while the different results for $O_{\rm {adapt}}$ as obtained
by the various codes may be considered as giving our estimation of the scheme
dependence.
Here {\it internal} must be understood as the estimate that one particular
code can produce by varying internally its options on the implementation
of radiative corrections. The corresponding error bars on the theoretical
predictions are in some cases very asymmetric, merely reflecting the specific
ideas or, even better, personal taste, beyond the choice of
preferred set-up. Clearly the way in which the different realizations
have been built into the codes is very indicative of the original
strategy. To give an example, we may note that the reason why some code
does not include the one-loop corrected axial-vector coupling of the $\z0$
to fermions in the definition of the $\rho$ parameter is because its
preferred set-up is the expanded solution. In some cases, the
theoretical uncertainty internally estimated by one code could turn out to be
large with respect to those of other codes. Basically, we do not attribute
any particular relevance to this fact, as the global indication should include
some sort of average among the codes. In the general discussion of the results
we have tried as much as possible to trace the roots of the phenomenon,
whenever it appears, and any further consideration should be left to the
potential users of our analysis.
Each point in the error bars has the same content of probability and the width
of the theoretical bands associated with each pseudo-observable should be
taken as an indication of the relevance of that quantity for the analysis of
the LEP data. Sometimes the theoretical error, which cannot be reduced unless
we come up with a full two-loop calculation, becomes of the same order as the
present or projected experimental error.
Even before entering into a full discussion of the results we may
anticipate the
predictions for $\stes^l$ by choosing some reference point, such as
$m_t = 175\,$GeV, $\hm = 300\,$GeV and $\als(\zm) = 0.125$.
The prediction is
 
\[\stes^l = \left\{
\begin{array}{ll}
{} &\mbox{} \\
0.23197^{+0.00004}_{-0.00007} & \mbox{BHM} \\
{} &\mbox{} \\
0.23200^{+0.00008}_{-0.00008} & \mbox{\tt LEPTOP} \\
{} &\mbox{} \\
0.23200^{+0.00004}_{-0.00004} & \mbox{\tt TOPAZ0} \\
{} &\mbox{} \\
0.23194^{+0.00003}_{-0.00007} & \mbox{\tt WOH} \\
{} &\mbox{} \\
0.23205^{+0.00004}_{-0.00014} & \mbox{\tt ZFITTER} \;. \\
{} &\mbox{}
\end{array}
\right. \]
 
\noindent
The corresponding width for the theoretical band is therefore $0.00011$,
$0.00016$,  $0.00008$,  $0.00010$ or $0.00018$ for {\tt BHM, LEPTOP, TOPAZ0,
WOH} and {\tt ZFITTER}.
This can be compared with what we obtain by combining the LEP
results on all the asymmetries: $\stes^l = 0.2321 \pm 0.0004$
\cite{lepwg3}-\cite{schaile}.
 
In order to present our results we have focused on the $W$ mass, on the
primary set of pseudo-observables chosen by the LEP collaborations
for fitting ---  $\gz, R_l, \afb^l$, and on the $b$- and
$c$-quark related charge asymmetries and ratios of partial widths.
 
\subsection{Experimental data and theoretical predictions}
 
To set the scene for our discussion we must first introduce, in
Table~\ref{ta3}, the relevant experimental data.
Clearly a detailed discussion of our results should take into account the whole
range of values for the input parameters but many of the relevant conclusions
can already be drawn by considering Tables~\ref{ta4}--\ref{ta7}, where we have
reported $\wm$~\cite{wm} and a sample of $11$ quantities as measured by the LEP
collaborations. For comparison we have fixed a reference point,
$m_t = 175\,$GeV, $\hm = 300\,$GeV and $\als = 0.125$, and reported
the predictions from {\tt BHM/LEPTOP/TOPAZ0/WOH/ZFITTER}, including the
estimated theoretical errors and the average.
The content of the tables should not be confused with a fit but should only
be taken as a first introduction to the theoretical results. Our reference
point is a mere consequence of the range indicated by CDF for the top quark
mass and of the most recent prediction for $\als(\zm)$ from LEP. As far as
$\hm$ is concerned, we must admit a high degree of arbitrariness.
As we have detailzed, the differences among the (theoretical)
central values for each quantity are basically (even though not totally) a
measure of the effect induced by a variation in the renormalization scheme.
As can be seen from Tables~\ref{ta4}--\ref{ta7} the ratios of the
maximal half-differences among the five codes over the experimental errors are:
 
\begin{eqnarray}
\Delta_c(\wm) &=&  2.5\tmt \nll
\Delta_c(\ge) &=&  6.7\tmt \nll
\Delta_c(\gz) &=&  3.9\tmt \nll
\Delta_c(R_l) &=&  1.0\tmo \nll
\Delta_c(R_b) &=&  3.3\tmt \nll
\Delta_c(R_c) &=&  2.0\tmth \nll
\Delta_c(\stes^l) &=&  1.4\tmo \nll
\Delta_c(\afb^l) &=&   5.6\tmt \nll
\Delta_c(\afb^b) &=&   7.8\tmt \nll
\Delta_c(\afb^c) &=&   2.5\tmt  \; .
\end{eqnarray}
 
\noindent
Another piece of information is obtained by looking at the theoretical
uncertainties as estimated internally by each code. A very conservative
attitude would be to report the half-difference between the global maxima and
minima among the five codes. By considering again the ratio with the
experimental errors we obtain
 
\begin{eqnarray}
\Delta_g(\wm) &=&  7.2\tmt \nll
\Delta_g(\ge) &=&  1.6\tmo \nll
\Delta_g(\gz) &=&  3.7\tmo \nll
\Delta_g(R_l) &=&  2.3\tmo \nll
\Delta_g(R_b) &=&  8.0\tmt \nll
\Delta_g(R_c) &=&  4.1\tmth \nll
\Delta_g(\stes^l) &=&  2.8\tmo \nll
\Delta_g(\afb^l) &=&   1.2\tmo \nll
\Delta_g(\afb^b) &=&   1.6\tmo \nll
\Delta_g(\afb^c) &=&   5.2\tmt \; .
\end{eqnarray}
 
\noindent
By comparing the various $\Delta_c$ and $\Delta_g$ we can obtain a rough
evaluation of the global theoretical error associated with the most
relevant quantities in the analysis of the LEP data. As far as the differences
among codes in their preferred(adapted) set-up is concerned, we can safely
conclude that the ratios of their predictions to the experimental errors are
usually less than $0.15$. However, the most important
message to be derived from this simple exercise is that very often the
theoretical uncertainty can be larger than what is to be expected on the
basis of a simple comparison of the results from different calculations.
 
A wider sample of results is shown in Figures~11--23, where, again, we
have fixed some reference point ---  $\hm = 300\,$GeV and $\als(\zm) = 0.125$
and where $100\,\gev < m_t < 250\,\gev$.
In every figure the corresponding experimental points(data), as they are given
in ref.~(\cite{lepwg3}), are shown at $m_t= 178\,$GeV.
The full collection of our results refers instead to $\hm = 60, 300, 1000\,$GeV
and to $\als(\zm) = 0.118, 0.125, 0.132${\footnote{A preliminary version of
the comparison among our results has been presented in~\cite{mt}}.
 
Here we discuss the main features of the comparison and for each quantity we
indulge in presenting an estimate of the {\it global uncertainty}
by roughly considering the half-difference between the maximum and minimum
among all predictions. Admittedly this is not a very rigorous procedure and
therefore it should be treated with due caution. It should be stressed again
that the $\Delta_c, \Delta_g$ factors presented above show the half-difference
of the predictions over the experimental errors. In the following discussion
we will mainly analyze two quantities:
 
\begin{description}
 
\item[{\bf $d_c, d_g$}]
--- the half-differences, either among central predictions or between the
maximum and the minimum among all predictions.
 
\end{description}
 
\noindent
It should be noted that we will not address in this section the question of
how the transformation from primordial distributions to the secondary
quantities will affect their precision.
In the following we have examined some of the pseudo-observables
at the standard reference point and tried to present the state of the art
for their theoretical predictions without including all sorts of parametric
uncertainties, but rather we have limited our discussion to the genuinely
theoretical ones. Below we discuss 13 pseudo-observables.
 
\begin{itemize}
 
\item{\bf $\wm$} \\
There is a certain spread in the predictions for increasing values of $m_t$
substantially independent of $\als$ and increasing for large $\hm$, with
the formation of two clusters, represented by {\tt BHM/WOH} and {\tt ZFITTER}
on one side and {\tt LEPTOP/TOPAZ0} on the other.
The maximum difference in central values is seen for $m_t = 250\,$GeV where
$d_c$ reaches a $10.5(12)\,$MeV for $\hm = 300(1000)\,$GeV, made even more
significant by the addional fact that the bands essentially do not overlap.
The situation improves for central values of $m_t$, where we see at most a
half-difference of $\approx 4.5\,$MeV even if the clustering is already evident.
 
\item{\bf $\ge$} \\
No pattern of any particular relevance is observed. Almost independent
of $\als$ we notice that for intermediate $m_t$ there is a substantial
agreement for all $\hm$, while for high $m_t$ the agreement is better at large
values of $\hm$, and for small $m_t$ it improves for low values of $\hm$.
All the error bars tend to become wider for increasing $m_t$ and $\hm$, with
the possible exception of {\tt TOPAZ0/ZFITTER}.
The largest half-difference among central values is for low $\hm$
and high $m_t$ where $d_c$ can reach $\approx 0.03\,$MeV.
A safe estimate of the overall theoretical error at $m_t = 175\,$GeV is of
about $0.03\,$~MeV
($0.030 < d_g < 0.031$ for the full range of $\als$ and $\hm$).
 
\item{\bf $\gz$} \\
Due to the final state QCD corrections $\gz$ is much more sensible to variations
in $\als$. However, we have verified that there is no substantial variation
among the codes as a function of $\als$, a sign that final state QCD
corrections are well under control. For instance we find $1.4\,\mev < d_g
< 1.5\,\mev$ for $0.118 < \als < 0.132$ at $m_t = 175\,$GeV and
$\hm = 300\,$GeV
{\footnote {This is not a trivial consequence of the fact that all five codes
use the same radiation functions, since their implementation is usually
different.}}.
For low $m_t$ the result of the comparison does not show any particular
pattern, while for high $m_t$ the agreement improves with high $\hm$, showing
again some sort of correlation between the two variables.
For intermediate $m_t$,
instead we find smaller differences around central values of $\hm$.
In general for $m_t > 150\div 175\,$GeV the various error bands, while growing,
have a tendency to overlap. The maximum deviation among codes is for high
$m_t$ and low $\hm$, where $d_c$ can reach approximately $0.6\,$MeV. For
$m_t = 175\,$GeV $\als = 0.125$ instead we get, as the global estimate of the
uncertainty, $1.3, 1.4$ and $1.4\,$MeV for $\hm = 60, 300$ and $1000\,$GeV
(with tiny variations in respect of $\als$).
 
\item{\bf $R_l$} \\
This quantity has a role of its own since quite often it is used for
extracting a precise determination of $\als(\zm)$. Indeed, up to
some degree of accuracy, the two variables are related by $\Delta\als
\approx \pi \Delta R/ R$ so that a difference of $0.01$ in $R_l$
is equivalent to an error of $0.002$ in $\als$.
There is some common trend in all our results for $R_l$, namely the {\tt BHM}
predictions always stay a little higher, while the other codes tend to
cluster, apart from the {\tt TOPAZ0} results, which, for very low $m_t$ and high
$\hm$ tend to converge towards those of {\tt BHM}. For $m_t > 175\,$GeV the
error bands tend to overlap so that each code has a central prediction within
the other error bands, again apart from the {\tt BHM} point, which sometimes is
fully contained only within the {\tt WOH} predictions and lies at the upper
boundary of the {\tt LEPTOP/ZFITTER} ones. Error bars are often very
asymmetric, expecially for {\tt WOH/ZFITTER}.
The maximal deviation is for high $m_t, \hm$, where $d_c$ may reach
$0.006$, whereas the global estimate of the uncertainty for $m_t = 175\,$GeV
gives $0.0085, 0.0090$ and $0.0095$ for $\als(\zm) = 0.118, 0.125$ and
$0.132$, independent of $\hm$. These values correspond to an error of
$0.002\div 0.003$ in the determination of $\als(\zm)$. We have also analyzed
in more detail the $\als$ dependence of the ratio $R_l$ for $m_t = 175\,$GeV
and $\hm = 300\,$GeV, including the case when QCD is switched off. Indeed,
a determination of $\als$ from $R_l$ is usually achieved by writing
$R_l = R_l(\als=0)(1 + \delta^{^{\rm {QCD}}})$ and by using the most updated
formulation of QCD corrections (see for instance Ref.~\cite{hmpq}).
In this way,
some relevance should also be attributed to a comparison of various predictions
for the ratio $R_l$, unfolded of QCD correction terms.
For $\als = 0$ the {\tt WOH} prediction is lower than the others, which cluster
around $19.946$ ({\tt WOH} is $-0.78\tmt$ below the average).
When QCD corrections become active, {\tt BHM} remains higher ($0.5\div
0.6\tmt$ above average) while the other codes form a cluster.
 
\item{\bf $R_b,R_c$}  \\
The ratios of the $b \bar b$ and $c \bar c$ partial widths to the total
hadronic width share some common features. The experimental errors are
$0.0020$ and $0.0098$ and the two quantities are $-38\%$ correlated.
Our central predictions for $R_b$ are all within $2\tmf$
and even the inclusion of the theoretical uncertainty only gives
$d_g \approx 6\tmf$. The overall theoretical error at $m_t = 175\,$GeV
is $1.7\tmf$. For $R_c$ the global uncertainty is well contained within
$6.0\tmfv$. $R_b$ shows some clustering which becomes more and more evident
for large $m_t$, with {\tt BHM} giving the higher prediction and {\tt
LEPTOP/TOPAZ0/WOH/ZFITTER} forming a lower cluster. The behavior of $d_g$ as
a function of $\als$ is practically flat.
 
\item{\bf $\stes^{\rm {lept}} $}  \\
The reported value of the leptonic effective weak mixing angle is the
average of all forward--backward and polarization asymmetries from LEP.
Therefore the analysis relies on the hypothesis that the peak forward--backward
asymmetry can simply be connected with the remaining asymmetries through
the use of the same $\stes^l$. For all values of $\hm$ and $\als$
the agreement is less satisfactory for low $m_t$ where {\tt ZFITTER}
remains on the higher side, {\tt BHM/WOH} form a lower cluster and {\tt
LEPTOP/TOPAZ0} are somewhere in between. The general trend is to have a
convergence of all codes for high $m_t$. The maximal half-difference among
central values is as for low $m_t$, where for all $\hm,\als$ we find
$d_c \approx 9.0\tmfv$.
For $m_t = 175\,$GeV the overall uncertainty is estimated to be $d_g\approx
1.1\div 1.4\tmf$ over the whole range $\hm-\als$.
For $\hm= 300\,$Gev and $\als= 0.125$ we find $d_g = 1.3, 1.1, 2.1\tmf$ for
$m_t = 100, 175, 250\,$GeV.
In view of the supposed relevance of this parameter and of its projected
experimental error we have to admit that the status of the theoretical
predictions
is far from satisfactory but totally related to the unknown higher order
effects. To give an example, we could say that the knowledge of the
sub-leading ${\ord}(\gf^2\zm^2 m_t^2)$ corrections to $\Delta r$ would
greatly improve the situation --- for instance for {\tt ZFITTER}, which
dominates the error.
 
\item{\bf $\stes^b$}  \\
From many points of view the situation is very similar to that described
for $\stes^l$. Let us remember that $\stes^b$ differs from $\stes^l$
because of flavour-dependent corrections, which are $m_t$-dependent and
not negligible. There is some kind of crossed behavior in our predictions,
with an agreement substantially better for intermediate values of
$m_t$ and deteriorating at the boundaries. For low $m_t$ the comparison is
similar to that for $\stes^l$: a higher prediction from {\tt ZFITTER},
a central cluster {\tt LEPTOP/TOPAZ0} and a lower one {\tt BHM/WOH}.
For high $m_t$ the highest prediction is from {\tt BHM}, with a lower cluster
of the other codes. The global estimate of the uncertainty for $m_t = 175\,$Gev
and $\als= 0.125$ is of $1.15(1.05,1.45)\tmf$ for $\hm = 60(300, 1000)\,$GeV,
with an uncertainty $\pm 0.1\tmf$ due to a variation in $\als$.
 
\item{\bf $\afb^l$}  \\
In presenting results for the leptonic forward--backward (peak) asymmetry,
as well as for any other leptonic asymmetries, we follow the indication
of the experimentalists who keep $\afb^l$ until the end in their standard
model fits, without necessarily identifying it directly with $\stes^l$.
We first start by analyzing the comparison among the central values.
Here the result of the comparison is rather good and essentially we
always start at low $m_t$ with two clusters --- a higher one containing
{\tt BHM/WOH} and a lower one containing {\tt TOPAZ0/ZFITTER} and with
{\tt LEPTOP} somewhere in between. There is a fast convergence of all results
for increasing $m_t$, expecially for high $\hm$. Typically for $m_t = 175\,$GeV
and $\hm = 300\,$GeV we find a half-difference of about $8.4(8.9,9.5)\tmfv$
for $\als = 0.118(0.125, 0.132)$.
When we come to the inclusion of the theoretical uncertainty it is immediately
evident that the previous comment still applies and the global error for
the standard reference point becomes $1.9 \div 2.0\tmf$ at  $\als = 0.118$,
$2.0 \div 2.2\tmf$ for $\als = 0.125$ and $2.0 \div 2.3\tmf$ for $\als = 0.132$
, where the variation with $\hm$ is illustrated. We will come back to $\afb^l$
and to the related uncertainty while discussing additional theoretical options
which have not been included in the working set.
 
\item{\bf $\alr$}  \\
Also for the left--right (peak) asymmetry there is some general behavior in
our predictions. At low $m_t$ we start with maximal disagreement, two clusters
with {\tt BHM/WOH} in the higher one and {\tt LEPTOP/TOPAZ0} in the central
one, to reach convergence of results for high $m_t$.
For low $m_t$ there is also a considerable spreading of the error bands.
Given the fact that the SLD measurement seems to require a much higher value
of $m_t$, we have considered the overall uncertainty for $m_t = 250\,$GeV with
the result of $1.82(1.59, 1.74)\tmth$ for $\hm = 60(300, 1000)\,$GeV
for $\als = 0.118$. These values become $1.84(1.61,1.74)\tmth$ and
$1.86(1.63,1.77)\tmth$ for $\als = 0.125$ and $0.132$.
 
\item{\bf $\afb^b, \afb^c$}  \\
Practically everything we state for the general behavior of $\alr$ can
be repeated for the $b$- and $c$- quark charge asymmetries. As for $\alr$,
the agreement is worst at low $m_t$ and there is a general convergence
at high $\hm$ when $m_t$ is growing. For low $\hm$ and large $m_t$
{\tt TOPAZ0} has the tendency to give a higher prediction.
The global uncertainty for $\afb^b$ at $m_t = 175\,$GeV is $6.2\div 7.4\tmf$.
There is no general agreement among codes on the definition of $\afb^b$.
In particular {\tt LEPTOP} and {\tt TOPAZ0} include mass effects into the
definition of this pseudo-observable.
For the $b$ quark, the effect of its non-vanishing mass leads to:
 
\begin{equation}
{\cal A}^b = \frac{2g_{_{V}}^b g_{_{A}}^b
}{[\frac{1}{2}(3-v^2)(g^b_{_{V}})^2+v^2(g^b_{_{A}})^2]} v \; ,
\label{20}
\end{equation}
 
\noindent
where $v$ is the $b$ quark velocity:
 
\begin{equation}
v = \sqrt{1 - \frac{4 {\bar m}^2_b(\zm^2)}{\zm^2}}\;,
\label{305}
\end{equation}
 
\noindent
where $\bar{m}_b(\zm^2)$ is the running $b$-mass in $\overline{MS}$-sheme
defined by Eq.(\ref{82}) below. However this is a very tiny effect. An estimate
of it, based on {\tt TOPAZ0} results, gives an absolute deviation between the
massless and massive definitions of $1.6, 1.0, 2.0\tmfv$ at $m_t = 100, 175,
250\,$GeV for $\hm = 300\,$GeV and $\als = 0.125$.
Similar conclusion can be drawn for $\afb^c$. The global uncertainty at
$m_t = 175\,$GeV is $d_g = 4.5\div 6.0\tmf$.
 
\item{\bf $\sigma_h$} \\
For the hadronic cross-section the theoretical error $d_g$ is always contained
in a range of $0.0075\div 0.0095\,$nb, with very little dependence on $\als$.
Indeed, for our standard reference point ($m_t = 175\,$GeV and $\hm = 300\,$GeV)
the largest half-difference among central values is $0.0065\,\nb, 0.0065\,\nb$
and $0.0070\,\nb$ for $\als = 0.118, 0.125$ and $0.132$. Our estimate of the
overall theoretical error at $m_t = 175\,$GeV is of about $0.009\,$nb and the
largest half-difference is seen at large $m_t$ and $\hm$ where it reaches
$0.0095\,$nb. Over the whole range of the parameters the prediction of {\tt WOH}
remains higher with a tendency for {\tt ZFITTER}, though
only at very low values of $m_t$, to slightly converge towards {\tt WOH}. In
any case, the {\tt WOH} error bands are sufficiently large to include the other
predictions or to overlap with the other bands.
 
\end{itemize}
 
\noindent
To summarize our results we have also presented in Table~\ref{ta8} the
largest half-differences between central values or between maximal and minimal
predictions among codes in the range $150\,\gev < m_t < 200\,\gev$,
$60\,\gev < \hm < 1\,\tev$ and $0.118 < \alpha_s < 0.125$.
In Tables~\ref{ta9} abd~\ref{ta10} we have fixed $m_t = 175\,$Gev and
illustrated
the largest half-differences among central values or those
between the maximum of the maxima and the minimum of the
minima among the five codes in two situations: $\als = 0.125$ fixed
and $60\,\gev < \hm < 1\,\tev$, or $\hm = 300\,\gev$ and $0.118 < \als < 0.125$.
In this way the separate contributions from $\hm$ or $\als$ to the theoretical
errors are indicatively given. A final but partial indication of our results
can be provided by computing some of the quantities which usually enter the
SM fits (at $m_t = 175\,$GeV) and by collecting all available
sources of error
 
\begin{eqnarray}
{\hbox{\tt BHM}} \quad \gz &=& 2497.4^{+0.9}_{-1.0}(th.)^{+7.9}_{-8.8}
(\hm,\als)
                    \pm 6.7\,\Delta\baral^{-1} \pm 0.8\,\Delta m_b
                           \,\mev\nll
{\hbox{\tt LEPTOP}} \quad \gz &=& 2497.2 \pm 1.1(th.)^{+8.0}_{-8.5}(\hm,\als)
                    \pm 6.8\,\Delta\baral^{-1} \pm 0.8\,\Delta m_b
                           \,\mev\nll
{\hbox{\tt TOPAZ0}}
\quad \gz &=& 2497.4^{+0.2}_{-0.9}(th.)^{+8.2}_{-9.1}(\hm,\als)
                    \pm 6.8\,\Delta\baral^{-1} \pm 0.8\,\Delta m_b
                           \,\mev\nll
{\hbox{\tt WOH}}
\quad \gz &=& 2497.4^{+1.5}_{-0.6}(th.)^{+8.0}_{-8.8}(\hm,\als)
                           \,\mev\nll
{\hbox{\tt ZFITTER}}
\quad \gz&=& 2497.5^{+0.6}_{-0.5}(th.)^{+7.9}_{-8.7}(\hm,\als)
                    \pm 6.8\,\Delta\baral^{-1} \pm 0.8\,\Delta m_b
                           \,\mev
\end{eqnarray}
 
\begin{eqnarray}
{\hbox{\tt BHM}} \quad \s0h &=& 41.436^{+0.006}_{-0.003}(th.)\pm 0.042
                           (\hm,\als)\pm 0.013\,\Delta\baral^{-1}
                           \pm 0.007\,\Delta m_b
                           \,\nb\nll
{\hbox{\tt LEPTOP}} \quad \s0h &=& 41.439^{+0.003}_{-0.004}(th.)
                           \pm 0.040(\hm,\als)
                           \pm 0.012\,\Delta\baral^{-1}
                           \pm 0.008\,\Delta m_b
                           \,\nb\nll
{\hbox{\tt TOPAZ0}}
\quad \s0h &=& 41.437^{+0.007}_{-0.002}(th.)^{+0.041}_{-0.040}
                               (\hm,\als)
                    \pm 0.012\,\Delta\baral^{-1} \pm 0.007\,\Delta m_b
                           \,\nb\nll
{\hbox{\tt WOH}} \quad \s0h &=& 41.449^{+0.000}_{-0.012}(th.)^{+0.041}_{-0.040}
                           (\hm,\als)
                           \,\nb\nll
{\hbox{\tt ZFITTER}}
\quad \s0h&=& 41.441^{+0.000}_{-0.005}(th.)^{+0.041}_{-0.040}
                           (\hm,\als) \pm 0.013\,\Delta\baral^{-1}
                               \pm 0.007\,\Delta m_b
                           \,\nb
\end{eqnarray}
 
\begin{eqnarray}
{\hbox{\tt BHM}} \quad R_l &=& 20.788^{+0.004}_{-0.008}(th.)^{+0.061}_{-0.060}
                           (\hm,\als)\pm 0.047\,\Delta\baral^{-1}
                           \pm 0.009\,\Delta m_b
                           \nll
{\hbox{\tt LEPTOP}} \quad R_l &=&
 20.780^{+0.006}_{-0.005}(th.)^{+0.061}_{-0.059}
                              (\hm,\als)
                           \pm 0.046\,\Delta\baral^{-1}
                           \pm 0.010\,\Delta m_b
                           \nll
{\hbox{\tt TOPAZ0}}
\quad R_l &=& 20.782^{+0.002}_{-0.005}(th.)^{+0.052}_{-0.059}
                               (\hm,\als)
                    \pm 0.046\,\Delta\baral^{-1} \pm 0.010\,\Delta m_b
                           \nll
{\hbox{\tt WOH}} \quad R_l &=& 20.780^{+0.013}_{-0.000}(th.)^{+0.062}_{-0.059}
                           (\hm,\als)
                           \nll
{\hbox{\tt ZFITTER}}
\quad R_l&=& 20.781^{+0.006}_{-0.001}(th.)^{+0.061}_{-0.060}
                               (\hm,\als) \pm 0.047\,\Delta\baral^{-1}
                               \pm 0.009\,\Delta m_b \; ,
\end{eqnarray}
 
\noindent
where we have allowed, as usual, $60\,\gev < \hm < 1\,\tev$ and $0.118 < \als <
0.132$.
Whenever available, the parametric uncertainties have been inserted.
It is assumed that $\Delta m_b$ is given in GeV.
 
\subsection{More on Theoretical Uncertainties}
 
There are some options on electroweak radiative corrections which, although
implemented in the codes, have not been used as working options for
producing our comparisons with the experimental data. The main argument
for this exclusion is related to their tendency to produce
results deviating sensibly from the average. As a matter of fact, this
is the visible consequence of their inclusion, but quite often we have
theoretical reasons against them. This short section is devoted to
a summary of those effects and of their eventual influence on the comparisons.
At the end one should not forget that the design and the implementation of
options is indeed something very peculiar to a given realization.
The following short considerations give another illustration of a very
important fact: theoretical uncertainties should be treated with due
caution, realizing that they contain, in any case, a large degree of
arbitrariness.
 
\subsubsection{\tt TOPAZ0}
 
With {\tt TOPAZ0}, the most striking effect is related
to an expansion of the non-leading terms to ${\ord}(\alpha)$, once bosonic
self-energies were not re-summed. Roughly speaking, we can assume that
a certain quantity $X$ is given by (see also 2.2.4)
 
\begin{equation}
X = X_0 + \frac{\alpha}{\pi}\,X_1 + {\ord}(\alpha^2) \;,
\label{coefx1}
\end{equation}
 
\noindent
where $X_0$, by construction, will include all the re-summed contributions.
As soon as we allow the two options,
 
\[X^2 = \left\{
\begin{array}{ll}
\left(X_0 + \frac{\alpha}{\pi}\,X_1\right)^2 \\
{}\\
X_0^2 + 2\,\frac{\alpha}{\pi}X_0X_1 \; ,
\end{array}
\right. \]
 
\noindent
and interpret the resulting difference as a theoretical uncertainty, then, under
some circumstances, we end up with errors considerably larger than
those presented above. The main problem is represented here by a clash between
accidental cancellations and gauge invariance (even the notion of cancellation
between fermionic and bosonic sector is gauge-dependent).
As it is well known, the bosonic self-energies are gauge-dependent and,
moreover,
the fermionic ones tend to dominate away from the intermediate $m_t$ region.
That is how {\tt TOPAZ0} allows, among other options, for a strict resummation
of the fermionic self-energies alone. Of course, one could just avoid
resummation altogether, but the option of expanding versus
non-expanding the remainders, with respect to the leading terms, still
applies and
the theoretical uncertainties would become, in this case, sensibly
$m_t$ dependent.
Finally, we stress once more that the identification of a gauge invariant
part of the bosonic self-energies is not at all a unique procedure and
therefore some degree of arbitrariness is always hidden in a global
resummation.
We admit that, this option allows for a nice construction of a small remainder,
in a situation, however, where we are working with one-loop
contributions and
higher order reducible ones. What is left out of our analysis is, in any case,
related to the two-loop irreducible terms, about which nothing is presently
known. To summarize we could say that in one case it is approximately the
`square' of the one-loop bosonic corrections that we use to estimate the
theoretical error, while in the other we can decide to re-sum part of it
into the leading term and make the remainder small, even though we are still
missing information about not-yet-computed higher orders and their effects,
which could make the smallness of the remainder inadequate.
 
The final reason why {\tt TOPAZ0} has excluded this option in presenting the
pseudo-observable Tables is therefore totally related to the abnormal
(as compared to the other codes) size of the errors for some of the quantities,
noticeably $\afb^l$. Here we say again that the expansion option is
bounded to produce larger errors whenever the remainders are not
(one way or another)
kept small and in some codes this effect is not seen, simply
because nothing equivalent to the expansion has been implemented. To give a
quick idea of the effect we present in Table~\ref{ta11} the shift in the central
values and in the error bands for some of the pseudo-observables.
Clearly the largest effect is seen for the leptonic forward--backward
asymmetry
where the theoretical error becomes comparable in size to the experimental
counterpart. Needless to say, when this option is activated, a large
fraction of the uncertainties become {\tt TOPAZ0} dominated.
Additional consequences will be introduced and discussed in the next chapter.
 
\subsubsection{\tt ZFITTER}
 
About {\tt ZFITTER}, one can also mention several peculiar
moments, related to the specific design of its options.
We begin with by making clear that nothing resembling what is described
in the previous section was observed.
To a large extent this is due to the fact that
the coefficient $X_1$ (see discussion in 2.2.4) happened to be small in the
framework of the {\tt ZFITTER} renormalization scheme.
Having accepted this statement, however, one should not conclude that
it has particular advantages over the other schemes.
The size of the coefficient is simply a numerical accident,
without any deep physical meaning.
 
However, {\tt ZFITTER} does  contain additional options,
which eventually were not
left among its {\it working options}. An example is given by the array of
expansions~(\ref{ifacr}). All four expansions have been implemented, and
it was noted that the third and fourth expansions enlarge the
theoretical uncertainty for some observables ($\wm$ and $\stes^l$),
roughly by factors of two and four correspondingly.
As it was pointed out in section 1.10.2, the third and the fourth
expansions
contradict to the conclusion of paper~\cite{sirlin84} about fermionic mass
singularities. It not surprising that this was the argument in favour of
excluding these options from the working set.
 
Another interesting example is related to the leading--remainder
splitting problem. Three variants of the resummation of the leading terms
in $\Delta \rho$, see Eq.~(\ref{rholead}), were implemented as different
sub-options:
 
\begin{itemize}
 
\item[1)] only the first term, i.e. $\Delta \rho^{\alpha}$, is re-summed;
 
\item[2)] all but X terms of~(\ref{rholead}) --- the content of square
 brackets is re-summed;
 
\item[3)] the whole expression~(\ref{rholead}) is re-summed ({\tt ZFITTER}
default).
 
\end{itemize}
 
\noindent
There is a noticeable increase in the uncertainty when we include
the second option with respect to a situation where only the
first and the third are retained among the working options.
One should emphasize, however, that this increase in the error is not
dramatic: for example for $\stes^l$ it amounts
to a nearly $m_t$ independent uncertainty,
of the order $5\times10^{-5}$. Examining the reasons for this
enhancement, it was revealed that for the second option the remainder
terms are about 5--10 times bigger compared to those for the first
and the third options. Moreover, they are only several times smaller
than the leading terms.
On the basis of this observation, the second option was termed {\it a
pathological} option and initially excluded from the working set,
since its effect contradicted to an accepted strategy. In playing
more with {\tt ZFITTER} options a striking property was observed. Some
of the {\tt ZFITTER} options do not possess the additivity property. As an
example, we mention that the `scale of remainder option' alone produces
nearly the same uncertainty as when it is applied `in conjunction' with
the following two other options:
 
\begin{itemize}
 
\item[-] the one dealing with expansion~(\ref{ifacr}), first two rows;
 
\item[-] the one dealing with the three above-mentioned variants of
resummation for $\Delta \rho$: $1)\div3)$.
 
\end{itemize}
As a consequence, the combined effect of all three options leads
practically to the same uncertainty, and the latter is independent from
the actual number of options included --- two or three $\Delta \rho$
resummation sub-options are left in the working set.
This eventually lead to a decision to retain all $1)\div3)$ among working
options.
 
\clearpage
 
\section{Realistic distributions}
 
To give predictions for pseudo-observables has not been our only task
and we have also devoted a noticeable effort in order to present the most
updated analysis for realistic observables. This process requires as
fundamental ingredients a QED {\it dresser} and therefore the comparisons
have been restricted to {\tt BHM, TOPAZ0} and {\tt ZFITTER}, since they
allow
for the treatment of QED diagrams involving the emission of real photons with
results which are dependent on energies and experimental cuts.
Roughly speaking we can distinguish between $s$-channel processes and
Bhabha scattering, ${\rm e}^+{\rm e}^-\to {\rm e}^+{\rm e}^-$.
For Bhabha scattering the commonly
accepted procedure is the so-called $t$-channel subtraction, where
the $s-t$ and $t-t$ contributions are subtracted from the data by using the
code {\tt ALIBABA}~\cite{ali}.
In the procedure the `most reasonable' values of $m_t$ and of $\hm$ are used
leading to additional sources of errors in the analysis. Given the present
situation we have also performed a comparison between {\tt ALIBABA} and
{\tt TOPAZ0} inspite of the fact that the electroweak library of
{\tt ALIBABA}
has not been constantly updated. Thus in the comparison will
emerge an intrinsic difference due to the improved electroweak and QCD
formulation of {\tt TOPAZ0}.
Our comparisons can be divided according to the following scheme:
 
\begin{itemize}
 
\item Fully extrapolated set-up for muonic and hadronic channel with the
possible inclusion of a cut on the invariant mass of the final fermion pair,
the so-called $s'$ cut ({\tt BHM, TOPAZ0, ZFITTER});
 
\item ${\rm e}^+{\rm e}^- \to \mu^+\mu^-$ (${\rm e}^+{\rm e}^-$,
$s$-channel for $40^{\circ} < \theta_{-} < 140^{\circ}$,
$\theta_{\rm {acoll}}^{\rm {max}} = 10^{\circ},25^{\circ}$,
and $E_{\rm {th}}(\mu^{\pm}) = 20\,$GeV
($E_{\rm {th}}(e^{\pm}) = 1\,$GeV), ({\tt TOPAZ0, ZFITTER});
 
\item ${\rm e}^+{\rm e}^- \to {\rm e}^+{\rm e}^-$ for
$40^{\circ} < \theta_-< 140^{\circ}$, $\theta_{\rm {acoll}}^{\rm {max}}
= 10^{\circ},25^{\circ}$, and $E_{\rm {th}}({\rm e}^{\pm}) = 1\,$GeV,
({\tt ALIBABA, TOPAZ0}).
 
\end{itemize}
 
\noindent
Moreover, we have fixed the following set of values for the c.m.energy:
$88.45, 89.45$, $90.20, 91.1887, 91.30$, $91.95, 93.00, 93.70\,$GeV.
All the results refer to a given reference point, $m_t = 175\,$GeV,
$\hm = 300\,$GeV and $\als(\zm) = 0.125$.
A first comment concerns the $s'$ cut, which should not be confused with
a cut on the invariant mass of the event after initial state radiation
(sometimes also used in the experiments).
For realistic observables we have avoided all references to any specific set
of {\it effective} formulas and to their realizations, the interested
reader having available the existing literature~\cite{ali}-\cite{real}.
Weak radiative corrections, depending on the assumptions of the electroweak
theory, have already been discussed from the point of view of the options that
arise in their practical implementation. Here we only mention that we have
fully propagated those theoretical uncertainties from the
pseudo-observables to the realistic ones with the result that ---
for the fully convoluted cross-sections and forward--backward
asymmetries --- the final results include a theoretical error bar.
In the following we present a short discussion of the main ingredients
entering the calculation of realistic observables and critically
compare some of the results obtained with {\tt BHM, TOPAZ0} and {\tt ZFITTER}.
 
\subsection{De-Convoluted Distributions}
 
To illustrate in more detail the construction of realistic observables
(RO) we start from the concept of de-convoluted quantities. For a given
process we construct $\xsf$($\xsb$), the forward(backward) kernel cross-section,
including electroweak corrections, and eventually the comprehensive of a
cut on the angular acceptance.
QCD corrections are included while all QED corrections are left out. After a
first comparison at this level we proceeded by introducing QED final-state
radiation (FSR), initial-state leptonic and hadronic pair production (PP),
initial--final-state QED interference (INT) and, finally, the kernel
distributions folded with initial-state QED radiation (ISR).
One should emphasize, however, that the INT contribution was simply
added --- at ${\ord}(\alpha)$ --- and was not folded with the ISR.
As far as the ISR is concerned we did not find it opportune to fully
review the
various treatments and implementations but have tried as much as possible
to illustrate the origin of possible discrepancies whenever they arise.
For instance, the differences that we find in the results of the various codes
are dominated by pure weak (and QCD) corrections around the region of the peak
and by QED radiation along the tails, where, however, the experimental
error is considerably larger.
Among the de-convoluted quantities, the most relevant are those computed at
$s = \zm^2$, which have an obvious counterpart in the pseudo-observables,
that we have already computed, $\afb^l$ and $\sigma^h$, which hereafter
will be characterized by an index $0$, as $\afb^{l,0}$ and $\s0h$.
There is, however, a noticeable difference between the two sets,
represented by
the interference of the $\z0-\gamma$ $s$-channel diagrams and by the presence
of imaginary parts in the formfactors, the latter being particularly relevant
for the leptonic forward--backward asymmetry.
 
As far as the propagation of electroweak uncertainties from
pseudo-observables to realistic ones is concerned we notice
that all three codes see an enhancement of the theoretical errors. For instance
for the standard reference point we denote with $\s0h$ the PO hadronic
(peak) cross-section and with $\sigma^h(\zm^2)$ the realistic one and get:
 
\begin{eqnarray}
{\hbox{\tt BHM}}     &\s0h& = 41.436^{+0.006}_{-0.003}\,\nb \quad
                \sigma^h(\zm^2) = 30.366^{+0.015}_{-0.002}\,\nb \nll
{\hbox{\tt TOPAZ0}}  &\s0h& = 41.437^{+0.007}_{-0.002}\,\nb \quad
                \sigma^h(\zm^2) = 30.375^{+0.016}_{-0.005}\,\nb \nll
{\hbox{\tt ZFITTER}} &\s0h& = 41.441^{+0.000}_{-0.005}\,\nb\quad
                \sigma^h(\zm^2) = 30.373^{+0.005}_{-0.002}\,\nb \; ,
\end{eqnarray}
 
\noindent
the enhancement factor being $1.9, 2.3$ and $1.4$ for the three codes
respectively. The reason behind this increase of the induced theoretical
error is that for $\s0h$, as defined in
Eq.~(\ref{sigh0}), we first compute $\ge, \gh$ and $\gz$ and then construct
the combination $\ge\gh/\gz^2$ without any further expansion in $\alpha$, while
for $\sigma^h (\zm^2)$ we include the expansion in the option set
(linearization of the cross-section), thus enlarging the error.
 
\subsection{Final-State Radiation}
 
A substantial difference exists between fully extrapolated RO and RO in the
presence of cuts.
This is illustrated bt the fact that without cuts we can simply
use the well known correction factor $1 + \frac{3}{4}Q_f^2\frac{\alpha}{\pi}$
for each partial channel and there is therefore no ambiguity in FSR. A
possible source of discrepancy can instead be introduced when cuts are present,
due to a different treatment of final-state higher-order QED effects. This can
lead to differences which in general depend on the experimental cuts required
and may grow for particularly severe cuts.
It has already been shown~\cite{topaz0} that two possible prescriptions,
--- completely factorized final-state QED correction versus factorized
leading-terms and non-leading contributions summed up ---
lead to differences in the cross-section for
Bhabha scattering of the order of $0.5\%$ far from the peak,
whereas the asymmetry is substantially left unchanged. To be more specific,
the final-state QED corrections amount to a leading correction term:
 
\begin{equation}
F_{\rm {cut}}^l (s) =
2{{\alpha} \over {\pi}} Q_f^2 \ln\left(1-\frac{s_0}{s} \right)
\left[\ln\left({ {s} \over {m_f^2} }\right) - 1 \right] \; ,
\end{equation}
 
\noindent
where $s_0$ represents a cut in the reduced invariant mass.
Renormalization group arguments suggest that such a leading term should
be exponentiated, which is of no practical importance at low thresholds but
could give sizeable effects at high thresholds. By defining
 
\begin{equation}
F_{\rm {cut,r}}^{\pm} (s) =
F_{\rm {cut},\alpha}^{\pm} (s) -F_{\rm {cut}}^l (s) \; ,
\end{equation}
 
\noindent
the leading term resummation can be implemented as follows:
 
\begin{equation}
F_{\rm {cut}}^{\pm} (s) = \exp \left[F_{\rm {cut}}^l (s) \right]
\left[ 1 + F_{\rm {cut,r}}^\pm (s)
- F_{\rm {cut,r}}^\pm (s) F_{\rm {cut}}^l (s) \right] \; ,
\end{equation}
 
\noindent
where spurious terms
$F_{\rm {cut}}^l (s) \times F_{\rm {cut,r}}^\pm (s)$ are confined at
least at ${\ord}(\alpha^3)$.
Indeed, several prescriptions for treating final-state correction are
possible, all equivalent at ${\ord}(\alpha)$. One reasonable recipe
could be to define the leading term in a different way. For instance, in the
presence of an acollinearity cut, the infrared logarithm could be defined as
 
\begin{equation}
l = \ln (1-x) \; ,
\end{equation}
 
\noindent
where $x$ is given by
 
\begin{equation}
x = {\rm max} (s_0/s, y_T) \; ,
\end{equation}
 
\noindent
$y_T$ being
 
\begin{equation}
y_T = {{1- \sin ( \zeta / 2)} \over {1+\sin (\zeta/2)}} \;,
\end{equation}
 
\noindent
and $\zeta$ the maximum acollinearity allowed. Another possibility would be
to exponentiate the full ${\ord}(\alpha)$ contribution
$F_{\rm {cut},\alpha}^\pm$, even though
there is no guarantee that the experimental cut-dependent terms do
exponentiate; in this case spurious terms appear already at ${\ord}(\alpha^2)$.
Alternatively, one could choose to factorize only a leading ${\ord}(\alpha)$
term and simply add the ${\ord}(\alpha)$ correction due to the acollinearity
cut (this simulates the choice in Ref.~\cite{ali}).
 
\subsection{Initial-State Pair Production and QED Interference}
 
Next we come to the inclusion of initial-state pair production in the realistic
distributions. 
A fermionic pair of four-momentum $q^2$ radiated from the ${\rm e}^+$
or ${\rm e}^-$ line gives a correction~\cite{kkks}:
 
\begin{eqnarray}
\sigma_{rm {pair}} &=&
\sigma^{S+V}_{\rm {pair}} + \sigma^{H}_{\rm {pair}} \; , \nll
\sigma^{S} &=& \int_{4m^2}^{\Delta^2}\,dq^2\int_{(\sqrt{s}-\Delta)^2}^{(
\sqrt{s}-\sqrt{q^2})}\,ds'\,{{d^2\sigma}\over{dq^2ds'}} \; , \nll
{{d\sigma^{H}}\over {dz}} &=& s\int_{4m^2}^{(1-\sqrt{z})^2s}\,dq^2\,
\,{{d^2\sigma}\over{dq^2ds'}} \;.
\end{eqnarray}
 
\noindent
Also for this term there are different treatments ---  we can exponentiate
the pair production according to the YFS formalism~\cite{jsm}
or the same pairs can be included at ${\ord}(\alpha^2)$.
In the end, however, we found a reasonable agreement among
the results of the
three codes relative to the specific inclusion of pair production.
The main features of this correction term are as follows.
 
\begin{itemize}
 
\item {\tt TOPAZ0/ZFITTER} employ the KKKS formulation~\cite{kkks} without
exponentiating the soft+virtual part, which is added linearly to the
cross-section. {\tt BHM}, instead employs the YFS formalism~\cite{jsm}.
The independence of the results from the soft--hard separator has also
been successfully investigated.
 
\item $\tau$-pairs are not included.
 
\item The lower limit of integration $z_{\rm {min}}$,
adopted for leptonic pair
production, is $0.25$ and the soft--hard separator $\Delta$ has been
fixed in the region where we see a plateau of stability.
 
\end{itemize}
 
Initial--final QED interference has been introduced in the calculations,
including the effects of hard photons. For a fully extrapolated set-up this
means that all photons, up to the maximum available energy, are taken
into account, while $s'$ cuts or energy thresholds and acollinearity cuts will
restrict the available phase space. In order to proceed step-by-step we have
introduced the procedure of comparing our results in the sequence
$\sigma_{_{F(B)}}(NN,NY,YN,YY)$, where the first argument in parenthesis
denotes inclusion or exclusion of PP, and the second refers to the
interference. In this way the relative influence of QED corrections has been
checked and kept under control.
To illustrate the trend, we present in Tables~\ref{ta12} and~\ref{ta13}
a comparison
for the hadronic cross-section and the muonic forward--backward asymmetry.
In particular, to continuously keep under control our comparisons,
we have introduced and analyzed the ratios
 
\begin{equation}
r\left({\hbox{C-{\tt TOPAZ0}}},{\hbox{conf}},O\right) =
{{dO \left({\hbox{C-{\tt TOPAZ0}}},{\hbox{conf}}\right)}\over
 {dO \left({\hbox{C-{\tt TOPAZ0}}},{\hbox{NN}}\right)}} \; ,
\end{equation}
 
\noindent
where $O = \sigma^h, \sigma^{\mu}, \afb^{\mu}$, conf = YN/NY/YY and
C = {\tt BHM,ZFITTER}. Moreover, $dO$ denotes the relative variation for
cross-sections and the absolute deviation for the asymmetry.
Whenever pair production or QED interference have similar effects,
the corresponding ratio assumes values of around $1$. When the ratio goes
to zero it is a signal that the apparent agreement does not reflect a
similar agreement in the NN quantities and is therefore due to
accidental compensations. Finally, if this ratio grows in modulus it
gives an indication that the agreement at the NN level is not respected
when PP or INT are introduced.
 
To illustrate this fact, consider Table~\ref{ta12}, which
refers to $\sigma^h$. We find, for instance,
that the YY cross-sections for
{\tt ZFITTER-TOPAZ0} agree in five digits at $\sqrt{s} = 89.45\, $GeV,
$90.20\,$GeV and $91.30\,$GeV, giving $10.042\,\nb, 17.992\,\nb$
and $30.514\nb$ respectively. Moreover, a closer look reveals that
 
\begin{eqnarray}
\sqrt{s} &=& 89.45\,\gev \qquad 10.067 - 10.068\,\nb    \nll
\sqrt{s} &=& 90.20\,\gev \qquad 18.039 - 18.040\,\nb    \nll
\sqrt{s} &=& 91.30\,\gev \qquad 30.590 - 30.590\,\nb    \nll
\end{eqnarray}
 
\noindent
for {\tt ZFITTER/TOPAZ0} in the NN configuration.
 
\subsection{Imaginary Parts of the Formfactors}
 
The presence of imaginary parts in the weak formfactors introduces
additional possibilities for the implementation of radiative corrections
with respect to pseudo-observables.
At the level of the kernel cross-sections and remembering the quite general
subdivision into leading parts and remainders introduced in section 2.2.1, we
can basically select two possible options:
 
\begin{itemize}
 
\item Imaginary parts confined in the remainders,
 
\item Imaginary parts inserted into the leading terms.
 
\end{itemize}
 
To illustrate the effect of imaginary parts we select $f = \mu$ and
consider the asymmetric part of the angular distribution for 
${\rm e}^+{\rm e}^- \to \mu^+\mu^-$.
If we denote with $\chig$ and $\chiz$ the corrected $\gamma-\gamma$
and $Z-Z$ propagators, this asymmetric part is proportional to
 
\begin{eqnarray}
\xsf-\xsb &\propto&{\cal R}e\left[\left(\ga^*\right)^2\chig\chiz^*\right] +
      \left[{\cal R}e\left(\gv\ga^*\right)\right]^2|\chiz|^2  +
{\hbox{boxes}}\nll
{} &=& {\cal R}e\ga^2 {\cal I}m \chig {\cal I}m \chiz
+ 2\,{\cal R}e \ga {\cal I}m \ga {\cal R}e \chig {\cal I}m \chiz \nll
{} && + \left({\cal R}e \ga \right)^2 \left({\cal R}e \gv\right)^2
+ {\ord}({\cal R}e \chiz)
+ {\ord}\left({\cal R}e \gv \times {\cal I}m^2\right)
+ {\ord}\left({\cal I}m^4\right) ,
\end{eqnarray}
 
\noindent
where ${\cal R}e \chiz$ is suppressed around the $\z0$ peak
and ${\cal R}e \gv$ is usually small.
 
\subsection{Initial-State QED Uncertainties}
 
An estimate of the theoretical uncertainty due to QED radiation
can also be derived, so that our results may contain two sources of
theoretical error, $\pm \Delta(EW) \pm \Delta(QED)$.
For instance such an estimate can be performed by using the
following algorithm. Since the main source of QED theoretical
uncertainty is the treatment of final-state radiation, namely the
use of a completely factorized formula versus a leading-log factorized
plus a non-log additive one, whenever the theoretical error is required,
we can run over the two possible options and return the corresponding
uncertainty. Moreover, when the large-angle Bhabha scattering is
considered, we must realize that the main approximation adopted by {\tt TOPAZ0}
is to treat the $t$ and $s+t$ contributions to the cross-section at the
leading logarithmic level. Thus the size of the convoluted $t$ and $s+t$
terms is computed as the difference between the full Bhabha prediction
and the pure $s$-channel one and the theoretical error is
estimated by assuming $1\%$ of the difference. In particular, this means
that the QED theoretical error depends on the detailed
experimental set-up, growing when tightening the experimental cuts
and, for the Bhabha case, when enlarging the angular acceptance to
smaller angles, since this increases the contribution of the non-$s$
terms to the Bhabha cross-section.
In the case of Bhabha scattering, the calorimetric measurement problem
arises if a cross-section not inclusive of the energy of the outgoing
fermion is considered. For thresholds of the order of $\approx 1\,$GeV
the effect is of order $0.01\div 0.02\%$, but for higher energy thresholds
($E_{\rm {th}}$) the contribution can grow considerably.
 
\subsection{Comparisons}
 
Let us now consider the question of de-convoluting the peak quantities in
order to extract pseudo-observables like $\s0h$ and $\afb^{l,0}$.
By definition the de-convoluted asymmetry does not include final-state QED
radiation, while the de-convoluted cross-sections include both QED and
QCD final-state corrections.
For $\sqrt{s} = \zm\,$GeV the three codes predict a hadronic cross-section
of $30.366\,\nb, 30.375\,$nb and $30.373\,$nb respectively. The corresponding
de-convoluted quantities --- no QED corrections --- are $41.400\,\nb,41.409\,$nb
and $41.402\,$nb which, in turn, means that for {\tt BHM}
the effect of extracting QED corrections amounts to $11.034\,$nb,
for {\tt ZFITTER} $11.029\,$nb, and for {\tt TOPAZ0}, also $11.034\,$nb
with a $0.045\%$ difference. If we introduce
 
\begin{equation}
\sigma^h = {\cal D}[\sigma^h]\,\left(1+\delta_{\rm {conv}}\right)
 \;,
\label{deconv}
\end{equation}
 
\noindent
where ${\cal D}$ denotes de-convolution, we get
 
\begin{equation}
\delta_{\rm {conv}}({\hbox{B,T,Z}}) = -0.2665\;,
\qquad -0.2665\;, \qquad -0.2664  \;.
\end{equation}
 
\noindent
From our previous tables we also find that for the same choice of input
parameters
 
\begin{equation}
\s0h({\hbox{B,T,Z}}) = 41.436\;, \qquad 41.437\;, \qquad 41.441\,\nb \;.
\nonumber
\end{equation}
 
\noindent
The corresponding differences as compared to
the de-convoluted observables, which are
$0.036\,\nb$, $0.028\,\nb$ and $0.039\,$nb, give an estimate of about
$0.011\,$nb for the uncertainty in the effect of the imaginary parts from the
weak formfactors. Coming now to the asymmetry we find:
 
\begin{equation}
\afb({\hbox{B,T,Z}}) = -0.00082\;, \qquad -0.00125\;, \qquad -0.00109
\nonumber
\end{equation}
 
\noindent
at $\sqrt{s} = \zm$, becoming after de-convolution,
 
\begin{equation}
{\cal D}\afb({\hbox{B,T,Z}}) = 0.0169\;, \qquad 0.0166\;, \qquad 0.0166
\;.
\nonumber
\end{equation}
 
\noindent
Thus the effect of de-convolution is $0.0177, 0.0179$ and $0.0177$ in {\tt BHM}
, {\tt TOPAZ0} and {\tt ZFITTER}, with a {\tt BHM/ZFITTER-TOPAZ0} difference
of $2.0\tmf$.
It must be noted that the quantity usually reported in the literature is the
de-convoluted peak asymmetry with a pure $\z0$ exchange. In this case our
predictions become
 
\begin{equation}
{\cal D}\afb({\hbox{B,T,Z}}) = 0.01544(0.01544)\;,
\quad 0.01536(0.01536)\;, \quad 0.01528(0.01531) \;,
\nonumber
\end{equation}
 
\noindent
where in parenthesis we have included the corresponding prediction for
$\afb^{l,0}$.
Therefore this additional and conventional filtering of the asymmetry brings
the {\tt BHM-TOPAZ0} difference to $8.5\tmfv$ and the
{\tt ZFITTER-TOPAZ0} difference to $-8.0\tmfv$. The various
combinations of de-convoluted asymmetries are presented in Table~\ref{ta14}.
The complete set of de-convoluted quantities is presented in Table~\ref{ta15},
where, around the peak, we see the largest difference among codes of
$0.04\%$ and of $0.02\%$ for $\sigma^{\mu}, \sigma^h$.
For the muonic cross-section we find at $\sqrt{s} = \zm$ $1.4785\,\nb,
1.4790\,\nb$ and $1.4794\,\nb$, while the corresponding de-convoluted
quantities are $2.0015\,\nb, 2.0019\,\nb$ and $2.0022\,\nb$ respectively.
Thus the effect of de-convolution in $\sigma^{\mu}$ is $-0.2613, -0.2611$ and
$-0.2611$ for {\tt BHM, TOPAZ0} and {\tt ZFITTER}.
An interesting question, which arises in this contest, is related to the
possibility of performing a real and significant test of the QED corrections
by comparing the predictions of different codes. We could define
 
\begin{equation}
\delta^i_{_{\rm {QED}}}(O^i) = {O^i \over {O^i_0}} -1
\end{equation}
 
\noindent
for a given observable $O$, with $O^i_0$ being the de-convoluted observable
as predicted by the $i$-th code.
For a given code this $\delta_{_{\rm {QED}}}$ represents the effect of
the convolution over an electroweak-corrected observable. It must however
be noted that the comparison of $\delta_{_{\rm {QED}}}$ of different codes does
not give an unambiguous information on QED corrections since we start
already from slightly different kernels. Let us quantify this statement.
Let $\Delta^i_{_{\rm {QED}}}$ be the absolute QED correction for some observable
$O$ as computed by the $i$-th code:

\begin{equation}
O^i = O^i_0 + \Delta^i_{_{\rm {QED}}}\; .
\end{equation}

\noindent
Then

\begin{equation}
\delta_{_{\rm {QED}}}(O^i) = {{\Delta^i_{_{\rm {QED}}}} \over {O^i_0}}\; ,
\end{equation}

\noindent
so that the difference between $\delta_{_{\rm {QED}}}$ of different codes can be
written as

\begin{equation}
\delta_{_{\rm {QED}}}(O^i) - \delta_{_{\rm {QED}}}(O^j) =
{{\Delta^i_{_{\rm {QED}}}} \over {O^i_0}}
- {{\Delta^j_{_{\rm {QED}}}} \over {O^j_0}}\; .
\end{equation}

\noindent
By adding and subtracting $\Delta^i_{_{\rm {QED}}} / O^j_0$ one obtains

\begin{equation}
\delta_{_{\rm {QED}}}(O^i) - \delta_{_{\rm {QED}}}(O^j) =
\Delta^i_{_{\rm {QED}}} {{O_0^j - O_0^i} \over {O_0^i O_0^j}} +
{{1} \over {O_0^j}} ( \Delta^i_{_{\rm {QED}}} - \Delta^j_{_{\rm {QED}}} )\; .
\end{equation}

\noindent
Now it is clear that the difference between $\delta_{_{\rm {QED}}}$
of different codes depends also on the differences in the pure weak and QCD
libraries of the codes.
In particular, for the leptonic asymmetries in
the $Z$ peak region, which are small, the first term in the
r.h.s. of the last equation becomes very large, so that
$\delta_{_{\rm {QED}}}(O^i) - \delta_{_{\rm {QED}}}(O^j)$
is weak- and QCD- dominated,
whereas it becomes QED-dominated far from the peak only.
 
Coming back to our original strategy we have made a constant effort to
understand the systematics inherent in the extraction of
pseudo-observables from the realistic distributions that the experiments
should consider as a theoretical uncertainty.
The main ingredient contributing to the systematics is of course
the QED radiation, inclusive of initial-final interference and of
initial-state pair-production. But in the extraction, some relevance must
also be attributed to the imaginary parts of the formfactors and to the
$\z0-\gamma$ interference. The latter may have some influence, since not all
the codes have the same splitting of the $\gamma f \bar f$ vertices among the
formfactors. We have already devoted some detailed discussion to the
differences among the convolutions around the peak. In Table~\ref{ta16},
we give all the remaining results, corresponding to the full set of
the energy points.
With due caution as to the correct interpretation of our comparison we
observe a somehow larger difference in the convolution around the tails with
respect to the peak region, notably $0.23\%$ at $\sqrt{s} = 88.45\,$GeV
and $0.59\%$ at $\sqrt{s} = 93.70\,$GeV. Since these are differences in
$\delta_{\rm {conv}}$ and not in the total prediction, even
$0.6\%$ away from the peak is quite reasonable.
 
Our results are presented in Figures 24 -- 37 where we have reported
$\sigma^{\mu}$ and $\afb^{\mu}$ for four different set-ups: fully extrapolated,
$s' > 0.5\,s$ and $40^{\circ} < \theta_- < 140^{\circ},
E_{\rm {th}} > 20\,$Gev, $\theta_{\rm {acoll}} < 10^{\circ},25^{\circ}$.
Also reported are $\sigma^h$ for two set-ups, fully extrapolated and $s' >
0.01\,s$ and the $s$ channel $\sigma^e, \afb^e$ where, however, $E_{\rm {th}}
> 1\,$GeV.
 
In all these figures we have also shown the deviations (relative for
cross-sections and absolute for the asymmetry) {\tt BHM-TOPAZ0} (when available)
and {\tt ZFITTER-TOPAZ0} with the corresponding theoretical error bars.
The choice of {\tt TOPAZ0} as a reference point is purely technical and
avoids the necessity of introducing an average among the codes. It emerges
from these comparisons that for a fully extrapolated set-up, the agreement for
the muonic cross-section for energies below the peak and around it is quite
reasonable, always taking the intrinsic error as a reference.
Given the reasonable agreement at the level of de-convoluted cross-sections
and once we have observed that the de-convolution is satisfactory for hadrons,
we come to the conclusion that for muons the low-$q^2$ region, where mass
effects may become relevant, gives the dominant difference in $\sigma^{\mu}$.
Indeed, a comparison for $s' > 0.5\,s$ shows a much better agreement,
giving $1.4391\,\nb, 1.4396\,\nb$ and $1.4397\,\nb$ for the three codes
(instead of $1.4785, 1.4790, 1.4794$). Since the corresponding de-convoluted
quantities are $1.9599\,$nb for {\tt BHM} and $1.9605\,$nb for {\tt TOPAZ0}
we conclude that de-convolutions amount to $-0.2657$ for both codes, while
for fully extrapolated they are $-0.2613$ versus $-0.2611$ with a $0.08\%$ of
relative difference.
Around the peak, the energy-dependence of the observables as predicted by
{\tt BHM} looks different --- a difference which should probably be related to
the way in which {\tt BHM} implements the effective coupling language.
For the hadronic cross-section we observe a consistent agreement among the
three codes. Finally, for the muonic asymmetry the agreement at the peak is
again quite reasonable, but the energy dependence again looks different, with
{\tt BHM/TOPAZ0} agreeing below the peak and {\tt TOPAZ0/ZFITTER} agreeing
above it.
To summarize the status of the comparison for extrapolated set-up or for
$s'$-cut we have from Figures 24 --- 29 the following.
 
\begin{itemize}
 
\item{\bf $\sigma^{\mu}$} \\
Around the peak the maximum deviation is $9\tmf\,$nb corresponding to
$0.06\%$, the {\tt BHM} predictions are always lower,
those of {\tt ZFITTER} higher, and {\tt TOPAZ0} is in-between.
On the high energy side the {\tt BHM} behavior
with energy tends to differ. For an $s'$-cut of $0.5\,s$ the maximum
deviation around the peak corresponds to $0.04\%$ and on the low-energy side
of the resonance we observe a rather remarkable agreement among all codes.
 
\item{\bf $\sigma^h$} \\
For the hadronic cross-section we have found an impressive agreement around
the peak, with $0.03\%$ of maximum relative deviation, which is only
slightly worse around the tails --- $0.08\%$ and $0.06\%$. The comparison
remains substantially unchanged for a low $s'$ cut.
 
\item{\bf $\afb^{\mu}$} \\
Here the agreement is quite reasonable at the peak, with an absolute deviation
of $4.3\tmf$ between {\tt BHM} and {\tt TOPAZ0}.
Not completely satisfactory is the energy dependence, which registers
a substantial disagreement with {\tt ZFITTER} giving higher predictions below
the peak and {\tt BHM} after it.
At least on the low-energy side the situation improves if an $s' = 0.5\,s$
cut is imposed.
 
\end{itemize}
 
\noindent
The agreement between {\tt TOPAZ0} and {\tt ZFITTER} remains rather remarkable
even when the geometrical acceptance is constrained, and as well, final-state
energies and the acollinearity angle are bounded with or without QED
initial--final interference.
For instance, for the $s$-channel leptonic cross-sections at $\sqrt{s} = \zm$
and $40^{\circ} <  \theta_- < 140^{\circ}, E_{\rm {th}}(\mu) > 20\, \gev,
E_{\rm {th}}(e) > 1\,$GeV, we find            
 
\begin{eqnarray}
\sigma^{\mu}, \qquad\theta_{\rm {acoll}} < 10^{\circ}
&{}& \qquad 0.9802  - 0.9801\,\nb \nll
\sigma^{\mu}, \qquad\theta_{\rm {acoll}} < 25^{\circ}
&{}& \qquad 0.9905  - 0.9900\,\nb \nll
\sigma^{e},   \qquad\theta_{\rm {acoll}} < 10^{\circ}
&{}& \qquad 0.9886  - 0.9884\,\nb  \nll
\sigma^{e},   \qquad\theta_{\rm {acoll}} < 25^{\circ}
&{}& \qquad 1.0012  - 1.0011\,\nb \; ,
\end{eqnarray}
 
\noindent
where the first entry is {\tt ZFITTER} and the second {\tt TOPAZ0}.
More generally, and remembering that we use $E_{\rm {th}} > 20\,$GeV for muons
and a lower cut of $1\,$GeV for electrons, we find from Figures 30 -- 37:
 
\begin{itemize}
 
\item{\bf $\sigma^{\mu,e}$} \\
for the muonic cross-section there is very good agreement for all energies
and $\theta_{\rm {acoll}} < 10^{\circ}$ --- agreement which slightly
deteriorates for $\theta_{\rm {acoll}} < 25^{\circ}$.
At the peak we have a relative difference of
$0.02\%, 0.05\%$ respectively for two above mentioned $\theta_{\rm {acoll}}$.
($0.03\%$ for a fully extrapolated set-up). For
$s$-channel electrons the agreement is everywhere of the same quality: in
particular at the peak we find a $0.02\%, 0.01\%$ of relative
{\tt TOPAZ0/ZFITTER} deviation --- for $\theta_{\rm {acoll}} < 10,25^{\circ}$.
Thus for $\sigma^l$ our agreement is not altered by introducing cuts.

\item{\bf $\afb^{\mu,e}$} \\
The two leptonic forward--backward asymmetries agree at the peak at the level
of $0.3, 4.3 \tmf$ for muons ($\theta_{\rm {acoll}} < 10,25^{\circ}$) and
$2.9, 0.1 \tmf$ for electrons.
For $\theta_{\rm {acoll}} < 10^{\circ}$ the agreement tends to
deteriorate at larger energies, reaching $9.6\tmf$ for muons and $ 8.7\tmf$
for electrons at $\sqrt{s} = 93.70\,$GeV, while remaining always very good
for $\theta_{acoll} < 25^{\circ}$.
Globally for $\afb^{\mu}$ the {\tt TOPAZ0-ZFITTER} comparison shows
for the fully extrapolated set-up a larger difference on the low-energy side 
of the resonance ($1.0\tmth$ at $\sqrt{s} = 88.45\,$GeV), a maximal agreement
over the whole range of energies for $\theta_{acoll} < 25^{\circ}$, and a larger
difference on the high-energy side for $\theta_{\rm {acoll}} < 10^{\circ}$
($9.6\tmf$ at $\sqrt{s} = 93.70\,$GeV).
 
\end{itemize}
 
\noindent
We illustrate the electroweak theoretical error by considering $\sigma^{\mu}$
at $\sqrt{s} = \zm$. In the four different set-ups considered --- fully
extrapolated, $s'$-cut of $0.5$ or $40^{\circ} < \theta_- < 140^{\circ},
E_{\rm {th}} > 20\,$GeV, and $\theta_{\rm {acoll}} < 10^{\circ}$ or
$< 25^{\circ}$ --- we find $0.095\%, 0.097\%, 0.066\%$ and $0.068\%$,
respectively, in the
relative deviation between the maximum and minimum predictions among the codes.
 
Finally, we illustrate the effect of different treatments of final state
QED radiation in the presence of severe kinematical cuts. By adopting
different strategies {\tt TOPAZ0} predicts for $\sigma^{\mu}$ at $\sqrt{s} =
\zm$ and $40^{\circ} < \theta_- < 140^{\circ},
E_{\rm {th}} > 20\,\gev, \theta_{\rm {acoll}} < 10^{\circ}$:
 
\begin{equation}
\sigma^{\mu} = 0.9801 - 0.9817\,\nb \; ,
\end{equation}
 
\noindent
while for a loose cut of $E_{\rm {th}} > 1\,$GeV we obtain
 
\begin{equation}
\sigma^{\mu} = 0.9892 - 0.9893\,\nb \; .
\end{equation}
 
\noindent
As already discussed, the $0.16\%$ difference at large $E_{\rm {th}}$ reduces
to a mere $0.01\%$ at low $E_{\rm {th}}$.
 
Coming now to the full Bhabha cross-section and forward--backward asymmetry,
we have considered in the following a comparison between the results
of {\tt TOPAZ0} and of {\tt ALIBABA}.
As already explained, the comparison is not fully consistent as it stands now,
because the electroweak and QCD libraries of {\tt ALIBABA} are not up-to-date.
Nevetheless, we show it because it gives an impression of the state-of-the-art.
For this particular comparison the input parameters are slightly
different from our default --- namely $m_t = 174\,$GeV and $\als = 0.124$ have
been used.
The {\tt ALIBABA-TOPAZ0} comparison for $\sigma^e$ is shown in
Tables~\ref{ta21} for $40^{\circ} < \theta_- < 140^{\circ}, E_{\rm {th}} > 1\,
\gev and \theta_{\rm {acoll}} < 10^{\circ}$.
For {\tt TOPAZ0} we have shown three sets of numbers, all including QED
interference, with the first two showing the effect of a different treatment of
QED final-state radiation --- I is the {\tt TOPAZ0} default, while II is the
{\tt ZFITTER}-like default, and III shows the effect of initial
state pair production, which is, however, not included in {\tt ALIBABA}.
It should be mentioned at this point that the insertion of pair production is
strictly valid in {\tt TOPAZ0} only for s-channel processes~\cite{topaz0} and
that it is approximate for full Bhabha.
The cross-section is shown for {\tt ALIBABA} with its numerical error,
while for {\tt TOPAZ0} we give first the electroweak theoretical error and
than the numerical one.
Expressing due caution in comparing the two codes we observe a relative
difference, of $0.05\%$ at the low-energy side of the resonance, which
becomes $0.003\div 0.004\%$ around it, with a visible deterioration at
high energies where we reach $0.86\%$. As already discussed~\cite{topaz0},
this is mainly due to a different treatment of higher-order QED
final-state corrections.
 
A more detailed comparison between {\tt ALIBABA} and
{\tt TOPAZ0} is shown in Tables~\ref{ta22}-\ref{ta29}. We have shown the
predictions for the full Bhabha cross-section and the forward-backward
asymmetry for two different values of $\theta_{\rm {acoll}}$ in
Tables~\ref{ta21}-\ref{ta24}. In  Tables~\ref{ta25}-\ref{ta26} we give the
relative deviation between the central values of {\tt ALIBABA-TOPAZ0} for full
Bhabha or for the $s$-channel alone. Next in Table~\ref{ta27} we give the
difference between full Bhabha and $s$-channel results both in {\tt ALIBABA}
and in {\tt TOPAZ0} with the relative contribution, i.e. $\delta =
\sigma/\sigma(s) -1$ and $\delta(A)/\delta(T)$.
In Table~\ref{ta28} we show the $s+t$, $s$ and $t$ forward-backward asymmetries.
Finally in Table~\ref{ta29} we give a rough estimate of the theoretical
uncertainty by considering the difference between maximal and minimal
predictions from the two codes. For {\tt ALIBABA} this takes into account the
numerical error alone while for {\tt TOPAZ0} we have added linearly the
electroweak and the numerical uncertainty. With due caution in the
interpretation we extract a $0.1\div 0.2 \%$ before and around the $Z$
resonance which becomes as large as $0.9\div 1.0 \%$ at higher energies.
 
\vfill
\pagebreak
 
\section{Basic Formulae for Electroweak Radiative Corrections \label{basics}}
 
In this section we give a more detailed description of the {\it realizations}
of the effective couplings $\gv$ and $\ga$.
 
\subsection{{\tt BHM/WOH} basics}
\subsubsection{Self-energies, propagators, and $\Delta r$}
 
\smallskip \noi
The radiative corrections to the photon-$Z$ propagator system
(considering only the transverse parts $\sim g_{\mu\nu}$)
can be obtained by inversion of the matrix
 
\beq
 ({\bf D}_{\m\nu})^{-1}\,=\,i\,g_{\m\nu}\,
 \left ( \begin{array}{cc}
       k^2+\sggr(k^2)  &  \sgzr(k^2)  \\
       \sgzr(k^2)     &  k^2-M_{_Z}^2+\szzr(k^2)
       \end{array}
                      \right )  \; ,
\label{prop}
\eeq
\noindent
with the renormalized self energies specified below, yielding
 
\beq
 {\bf D}_{\m\nu}\,=\,-i\,g_{\m\nu}\,
 \left ( \begin{array}{cc}
        D_{_{\g}}     &   D_{_{\g Z}}  \\
        D_{_{\g Z}}   &   D_{_Z}
       \end{array}
                    \right)  \; ,
\label{prop1}
\eeq
\noindent
where ($s=k^2$)
 
\bea
 D_{_{\g}}(s) & = & \frac{1}{\irg \,-\,\frac{[\sgzr(s)]^2}{\irz}} \;, \nn  \\
           &   &            \nn \\
 D_{_Z}   (s) & = & \frac{1}{\irz\, -\,\frac{[\sgzr(s)]^2}{\irg}} \; \nn  \\
           &   &            \nn \\
 D_{_{\g Z}}(s) & = & -\,\frac{\sgzr(s)}
                {[\irg]\,[\irz]\,-\,[\sgzr(s)]^2} \; .
\label{prop2}
\eea
 
\medskip
\noi
The building blocks of Eq.~(\ref{prop}) are the renormalized self-energies
$\hat{\Sigma}$, which are decomposed into unrenormalized ones $\Sigma$
and counter terms, as follows:
 
\bea
\sggr(k^2) & = & \sg(k^2)\,+\,\dz_2^{\g}\,k^2     \;, \nn \\
\szzr(k^2) & = & \sz(k^2) - \delta M_{_Z}^2 \,+\,
                \dz_2^Z \, (k^2-M_{_Z}^2)         \;, \nn \\
\swwr(k^2) & = & \sw(k^2) - \delta M_{_W}^2 \,+\,
                \dz_2^W \, (k^2-M_{_W}^2)         \;, \nn  \\
\sgzr(k^2) & = & \sgz(k^2) \,-\,\dz_2^{^{\g Z}}\,k^2 \,+\,
                (\dz_1^{^{\g Z}} - \dz_2^{^{\g Z}})\,M_{_Z}^2 \; .
\label{rself}
\eea
 
In the last line the abbreviations ($i=1,2$)
 
$$
 \dz_i^{^{\g Z}} \,=\, \frac{c_{_W} s_{_W}}{c_{_W}^2-s_{_W}^2}\,
                  (\dz_i^{^Z} - \dz_i^{^{\g}})
$$
\noindent and
 
$$
 s_{_W}^2 = 1-M_{_W}^2/M_{_Z}^2, \;\;\;
 c_{_W}^2 = 1-s_{_W}^2
$$
\noindent were used.
 
\smallskip
The self-energies $\Sigma^{ij}$ in (\ref{rself}) are the sum
of the electroweak one-loop diagrams \cite{bhm}, completed in the quark loops
by the ${\cal O} (\alpha\alpha_s)$ two-loop QCD-electroweak contributions.
 
\bigskip
The mass counter terms for $W$ and $Z$ follow  from the on-shell conditions
for the $W$ and $Z$ propagators:
 
\bea
\delta M_{_W}^2 & = & \real\,\Sigma^{_{WW}}(M_{_W}^2) \; ,   \nn \\
\delta\mz  & = &
\real\,\left\{ \sz(\mz)\,-\,
 \frac{\left[\sgzr(\mz)\right]^2}{\mz+\sggr(\mz)} \right\} \; .
\eea
 
The other renormalization constants in (\ref{rself}) are given by the
following set of equations:
 
\bea
 \delta Z_2^{^{\gamma}} & = & -\Pi^{^{\gamma}}(0)\,\equiv\,
       -\frac{\partial\sg}{\partial k^2}(0)                     \;, \nn \\
 \delta Z_1^{^{\gamma}} & = & -\Pi^{^{\gamma}}(0)\,-\,
   \frac{s_{_W}}{c_{_W}}\,\frac{\Sigma^{^{\gamma Z}}(0)}{M_{_Z}^2}
\;, \nn \\
 \delta Z_2^{^Z} & = & -\Pi^{^{\gamma}}(0)\,-\,
   2\,\frac{c_{_W}^2-s_{_W}^2}{s_{_W}c_{_W}}\,
   \frac{\Sigma^{^{\gamma Z}}(0)}{M_{_Z}^2}\,+\,
   \frac{c_{_W}^2-s_{_W}^2}{s_{_W}^2}\,\left (
   \frac{\delta M_{_Z}^2}{M_{_Z}^2}-\frac{\delta M_{_W}^2}{M_{_W}^2} \right)
\;, \nn\\
 \delta Z_1^{^Z} & = & -\Pi^{^{\gamma}}(0)\,-\,
   \frac{3c_{_W}^2-2s_{_W}^2}{s_{_W}c_{_W}}\,\frac{\Sigma^{^{\gamma Z}}(0)}
{M_{_Z}^2}\,
   +\,\frac{c_{_W}^2-s_{_W}^2}{s_{_W}^2}\,\left (
   \frac{\delta M_{_Z}^2}{M_{_Z}^2}-\frac{\delta M_{_W}^2}{M_{_W}^2} \right)
\;, \nn\\
 \delta Z_2^{^W} & = & -\Pi^{^{\gamma}}(0)\,-\,
   2\,\frac{c_{_W}}{s_{_W}}\,\frac{\Sigma^{^{\gamma Z}}(0)}{M_{_Z}^2}\,+\,
   \frac{c_{_W}^2}{s_{_W}^2}\, \left (
   \frac{\delta M_{_Z}^2}{M_{_Z}^2}-\frac{\delta M_{_W}^2}{M_{_W}^2} \right)
\;, \nn\\
 \delta Z_1^{^W} & = & -\Pi^{^{\gamma}}(0)\,-\,
   \frac{3-2s_{_W}^2}{s_{_W}c_{_W}}\,\frac{\Sigma^{^{\gamma Z}}(0)}{M_{_Z}^2}
\,+\,
   \frac{c_{_W}^2}{s_{_W}^2}\, \left (
   \frac{\delta M_{_Z}^2}{M_{_Z}^2}-\frac{\delta M_{_W}^2}{M_{_W}^2} \right)\,.
\label{counterm}
\eea
 
The photon vacuum polarization $\Pi^{^{\gamma}}(0)$ contains as its light
fermion part the quantity $\Delta\alpha$ discussed in section 1.2.
The last two constants $\delta Z_i^{^W}$ are not independent but are linear
combinations of $\delta Z_i^{^{\gamma}}$ and $\delta Z_i^{^Z}$.
They are given here for completeness.
 
\medskip
By the presence of the $(\sgzr)^2$ term on the r.h.s. the equations
(\ref{counterm}) are non-linear equations.
It is, however, straightforward to solve them for the renormalization
constants in terms of the unrenormalized quantities.
 
\smallskip
The higher-order irreducible contributions to the $\rho$-parameter, as far as
available, are built in by means of substituting
 
\beq
 \dmz-\dmw \, \ra \, \dmz-\dmw + \dro^{(\rm {HO})}
\eeq
\noindent
for the r.h.s of (\ref{counterm}), with
 
\bea
\dro^{(\rm {HO})} & = & 3\, \bar{x}_t \left[
                 1 +      x_t \, \dro^{(2)}(\xi)
                 + \delta^{\rm {QCD}}_{(3)} \right]\, ,
\eea
\beq
 \bar{x}_t  =  \frac{\alpha}{16\pi s_{_W}^2 c_{_W}^2} \,
                 \frac{m_t^2}{\mz} , \;\;\;
       x_t  =  \frac{\Gmu m_t^2}{8\pi^2 \sqrt{2}},
                      \;\;\;
       \xi \; = \;  \frac{m_t^2}{M_{_H}^2} \, ,\nn
\eeq
\noindent
comprising the two-loop electroweak and three-loop QCD contributions.
For the functions $\dro^{(2)}$ and $\delta^{\rm {QCD}}_{(3)}$
see section 1.10.2 and 2.3.
 
\smallskip \noi
Around the $Z$ peak, the $Z$ propagator  has the form
 
\bea
D_{_Z}(s)
    & \simeq & \frac{1}{1+\Pizr(\mz)} \;
          \frac{1}{s-\mz+\,i\,\frac{s}{M_{_Z}} \G_{_Z}} \; ,
\eea
\noindent
with
 
\bea
\Pizr(\mz)
   & = & \real\, \frac{d\,\szr}{d s}(\mz)  \; , \nn  \\
 \szr(s)  & = & \szzr(s)- \frac{\mixs}{\irg}  \, .
\label{piz}
\eea
 
The $Z$  width $\G_{_Z}$ is calculated from the effective coupling
constants given below in Eq.\ (\ref{effva}) together with the
general formulae of section 1.10.1.
 
\bigskip  \noi
The vector boson masses $M_{_W}, M_{_Z}$ are correlated by the Fermi constant
 
\beq
\Gmu = \frac{\pi\al}{\sqrt{2}M_{_W}^2s_{_W}^2} \, \frac{1}{1-\Dr}
= \frac{\pi\al}{\sqrt{2}M_{_Z}^2c_{_W}^2s_{_W}^2} \, \frac{1}{1-\Dr}\;.
\label{wzcorr}
\eeq
 
The quantity
$\Dr(\alpha,M_{_Z},M_{_W},m_t,M_{_H})$ has the following representation:
 
\bea
\Dr & = & \Pi^{^{\g}}(0) -\frac{c_{_W}^2}{s_{_W}^2} \left(\dmz-\dmw\right)
          +\frac{\sw(0)-\dmmw}{M_{_W}^2}   \nn    \\
    &   &+ 2\,\frac{c_{_W}}{s_{_W}}\,\frac{\Sigma^{^{\g Z}}(0)}{M_{_Z}^2} \, +\,
\frac{\al}{4\pi s_{_W}^2} \left( 6+
\frac{7-4 s_{_W}^2}{2 s_{_W}^2} \log c_{_W}^2   \right) \, .
\label{deltar}
\eea
 
The last term is the sum of the box contributions and renormalized
vertex corrections to the muon decay amplitude after removing the
Fermi-model-like virtual photonic corrections. Due to the inclusion of the
higher-order reducible and irreducible terms the way of writing $\Delta r$ in
the denominator of (\ref{wzcorr}) automatically takes account of the proper
resummation of the leading $\Delta\alpha$ and $\dro$ terms and some of the
sub-leading terms as discussed in sub-section 1.10.2.
 
\subsubsection{Vertex corrections}
 
\smallskip \noi
The vertex corrections can be summarized in terms of $s$-dependent vector and
axial vector form factors if the masses $m_f$ of the external fermions are small
compared to $M_{_W}$, both for the electromagnetic and the weak NC vertex.
In our terminology, `vertex corrections' denote the renormalized
$\g(Z)ff$ three-point functions in  one-loop order, together with the finite
wave function renormalizations for external fermions.
 
\smallskip
In contrast to the propagator corrections, the vertex corrections are
not universal and depend on the fermion species.
For this reason we have to list them separately for $\nu,\, e,\,
u, and  d$ type fermions. In addition, the $b$ quark is exceptional due
to the virtual top contributions in the vertex.
 
\bigskip   \noi
 
Our terminology is as follows.
 
\smallskip \noi
$F_{V,A}^{Zf}$ and $F_{V,A}^{\g f}$ denote the IR finite weak (without 
the virtual photon diagrams) form factors for the $Zff$ and $\g ff$
vertex which, together with the lowest order terms, yield the dressed vertices:
 \\[0.1cm]
 
\bea
\Gr^{Zff} & = & i\,\frac{e}{2s_{_W}c_{_W}}\,\gamu \left\{
 v_f+ F_V^{Zf}(s) - \,\gafi \left[
 a_f+ F_A^{Zf}(s) \right] \right\} \, ,    \nn \\
  &     &   \nn \\
\Gr^{\g ff} & = & -i\,e\,Q_f\, \gamu
          -\,i\,e\,\gamu \left[ F_V^{\g f}(s) - F_A^{\g f}(s)\,\gafi
           \right]  \, .
\label{vertex}
\eea
 
\bigskip \noi
The lowest order coupling reads:
 
\beq
 v_f = I^{(3)}_f -2 Q_f s_{_W}^2\;, \;\;\;\;\;
 a_f = I^{(3)}_f\, ,
\eeq
\noindent
and the weak form factors in (\ref{vertex}) are explicitly given by
the following set of formulae.
 
\smallskip
\noi  {\bf Neutral current vertex:}
 
\smallskip \noi
neutrinos:
\bea
 F_V^{Z\nu} & = & F_A^{Z\nu}   \\
            & = & \alpi                   \left[
                  \frac{1}{8c_{_W}^2 s_{_W}^2} \Lambda_2(s,M_{_Z})
                  +\frac{2s_{_W}^2-1}{4s_{_W}^2}\Lambda_2(s,M_{_W})
                  +\frac{3c_{_W}^2}{2s_{_W}^2}\Lambda_3(s,M_{_W}) \right] 
\; , \nn
\eea
charged fermions:
\bea
 F_V^{Zf} & = & \alpi \left[
          \frac{v_f(v_f^2+3a_f^2)}{4s_{_W}^2c_{_W}^2} \Lambda_2(s,M_{_Z})
                \,+\,F_L^f \right]  \; ,    \nn \\[0.1cm]
 F_A^{Zf} & = & \alpi \left[
          \frac{a_f(3v_f^2+a_f^2)}{4s_{_W}^2c_{_W}^2} \Lambda_2(s,M_{_Z})
                \,+\,F_L^f \right]  \; ,
\label{zvertex}
\eea
\noindent
with
\bea
  F_L^{\ell} & = & \frac{1}{4s_{_W}^2}\,\Lambda_2(s,M_{_W})-
  \frac{3c_{_W}^2}{2s_{_W}^2}\,\Lambda_3(s,M_{_W}) \; ,   \nn \\
F_L^u & = & -\frac{1-\frac{2}{3}s_{_W}^2}{4s_{_W}^2}\,\Lambda_2(s,M_{_W})+
  \frac{3c_{_W}^2}{2s_{_W}^2}\,\Lambda_3(s,M_{_W}) \; ,   \nn \\
F_L^d & = & \frac{1-\frac{4}{3}s_{_W}^2}{4s_{_W}^2}\,\Lambda_2(s,M_{_W})-
  \frac{3c_{_W}^2}{2s_{_W}^2}\,\Lambda_3(s,M_{_W})\; .
\eea
 
\medskip
\noi
     {\bf Electromagnetic vertex: }
\bea
 F_V^{\g f} & = & \alpi \left[
          \frac{Q_f(v_f^2+a_f^2)}{4s_{_W}^2c_{_W}^2} \Lambda_2(s,M_{_Z})
                \,+\,G_L^f \right]  \; ,    \nn \\[0.1cm]
 F_A^{\g f} & = & \alpi \left[
          \frac{Q_f\,2v_f a_f}{4s_{_W}^2c_{_W}^2}\,\Lambda_2(s,M_{_Z})
                \,+\,G_L^f \right]  \; ,
\label{emvertex}
\eea
\noindent
with
\bea
  G_L^{\ell} & = &
  - \frac{3}{4s_{_W}^2}\,\Lambda_3(s,M_{_W}) \; , \nn \\
G_L^u & = & -\frac{1}{12s_{_W}^2}\,\Lambda_2(s,M_{_W})+
  \frac{3}{4s_{_W}^2}\,\Lambda_3(s,M_{_W})   \; , \nn \\
G_L^d & = & \frac{1}{6s_{_W}^2}\,\Lambda_2(s,M_{_W})-
  \frac{3}{4s_{_W}^2}\,\Lambda_3(s,M_{_W})   \; .
\eea
 
\smallskip
\noi
 
The functions $\Lambda_2$, $\Lambda_3$ have the form,
 
\bea
  \Lambda_2(s,M) &  = &  -\frac{7}{2}-2w-(2w+3)\log(-w) \nn \\
                 &    & +
  2(1+w)^2\,\left[\dlog\left(1+\frac{1}{w}\right) -\frac{\pi^2}{6}
  \right]                       \; ,   \nn \\
                 &   &   \nn \\
  \Lambda_3(s,M) & = & \frac{5}{6}-\frac{2w}{3}+
  \frac{(2w+1)}{3}\sqrt{1-4w}\log(x)
   +\frac{2}{3}w(w+2) \log^2(x) \; ,
\eea
\noindent
with
 
$$
  w=\frac{M^2}{s+i\veps}, \;\;\;
  x = \frac{\sqrt{1-4w}-1}{\sqrt{1-4w}+1} \, .
$$
 
\bigskip  \noi
The functions $F^d_L$ and $G_L^d$ cannot be used for $b$ quarks.
The full expressions for $F_L^b,\, G_L^b$ can be found, for example, in the
second of Ref.\ \cite{bhm}
\footnote{ Note that the normalization is different:
$$ F_L^b(\mbox{this report}) = 2 s_{_W} c_{_W}\, F_L^b(
  \mbox{Ref.\ \cite{bhm}}).$$ }.
 
\subsubsection{$\epmf$ amplitudes}
 
\smallskip \noi
Around the $Z$ resonance, the amplitude for $\epmf$  can be cast into a form
close to the lowest order amplitude:
 
\beq
 A(\epmf) \,=\, A_{\g} \, +\, A_{_Z} \,+\, (box) \; ,
\eeq
\noindent
where $A_{\g}$ denotes the dressed photon, $A_Z$ the dressed
$Z$ exchange amplitude, and $(box)$ summarizes the terms from the
massive box diagrams, which, however, can be neglected around the $Z$.
 
\smallskip \noi
The dressed {\bf photon exchange} amplitude can be written in
the following way:
 
\beq
A_{\g} \,=\, \frac{e^2}{1+\Pigr(s)} \;\frac{Q_eQ_f}{s} \;
\left[ (1+\fvge)\gamu-\fage\gamu\gafi \right] \otimes
\left[ (1+\fvgf)\gimu-\fagf\gimu\gafi \right]           \, .
\eeq
$\Pigr$ is the subtracted vacuum polarization
$\Pigr(s) = \Pig(s) -\Pig(0)$ with
 $$
\Pig(s) = \frac{\sg(s)}{s} \, .
$$
The vertex form factors $F_{V,A}^{\gamma f}$ from Eq.\ (\ref{emvertex})
are evaluated for $s=\mz$.
 
\bigskip \noi
The {\bf $Z$ exchange} amplitude without the box diagrams
factorizes as follows:
\bea
A_{_Z} & = & \sqrt{2} \Gmu\mz \; (\rho_e\rho_f)^{1/2} \;     \nn \\
 &   & \times \; \frac
{ \left[\gamu\left(I^{(3)}_e-2Q_e s_{_W}^2\kappa_e\right)
-I^{(3)}_e\gamu\gafi\right]\otimes
  \left[\gimu\left(I^{(3)}_f-2Q_f s_{_W}^2\kappa_f\right)
-I^{(3)}_f\gimu\gafi\right] }
  {s-\mz \,+\,i\,\frac{s}{\mz} \; M_{_Z} \Gamma_{_Z} } \; .
\label{az}
\eea
The weak corrections appear in terms of fermion-dependent
form factors $\rho_f$ and $\kappa_f$ in the coupling constants
and in the width in the denominator.
 
\subsubsection{Effective neutral current couplings}
 
\smallskip \noi
The factorized amplitude (\ref{az}) allows us to define
NC vertices at the $Z$ resonance
with effective coupling constants
$g_{_{V,A}}^f$, synonymously with the use of $\rho_f, \kappa_f$:
\bea
 J_{\m}^{NC} & = & \left( \sqrt{2}\Gmu\mz \rho_f \right)^{1/2}
\left[ \left(I^{(3)}_f-2Q_fs_{_W}^2\kappa_f\right)\gamu
  -I^{(3)}_f\gamu\gafi \right] \nn\\
  & = & \left( \sqrt{2}\Gmu\mz \right)^{1/2} \,
  [g_{_V}^f \,\gamu -  g_{_A}^f \,\gamu\gafi]  \, .
\label{effc}
\eea
The effective couplings read as follows:
\bea
 g_{_V}^f & = &  \left[ v_f +2s_{_W} c_{_W} \, Q_f\, \Pgzrz
             + \fvzf \right] \left[ \frac
              {1-\Dr}{1+\Pizr(\mz)} \right]^{1/2}, \nn \\
 g_{_A}^f & = &  \left[ a_f
             + \fazf \right] \left[ \frac
              {1-\Dr}{1+\Pizr(\mz)} \right]^{1/2} \, .
\label{effva}
\eea
The building blocks are:
\begin{itemize}
\item
 $\Dr$ from Eq.\ (\ref{deltar})
\item
 $\Pizr(\mz)$ from Eq.\ (\ref{piz})
\item
 $\Pgzrz = \sgzr(\mz)/\mz$ from Eq.\ (\ref{rself})
\item
 $F_{V,A}^{Zf}$ from Eq.\ (\ref{zvertex}) for $s=\mz$.
\end{itemize}
For given $m_t,M_{_H}$ the values of $s_{_W}^2$ respectively \ $M_{_W}$ are
chosen such that Eq.\ (\ref{wzcorr}) is fulfilled.
 
\smallskip
For the $b$ quark couplings the next-order leading corrections
$\sim \Gmu^2 m_t^4, \; \sim \alpha_s \Gmu m_t^2$ are taken
into account by performing in the one-loop expression $F_L^b$
the following substitution:
 
\beq
 F_L^b \, \ra \, F_L^b - \frac{\alpha}{16\pi s_{_W}^2 c_{_W}^2} \;
                         \frac{m_t^2}{\mz} + \Delta\tau_b \; ,
\eeq
with
\bea
  \Delta\tau_b & = & x_t \left[ 1 + x_t\, \tau^{(2)}(\xi)
              - \alpha_s(m_t) \frac{\pi}{3} \right]\; .
\eea
 
The function $\tau^{(2)}$ is taken from Refs. \cite{barb,ftj}, and the QCD
correction term from Ref.~\cite{ftjr}.
 
\smallskip \noi
The alternative form factors $\rho$ and $\kappa$ can then be obtained via
 
\beq
 \frac{g_{_V}^f}{g_{_A}^f} = 1 - 4 \mid Q_f\mid \kappa_f s_{_W}^2, \;\;\;\;\;
 \left( \frac{g_{_A}^f}{a_f} \right)^2 = \rho_f \,  .
\eeq
 
\noi
Due to the imaginary parts of the self energies and vertices,
the form factors and the effective couplings, respectively, are
complex quantities. The effective mixing angles are calculated from the real
parts according to
 
\beq
 \sin^2\theta^f_{\rm {eff}} = \frac{1}{4\mid Q_f\mid} \left(
              1-\frac{\real\, g_{_V}^f}{\real\, g_{_A}^f} \right) \, .
\eeq
 
\subsection{{\tt LEPTOP} basics}
\subsubsection{Electroweak loops for hadron-free observables:
 functions $V_i$ }
 
For hadron-free observables we write the result of one-loop electroweak
calculations in the form suggested in ref. \cite{2}:
\begin{eqnarray}
M_{_W}/M_{_Z} &=& c+\frac{3c}{32\pi s^2(c^2-s^2)}
\baral V_m(t,h) \; , \nonumber \\
g^l_{_A} &=& -\frac{1}{2}-\frac{3}{64\pi s^2c^2}\baral V_A(t,h) \; ,
\nonumber \\
R &=& g^l_{_V}/g^l_{_{A}} =
1- 4s^2 + \frac{3}{4\pi(c^2 - s^2)}\baral V_R(t,h) \; ,
\nonumber
\\
g^{\nu} &=& \frac{1}{2} + \frac{3}{64\pi s^2c^2}\baral V_{\nu}(t,h) \; ,
\label{31}
\end{eqnarray}
where $t=m_t^2/M_{_Z}^2$, $h= M_{_H}^2/M_{_Z}^2$
and functions $V_i(t,h)$ are normalized by the condition,
\begin{equation}
V_i(t,h)\simeq t \; ,
\label{32}
\end{equation}
at $t\gg 1$.
 
Each function $V_i$ is a sum of five terms \cite{2}, \cite{3}
\begin{equation}
V_i(t,h) = t + T_i(t)+H_i(h)+C_i+\delta V_i(t,h) \; .
\label{33}
\end{equation}
 
The functions $t + T_i(t)$ are due to  $(t,b)$ doublet contribution
to self-energies  of the vector bosons, $H_i(h)$ is due to
 $W^{\pm},Z$ and
$H$ loops, the constants $C_i$ include light fermion
 contribution both to
self-energies, vertex and box diagrams.
 
To give explicit expressions for $T_i(t)$ and $H_i(h)$
 it is convenient to
introduce three auxiliary functions $F_t(t)$, $F_h(h)$
 and $F'_h(h)$ (see subsection 4.2.5 for their expressions).
 
The equations for $T_i(t)$ and $H_i(h)$ have the form \cite{2}:
$$
\underline{i = m}
$$
\ba
T_m(t) &=& \left(\frac{2}{3} - \frac{8}{9}s^2\right)\ln t -
 \frac{4}{3} + \frac{32}{9} s^2
+ \frac{2}{3}\;(c^2 - s^2)\left(\frac{t^3}{c^6} -
 \frac{3t}{c^2} + 2\right) \ln \mid 1 - \frac{c^2}{t} \mid    \nonumber \\
&&+\; \frac{2}{3} \frac{c^2-s^2}{c^4} t^2 +
      \frac{1}{3} \frac{c^2 - s^2}{c^2} t
  + \left(\frac{2}{3} - \frac{16}{9} s^2 -
    \frac{2}{3} t - \frac{32}{9} s^2t\right) F_t(t)\;,
\nonumber \\
H_m(h) &=& -\frac{h}{h-1} \ln h + \frac{c^2 h}{h-c^2}
\ln \frac{h}{c^2} -
\frac{s^2}{18c^2} h - \frac{8}{3}s^2 +
  +\; \left(\frac{h^2}{9} -\frac{4h}{9} + \frac{4}{3}\right)F_h(h)
\nonumber \\
&&-(c^2 - s^2)\left(\frac{h^2}{9c^4} - \frac{4}{9}
   \frac{h}{c^2} + \frac{4}{3}\right)F_h\left(\frac{h}{c^2}\right)
+ (1.1205 - 2.59\delta s^2)\;,
\label{34}
\ea
where $\delta s^2 = 0.23117 - s^2$~.
$$
\underline{i = A}
$$
\ba
T_A(t) &=& \frac{2}{3} - \frac{8}{9} s^2 + \frac{16}{27} s^4 -
 \frac{1 - 2tF_t(t)}{4t-1}
\nonumber \\
&&+\;\left(\frac{32}{9} s^4 - \frac{8}{3} s^2 - \frac{1}{2}\right)
\left[ \frac{4}{3}tF_t(t) -
\frac{2}{3}(1+2t)\frac{1-2tF_t(t)}{4t-1}\right] \; ,
\nonumber \\
H_A(h) &=& \frac{c^2}{1-c^2/h}\ln \frac{h}{c^2} -
 \frac{8h}{9(h-1)}\ln h 
+\;\left(\frac{4}{3} - \frac{2}{3}h+ \frac{2}{9}h^2\right)F_h(h)
\nonumber \\
&&- \left(\frac{4}{3} - \frac{4}{9}h + \frac{1}{9} h^2\right) F'_h(h)
  -\frac{1}{18}h + (0.7751 + 1.07\delta s^2) \;.
\label{35}
\ea
$$
\underline{i = R}
$$
\ba
T_R(t) &=& \frac{2}{9}\ln t + \frac{4}{9} -
\frac{2}{9}(1+ 11t)F_t(t)\;,
\nonumber \\
H_R(h) &=& -\frac{4}{3} - \frac{h}{18} + \frac{c^2}{1 - c^2/h}\ln
\frac{h}{c^2}
+ \left(\frac{4}{3} - \frac{4}{9}h + \frac{1}{9}h^2\right)F_h(h)
\nonumber \\
&& +\; \frac{h}{1-h}\ln h + (1.3590 + 0.51\delta s^2) \; .
\label{36}
\ea
$$
\underline{i=\nu}
$$
\ba
T_{\nu}(t) &=& T_A(t) \; ,
\nonumber \\
H_{\nu}(h) &=& H_A(h) \; .
\label{37}
\ea
 
\noindent
The constants  $C_i$ are rather complicated  functions of
 $\sin^2 \theta$
and we present their numerical values near $s^2 = 0.23117$:
 
\begin{equation}
C_m = -1.3500 + 4.13 \; \delta s^2      \; ,
\label{501}
\end{equation}
\begin{equation}
C_A = -2.2619 - 2.63 \; \delta s^2      \; ,
\label{502}
\end{equation}
\begin{equation}
C_R = -3.5041 - 5.72 \; \delta s^2      \; ,
\label{503}
\end{equation}
\begin{equation}
C_{\nu} = -1.1638 - 4.88 \; \delta s^2  \; .
\label{504}
\end{equation}
 
\subsubsection{Corrections $\delta V_i$ }
 
Functions $\delta V_i(t,h)$ in Eq. (\ref{33})
 are small corrections to
$V_i$. They can be presented as a sum of five
 terms $\delta_k V_i (k=1 \div 5)$:
 
\begin{enumerate}
 
\item{Corrections to polarization of electromagnetic vacuum due to W boson loop,
$\delta_{_W}\alpha$, and t-quark loop, $\delta_t\alpha$,
are traditionally not included in the running of $\alpha(q^2)$.
We also prefer to consider them together with electroweak corrections.
 This is
especially reasonable because $W$ contribution $\delta_{_W}\alpha$ is
gauge-dependent, while $\delta\alpha_t$ is negligibly small. Here
and for all other electroweak corrections we use the 't Hooft--Feynman
 gauge.
The corrections $\delta_{_W}\alpha$ and $\delta_t\alpha$
were neglected in Ref.  \cite{2} and were introduced in Ref. \cite{7}:
 
\ba 
\delta_1 V_m(t,h) &=& -\frac{16}{3} \pi s^4
\frac{1}{\alpha}(\delta_{_W}\alpha + \delta_t \alpha) = -0.055 \; , \\
\label{405} 
\delta_1 V_R(t,h) &=&
- \frac{16}{3} \pi s^2 c^2 \frac{1}{\alpha}(\delta_{_W}\alpha
+ \delta_t\alpha) = -0.181 \; , \\
\label{38}
 \delta_1 V_A(t,h) &=& \delta_1 V_{\nu}(t,h) = 0\;, 
\label{306}
\ea
where

\ba
\frac{\delta_{_W}\alpha}{\alpha} &=& \frac{1}{2\pi} \left[(3 + 4c^2)\left(1
-\sqrt{4c^2-1} \arcsin \frac{1}{2c}\right) - \frac{1}{3}\right]
= 0.0686 \; ,  \\
\label{307}
\frac{\delta_t\alpha}{\alpha} &=&
-\frac{4}{9\pi}\left[(1+2t)F_t(t) - \frac{1}{3}\right] \simeq -
\frac{4}{45\pi}\frac{1}{t} + ... \simeq - 0.00768 \; .  
\label{308}
\ea
 
(Here and in Eqs.~(\ref{309} -- \ref{319}) we use $m_t = 175$
GeV for numerical estimates.)}
 
\item{Corrections of the order of
$\baral \hat{\alpha}_s$, due to the gluon exchange
in the quark electroweak loops \cite{zbb} (see also Ref.~\cite{3}),
$\delta_2 V_i = \delta_2^q V_i + \delta_2^t V_i$.
For the two generations of light quarks $(q = u,d,s,c)$ this gives:
 
\ba
\delta^q_2 V_m(t,h) &=&
2 \; \left\{\frac{4}{3}
\left[\frac{\hat{\alpha}_s(M_{_Z})}{\pi}\right](c^2-s^2)\ln c^2 \right\}
= \left[\frac{\hat{\alpha}_s(M_{_Z})}{\pi}\right](-0.377) \;,
\label{309}
\\
\delta^q_2 V_A(t,h) &=&
 2 \; \left\{\frac{4}{3}\left[\frac{\hat{\alpha}_s(M_{_Z})}{\pi}\right]
\left(c^2-s^2 + \frac{20}{9} s^4\right)\right\} =
\left[\frac{\hat{\alpha}_s(M_{_Z})}{\pi}\right](1.750),
\label{310}
\\
\delta^q_2 V_R(t,h) &=& 0  \; ,
\label{311}
\\
\delta_2^q V_{\nu} (t,h) &=& \delta_2^q V_A (t,h) \;.
\label{3111}
\ea
 
\noindent
  The results of calculations for the third generation are given
  by rather complicated functions of the top mass and $s^2$. Here
  we present an expansion for large values of $t$ and for fixed value of
  $s^2$ = 0.23117:
 
\ba
 \delta^t_2 V_m(t,h) &=&
 \left[\frac{\hat{\alpha}_s(m_t)}{\pi}\right]
 \left[-2.86\;t + 0.46\ln t - 1.540 - \frac{0.68}{t} - \frac{0.21}{t^2}\right]
\nonumber \\
 &=&\frac{\hat{\alpha}_s(m_t)}{\pi}(-11.67) \; ,
\label{312}
\\
\delta^t_2 V_A(t,h) &=&
 \left[\frac{\hat{\alpha}_s(m_t)}{\pi}\right]
 \left[-2.86\;t + 0.493 - \frac{0.19}{t} - \frac{0.05}{t^2}\right]
\nonumber \\
&=& \frac{\hat{\alpha}_s(m_t)}{\pi}(-10.10) \; ,
\label{313}
\\
\delta^t_2
V_R(t,h) &=&
 \left[\frac{\hat{\alpha}_s(m_t)}{\pi}\right]
 \left[-2.86\;t + 0.22 \ln t - 1.513 - \frac{0.42}{t} -
 \frac{0.08 }{t^2}\right]
\\
&=& \frac{\hat{\alpha}_s(m_t)}{\pi}(-11.88) \; ,
\label{314}
\\
\delta_2^t V_{\nu} (t,h) &=& \delta_2^t V_A (t,h) \;.
\label{3141}
\ea
 
\noindent
As these formulas are valid for $m_t > M_{_Z}$,
 in order to go to the region
$m_t < M_{_Z}$ we either put $\delta^t_2 V_i = 0$
 or use a massless limit in which
$\delta^t_2 V_i = \frac{1}{2}\delta^q_2 V_i$.
 In any case, this region gives
a tiny contribution to the global fit.}
 
\item{Corrections of the order of $\baral \hat{\alpha}
^2_s$ were calculated for the leading term $\baral
\hat{\alpha}^2_s t$ only~\cite{afmt}
\begin{equation}
\delta_3 V_i(t,h) \simeq -(2.1552 - 0.18094 \; N_f)
 \hat{\alpha}^2_s(m_t)t
\simeq -1.250 \; \hat{\alpha}^2_s (m_t)t = -0.06
\label{315}
\end{equation}
 
\noindent
for $N_f = 5$ light flavours. [For the numerical estimate we use
$\hat{\alpha}_{_S}(M_{_Z})=0.125$.]}
 
\item{The leading correction of the order
$\baral ^2 t^2$, which originates from
the second-order Yukawa interaction,
was calculated in Refs.~\cite{barb}-\cite{ftj}:

\begin{equation}
\delta_4 V_i(t,h) =
-\frac{\baral }{16\pi s^2c^2} A\left(\frac{h}{t}\right) \; t^2 \; ,
\label{316}
\end{equation}
 
\noindent
where function $A(M_{_H}/m_t)$ can be found in
Refs.~\cite{barb},\cite{ftj}. For
$m_t = 175$ GeV and $M_{_H} = 300$ GeV one has
 $A = 8.9$ and $\delta_4V_i(t,h) = -0.11$.}
 
\item{In the second order in electroweak interactions quadratic dependence 
appears on the Higgs mass \cite{19}

\ba
\delta_5 V_m &=&
 \frac{\bar{\alpha}}{24\pi}\left(\frac{h}{c^2}\right)
\; 0.747 = 0.0011 \; ,
\label{317}
\\
\delta_5 V_A &=&
 \frac{\bar{\alpha}}{24\pi}\left(\frac{h}{s^2}\right)
\; 1.199 = 0.0057 \; ,
\label{318}
\\
\delta_5 V_R &=&
-\frac{\bar{\alpha}}{24\pi}\left(\frac{h}{c^2}\right)\frac{c^2 -s^2}{s^2}
\; 0.973 =-0.0032 \; .
\label{319} 
\ea
 
(Here for numerical estimates we use $m_H = 300$ GeV.)}
 
\end{enumerate}
 
\subsubsection{Hadronic decays of $Z$ boson}
 
For the partial width of the $Z$ decay into a pair of quarks
$q\bar{q}$ ($q = u, d, s, c, b$), we use the equation
$$
\Gamma_q = 12\;\left[(g^q_{_V})^2 R^q_{_V}+(g^q_{_A})^2 R^q_{_A}\right]\;
\Gamma_0\;,
$$
\noindent
where the final state QED and QCD radiative functions $R^q_{_V}$ and
$R^q_{_A}$ are given by Eqs. (\ref{78}) and (\ref{79}) respectively,
and $\Gamma_0$
is defined by Eq. (\ref{B}). The electroweak radiative corrections are
included in $g^q_{_V}$ and $g^q_{_A}$:
 
\begin{equation}
g^q_{_A} =
 I^{(3)}_q \left[1+\frac{3\baral }{32\pi s^2 c^2}V_{Aq}(t,h)\right]\;,
\label{320}
\end{equation}
\begin{equation}
g^q_{_V}/g^q_{_A} = 1-4|Q_q|s^2 +\frac{3|Q_q|}{4\pi(c^2
-s^2)}\baral V_{Rq}(t,h) \; .
\label{321}
\end{equation}
 
\noindent
The functions $V_{Aq}(t,h)$ and $V_{Rq}(t,h)$
 in the one-loop electroweak
approximation are related to the functions $V_A(t,h)$ and $V_R(t,h)$
from leptonic decays~\cite{5}:
 
\ba
V_{Au}(t,h)&=&V_{Ac}(t,h) = V_{A}(t,h) + \frac{128\pi s^3 c^3}{3 \bar{\alpha}}
 (F_{Al}+F_{Au}) \; ,
\label{322}
\\
V_{Ad}(t,h)&=&V_{As}(t,h) = V_{A}(t,h) + \frac{128\pi s^3 c^3}{3 \bar{\alpha}}
 (F_{Al}-F_{Ad}) \; ,
\label{323}
\\
V_{Ru}(t,h)&=&V_{Rc}(t,h) = V_{R}(t,h)+ \frac{16\pi sc(c^2 - s^2)}{3
 \bar{\alpha}}
\\
&&\times [F_{Vl}-(1-4s^2)F_{Al} + \frac{3}{2}(-(1-\frac{8}{3}s^2 )F_{Au}+F_{Vu})]
\; ,
\label{324}
\nonumber \\
V_{Rd}(t,h)&=&V_{Rs}(t,h) = V_{R}(t,h)+ \frac{16\pi sc(c^2 - s^2)}{3
 \bar{\alpha}}
\\
&&\times [F_{Vl}-(1-4s^2)F_{Al} + 3((1-\frac{4}{3}s^2 )F_{Ad}-F_{Vd})] \; ,
\label{325}
\ea
where $F_{Vl} = 0.00197\;,\;\;\; F_{Al} = 0.00186\;,\;\;\;
F_{Vu} = -0.00169\;,\;\;\; F_{Au} = -0.00165\;,\;\;\; F_{Vd} = 0.00138\;,
\;\;\; F_{Ad} = 0.00137$~\cite{5}.
The difference $V_{i,q} - V_i$
 is due to different electroweak corrections
to the vertices $Zq\bar{q}$ and $Zl\bar{l}$
 ($s^2$= 0.23117 is assumed: formulas used in the code have
explicit $s^2$ dependence).
 
The oblique corrections of the order of $\hat{\alpha}_s$,
$\hat{\alpha}_s^2 t$ and $\baral t^2$ to the $V_{Aq}(V_{Rq})$ are the same
as in the case of $V_A$ and $V_R$. But for $Z$ boson decay into pair
$q\bar{q}$ there are additional $\hat{\alpha}_s$ corrections to the
vertices that have not yet been calculated.
 This brings additional uncertainty into the theoretical accuracy.
 
\subsubsection{Specific features of the decay $Z\to b\bar{b}$ }
 
For $Z\to b\bar{b}$ decays we have
 to take into account corrections to the $Z\to b\bar{b}$ vertex which
depend on $t$ \cite{aasse}, \cite{ftjr}.

\ba
V_{Ab}(t,h) &=& V_{Ad}(t,h)
 - \frac{8 s^2 c^2}{3(3-2s^2)}\left[\phi(t)+\delta \phi(t)\right] \; ,
\\
V_{Rb}(t,h) &=& V_{Rd}(t,h)
 - \frac{4 s^2 (c^2-s^2)}{3(3-2s^2)}\left[\phi (t) + \delta \phi (t)\right]\; .
\ea

For $\phi(t)$ we use the following expansion \cite{aasse}

\ba
\phi(t)&=&\frac{3-2s^2}{2s^2 c^2}\Biggl\{t+c^2\Biggl[2.88
ln\left(\frac{t}{c^2}\right) -6.716
  +\frac{1}{t}\left(8.368\;c^2\ln\left(\frac{t}{c^2}\right)-3.408\;c^2\right)
\nonumber \\&&+\;
 \frac{1}{t^2}\left(9.126\;c^4\ln\left(\frac{t}{c^2}\right)+2.26 \;c^4\right)
+\frac{1}{t^3}\left(4.043\;c^6\ln\left(\frac{t}{c^2}\right)+7.41 \;c^6\right)
\nonumber \\&&+\;
...\Biggr] \Biggr\} \; ,
\label{402}
\ea

and for $\delta\phi(t)$ we use the leading approximation calculated
in Refs.~\cite{ftjr} and~\cite{barb},~\cite{ftj}

\begin{equation}
\delta\phi(t) = \frac{3-2s^2}{2s^2c^2}\left
\{-\frac{\pi^2}{3}\left[\frac{\hat{\alpha}_s(m_t)}{\pi}\right]\;t 
+\frac{1}{16s^2c^2}\left(\frac{\bar{\alpha}}{\pi}\right)t^2
\tau_b^{(2)}\left(\frac{h}{t}\right)\right\} \; ,
\label{403}
\end{equation}
 
\noindent
where function
 $\tau_b^{(2)}$ has been calculated in Ref.~\cite{ftj}. 

For $m_t= 175$ GeV and  $M_{_H} = 300 $ GeV
$$ \tau_b^{(2)} = 1.245\;\; .
$$
 
\underline{Asymmetries}
are calculated with the loop corrected values of $g_{_V}$ and $g_{_A}$.

\subsubsection{Appendix: Auxiliary functions $F_t$ and $F_h$}
\vspace{0.5cm}
 
\begin{equation}
F_t(t) = \left\{ \begin{array}{ccc}
2(1-\sqrt{4t-1} \arcsin \frac{1}{\sqrt{4t}}) & , & 4t > 1 \\
 & & \\
2(1-\sqrt{1-4t} \ln \frac{1+\sqrt{1-4t}}{\sqrt{4t}}) & , & 4t < 1
\end{array}
\right.
\end{equation}
 
\begin{equation}
F_h(h) = \left\{ \begin{array}{ccc}
1+(\frac{h}{h-1}-\frac{h}{2})\ln~ h + h\sqrt{1-\frac{4}{h}}
\ln\left(\sqrt{\frac{h}{4}-1} +\sqrt{\frac{h}{4}}\right) & , & h > 4 \;\;, \\
 & & \\
1+(\frac{h}{h-1}-\frac{h}{2})\ln~ h - h\sqrt{\frac{4}{h}-1}
\arctan \sqrt{\frac{4}{h}-1} & , & h < 4 \;\;,
\end{array}
\right.
\end{equation}
 
\begin{equation}
F'_h(h) = \left\{ \begin{array}{ccc}
-1+\frac{h-1}{2}\ln~ h
 +(3-h)\sqrt{\frac{h}{h-4}}\ln\left(\sqrt{\frac{h}{4}-1}
+\sqrt{\frac{h}{4}}\right) & , & h > 4  \;\;, \\
 & & \\
-1+\frac{h-1}{2}\ln~ h
+(3-h)\sqrt{\frac{h}{4-h}}\arctan\sqrt{\frac{4}{h}-1}& , & h < 4 \;\;.
\end{array}
\right.
\end{equation}
 
\subsection{{\tt TOPAZ0} basics}
The {\it realization} given in Ref.~\cite{myr} describes the coupling of the
$\z0$ as:
 
\begin{equation}
\srt\,(\gf\rhoz)^{1/2}\zm \gamma^{\mu} \left[ \i3f - 2 Q_f\shat^2 +
\dgvf + \left(\i3f + \dgaf\right)\gfd\right] \;.
\end{equation}
 
\noindent
Before giving the specific expression of the various quantities entering the
previous equation, we stress that our metric is such that a time-like momentum
squared is negative. Next we decompose the unrenormalized vector boson
self-energies as:
 
\begin{eqnarray}
S_{_{\gamma\gamma}}\ap2 &=& {{g^2\stws}\over {\osp2}}\,\Pgg\ap2 p^2 \;,  \nll
S_{_{ZZ}}\ap2 &=& {{g^2}\over {\osp2\ctws}}\,\Szz\ap2 \; ,  \nll
S_{_{Z\gamma}}\ap2 &=& {{g^2\stw}\over {\osp2\ctw}}\,\Szg\ap2 \; ,  \nll
S_{_{WW}}\ap2 &=& {{g^2}\over {\osp2}}\,\Sww\ap2 \; ,  \nll
\Szz\ap2 &=& \Stt\ap2 - 2\,\stws\Stg\ap2 + \stwf\Pgg\ap2 p^2 \; ,  \nll
\Szg\ap2 &=& \Stg\ap2 - \stws\Pgg\ap2 p^2 \; ,
\label{topazse}
\end{eqnarray}
 
\noindent
where $\theta$ denotes the bare mixing angle.
After a re-diagonalization in the neutral sector, which makes 
$S_{_{Z\gamma}}(0)=0$ 
in the $\xi = 1$ gauge and replaces $\Sww,\Stt$ and $\Stg$ with
$\Sww + 4\Gamma, \Stt + 4\Gamma$ and $\Stg + 2\Gamma$
with $\Gamma = M^2B_0(0,M^2,M^2)$ --- $M$ being the bare $W$ mass ---
we consider the ultraviolet and infrared
finite corrections to the $\mu$-decay $\delta_{_G}$ and introduce
 
\begin{equation}
\Swwg = \Sww(0) + {{\sqrt{2}\pi\alpha}\over {\gf}}\,\delta_{_G} \; ,
\end{equation}
 
\noindent
which has the virtue of being gauge invariant, a property not satisfied by
$\Sww(0)$ alone. The extra term induced by the diagonalization is
gauge invariant by construction.
From now on we will denote the fermionic (bosonic) contributions to a given
quantity $A$ with the notation $A^{\rm {ferm}}(A^{\rm {bos}})$.
With these quantities we define
 
\begin{eqnarray}
\rhoz &=& {1\over {{1 + {{\gf\zm^2}\over {2\srt\pi^2}}\,\Sigma^{\rm {ferm}} +
{\mbox{h.o.}}}}} \; ,  \nll
\zm^2 \Sigma &=& \Swwg - {\cal R}e\Stt(\zm^2) 
+ {\cal R}e\Stg(\zm^2) + s^2c^2\zm^2 \Pf^{\rm {bos}}(\zm^2) \; ,
\label{rhodef}
\end{eqnarray}
 
\noindent
where
 
\begin{equation}
s^2c^2 = {{\pi\alpha(\zm)}\over {\sqrt{2}\gf\zm^2}} \; ,
\end{equation}
 
\noindent
and $\Pf(\zm^2) = {\cal R}e\Pgg(\zm^2) - \Pgg(0)$.
Note that $\Sigma$ is ultraviolet finite. Next we can introduce $\shat^2$ as:
 
\begin{equation}
\shat^2 = \frac{1}{2}\,\left[\, 1 - \sqrt{1 - \frac{4\,\pi\alpha(\zm)}
{\srt\,\gf\zm^2\rhozr}}\,\,\right] \; ,
\label{cdshat}
\end{equation}
 
\noindent
and give the definition of $\dgvf,\dgaf$,
 
\begin{eqnarray}
\dgvf &=& \frac{\alpha}{4\pi}\left[ {{2\fvf - \frac{1}{2} v_f \Delta\Pi_{_Z}}
\over {c^2s^2}} - 2 Q_f\Delta s^2 \right] \;,  \nll
\dgaf &=& \frac{\alpha}{4\pi}\left[ {{2\faf - \frac{1}{2} \i3f \Delta\Pi_{_Z}}
\over {c^2s^2}} \right] \; ,
\end{eqnarray}
 
\noindent
where $v_f = \i3f -2 Q_f s^2$,  and
 
\begin{eqnarray}
\zm^2 \Delta\Pi_{_Z} &=& {\cal R}e\left\{\left(\Swwg\right)^{\rm {bos}}
- \Stt^{\rm {bos}}(\zm^2)
- \Stg^{\rm {ferm}}(\zm^2) - \zm^2\Sigma'_{_{33}}(\zm^2)\right. \nll
 & &+ \; \left. 2 s^2\left[\Stg(\zm^2) + \zm^2\Sigma'_{_{3Q}}
(\zm^2)\right] + s^4\zm^4\Pi'_{_{\gamma\gamma}}(\zm^2)\right\} \; .
\label{cdpiz}
\end{eqnarray}
 
\noindent
Here $f'$ stands for $-df/dp^2$. The $F$ refers to flavour-dependent vertex
corrections.
 
\begin{itemize}
 
\item If no resummation of bosonic self-energies is performed we have
 
\begin{equation}
\rhozr = \rhoz, \quad {\hbox{and}} \quad
\Delta s^2 = {{\Sigma^{\rm {bos}}}\over {c^2 - s^2}} \; ;
\label{nores}
\end{equation}
 
\item otherwise $\rhozr$ has the same structure of $\rhoz$ with 
$\Sigma^{\rm {ferm}}$ replaced by $\Sigma_{_R}$:
 
\begin{eqnarray}
\rhozr &=& {1\over {{1 + {{\gf\zm^2}\over {2\srt\pi^2}}\,\Sigma_R +
{\mbox{h.o.}}}}} \; ,  \nll
\Sigma_R &=& \left[\Sigma^{tot}  - (c^2 - s^2)\,
{{{\cal R}e\Szg(\zm^2)}\over \zm^2}\right]_{\msb} \; ,
\label{myres}
\end{eqnarray}
 
\noindent and
 
\begin{equation}
\zm^2 \Delta s^2 = \left[{\cal R}e\Szg(\zm^2)\right]_{\msb} \;.
\end{equation}
 
\end{itemize}
 
\noindent
$\Delta\Pi_{_Z}$ is the residual $\z0$ wave function factor which obtains by
writing the $\z0$ propagator $\chiz$ as
 
\begin{eqnarray}
{\cal R}e\chiz^{-1} &=& \left(1 + {{\gf\zm^2}\over {2\sqrt{2}\pi^2}}\Pi_{_Z}
\right)\,(p^2 + \zm^2) \; ,  \nll
\zm^2\Pi_{_Z} &=& \Swwg - {1\over {p^2+\zm^2}}\,{\cal R}e\left[{p^2\over\zm^2}
\Szz(\zm^2) + \Szz(p^2)\right] \; , \nll
\Pi_{_Z} &=& \Sigma^{\rm {ferm}} + \Delta\Pi_{_Z} \; .
\end{eqnarray}
 
The resummation operates on a gauge invariant quantity since it can be
proved that
 
\begin{equation}
\left[\zm^2\,{{\Sigma}\over {c^2 - s^2}} - {\cal R}e\Szg\left(\zm^2\right)
\right]^{\rm {bos}} = {1\over {c^2 - s^2}}\,\left[\Swwg - s^2c^2\zm^2\Pgg(0) -
{\cal R}e\Szz(\zm^2)\right]^{\rm {bos}},
\end{equation}
 
\noindent
and the sum of the terms in the last parenthesis is automatically
gauge invariant.
The same quantity, however, is not ultraviolet-finite and therefore has to
be strictly understood in the $\msb$ sense at the scale $\mu = \zm$.
The higher-order terms in Eq.~(\ref{rhodef}) are given by:
 
\begin{eqnarray}
{\hbox{h.o.}} &=&{{\gf\zm^2}\over {2\pi^2}}\Delta\Sigma_2
+ \frac{1}{2}\,\frac{\als(m_t)}{\pi}
\left(1 + \frac{\pi}{3}\right) m_t^2\left[1 + {\ord}\left(\frac{\zm^2}{m_t^2}
\right)\right] \nll
& & + \; \frac{\als(\zm)}{\pi}\,\Delta\Sigma_{lq} +
10.55\,\left(\frac{\als(m_t)}{\pi}\right)^2 \; ,
\end{eqnarray}
 
\noindent
where $\Delta\Sigma_2$ is the two-loop factor proportional to
$m_t^4/\zm^4$, $\Delta\Sigma_{lq}$ denotes the $\ord(\alpha\als)$ light quark
contribution and the last factor is the three-loop correction computed for
six active flavours. In the special case of the $Z \to b \bar b$ width we
include
the well known correction factor~\cite{barb}, which modifies the vector and
axial-vector couplings into $1 - {4}/{3}\shat^2 + \tau$ and $1 + \tau$.
 
To illustrate the internal structure of the renormalization procedure we use
a well defined and generalizable example. The three bare parameters of the
MSM Lagrangian are related to experimental data by one-loop relations. One
example of a solution concerns the bare mixing-angle
 
\begin{eqnarray}
\sin^2\theta &=& s^2 + {\alpha\over {4\pi}}\,s_1 + \left({\alpha\over
 {4\pi}}\right)^2
s_2  \; , \nll
s_1 &=& {1\over {c^2-s^2}}\Sigma - {1\over \zm^2}{\cal R}e\Szg(\zm^2) \; , \nll
s_2 &=& {1\over {(c^2-s^2)^3}}\Sigma^2 - \Pgg(0)s_1 \; .
\end{eqnarray}
 
\noindent
For the $\z0 \to \barf f$ amplitude the self-energy corrections give rise
to
 
\begin{equation}
\gamma^{\mu}\left[ \i3f -2Q_fV_f + \i3f\gamma^5\right] \; .
\end{equation}
 
\noindent
Thus:
 
\begin{eqnarray}
V_f &=& \sin^2\theta + {\alpha\over {4\pi}}\,{1\over \zm^2}\,
{{{\cal R}e\Szg(\zm^2)}\over {1 - {\alpha\over {4\pi}}\Pf}} +
{\ord}(\alpha^2)  \nll
{} &=& s^2 + {\alpha\over {4\pi}}{\Sigma\over{c^2-s^2}} +
\left({\alpha\over {4\pi}}\right)^2\left[
{{\Sigma^2}\over{(c^2-s^2)^3}} + {{\Sigma\Pf}\over{c^2-s^2}}\right] +
{\ord}(\alpha^3)  \nll
{} &=& s^2 + {\alpha(\zm)\over {4\pi}}\,{{\Sigma}\over{c^2-s^2}} +
\left[{\alpha(\zm)\over {4\pi}}\right]^2{{\Sigma^2}\over {(c^2-s^2)^3}} +
{\ord}(\alpha^3) \; .
\end{eqnarray}
 
\noindent
After this result we may perform a partial resummation ---
 
\begin{equation}
V_f = \shat^2 + {\alpha(\zm)\over {4\pi}}\,{1\over{c^2-s^2}}
\Sigma^{\rm {bos}} + \dots
\end{equation}
 
\noindent
As a final comment let us consider the term proportional to the square of
the $\z0-\gamma$ transition in the propagators. It would contribute to
$\Pi_{_Z}$ with
 
\begin{eqnarray}
\Delta\Pi_{_Z} &\to& \Delta\Pi_{_Z} - {\gf\over {2\pi^2}}s^2c^2\,
{p^2\over {\zm^2(p^2+\zm^2)}}\,\left[ {\cal R}e\Szg(\zm^2) +
{{\zm^2}\over p^2}{\cal R}e\Szg(p^2)\right]^2  \nll
{} &=& -{\gf\over {2\pi^2}}s^2c^2{p^2\over \zm^2}(p^2+\zm^2)\left[
-{\cal R}e\Sigma'_{Z\gamma}(\zm^2) + {{{\cal R}e\Szg(\zm^2)}\over \zm^2}
+ \dots\right]^2 \; ,
\end{eqnarray}
 
\noindent
where $\dots$ indicates terms of ${\ord}\left(p^2+\zm^2\right)$.
Thus the additional term, being ${\ord}(p^2+\zm^2)$, does not contribute
to the $\z0$ wave function renormalization factor.
It should be noted that there is an easy dictionary of translation with other
realizations, for instance:
 
\begin{eqnarray}
\rho_f &=& 4\,\rhoz \left(\i3f + \dgaf\right)^2 \; ,  \nll
2 Q_f s_W^2 \kappa_f &=& \i3f\,{{2 Q_f\shat^2 + \dgaf - \dgvf}\over
{\i3f + \dgaf}} \; .
\end{eqnarray}
 
\noindent
Once electroweak corrections are included in the formulation,
{\tt TOPAZ0} implements initial state QED corrections by a convolution on the
weakly corrected kernel distributions with a radiator function, or with
structure functions --- depending on the experimental set-up. Resummation of
soft photon effects and hard photon emission up to ${\ord}(\alpha^2)$
is taken into account, and final state QED radiation with realistic cuts, QED
initial--final interference, and initial state leptonic and hadronic pair
production are also included.
In the next two appendices further details are given for pure electroweak
corrections.
 
\subsubsection{Appendix 1: The self-energies}
 
Starting from the decomposition of Eq.~(\ref{topazse}) for the vector boson
self-energies we give their general expression in terms of two-point scalar
form factors~\cite{ossch1}. In the following, the first argument, $p^2$,
is always left understood:
 
\begin{eqnarray}
           \Pgg &=& \frac{2}{3} - 12\,B_{21}(\wm,\wm) +  7\,B_0(\wm,\wm) \nll
           & & + \; 4\sum_g\,\left[
            B(m_l,m_l) + \frac{4}{3}\,B(m_u,m_u) + \frac{1}{3}\,
            B(m_d,m_d) \right] \; , \nll
           \Stg &=& p^2\,Biggl\{\frac{2}{3} - 10\,B_{21}(\wm,\wm) +
            \frac{13}{2}\,B_0(\wm,\wm) \nll
           & & + \; \sum_g\,\left[
            B(m_l,m_l) + 2\,B(m_u,m_u) + B(m_d,m_d) \right] Biggr\}
            -2\,\wm^2B_0(\wm,\wm) \; , \nll
           \Stt &=& p^2\,\Pi_{_{33}} + \sigma_{_{33}} \; , \nll
           \Sww &=& p^2\,\left[\Pi^0_{_{WW}}+s^2\Pi^1_{_{WW}}\right] +
                    \left[\sigma^0_{_{WW}}+s^2\sigma^1_{_{WW}}\right] \; , \nll
           \Pi_{_{33}} &=& \frac{2}{3} -9\,B_{21}(\wm,\wm) + \frac{25}{4}\,
           B_0(\wm,\wm)
           - B_{21}(\zm,\hm) \nll
           & & - \; B_1(\zm,\hm) - \frac{1}{4}\,B_0(\zm,\hm) \nll
           & & + \; \frac{1}{2}\sum_g\,\left[ B(m_l,m_l) + B(m_{\nu},m_{\nu})
           + 3\,B(m_u,m_u) + 3\,B(m_d,m_d) \right] \; , \nll
           \sigma_{_{33}} &=& -2\,\wm^2B_0(\wm,\wm) + \frac{1}{2}\,\zm^2B_1(\zm,
           \hm) + \frac{5}{4}\,\zm^2B_0(\zm,\hm) \nll
           & & - \;  \frac{1}{2}\,\hm^2
           B_1(\zm,\hm) - \frac{1}{4}\,\hm^2B_0(\zm,\hm)  \nll
           & & - \; \frac{1}{2}\sum_g\,\left[
           m_{\nu}^2B_0(m_{\nu},m_{\nu}) + m_l^2B_0(m_l,m_l) \right. \nll
           & & \; \left.+3\,m_u^2B_0(m_u,m_u) + 3\,m_d^2B_0(m_d,m_d) \right]\;,
\nll
           \sigma^0_{_{WW}} &=& \frac{9}{2}\,(\zm^2 - \wm^2)B_1(\zm,\wm) +
           \frac{1}{4}\,(13\zm^2 - 21\wm^2)B_0(\zm,\wm) \nll
           & & + \; \frac{1}{2}\,
           (\wm^2 - \hm^2)B_1(\wm,\hm) + \frac{1}{4}\,(5\wm^2 - \hm^2)
           B_0(\wm,\hm)  \nll
           & & + \; \sum_g\,\left[ (m_l^2 - m_{\nu}^2)B_1(m_{\nu},m_l)
           - m_{\nu}^2B_0(m_{\nu},m_l) +3\,(m_d^2 - m_u^2)B_1(m_u,m_d)
           \right. \nll
           & & \; \left.- 3\,m_u^2B_0(m_u,m_d) \right] \; , \nll
           \sigma^1_{_{WW}} &=& 2\,(\wm^2 - \zm^2)\,\left[
           2B_1(\zm,\wm) + B_0(\zm,\wm) \right]  -2\,\wm^2\left[
           2\,B_1(0,\wm) + B_0(0,\wm)\right], \nll
           \Pi^0_{_{WW}} &=& \frac{2}{3} - 9B_{21}(\zm,\wm) - 9B_1(\zm,\wm)
           + \frac{7}{4}\,B_0(\zm,\wm) \nll
           & & - \;  B_{21}(\wm,\hm) - B_1(\wm,\hm) -
           \frac{1}{4}\,B_0(\wm,\hm) \nll
           & & + \; 2\sum_g\,\left[ B_{21}(m_{\nu},m_l) + B_1(m_{\nu},m_l)
           + 3\,B_{21}(m_u,m_d) + 3\,B_1(m_u,m_d) \right] \; , \nll
           \Pi^1_{_{WW}} &=& 8\,\left[B_{21}(\zm,\wm) + B_1(\zm,\wm) -
           B_{21}(0,\wm) -  B_1(0,\wm)\right]  \nll
           &+& 2\,\left[B_0(0,\wm) - B_0(\zm,\wm)\right] \; ,
\end{eqnarray}
 
\noindent
where $B = 2\,B_{21} - B_0$ and the sum is over the fermionic generations.
The factor $\Delta$ is given by $\Delta = - 2/(n-4) + \gamma_{_E} - \ln4\pi$.
The functions $\chi(x),\; G_n(y)$ are defined by~\cite{gv}
 
\begin{eqnarray}
\chi(x) &=& -p^2x^2 + (p^2+m_2^2-m_1^2)x + m_1^2 \; , \nll
G_n(y) &=& \int_0^1\,dx x^{n-1}\,\ln\,(x-y) \; .
\end{eqnarray}
 
\noindent
In terms of $\chi$ (where $m^2 \to m^2 -i\epsilon$) we have
 
\begin{eqnarray}
B_0 &=& \Delta - \int_0^1\,dx \ln\chi \; ,  \nll
B_1 &=& -\frac{1}{2}\Delta + \int_0^1\,dx x\ln\chi \; ,  \nll
B_{21} &=& \frac{1}{3}\Delta - \int_0^1\,dx x^2\ln\chi \; ,
\end{eqnarray}
 
\noindent
and the corresponding integrals can be written in terms of the $G_n$-functions,
for which we write recursion formulae to be worked upwards or downwards,
according to the magnitude of $y$.
 
The $\ord(\alpha\als)$ contributions to $\Pgg \dots \Sww$ are computed
according to the formulation of Kniehl (see Ref.~\cite{aasse}). For
instance,
 
\begin{eqnarray}
\Pgg &=& \Pgg + \Pgg^{\alpha\als},  \nll
\Pgg^{\alpha\als} &=& \frac{64}{9}\; {{\alpha_s(m_t)}\over {\pi}}\;
{{m_t^2}\over {p^2}}\,\left[r X + V_1(r)\right] \; ,
\end{eqnarray}
 
\noindent
where $r = -{1}/{4}\;{p^2}/{m_t^2}$ and the functions $X$ and $V_1$
are explicitly given in Ref.~\cite{aasse}.
 
\subsubsection{Appendix 2: The $Zf\bar f $ vertices}
 
In order to introduce the vertex contributions $F_{_{V,A}}^f$ we first present
the fermion wave function renormalization factors ($f\not= b$):
 
\begin{eqnarray}
W_{_V}^f &=& -\frac{1}{32}\,\left[\left(\i3f - 2Q_f\,s^2\right)^2 + 1\right]\,
             F_{_Z} - \frac{1}{8}\,c^2\,F_{_W} \; ,  \nll
W_{_A}^f &=& -\frac{1}{16}\,\left(1 - 8\i3f Q_f\,s^2\right)\,F_{_Z} -
             \frac{1}{8}\,c^2\,F_{_W},
\end{eqnarray}
 
\noindent
where $F_{_X} = \Delta - \ln\left(M_{_X}^2\right) - \frac{1}{2}$. With them
we can write ($f\not=b$):
 
\begin{eqnarray}
F_{_V}^f &=& W_{_V}^f v_f + W_{_A}^f\i3f -2\,v_f\left(v_f^2+\frac{3}{4}\right)
             F_1^f + 4\,\i3f c^2 F_2^f -c^4\i3f F_3^f \; ,  \nll
F_{_A}^f &=& W_{_V}^f \i3f + W_{_A}^f v_f -2\,\i3f\left(3\,v_f^2+\frac{1}{4}
             \right) F_1^f + 4\,\i3f c^2 F_2^f - c^4\i3f F_3^f \; ,
\end{eqnarray}
 
\noindent
where again $v_f = \i3f - 2\,Q_f s^2$. The functions $F_i^f$ are given in
terms of three-point scalar formfactors~\cite{ossch1}. For instance,
 
\begin{eqnarray}
F_1^e &=&  -\frac{1}{4}\,C_{24}(m_e^2,\zm^2,m_e^2) - \frac{1}{8}\,p^2\,
           \left[C_{11}(m_e^2,\zm^2,m_e^2) + C_{23}(m_e^2,\zm^2,m_e^2)\right]
           + \frac{1}{8} \; ,  \nll
F_2^e &=&  -\frac{1}{4}\,C_{24}(m_{\nu}^2,\wm^2,m_{\nu}^2) - \frac{1}{8}\,p^2\,
           \left[C_{11}(m_{\nu}^2,\wm^2,m_{\nu}^2) +
            C_{23}(m_{\nu}^2,\wm^2,m_{\nu}^2)\right]
           + \frac{1}{8} \; ,  \nll
F_3^e &=&  -6\,C_{24}(\wm^2,m_{\nu}^2,\wm^2) - p^2\,\left[
           C_0(\wm^2,m_{\nu}^2,\wm^2) + C_{11}(\wm^2,m_{\nu}^2,\wm^2)\right.
           \nll
           & & + \; \left. C_{23}(\wm^2,m_{\nu}^2,\wm^2)\right]
           + 1 + \Delta - \ln\,(\wm^2) \; .
\end{eqnarray}
 
\noindent
In {\tt TOPAZ0} the most general (arbitrary internal and external masses) two-,
three- and four-point scalar functions are available~\cite{tv}.
For $b \bar b$ final states the expression for vertices contains additional
$m_t$-dependent terms which can be found in Ref.~\cite{myr}.
 
\subsection{{\tt ZFITTER} basics}
Here, we introduce explicit expressions for
$\Delta r$ and the weak form factors of $Z$ decay and of fermion pair
production process $\epmf$
 $\rho$, $\kappa$ to order (1loop,$\alpha$) and
order (2loop,$\alpha\alpha_s$) as used in Eqs.(\ref{romus})-(\ref{kappafs})
and~(\ref{drrem})-(\ref{kaprem}) of the main text.
\subsubsection{Muon life-time}
The virtual, non-photonic one-loop corrections to the muon life-time are:
\ba
\Delta r^{\mathrm{1loop},\alpha}
&=& \frac{\alpha}{4\pi}
\Biggl\{
-\frac{2}{3} \left( 1+2 \sum_f Q_f^2 \ln \frac{m_f^2}{M_{_W}^2}\right)
+ \frac{R}{(1-R)^2}\left[{ W}(M_{_W}^2) - { Z}(M_{_Z}^2)\right]
\nll &&+~\frac{1}{1-R}
\left[{ W}(0) - { W}(M_{_W}^2) -\frac{5}{8}R(1+R)+\frac{11}{2}+
\frac{9}{4}\frac{R}{1-R}\ln R\right]
\Biggr\}.
\label{zfbdr}
\ea
Here and in the following sections, we use the abbreviations,
\ba
R = \frac{M_{_W}^2}{M_{_Z}^2}\;,~~~~~
r_{_W} = \frac{M_{_H}^2}{M_{_W}^2}\;,~~~~~
r_{_Z} = \frac{M_{_H}^2}{M_{_Z}^2}\;.
\label{ssss2}
\ea
The $\Delta r$ was introduced in (E.8) and (F.3) of~\cite{aa2}.
The $t$ mass dependent terms may be found in the appendix of~\cite{bbb}.
\subsubsection{Partial widths of the $Z$ boson}
The two form factors for each fermionic partial width of the $Z$ boson
are in one loop order and in the approximation of vanishing external
fermion masses~\cite{bbb}:
\ba
\rho_f^{\mathrm{1loop},\alpha}
&=& 1+\frac{\alpha}{4\pi(1-R)}
\Biggl\{{ Z}(M_{_Z}^2) + { Z}^{^F}(M_{_Z}^2) - { W}(0)
\nl &&
+~\frac{5}{8} R (1+R) - \frac{11}{2} - \frac{9R}{4(1-R)} \ln R + u_f
+\delta\rho_{ct,f}^t\Biggr\} \; ,
\label{zfrho}
\ea
\ba
\kappa_f^{\mathrm{1loop},\alpha}
&=& 1+\frac{\alpha}{4\pi(1-R)}
\Biggl\{ \frac{R}{1-R}\left[{ Z}(M_{_Z}^2) - { W}(M_{_W}^2) \right]
+~{ M}(M_{_Z}^2)
\nl &&
+~\frac{(1-R)^2}{R} Q_f^2 \left[ {V}_{_{1Z}}(M_{_Z}^2) +\frac{3}{2} \right]
- \frac{1}{2} \left[ u_f + \delta\rho_{ct,f}^t \right] \Biggr\} \; ,
\label{zfk}
\ea
\ba
u_f &=& \frac{1}{2R}\left[1-6|Q_f|(1-R)+12 Q_f^2(1-R)^2\right]
        \left[ {V}_{_{1Z}}({M}_{_Z}^2) +\frac{3}{2} \right]
+~ \biggl[1-2R
\nl &&-~2|Q_f|(1-R)\biggr] \left[ { V}_{_{1W}}({M}_{_Z}^2) +\frac{3}{2} \right]
+ 2R \left[ { V}_{_{2W}}({M}_{_Z}^2)+\frac{3}{2} \right] \; .
\label{zfuf}
\ea
\subsubsection{Auxiliary functions}
All $W$ and $Z$ boson self-energy functions
are sums of bosonic and fermionic parts, e.g.
\begin{equation}
W(0) = W_b(0) + W_f(0) \; .
\label{ssss9}
\end{equation}
The bosonic parts are given in appendix~A
of~\cite{aa2}\footnote{One should eliminate there the
(approximate) fermionic parts proportional to Tr$Q_f^2$ and
$N_f$.}
\ba
W_b(0) &=&  \frac{5R(1+R)}{8}  - \frac{17}{4} + \frac{5}{8R} -\frac{\rW}{8}
+\left[ \frac{9}{4} +\frac{3}{4R} - \frac{3}{1-R}\right]\ln R
+ \frac{3\rW}{4(1-\rW)} \ln \rW\; , \nonumber \\
\label{Ol_W0}  \\
\label{Ol_W1}
W_b(M_{_W}^2) &=& - \frac{157}{9} + \frac{23}{12R} + \frac{1}{12R^2}
           - \frac{\rW}{2} +\frac{\rW^2}{12}
+\frac{1}{R}\left[ - \frac{7}{2} + \frac{7}{12R}
           + \frac{1}{24R^2}\right] \ln R
\nll   &&
+ \rW \left( -\frac{3}{4} + \frac{\rW}{4} - \frac{\rW^2}{24} \right) \ln \rW
+\left( \frac{1}{2}   - \frac{\rW}{6} + \frac{\rW^2}{24} \right)
            \frac{L_{_{WH}}(M^2_{_W})}{M^2_{_W}}
\nll  &&
+~\left(- 2R - \frac{17}{6} + \frac{2}{3R} + \frac{1}{24R^2} \right)
\frac{ L_{_{WZ}}(M^2_{_W})}{M^2_{_W}} \;,
\\
\label{Ol_Z1}
Z_b(M_{_Z}^2) &=&
-8R^2 -\frac{34R}{3} + \frac{35}{18}\left(1 + \frac{1}{R} \right)
 -\frac{\rW}{2} + \frac{\rZ^2}{12R} + \rW \left( -\frac{3}{4} + \frac{\rZ}{4}
 - \frac{\rZ^2}{24} \right) \ln \rZ
\nll &&
  + \frac{5\ln R}{6R} +\Biggl( \frac{1}{2}
-~\frac{\rZ}{6}+ \frac{\rZ^2}{24} \Biggr) \frac{L_{_{ZH}}(M^2_{_Z})}{M_{_W}^2}
\nll &&
+\left[-2R^3 -\frac{17}{6}R^2 + \frac{2}{3}R   + \frac{1}{24} \right]
    \frac{{L}_{_{WW}}(M^2_{_Z})}{M^2_{_W}} \; ,
\\
Z^{^F}_b(M_Z^2) &=&
-4R^2+ \frac{17R}{3}-\frac{23}{9}+\frac{5}{18R}-\frac{\rW}{2}
+\frac{\rW\rZ}{6}
+\rW\left(-\frac{3}{4}+\frac{3\rZ}{8}-\frac{\rZ^2}{12} \right) \ln \rZ
\nll
&&-~\frac{1}{12R}\ln R + \frac{\ln \rZ}{2R}
+\left(-R^3+\frac{7R^2}{6}-\frac{17R}{12}-\frac{1}{8}
\right)\frac{L_{_{WW}}(\zm^2)}{\wm^2}
\nll
&&+~\left[\frac{1}{2}-\frac{5\rZ}{24}+\frac{\rZ^2}{12}+\frac{1}{2(\rZ-4)}
\right] \frac{ L_{_{ZH}} (\zm^2) } {{\wm}^2} \; ,
\ea
Function $L_{_{V_1V_2}}(s) \equiv
L(-s;M_{_{V_1}}^2,M_{_{V_2}}^2)$ is defined in~(2.14) of Ref.~\cite{aa1} and
${\cal F}_n$ and  \\
${\bf I}_n(Q^2;M_1^2,M_2^2)$ in appendices~C and~D of Ref.~\cite{ccc}.
The fermionic corrections are:
\ba
Z_f(s) &=& \frac{1}{2R} \sum_{f} N^f_c
\left[ \frac{-s}{M_{_Z}^2} \left(1+v_f^{2} \right)
{\bf I}_3(-s;m_f^2,m_f^2) + \frac{m_f^2}{M_{_Z}^2}
{\bf I}_0(-s;m_f^2,m_f^2) \right] \; ,
\label{ssss12}
\\
W_f(s) &=& \sum_{f=(f_u,f_d)} N^f_c
\left[ 2\frac{-s}{M_{_W}^2} {\bf I}_3(-s;m_{f_u}^2,m_{f_d}^2)
+ \frac{m_{f_u}^2}{M_{_W}^2} {\bf I}_1(-s;m_{f_u}^2,m_{f_d}^2)
\right.
\nl &&
\left.
+~\frac{m_{f_d}^2}{M_{_W}^2} {\bf I}_1(-s;m_{f_d}^2,m_{f_u}^2) \right],
\label{ssss13}
\\   
\label{zfzft}
Z^{^F}_f(\zm^2) &=& -\sum_f N^f_c
\Biggl\{\frac{r_f}{2} \left[ 1- r_f \wm^2 {\cal F}(-\zm^2,m_f^2,m_f^2) \right]
+\frac{1}{6R} \left(1+v_f^2\right)
\\  &&
\times~\left[ \frac{1}{2}\ln(r_fR) + r_fR + (-\frac{1}{4R} + \frac{r_f}{2}
-r_f^2R) \wm^2 {\cal F}(-\zm^2,m_f^2,m_f^2) \right] \Biggr\} \; ,
\nonumber
\ea
with $r_f=m_f^2 / \wm^2$ and the color factor $N^f_c$.
The $\gamma Z$-mixing function is:
\ba
M(s) &=& 2 \sum_f N^f_c |Q_f| v_f {\bf I}_3(-s;m_f^2,m_f^2) \; .
\label{ssss8}
\ea
 
The vertex functions are in the limit of vanishing fermion masses:
\ba
\label{ssss3} 
{V}_{_{1V}}(s) &=& -\frac{7}{2}-2 R_{_V}-(3+2R_{_V})\ln(-\tilde{R}_{_V} )
+ 2(1+R_{_V})^2 \left[ \mbox{Li}_2(1+\tilde{R}_{_V})-\mbox{Li}_2(1)
\right] \; ,
\nl
\\  
V_{_{2W}}(s) & = & -\frac{1}{6}-2R_{_W}-\left(\frac{7}{6}+R_{_W}\right)
\frac{L_{_{WW}}(s)}{s}
+2R_{_W} \left( R_{_W} +2 \right){\cal F}_3(s,M_{_W}^2) \; ,
\label{ssss4}
\ea
with
\ba
\tilde{R}_{_V}&=& R_{_V} -i\gamma_{_V}\;,~~~~
\gamma_{_V} = \frac{M_{_V} \Gamma_{_V}}{s}\;,
~~~~R_{_V} = \frac{M_{_V}^2}{s}\;.
\ea
The additional $t$ mass corrections to the $Zb\bar b$ vertex and
the counter term are at the $Z$ resonance\footnote{
The net one loop finite $t$ mass effect from the two vertices and
the counter term is taken into account in {\tt ZFITTER}
by a variable {\tt VTB}.}:
\ba
\label{zfv1t}
V_{_{1W}}^t(\zm^2) &=& \frac{1}{R}\int_0^1dy
\Biggl\{\left[ \frac{1}{2} - 3y(1-y) \right]\ln|r_1|
+2rR\ln|r_2| -rR +(2+R)
\nll &&
\times~\left[ {\bar F}_1(r)-{\bar F}_1(0)\right]-\left(\frac{3}{2}+R\right)
\left[ {\bar F}_2(r)-{\bar F}_2(0)\right]+\frac{1}{2}rR(2+R){\bar F}_2(r)
\nll &&
-~\frac{2rR(1-R)}{(1-4R)}
\left[1+\ln|r_2| - 4{\bar F}_1(r) + \frac{1-r}{2} {\bar F}_2(r)\right]
\Biggr\} \; ,
\\   
\label{zfv2t}
V_{_{2W}}^t(\zm^2) &=& \frac{1}{R}\int_0^1 dy
\Biggl\{-(2+R)\left[ {\bar F}_2(r)-{\bar F}_2(0)\right]
+r\Biggl[2R\ln|r_3| +\frac{1-2R}{4} (\ln|r_4|-1)
\nll &&
+~\frac{1}{2} (r-2rR-4R-4)
F_1(r) + \frac{1}{4}(1-r+2R+2rR)F_2(r)\Biggr] \Biggr\} \; ,
\\
\label{zfdrhbt}
\delta\rho_{ct,b}^t &=&
-\frac{r(1+2R)}{6(1-r)}\left[ \frac{1}{2} (5r-11)+\frac{3r(r-2)}{1-r}\ln r
\right] \; ,
\ea
with $r=r_t$ and
\ba
\begin{array}{rclrcl}
r_1&=&\frac{\displaystyle rR}{\displaystyle y(1-y)} -1\;,
& 
r_2&=&r-(1-y)\frac{\displaystyle y}{\displaystyle R}\;,
\nll 
r_3 &=& y+(1-y)r\;,
&   
r_4 &=& 1-(1-y)\frac{\displaystyle y}{\displaystyle R}\;,
\nll   
F_i(r) &=& f_i(1,r)\;,
&   
{\bar F}_i(r) &=& f_i(r,1)\;, \hspace{1.5cm} i=1,2\;,
\end{array}
\ea
\ba
f_1(a,b) &=& \frac{1}{a-b-y/R} \ln \frac{a-y(1-y)/R}{ay+b(1-y)} \; ,
\nll      
f_2(a,b) &=& \frac{1-y}{a-b-y/R}-\left(b+\frac{y^2}{R}\right)f_1(a,b)\;.
\ea
A completely analytical expression, valid at arbitrary $s$
may be found in~\cite{mannriemann}.
In the next section, we will need in addition the photonic self energy
function $A(s)$, and $D_Z(s)$:
\ba
A(s) &=& \frac{34}{3} +8R_{_W} +\left( \frac{17}{6} + 2R_{_W} \right)
\frac{L_{_{WW}}(s)}{s} \; ,
\label{ssss5}
\\  
D_{_{Z,b}}(s) &=&
\frac{34R}{3} - \frac{35}{18} - \frac{4}{9R} -\frac{\ln R}{12R}
-2R^2\frac{L_{_{WW}}(s)}{s}
+~\left(-2R^2 -\frac{17R}{6} +\frac{2}{3} + \frac{1}{24R} \right)
\nl &&
\times~\frac{L_{_{WW}}(s)-{\cal R}e L_{_{WW}}(M_{_Z}^2)}{-s+M_{_Z}^2}
+~\frac{1}{R} \Biggl\{ \frac{R_{_Z}}{12} \left(1-r_{_Z} \right)^2
+ \Bigl[ 1+ \left(1-r_{_Z} \right)
\nl &&
\times~\left( 10 - 5r_{_Z} + r_{_Z}^2 \right) R_{_Z}
+~\left( 1-r_{_Z} \right) ^3 R_{_Z}^2 \Bigr] \frac{\ln r_{_Z}}{24}
+~\biggl[ 11-4r_{_Z} +r_{_Z}^2
\nl &&
+~\left(1-r_{_Z} \right) R_{_Z} \biggr] \frac{L_{_{ZH}}(s)}{24s}
+ \left(\frac{1}{2} -\frac{r_{_Z}}{6}+ \frac{r_{_Z}^2}{24} \right)
\frac{ L_{_{ZH}}(s)-{\cal R}e L_{_{ZH}}(M_{_Z}^2)}{-s+M_{_Z}^2}\Biggr\},
\label{ssss10}
\\
D_{_{Z,f}}(s) &=& \frac{M_{_Z}^2}{M_{_Z}^2 - s}\left[ Z_f(s)
- {\cal R}e Z_f(M_{_Z}^2) \right] \; .
\label{ssss11}
\ea
\subsubsection{Form factors of the process $\epmf$}
The virtual, non-photonic corrections to fermion pair
production, including Bhabha scattering,
may be described by four form factors
per production channel as is indicated in Eq.~(\ref{meff}).
The contributions from the $ZZ,WW$ box diagrams are vanishingly small
at the $Z$ peak. Leaving them out\footnote{
The interested reader may find expressions for $WW,ZZ$ boxes in
Ref.~\cite{ccc} and in the subroutine {\tt ROKANC($\ldots,t-s,-s,-t,\ldots$)}
of {\tt ZFITTER} for the s channel ($t>0$).
For the t channel (used in Bhabha scattering) the call is
{\tt ROKANC($\ldots,s,t,u,\ldots$)}.}, it is~\cite{ccc}:
\ba
\rho_{ef}(s)^{\mathrm{1loop},\alpha} &=&
1 + \frac{\alpha}{4\pi(1-R)} \Biggl\{ { Z}(\zm^2) - { W}_f(0)
+\frac{19}{12}-\frac{5}{8R} +\frac{r_{_W}}{8}
+\frac{3}{4} \Biggl[-\frac{r_{_W}\ln r_{_W}}{1-r_{_W}}
\nl &&
+~\left(\frac{1}{1-R} - \frac{1}{R} \right) \ln R \Biggr]
+ \frac{5 L_{_{WW}}(s)}{6s} +D_{_Z}(s) + 2 R { V}_{_{2W}}(s)
+ \biggl[-2R +\frac{1}{2}
\nl &&
+~\frac{1}{4}\left(v_e + v_f\right) \biggr] { V}_{_{1W}}(s) + \frac{1}{8R}
\left[1+\frac{3}{2}\left(v_e^{2} + v_f^{2}\right)\right]
{V}_{_{1Z}}(s) + \frac{1}{2} \delta\rho_{ct,f}^t \Biggr\}\; ,
\label{zfref}
\ea
\ba
\kappa_f(s)^{\mathrm{1loop},\alpha} &=&
1 + \frac{\alpha}{4\pi(1-R)} \Biggl\{ \frac{R}{1-R} \left[
{ Z}(\zm^2) - { W}(\wm^2) \right]
\\
&& -~\frac{4}{9} - \frac{L_{_{WW}}(s)}{12 s} -{ M}(s)
-R A(s) + \left(R_{_W}-2R\right){ V}_{_{2W}}(s)
\nl &&
+~\left[2R-|Q_f|-\frac{1}{4}\left(v_e +v_f\right)
+\left(|Q_f|-1\right) R_{_Z} \right] { V}_{_{1W}}(s)
\nl &&
+~\left[-\frac{1}{8R}v_e\left(1+v_e\right)
-~\frac{1}{2}|Q_f|v_f\left(1-R_{_Z}\right) \right] { V}_{_{1Z}}(s)
-~\frac{1}{2}\delta\rho_{ct,f}^t
\Biggr\} \; ,
\nonumber
\label{ssss0}
\ea
\ba
\kappa_{ef}(s)^{\mathrm{1loop},\alpha} &=&
1 + \frac{\alpha}{4\pi(1-R)}\Biggl\{ \frac{2R}{1-R} \left[
{ Z}(\zm^2) - { W}(\wm^2) \right]
-\frac{8}{9} - \frac{ L_{_{WW}} (s) }{ 6s }
\nl &&
-~2{ M}(s) -\frac{4}{3} R -2R { V}_{_{2W}}(s)
-~\left( R + R_{_W} \right) \left[ A(s) -\frac{2}{3} \right]
\nl &&
+~\left[ 2R - \frac{1}{2} - \frac{1}{4}\left(v_e +v_f\right) \right]
{ V}_{_{1W}}(s)
+~\biggl[ -\frac{1}{8R} -\frac{3}{16R} \left( v_e^{2} + v_f^{2} \right)
\nl &&
+~\frac{1-R}{R} \left( Q_e^2 + Q_f^2 \right) \left( 1-R_{_W} \right)
\biggr] {V}_{_{1Z}}(s)
-~\frac{1}{2}\delta\rho_{ct,f}^t \Biggr\} \; .
\label{ssss1}
\ea
\subsubsection{The ${\cal O}(\alpha \alpha_s)$ corrections
            to electroweak observables \label{rkqcd} }
Here we follow presentation of Ref.~\cite{ddd},
based on the second Ref.~\cite{aasse}.
 
\ba
\Delta \alpha^{\mathrm{2loop},\alpha\alpha_s}
&=&
\frac{\alpha\alpha_s}{3 \pi^2 } Q^2_t \frac{m_t^2}{Q^2}
{\cal R}e \Biggl\{\Pi_t^{^{VF}} (Q^2)+\frac{45}{4}\frac{Q^2}{m^2_t}
\Biggr\} \; ,
\label{alqcd} \\
%
\Delta r^{\mathrm{2loop},\alpha\alpha_s} &=&
\frac{\alpha\alpha_s}{12 \pi^2 s_{_W}^2}
\frac{m_t^2}{M_{_W}^2} {\cal R}e \Biggl\{ -4s_{_W}^2
\left[ Q_b^2 \Pi_b^{^F}(M_{_W}^2) + Q_t^2 \Pi_t^{^{VF}}(M_{_W}^2) \right]
\nl  &&
+~\frac{R}{4s_{_W}^2} \left[ \left( v_b^2 +1 \right) \Pi_b^{^F}(M_{_Z}^2)
+ v_t^2  \Pi_t^{^{VF}}(M_{_Z}^2) +\Pi_t^{^{AF}}(M_{_Z}^2) \right]
\nl &&
-~\frac{1}{s_{_W}^2} \left(c_{_W}^2 - s_{_W}^2 \right) \Pi_t^{^{WF}} (M_{_W}^2)
- \Pi_0^{^{WF}} \Biggr\} \; ,
\label{drqcd}
%
%
\\
\Delta \kappa^{\mathrm{2loop},\alpha\alpha_s} &=&
\frac{\alpha\alpha_s}{12 \pi^2 s_{_W}^2}
\frac{m_t^2}{M_{_W}^2}{\cal R}e \Biggl\{ -\frac{R}{4s_{_W}^2}
\left[ \left( v_b^2 +1 \right) \Pi_b^{^F}(M_{_Z}^2) + v_t^2
\Pi_t^{^{VF}}(M_{_Z}^2) + \Pi_t^{^{AF}}(M_{_Z}^2) \right]
\nl & &
 +~\frac{c_{_W}^2}{s_{_W}^2}
\Pi_t^{^{WF}}(M_{_W}^2) - \frac{M_{_W}^2}{Q^2}
\left[ v_b|Q_b| \Pi_b^{^F}(Q^2)+
       v_t|Q_t| \Pi_t^{^{VF}}(Q^2) \right] \Biggr\} \; ,
\\
%
%
\Delta \rho^{\mathrm{2loop},\alpha\alpha_s} & = &
\frac{\alpha\alpha_s}{48 \pi^2 s_{_W}^2}
\frac{m_t^2}{M_{_W}^2}
{\cal R}e
\Biggl\{ -
\frac{Q^2}{m_t^2}
\left( v_b^2 +1 \right)
\frac{ \ln \left( Q^2 / M_{_Z}^2 \right)}{1-Q^2/M_{_Z}^2}
+ 4\Pi_0^{^{WF}}
\nl &&
+~
\Biggl[ \frac{1}{1-Q^2/M_{_Z}^2} \Biggl(
v_t^2 \left[ \Pi_t^{^{VF}}(M_{_Z}^2) -\Pi_t^{^{VF}}(Q^2) \right]
\nl &&
+~\Pi_t^{^{AF}}(M_{_Z}^2) - \Pi_t^{^{AF}}(Q^2) \Biggr)
-v_t^2 \Pi_t^{^{VF}}(M_{_Z}^2) -\Pi_t^{^{AF}}(M_{_Z}^2) \Biggr]
\Biggr\} \; ,
\ea
and, when $Q^2 = M_{_Z}^2$,
%
\ba
\Delta \rho^{\mathrm{2loop},\alpha\alpha_s}
& = &
\frac{\alpha\alpha_s}{48 \pi^2 s_{_W}^2}
\frac{m_t^2}{M_{_W}^2}
{\cal R}e\Biggl\{\frac{M_{_Z}^2}{m_t^2}
\left(v_b^2 +1 \right)
+
\Biggl[ v_t^2 \Biggl( M_{_Z}^2 \frac{d\Pi_t^{^{VF}}(Q^2)}{dQ^2}
{|_{Q^2=M_{_Z}^2}}
\\ & &
-~\Pi_t^{^{VF}}(M_{_Z}^2) \Biggr)
+
M_{_Z}^2 \frac{d\Pi_t^{^{AF}}(Q^2)}{dQ^2} {|_{Q^2=M_{_Z}^2}}
- \Pi_t^{^{AF}}(M_{_Z}^2) \Biggr]  +
4\Pi_0^{^{WF}} \Biggr\} \; .
\nonumber
\ea
Here, the two-loop functions are:
%
\ba
\label{ptvf}
\Pi_t^{^{VF}}(Q^2) &=&
\frac{1}{\alpha} \Biggl\{ \frac{55}{4} -26 \alpha + 3\left(1+x_{\alpha}
\right) \left(1-6\alpha \right) f_{\alpha} -2\left[\alpha \left(
2x_{\alpha}^2 - 3x_{\alpha} + 2 \right) +2x_{\alpha} \right] f_{\alpha}^2
\nl & &
+~ 2\left( 4\alpha^2 -1 \right) \mathrm{I}_{\alpha}
+4\alpha\left(2\alpha -1 \right) \left( 4\alpha+1 \right)
\mathrm{J}_{\alpha} \Biggr\} \; , \\
%
%
\Pi_t^{^{AF}}(Q^2) &=&
\frac{1}{\alpha} \Biggl\{ \frac{55}{4} - \frac{19}{2}\alpha + 12\alpha^2
+ 3\left( 1+x_{\alpha} \right) \left( 1+12\alpha +4\alpha^2 \right)
f_{\alpha}
+2\Bigl[2x_{\alpha}\left(3\alpha^2-1\right)
\nl && 
+~\alpha \left( 7x_{\alpha}^2 -3x_{\alpha} + 7 \right) \Bigr] f_{\alpha}^2
-~ 2\left(1+2\alpha \right) \left(1+4\alpha \right) \mathrm{I}_{\alpha}
-4\alpha \left(1+4\alpha \right)^2 \mathrm{J}_{\alpha} \Biggr\}\; ,\nl\\
%
%
\label{ptwf}
\Pi_t^{^{WF}}(Q^2) &=&
\frac{1}{4\alpha} \Biggl\{ 55 - \frac{71}{2}\alpha - 10 \alpha^2
-8\alpha G(x_b) + 2 \left(6 + 9\alpha -5\alpha^2 \right) f_b
\nl &&
+~ 2 \left[ 4x_b \left( \alpha^2 -\alpha -1 \right) + \alpha \left(
5-4\alpha \right) \right] f_b^2
\nl & &
+~4\left( \alpha -2 \right) \left( \alpha + 1 \right)^2 \mathrm{I}_b
+8\alpha \left( \alpha - 2 \right) \left( \alpha + 1 \right)
\mathrm{J}_b \Biggr\} \; , \\
%
%
\frac{d\,\Pi_t^{^{VF}}(Q^2)}{dQ^2}
&=&
-\frac{1}{m_t^2} \left[ \frac{43}{4} +18\alpha + \left(
10\alpha +3 \right) \left( 1+x_{\alpha} \right) f_{\alpha}
+2\left( 5\alpha -2 \right) x_{\alpha}f_{\alpha}^2 \right.
\nl & &                                            \left.
-~8 \left(2\alpha - 1 \right) x_{\alpha}^2 f_{\alpha}^2(x_{\alpha}^2)
-2 \left(4\alpha^2 + 1 \right) \mathrm{I}_{\alpha}
-8 \alpha^2 \left( 1+4\alpha \right) \mathrm{J}_{\alpha} \right] \; ,\\
%
%
\frac{d\,\Pi_t^{^{AF}}(Q^2)}{dQ^2}
&=&
-\frac{1}{m_t^2} \Biggl[ \frac{43}{4} -12\alpha \left(
2\alpha + 3 \right) - \left(
24\alpha^2 + 26\alpha
-3 \right)
\left(1+x_{\alpha}\right) f_{\alpha}
\nl & & 
-~2\Bigl( 12 \alpha^2
+19\alpha
+2 \Bigr)
 x_{\alpha}f_{\alpha}^2
+ 8 \left( 1+4 \alpha \right) x_{\alpha}^2 f_{\alpha}^2(x_{\alpha}^2)
+~2 \left( 8\alpha^2 - 1 \right) \mathrm{I}_{\alpha}
\nl & & 
+~16 \alpha^2 \left( 1+ 4\alpha \right) \mathrm{J}_{\alpha} \Biggr] \; ,
\ea
where
\ba
\begin{array}{rclrclrcl}
x_{\alpha} &=& \frac{ \displaystyle 2\alpha}{\displaystyle
1+2\alpha + \sqrt{1+4\alpha}}\;,
&  
\alpha &=& -\frac{\displaystyle m_t^2}{\displaystyle Q^2}\;,
&   
x_b &=& \frac{\displaystyle\alpha}{\displaystyle 1+\alpha}\;,
\nl 
\mathrm{I}_{\alpha} &=& F(1) + F(x_{\alpha}^2)- 2F(x_{\alpha})\;,
&  
\mathrm{I}_b &=& F(1)-F(x_b)\;,
&   
f_{\alpha} &=& \frac{\displaystyle\ln x_{\alpha}}{\displaystyle
1-x_{\alpha}}\;,
\\   
\mathrm{J}_{\alpha} &=&\frac{\displaystyle 1-x_{\alpha}}{
\displaystyle\left(1+x_{\alpha}\right)\alpha}
\left[ G(x_{\alpha}^2) - G(x_{\alpha}) \right]\;,
&   
\mathrm{J}_b &=&-\frac{\displaystyle 1}{\displaystyle 2\alpha}G(x_b)\;,
&    
f_b &=& \frac{\displaystyle\ln x_b}{\displaystyle 1-x_b}\;,
\end{array}
\ea
\ba
F(x) = \int_0^x dy \left(\frac{\ln y}{1-y}\right)^2 \ln \frac{x}{y}\;,
\hspace{1.0cm}
G(x) = \int_0^x dy \left( \frac{\ln y}{1-y}\right)^2.
\ea
The $\Pi_b^{^F}(Q^2)$ is a common asymptotic expression for
functions (\ref{ptvf}--\ref{ptwf}) for
$m_t \to 0$:
\begin{equation}
\alpha \Pi_b^{^F}(Q^2) = \left[\alpha \Pi^{^{VF}}\right]_{m_t=0}
=\left[\alpha \Pi^{^{AF}}\right]_{m_t=0}
=\left[\alpha \Pi^{^{WF}}\right]_{m_t=0}
= \frac{55}{4} - 12 \zeta (3) + 3 \ln \alpha ,
\end{equation}
with $\zeta(3)=1.0202...$  Further,
\ba
\Pi_0^{^{WF}} = \lim_{Q^2\rightarrow0}\Pi_0^{^{WF}}(Q^2)
= -\frac{1}{4} \left[ \frac{105}{2} + 2\pi^2 \right]\;.
\ea
%
\section{Description of options in different approaches \label{pecul}}
 
\subsection{{\tt BHM} options}
 
In this subsection
we describe the main options implemented in the version
of {\tt BHM} used for the present study to estimate the theoretical
uncertainties.
Out of the actual different user-accessible flags and their possible values,
only the ones used in the present analysis are discussed. The rest
of the flags and/or values either have a negligible effect on the predictions
or exist only for testing purposes.
 
\begin{itemize}
 
\item[$\bullet$]{\tt IRES=0,1,2} \\
The resummation of the one-loop one-particle irreducible contributions
in $\Delta r$ and, in general, in the whole set of self-energies is implemented
in {\tt BHM} in three different ways. 
Tsese can be considered representative of
at least two rather different philosophies. The basic difference
comes from the treatment of the $(\sgzr)^2$ mixing term.
If {\tt IRES=0 }, then the resummation comes from the resolution of the
renormalization equations keeping that term (see for instance
W.Hollik in Ref.~\cite{bhm}).
In this case, not only leading top and Higgs terms are re-summed but
also non-leading ones. This is the default working option.
If  {\tt IRES=1,2 }, then the renormalization equations are solved keeping
strictly one-loop contributions and the inclusion of higher-order terms coming
from one-loop one-particle irreducible diagrams is explicitly done for
$\Delta r$ and for each self-energy. If {\tt IRES=1 }, the prescription
used to treat the top, Higgs and remainder terms follows the suggestions of
Halzen, Kniehl and Stong \cite{halzknis}, whereas if {\tt IRES=2 }, then it
follows the suggestions from S.Fanchiotti and A.Sirlin \cite{fansir}. Both
differ in the detailed treatment of the Higgs and remainder terms.
 
\item[$\bullet$]{\tt ITWO=1,2}   \\
This flag allows the choice of scale in vertex corrections. The dominant
effect happens in the b-vertex. If {\tt ITWO=1 }, then the $\alpha(0)$ scale is
used whereas, if {\tt ITWO=2}, then the $G_{\mu}$ scale is used. The later is
the default working option.
 
\item[$\bullet$]{\tt IFAC=0,1,2}   \\
This flag allows the choice of different factorization schemes for the
final-state corrections.
If {\tt IFAC=0} then no factorization at all is applied for weak vertex,
QED and QCD final-state corrections. That means that these three kinds of
corrections are independently applied to the vacuum-polarization dressed
amplitudes. This is the default working option. If {\tt IFAC=1}, then
QED corrections are applied on top of weak vertex ones, and if {\tt IFAC=2},
then QCD corrections are also applied on top of weak vertex and QED ones.
 
\item[$\bullet$]{\tt IQCD=3,4}
This flag allows the choice of different treatments of the QCD corrections to
electroweak loops. Of its possible values, only two have been
used in the present study: if {\tt IQCD=3}, then the exact AFMT correction
is implemented, whereas if {\tt IQCD=4}, then the Sirlin's scale $\xi=0.248$
is used. These two approaches have already been discussed elsewhere in
the text.
 
\end{itemize}
\noindent
To summarize, {\tt BHM} runs in $3 \times~2 \times 3$ electroweak
$\times 2$ QCD options. We have added Table~\ref{ta17}, showing
the effect of the working options of {\tt BHM} on theoretical errors.
 
\subsection{{\tt LEPTOP} options}
 
Several options to the preferred formulas used in
{\tt LEPTOP} have been chosen and
variations of these formulae have been made.
 Usually each option consists
of the addition of an extra term corresponding
 to a rough guess of the value
of the uncalculated higher-order terms. We then make all possible
combinations of these options --- $2^n$ in total, where $n$ is the number
of options.  Among all these $2^n$ combinations we locate those yielding
the minimum and the maximum values of the observables and take as the
estimate of the theoretical errors their deviations from the central values.
 
Theoretical uncertainties come from two different sources: 1) not
yet calculated Feynman diagrams, 2) not yet calculated terms in a given
diagram. The first source is represented in {\tt LEPTOP} by the gluonic
corrections to the electroweak vertices of light quarks,
 $F_{iq}$, where
$q=u,d,s,c$;~~$i=V,A$. In this case as a basis for options
 a crude estimate
of these corrections is used:
 
\begin{equation}
\delta F_{iq} = \frac{\hat{\alpha}_{_S}}{\pi}F_{iq}\;.
\label{K1}
\end{equation}
 
The second source is represented by non-leading terms
 of a) higgs and b)
$\alpha_{_S}^2$ corrections to the $t\bar{t}$
 loop in $Z$ boson self energy and
$t-b$ loop in $W$ boson self energy denoted by
 $\delta V_i^{t^2}$ and $\delta
V_i^{\alpha_{_S}^2}$ respectively and c) by the $t\bar{t}$ vertex
contribution to $Z\to b\bar{b}$ decay denoted by $\delta\phi^{t^2}$.
 
In cases a) and b) the missing non-leading terms are
 estimated by multiplying
the corresponding leading terms by a factor $2/t$,
 where $t = (m_t/M_{_Z})^2$.
In the case c) the leading correction itself
 is so small that it is taken as
a measure of uncertainty. Thus the {\tt LEPTOP} uncertainties are:
 
\begin{equation}
\Delta V_i^{t^2} = (2/t)\delta V_i^{t^2} = (2/t)
\left(-\frac{\bar{\alpha}}{\pi}\frac{A(M_{_H}/m_t)t^2}{16s^2c^2}\right)
\label{K2}
\end{equation}
 
\noindent
in accord with Eq. (\ref{316});
 
\begin{equation}
\Delta V_i^{\alpha_{_S}^2} = (2/t)\delta V_i^{\alpha_{_S}^2} =
(2/t)(-1.25 \hat{\alpha}_{_S}^2(m_t)t)
\label{K3}
\end{equation}
 
\noindent
in accord with Eq. (\ref{315});

\begin{equation}
\Delta \phi^{\alpha_s^2} =-1.37 \frac{3 -2s^2}{2s^2c^2}\frac{\pi}{3}
\hat \alpha^2_s(m_t) t
\label{dphias2}
\end{equation}
in accord with the first term in Eq. (\ref{403});

\begin{equation}
\Delta\phi^{t^2}=\delta\phi^{t^2}=\frac{3-2s^2}{2s^2c^2}
\frac{\bar{\alpha}t^2}{16\pi s^2c^2}\tau_b^{(2)}
\label{K4}
\end{equation}
 
\noindent
in accord with the second term in Eq. (\ref{403}),
as $\Delta \phi^{\alpha^2_s}$
 is much larger then $\delta \phi^{t^2}$,
the latter will be neglected;
 
\begin{equation}
\Delta F_{iq}^{\alpha_S}=\delta F_{iq}^{\alpha_S}
=\frac{\hat{\alpha}_{_S}}{\pi}F_{iq}
\label{K5}
\end{equation}
 
\noindent
in accord with Eq. (\ref{K1}).
 
  Explicit calculations \cite{yr89, 5}  give:
 
\begin{equation}
F_{Vu} =-0.00169 \;, \;\;\;\; F_{Au} = -0.00165 \; ,
\label{K6}
\end{equation}
 
\begin{equation}
F_{Vd} = 0.00138 \;, \;\;\;\; F_{Ad} = 0.00137  \; .
\label{K7}
\end{equation}
 
Note that the uncertainties $\Delta F_{iq}^{\alpha_{_S}}$
 produce corresponding
uncertainties in electroweak corrections $\delta\Gamma_q$ to
the partial widths $Z\to q\bar{q}$:
 
\begin{equation}
\delta\Gamma_q =
 24sc\Gamma_0 2 I^{(3)}_q [(1-4|Q_q|s^2)F_{Vq}+F_{Aq}] \; .
\label{K8}
\end{equation}
 
With above numbers
 
\begin{equation}
\Delta\Gamma_u = \Delta \Gamma_c =
 -1.9\left(\frac{\alpha_{_S}}{\pi}\right) {\mbox {MeV}} \simeq
-0.08 \; {\mbox {MeV}} \; ,
\label{K9}
\end{equation}
 
\begin{equation}
\Delta\Gamma_d = \Delta \Gamma_s  =
 -2.0\left(\frac{\alpha_{_S}}{\pi}\right) {\mbox {MeV}} \simeq
-0.08 \; {\mbox {MeV}} \; .
\label{K10}
\end{equation}
 
So that the sum over four light quarks is
 
\begin{equation}
\Sigma_1^4\Delta\Gamma_q = -0.3 \; {\mbox {MeV}} \; .
\label{K11}
\end{equation}
 
Now the options of {\tt LEPTOP} can be formulated:
 
\begin{itemize}
 
\item Options 1,2
 
Uncertainty $\Delta V_i^{t^2}$ given by Eq.~(\ref{K2}) is added to
(option 1), or subtracted from (option 2) the functions
$V_i$, $i=m,A,R$. \\
 
\item Options 3,4
 
Uncertainty $\Delta V_i^{\alpha^2_{_S}}$ given by
Eq.(\ref{K3}) is added to (option 3), or subtracted from
 (option 4) the
function $V_i$, $i=m,A,R$. \\
 
\item Options 5,6
 
Uncertainties $\Delta \Gamma_q (q=u, d, s, c)$ given by 
Eqs.~(\ref{K9}),~(\ref{K10})
are added to (option 5) or subtracted from (option 6) the partial
widths $\Gamma_q$. \\
 
\item Option 7

Uncertainty $\Delta \phi^{\alpha^2_s}$ is added to the function
$\phi(t)$ given by Eq. (\ref{402}). \\
 
\end{itemize}
 
Theoretical uncertainties for observables,
 assuming $\hm = 300$ GeV$,\;\; m_t
= 175 $ GeV, $\;\; \hat{\alpha}_{_S}(\zm) = 0.125$ are shown
in Table~\ref{ta18}.

\subsection{{\tt TOPAZ0} options}

In this subsection we describe the options implemented in {\tt TOPAZ0} version
2.0 for studying the theoretical uncertainties. A general comment is in order
here.
Some of the {\tt TOPAZ0} electroweak options have been originally designed to
produce a conservative estimate of the uncertainty. If nothing stands against
a certain option then we accept it, even if it goes against our own philosophy.
 
\begin{itemize}
 
\item{\tt OU0.EQ.'N'('Y')}
 
This is the primordial option since it controls the partial resummation of
bosonic self-energies. If {\tt OU0.EQ.'N'}, then in $\shat^2$ and $\Delta s^2$,
we use:
 
\begin{eqnarray}
\Sigma &\to& \Sigma^{\rm {ferm}} \;, \nll
\Delta s^2 &=& {{\Sigma^{\rm {bos}}}\over {c^2 - s^2}} \; ,
\end{eqnarray}
 
\noindent
i.e., all bosonic self-energies are expanded. Otherwise for {\tt OU0.EQ.'Y'}
we use:
 
\begin{eqnarray}
\Sigma &\to& \Sigma_{_R} = \Sigma^{\rm {tot}} - (c^2 - s^2)\,
{{{\cal R}e\Szg(\zm^2)}\over \zm^2} \; , \nll
\zm^2 \Delta s^2 &=& {\cal R}e\Szg(\zm^2) \; .
\end{eqnarray}
 
\noindent
For {\tt OU0.EQ.'Y'} {\tt TOPAZ0} will automatically select {\tt OU2 = 'N'}
and {\tt OU3 = 'Y'}.
 
\item{\tt OU1.EQ.'N'('Y')}
 
The default choice requires that in
 
\begin{eqnarray}
\dgvf &=& \frac{\alpha}{4\pi}\left[ {{2\fvf - \frac{1}{2} v_f \Delta\Pi_{_Z}}
\over {c^2s^2}} - 2 Q_f\Delta s^2 \right] \; , \nll
\dgaf &=& \frac{\alpha}{4\pi}\left[ {{2\faf - \frac{1}{2} \i3f \Delta\Pi_{_Z}}
\over {c^2s^2}} \right] \; ,
\label{defg}
\end{eqnarray}
 
everything must be expanded in terms of $\alpha\equiv\alpha(0)=1/137.036\dots$
while for {\tt OU1.EQ.'Y'} the parameter expansion is selected as $\alpha
\equiv \alpha(\zm)$.
 
\item{\tt OU2.EQ.'N'('Y')}
 
This option deals with the so-called problem of expansion. The default for
{\tt TOPAZ0} is the expanded solution where, for instance, a $\z0$ partial
width is computed according to
 
\begin{eqnarray}
\Gamma_f &=& {{\gf\zm^3}\over {6\srt\pi}}\,
N_c^f \rhoz\,\left[\vhat_f^2 + \frac{1}{4} + 2 \vhat_f\dgvf +2\i3f\dgaf\right]
\; , \nll
\vhat_f &=& \i3f - 2 Q_f\shat^2 \; ,
\end{eqnarray}
 
\noindent
where for the moment we assume that there is no final-state QED + QCD radiation ,
factor, whose treatment will be explained by one of the following options.
If, instead, {\tt OU2.EQ.'Y'}, then the electroweak corrected vector and
axial-vector couplings are squared numerically.
 
\item{\tt OU3.EQ.'Y'('N')}

Two different procedures are introduced for dealing with the physical Higgs
contribution to the self-energies. If {\tt OU3.EQ.'Y'}, then by working in
a $\msb$ environment with a scale $\mu$ set to $\zm$ we extract from
$\Sigma^{\rm {bos}}$ the physical Higgs contribution, $\Sigma^{^H}$, and
redefine
 
\begin{equation}
\Sigma^{\rm {ferm}} \to \Sigma^{\rm {ferm}} + [\Sigma^{^H}]_{\msb} \; ,
\end{equation}
 
\noindent
where $\Sigma^{H}$ is subject to no additional expansion, such as leading
or sub-leading behavior with $\hm$. Of course, {\tt OU3.EQ.'N'} leaves
the Higgs contribution expanded as for any other bosonic contribution.
This option reflects and partially illustrates one of the defining rules
of {\tt TOPAZ0} --- a certain reluctance to accept the isolation and a
different treatment for something which can be considered as the leading
part of some quantity only for {\it extraordinary} values of some of the
parameters.
 
\item{\tt OU4.EQ.'N'('Y')}

When we consider the mass corrections to the $\barb b$ partial decay rate
there will be something like
 
\begin{equation}
-6\frac{m_b^2}{\zm^2}\,\left[ 2\i3f\dgaf + \left(\dgaf\right)^2\right] \; .
\end{equation}
 
\noindent
In these mixed corrections there is an additional uncertainty connected
with the choice of $m_b$ --- i.e., pole mass or running mass.
 
\item{\tt OU5.EQ.'N'('Y')}

According to the strategy that all non-leading and gauge-variant quantities
should be expanded, in Eq.(~\ref{defg}) we use as the zero order approximation,
 
\begin{equation}
s^2 = \frac{1}{2}\,\left[\, 1 - \sqrt{1 - \frac{4\,\pi\alpha(\zm)}
{\srt\,\gf\zm^2}}\,\,\right] \; .
\end{equation}
 
\noindent
However, since perturbation theory rearranges itself in such a way that
the expansion of $\stes^l$ starts with $\shat^2$, i.e.,
 
\begin{equation}
\stes^l = \shat^2 + \frac{1}{2}\,\dgvl + \frac{1}{2}\,\left(4\,\shat^2 -
1\right)\dgal \; ,
\end{equation}
 
\noindent
we have envisaged the possibility of reorganizing the structure of the
pseudo-obser- vables such that to all orders the bare weak mixing
angle has an expansion starting with $\shat^2$.
Combined with {\tt OU1.EQ.'N'('Y')} this option tells us that the expansion
parameter, formally $\alpha/(4\pi s^2c^2)$, can be set to
 
\begin{equation}
{{\gf\zm^2}\over {2\srt\pi^2}}\,\times\,\left\{1 ; \rhozr ; 1 - \Delta\alpha ;
\rhozr(1 - \Delta\alpha)\right\} \; .
\end{equation}
 
\item{\tt OU6.EQ.'N'('Y')}

This option deals with factorization of electroweak corrected kernels
and QCD radiation. For instance, if {\tt OU2.EQ.'N'} --- the expanded 
solution --- we still distinguish between a non-factorized solution,
 
\begin{equation}
\Gamma_f = {{\gf\zm^3}\over {6\srt\pi}}\,
N_c^f \rhoz\,\left[\vhat_f^2R^f_{_V} + \frac{1}{4}
R^f_{_A} + 2\vhat_f\dgvf + 2\i3f\dgaf\left(1-6\frac{m_f^2}{\zm^2}\right)\right]
\; ,
\end{equation}
 
\noindent
and a fully factorized solution. A special treatment is of course devoted to
$b \bar b$ final state in order to reproduce the FTJR correction term.
 
\item {\tt OU7.EQ.'N'('Y')}

Higher order QCD corrections to $\rhoz$ are implemented with the exact
AFMT term or by subtracting from $\Sigma^{\rm {ferm}}$ the leading $m_t$
term and by replacing it with the corresponding one evaluated at a scale of
$\xi = 0.248 m_t$.
 
\end{itemize}
 
\noindent
To summarize {\tt TOPAZ0} runs in $2^4$ or $2^6$ electroweak $\times 2$ QCD
options.
While we have devoted a special section to the effect of {\tt OU0.EQ.'N'} and
{\tt OU2.EQ.'Y'}, here we present a short Table~\ref{ta19} to indicate the
increasing spread in the theoretical errors while the various flags are
switched on.
 
\subsection{{\tt ZFITTER} options}
 
In this subsection we describe some electroweak and QCD
options implemented in {\tt ZFITTER} (version 4.9).
Simultaneously, we give a description of the flags, implemented in version
4.9 for studying the theoretical uncertainties.

\begin{itemize}
 
\item[$\bullet$]{\tt OZ1: IAMT4=3,2,1} \\
The realization of leading and remainder terms, given by
formulae~(\ref{romus}--\ref{c2qcd}), is the default, {\tt IAMT4=3}.
If {\tt IAMT4=2}, then $X$ is not included in the leading terms, it remains
a part of remainders~\cite{halzkni}. If {\tt IAMT4=1}, then both $X$ and
${\mathrm 2loop}$-$ \alpha \alpha_s$ terms are in the remainders.
In the last case the
${\mathrm 3loop}$-$ \alpha \alpha_s^2$ term is also placed in remainders.
In the result of numerical investigations it was revealed
that for {\tt IAMT4=2} the remainder terms in $\rho_f$ and $\kappa_f$ are
not small. Since
this contradicts our philosophy to keep remainder terms small, one could
exclude it from the set of working options. However, it was found that
the difference between {\tt IAMT4=3 and 1} is rather small, therefore this
option doesn't sizably influence the theoretical errors.
 
\item[$\bullet$]{\tt OZ2: IHIGS=0,1}   \\
This option governs the resummation of the leading Higgs contribution in
$\Delta r$
\begin{equation}
\Delta r^{\rm {Higgs}} \simeq \frac{\srt G_{\mu}M^2_{_W}}{4\pi^2} \;
     \frac{11}{12}\left(\ln\frac{M^2_{_H}}{M^2_{_W}} - \frac{5}{6} \right)\;,
               \;\;\;\;\;\; M_{_H} \gg M_{_W} \; .
\label{leadhigs}
\end{equation}
If {\tt IHIGS=0} (the default), then the Higgs contribution is not re-summed.
If {\tt IHIGS=1} and if 
\begin{equation}
\ln\frac{M^2_{_H}}{M^2_{_W}} - \frac{5}{6} \geq 0  \;,
\end{equation}
then it is re-summed ---
i.e., it is extracted from remainders with the scale $\alpha/s^2_W$
and put to the leading terms with the scale $G_{\mu}$, as in~(\ref{leadhigs}).
We observed, that $10/12$ of $\Delta r^{\rm {Higgs}}$ is contained in $X$.
Therefore, if {\tt IAMT4=3}, then only $1/12$ of it is additionally re-summed.
The influence of this option on theoretical errors was found to be tiny.
For this reason, the Higgs resummation has not been implemented in
$\rho_f$ and $\kappa_f$.
 
\item[$\bullet$]{\tt OZ3: ISCRE=0,1,2} \\
This option defines the scale of the remainder terms. If {\tt ISCRE=0}
(the default),
then the scale is $\alpha/s^2_W$, if {\tt ISCRE=2} it is $G_{\mu}$.
(More precisely, $G_{\mu}$ in $\rho_f$ and $\kappa_f$ and
$G_{\mu}(1-\Delta\alpha)$ in $\Delta r$.) For {\tt ISCRE=1}, the scale
of the remainder in $\Delta r$ is set equal to
$G_{\mu}(1-\Delta r_{_L})$, which
was not included in the set of working options, since its influence is much
smaller then the previous one's. The effect of the variation of the scale of
the remainder term on the theoretical bands was found to be dominating.
Especially influential is the scale variation in $\Delta r$,
which introduces terms of the order
$c^2_{_W}/s^2_{_W}\Delta \rho\cdot\Delta r_{\rm {rem}}$.
This illustrates the importance in the calculation of the next-to-leading
term of the order ${\cal O}(G^2_{\mu}m^2_t\zm^2)$.
 
\item[$\bullet$]{\tt OZ4: IFACR=0,1,2,3}    \\
It realizes four subsequent expansions of $\Delta r$
as they are given in~(\ref{ifacr}). The first, fully non-expanded option,
is the default. Only the first expansion (the second row)
was retained in the set of working options. The last two were excluded, since
they contradict Sirlin's theorem on mass singularities. This leads to a
visible spread of theoretical bands, although one much smaller than that of
the previous option.
 
\item[$\bullet$]{\tt OZ5: IFACT=0,1,2,3,4,5}\\
It realizes six subsequent expansions for $\rho_f$ and $\kappa_f$.
The default, {\tt IFACT=0}, corresponds to the non-expanded
realizations~(\ref{rofs}--(\ref{kappafs}).
The four first options for $\rho_f$, {\tt IFACT=0,1,2,3,} are exactly
the same as are given in~(\ref{ifacr}),
while $\kappa_f$ is expanded for {\tt IFACT=1,2,3}:

\begin{equation}
\kappa_f = \kappa_{_L} + \Delta \kappa_{f,\mathrm{rem}}
  = 1 + \frac {c^{2}_{_W}}{s^{2}_{_W}} \Delta \rho_{_X}
                                  + \Delta \kappa_{f,\mathrm{rem}} \; .
\label{kappal}
\end{equation}
 
\noindent
For {\tt IFACT=4,5}, we linearize the full expression~(\ref{a}).
Introducing
\begin{equation}
\rho_{_L} = \frac{1} { 1 - \Delta \rho } \; ,
\label{rofl}
\end{equation}
\begin{equation}
{\bar g}^f_{_{VL}} = 1 - 4|Q_f|s^2_{_W} \kappa_{_L} \; ,
\label{vefl}
\end{equation}
see~(\ref{rofs}) and~(\ref{rholead}), we have for {\tt IFACT=4},
\begin{eqnarray}
\Gamma_{f} &=& \frac{\gf\zm^3}{24\srt\,\pi}\,N_c^f\,
\Biggl\{ \left(\rho_{_L} + \Delta\rho_{f,{\mathrm rem}}\right)
        \left[ \left({\bar g}^f_{_{VL}}\right)^2 R^f_{_V}\,+\,R^f_{_A}\right]
\nll
  & & + \; \rho_{_L} \left[-8s^2_{_W} {\bar g}^f_{_{VL}} \;
       \Delta\kappa_{f,{\mathrm rem}} \; R^f_{_V} \right] \Biggr\} \; .
\label{gamzl}
\end{eqnarray}
Realizing, that
\begin{equation}
R^f_{_{V,A}} = 1 + \Delta R^f_{_{V,A}} \; ,
\end{equation}
for {\tt IFACT=5} we finally end up with a fully expanded equation for
the partial widths:
\begin{eqnarray}
\Gamma_{f} &=& \frac{\gf\zm^3}{24\srt\,\pi}\,N_c^f\,
\Biggl\{\rho_{_L} \left[ ({\bar g}^f_{_{VL}})^2\,R^f_{_V}\,+\,R^f_{_A}\right]
    + \Delta\rho_{f,{\mathrm rem}} \left[ ({\bar g}^f_{_{VL}})^2+1 \right]
\nll
  & & + \; \rho_{_L} \left(-8s^2_{_W} {\bar g}^f_{_{VL}} \;
       \Delta\kappa_{f,{\mathrm rem}}  \right) \Biggr\} \; .
\label{gamzfl}
\end{eqnarray}
It was found that the spread of error bands gradually grows while coming
from {\tt IFACT=0} to {\tt IFACT=5}. For this reason, only these two limiting
cases were left among the working options. This option was found to be rather
influential.
 
\item[$\bullet$]{\tt OZ6: ISCAL=0,1,2,3,4}  \\
is the only QCD option.
At the default, {\tt ISCAL=0}, we implemented the exact AFMT correction.
For {\tt ISCAL=2,1,3} we implemented the $\xi$-factor as given
in Ref.~\cite{kyr}. Finally, for {\tt ISCAL=4},
Sirlin's scale $\xi=0.248$ (see Ref.\cite{sir})
was implemented. Only {\tt ISCAL=0,4} were left among working options.
\end{itemize}
\noindent
To summarize, {\tt ZFITTER} runs in $2^5$ electroweak $\times\;2$
QCD options
{\footnote {The option {\tt OZ5, IFACT=5} also may be ranked among
QCD options, since it simulates missing terms of the order
${\cal O}(\alpha\alpha_{_S})$.
}}.
Table~\ref{ta20} has been added to show the effect of the working options of
{\tt ZFITTER} on theoretical errors.
 
\section*{Acknowledgements}
The authors acknowledge G.~Altarelli,
K.~Chetyrkin,
G.~Degrassi,
J.~Ellis,
S.~Fanchiotti,
D.~Haidt,
A.~Kataev,
B.~Kniehl,
H.~K\"uhn,
A.~Kwiatkowski,
S.~Larin,
A.~Sirlin,
O.~Tarasov
for useful discussions.
We are very much indebted to A.D.~Schaile for a critical reading
of the manuscript and useful suggestions.
The authors of the codes which have presented flowcharts
are grateful to V.L.~Telegdi for insistence on and help in preparing
them.
 

\newpage
 
\section*{Figures}
\addcontentsline{toc}{section}{Figures}
\subsection*{ Pseudo-observables}
\addcontentsline{toc}{subsection}{Pseudo-observables}
\vspace*{1cm}
 
\begin{center}
\mbox{\epsfig{file=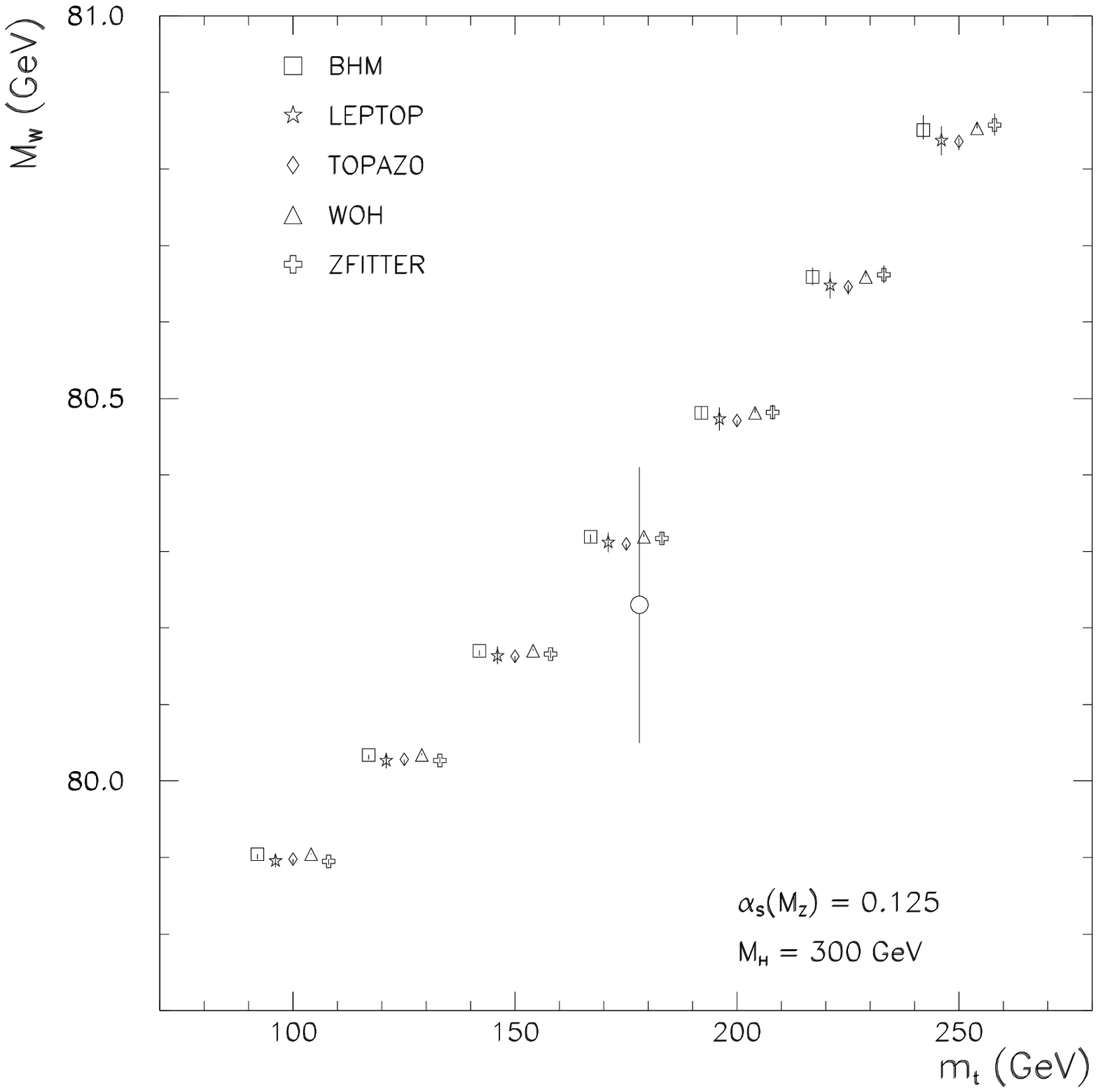,width=15cm,height=15cm}}
\end{center}
Figure 11: The {\tt BHM, LEPTOP, TOPAZ0, ZFITTER, WOH} predictions for
$\wm$, including an estimate of the theoretical error 
as a function of $m_t$,
for $\hm = 300\,$GeV and $\hat{\als} = 0.125$.
 
\newpage\vspace*{2cm}
\begin{center}
\mbox{\epsfig{file=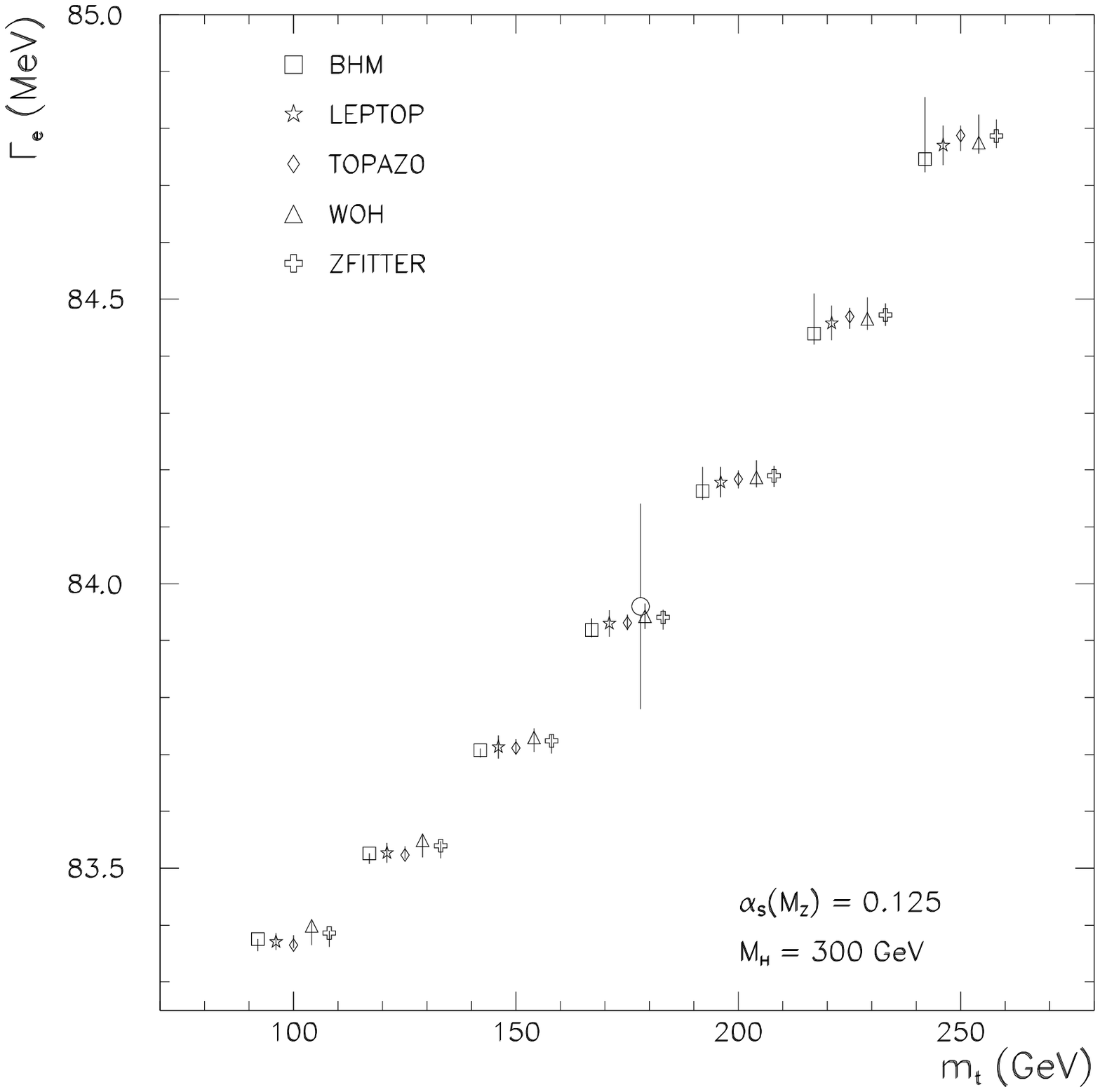,width=15cm,height=15cm}}
\end{center}
Figure 12: The {\tt BHM, LEPTOP, TOPAZ0, ZFITTER, WOH} predictions for
$\ge$, including an estimate of the theoretical error 
as a function of $m_t$,
for $\hm = 300\,$GeV and $\hat{\als} = 0.125$.
 
\newpage\vspace*{2cm}
\begin{center}
\mbox{\epsfig{file=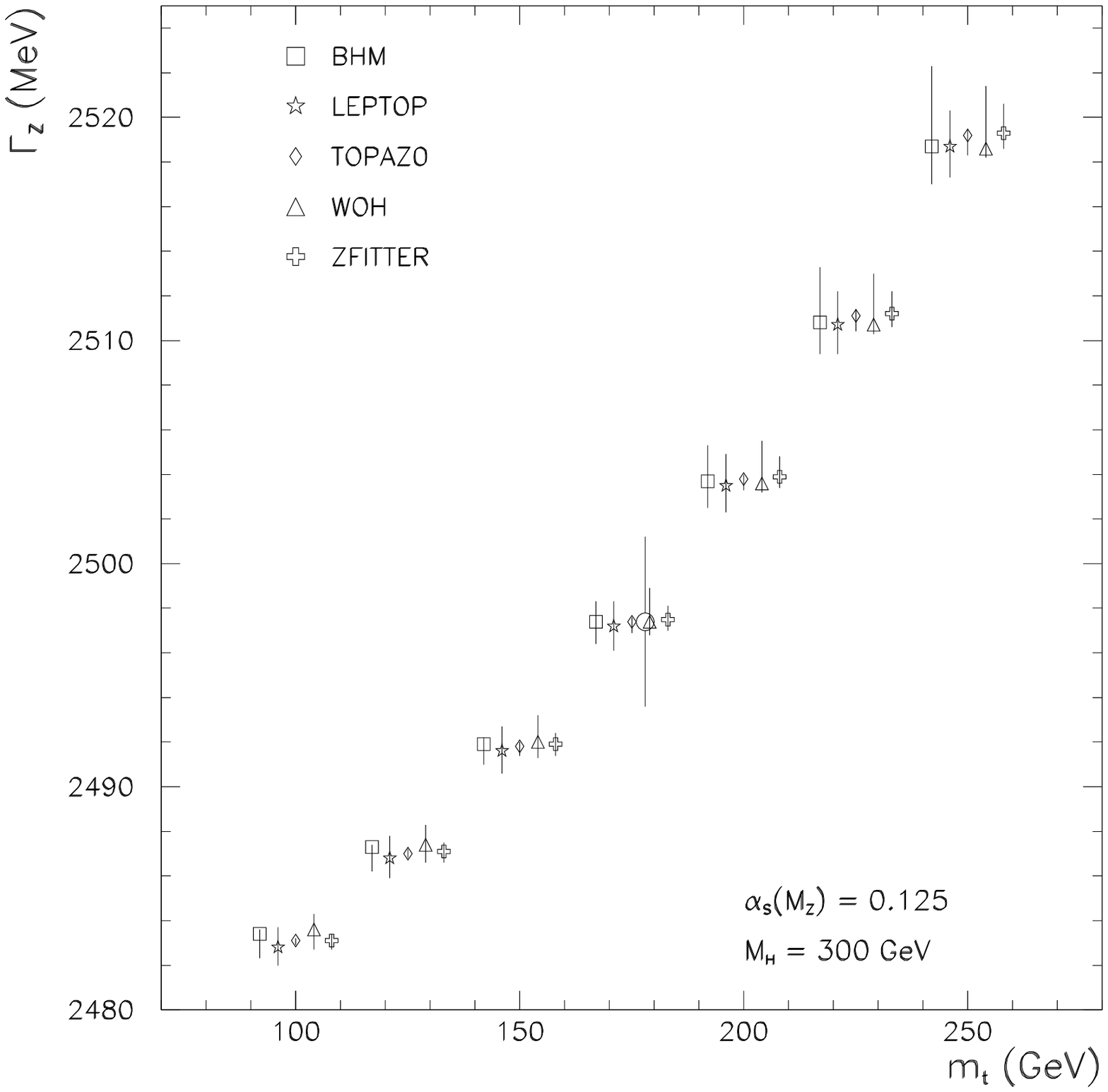,width=15cm,height=15cm}}
\end{center}
Figure 13: The {\tt BHM, LEPTOP, TOPAZ0, ZFITTER, WOH} predictions for
$\gz$, including an estimate of the theoretical error 
as a function of $m_t$,
for $\hm = 300\,$GeV and $\hat{\als} = 0.125$.
 
\newpage\vspace*{2cm}
\begin{center}
\mbox{\epsfig{file=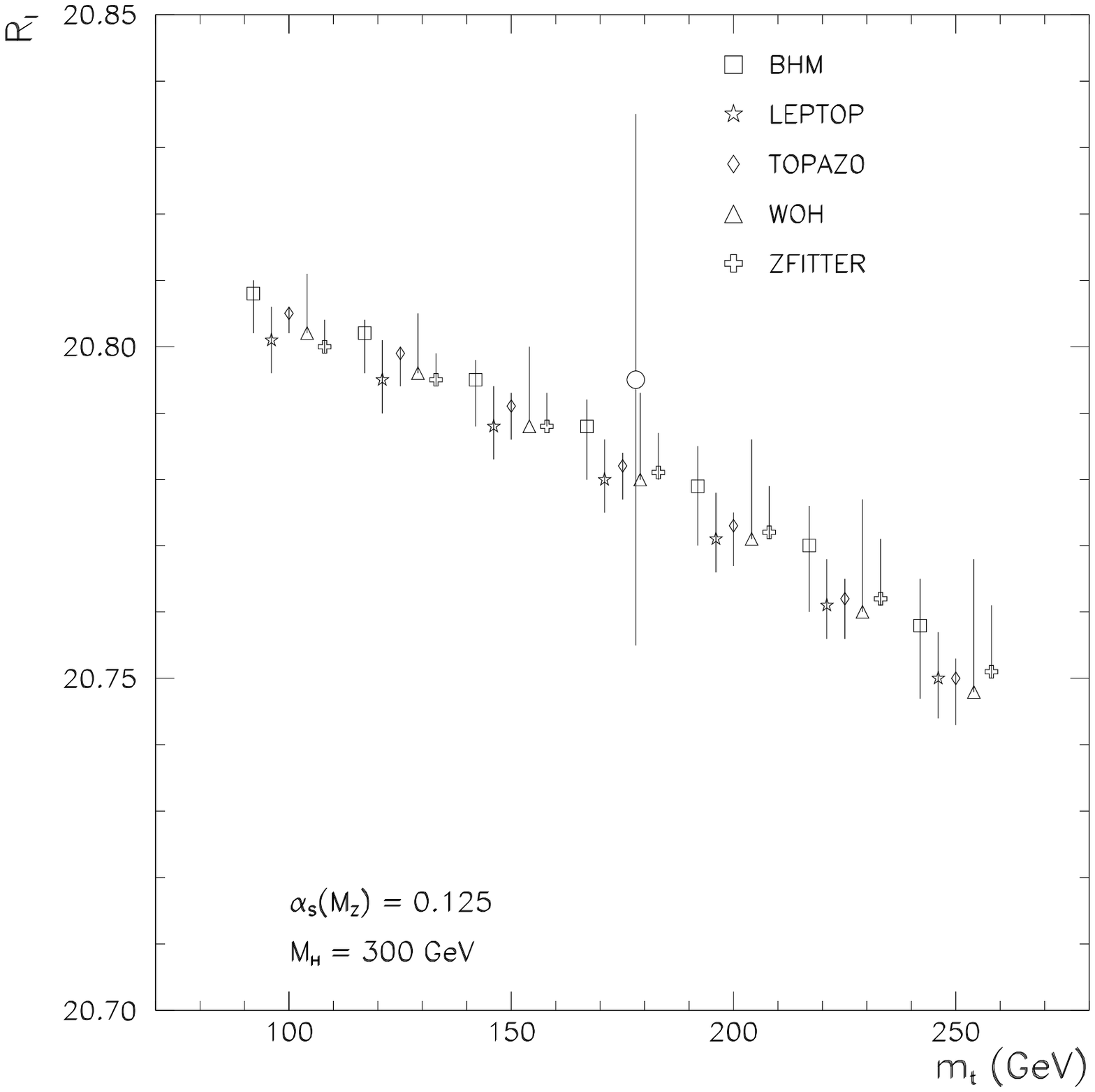,width=15cm,height=15cm}}
\end{center}
Figure 14: The {\tt BHM, LEPTOP, TOPAZ0, ZFITTER, WOH} predictions for
$R_l$, including an estimate of the theoretical error 
as a function of $m_t$,
for $\hm = 300\,$GeV and $\hat{\als} = 0.125$.
 
\newpage\vspace*{2cm}
\begin{center}
\mbox{\epsfig{file=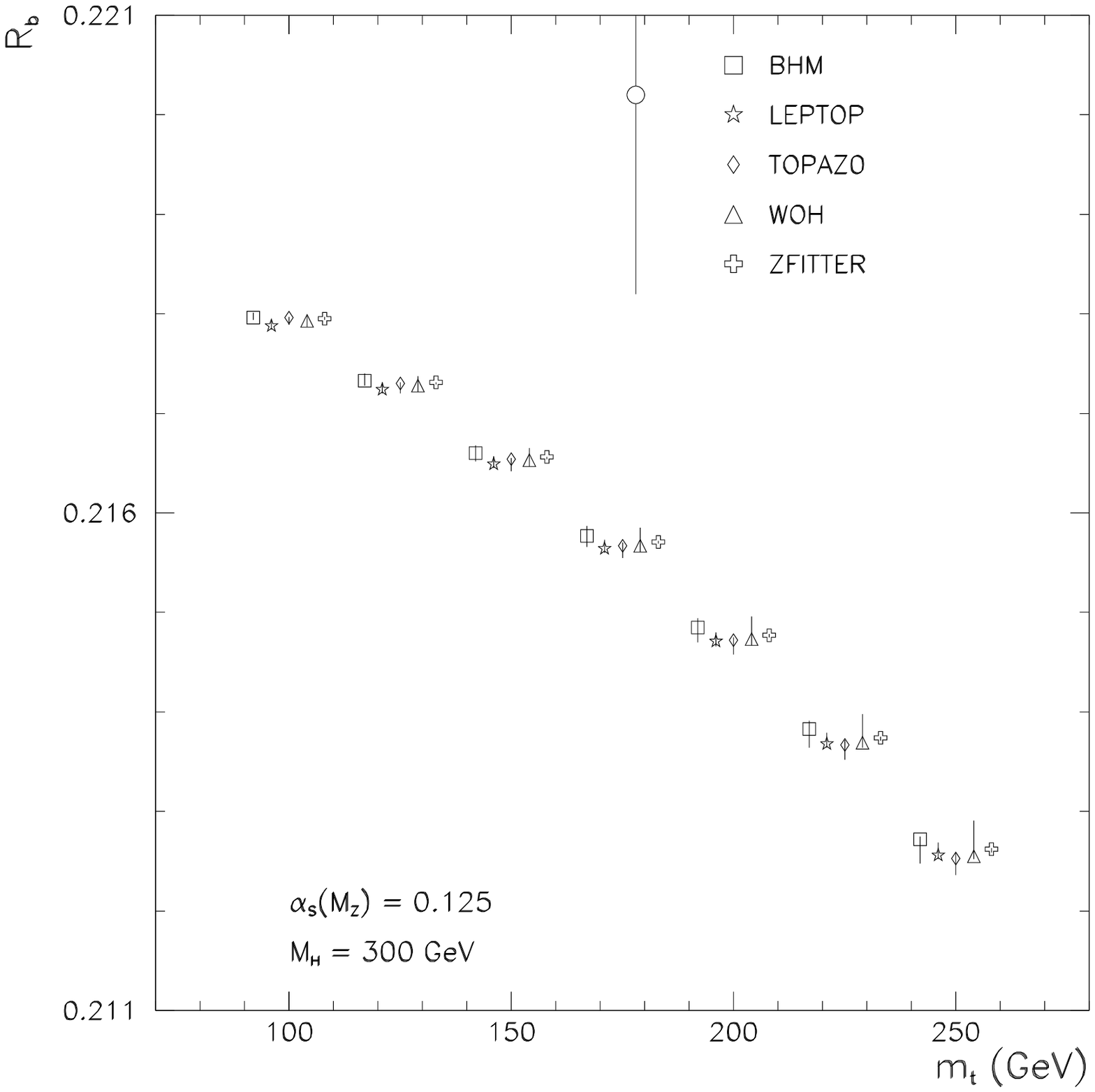,width=15cm,height=15cm}}
\end{center}
Figure 15: The {\tt BHM, LEPTOP, TOPAZ0, ZFITTER, WOH} predictions for
$R_b$, including an estimate of the theoretical error 
as a function of $m_t$,
for $\hm = 300\,$GeV and $\hat{\als} = 0.125$.
 
\newpage\vspace*{2cm}
\begin{center}
\mbox{\epsfig{file=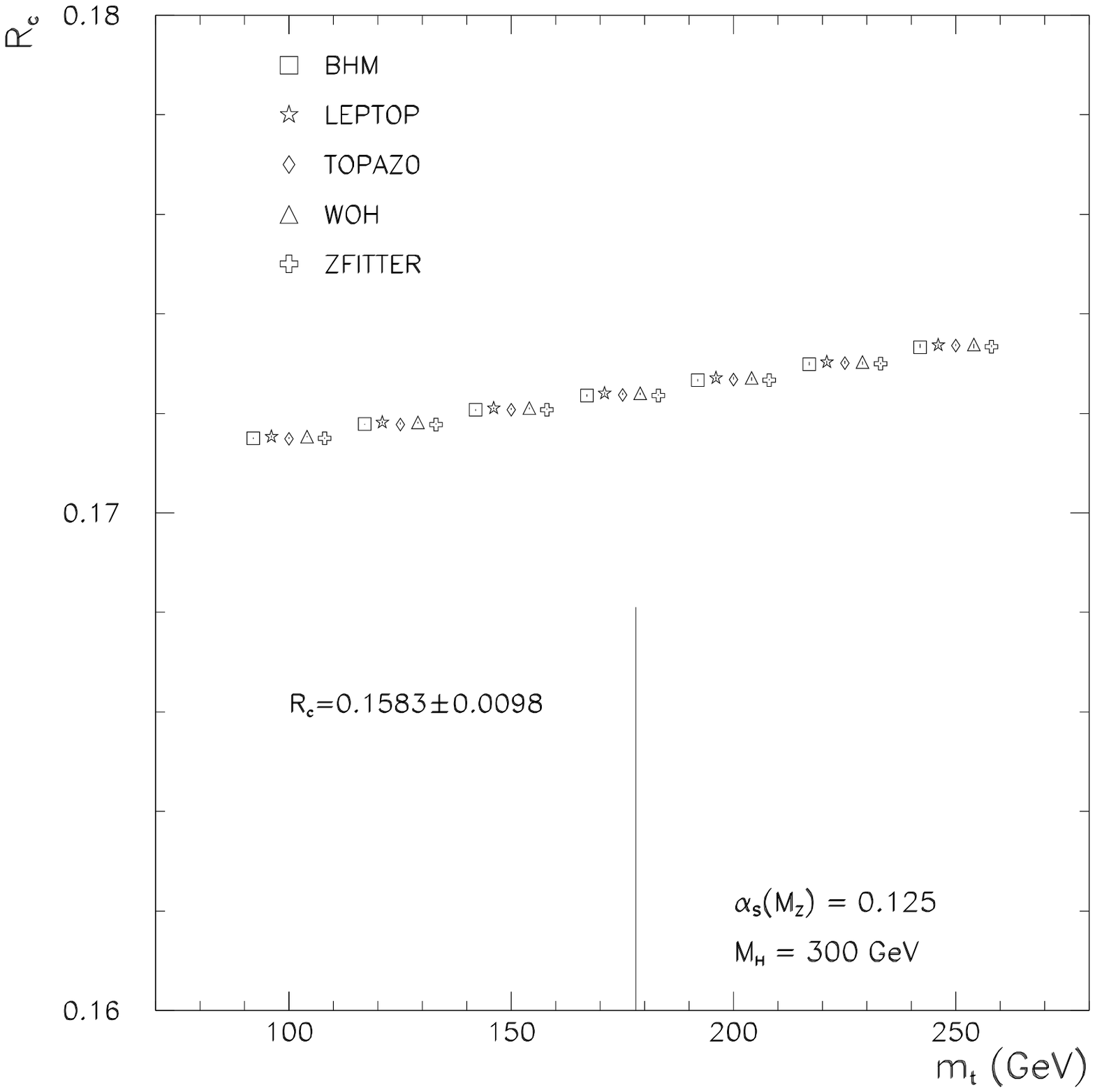,width=15cm,height=15cm}}
\end{center}
Figure 16: The {\tt BHM, LEPTOP, TOPAZ0, ZFITTER, WOH} predictions for
$R_c$, including an estimate of the theoretical error 
as a function of $m_t$,
for $\hm = 300\,$GeV and $\hat{\als} = 0.125$.
 
\newpage\vspace*{2cm}
\begin{center}
\mbox{\epsfig{file=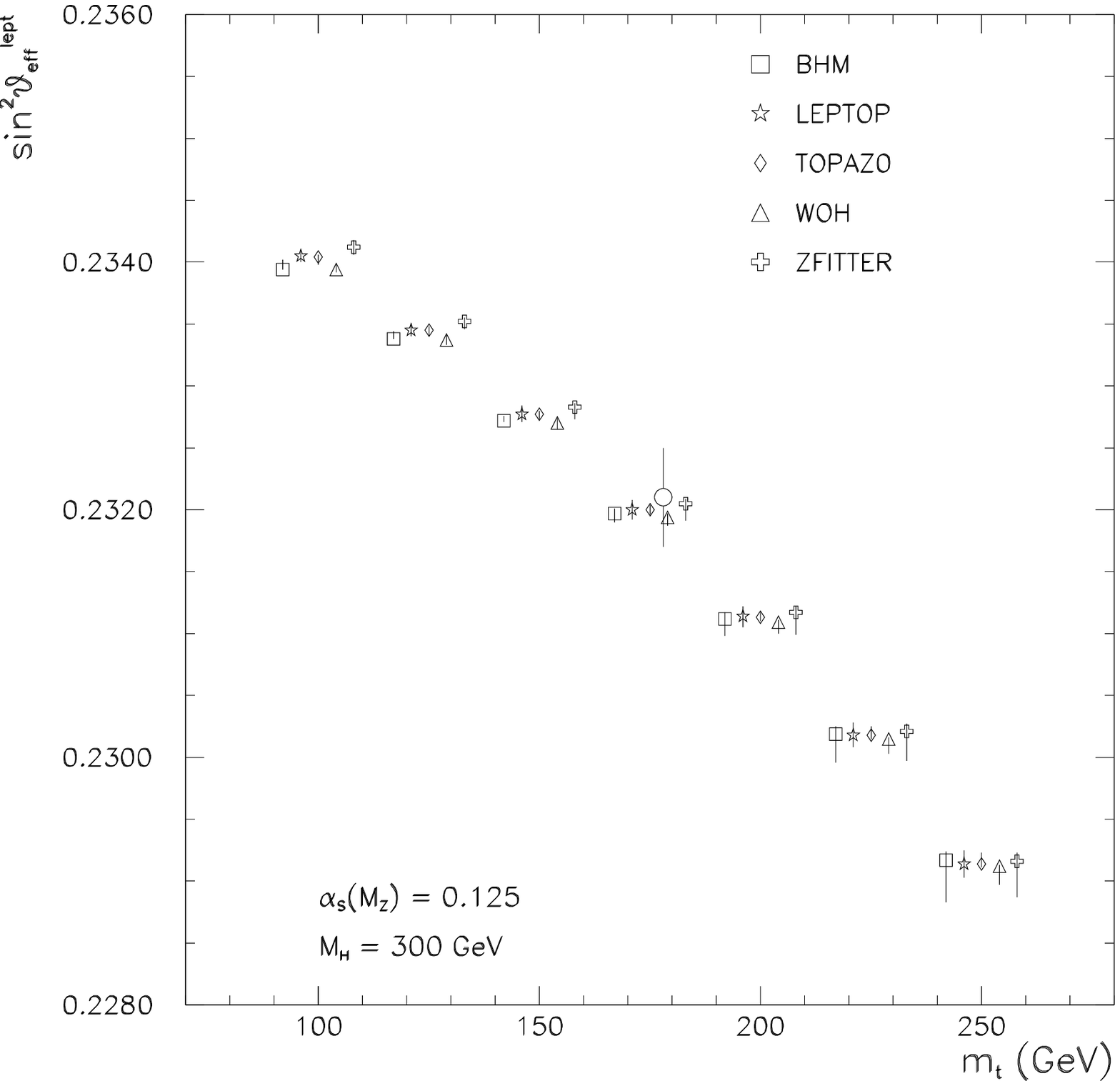,width=15cm,height=15cm}}
\end{center}
Figure 17: The {\tt BHM, LEPTOP, TOPAZ0, ZFITTER, WOH} predictions for
$\stes^l$, including an estimate of the theoretical error 
as a function of $m_t$,
for $\hm = 300\,$GeV and $\hat{\als} = 0.125$.
 
\newpage\vspace*{2cm}
\begin{center}
\mbox{\epsfig{file=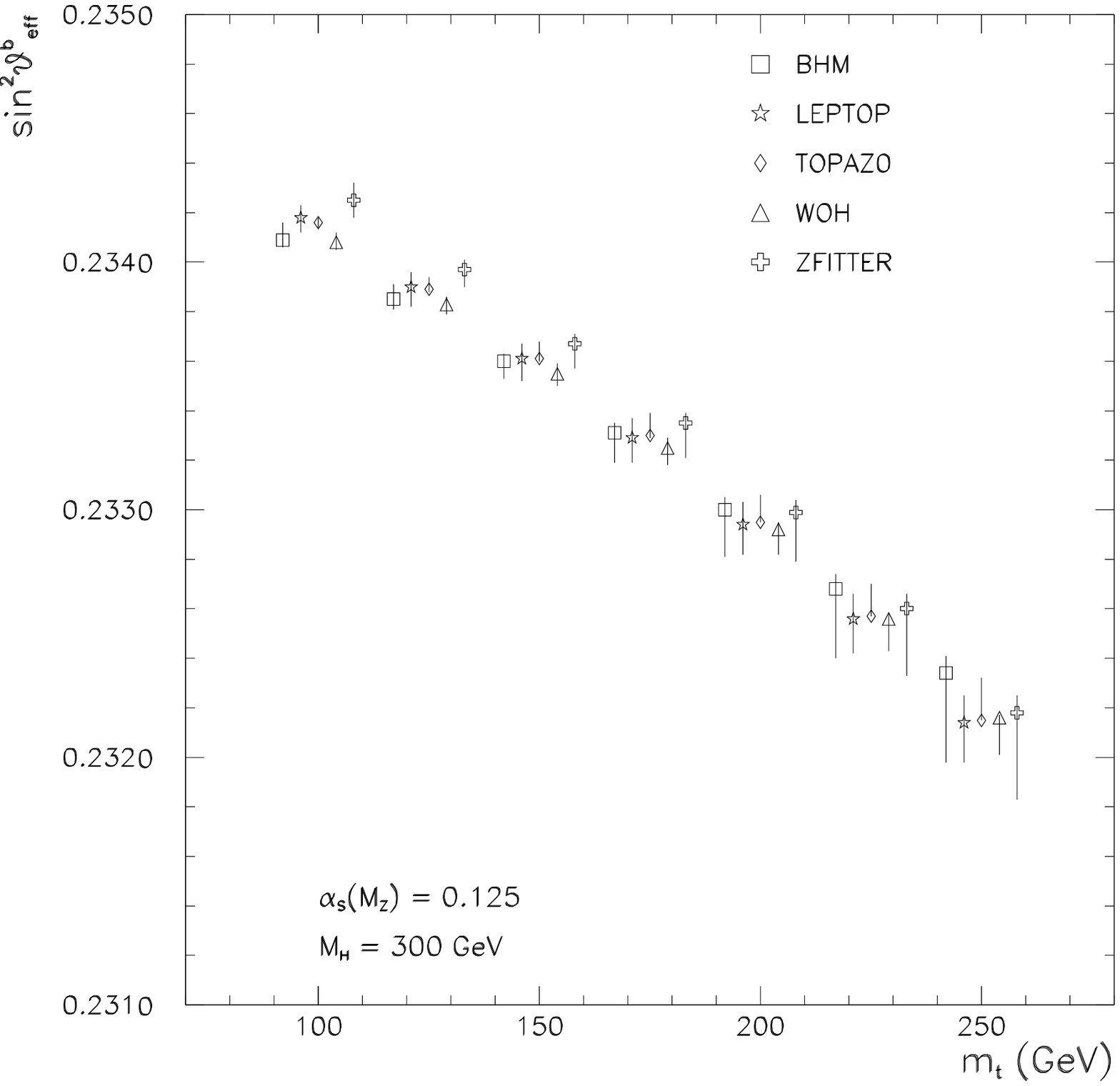,width=15cm,height=15cm}}
\end{center}
Figure 18: The {\tt BHM, LEPTOP, TOPAZ0, ZFITTER, WOH} predictions for
$\stes^b$, including an estimate of the theoretical error 
as a function of $m_t$,
for $\hm = 300\,$GeV and $\hat{\als} = 0.125$.
 
\newpage\vspace*{2cm}
\begin{center}
\mbox{\epsfig{file=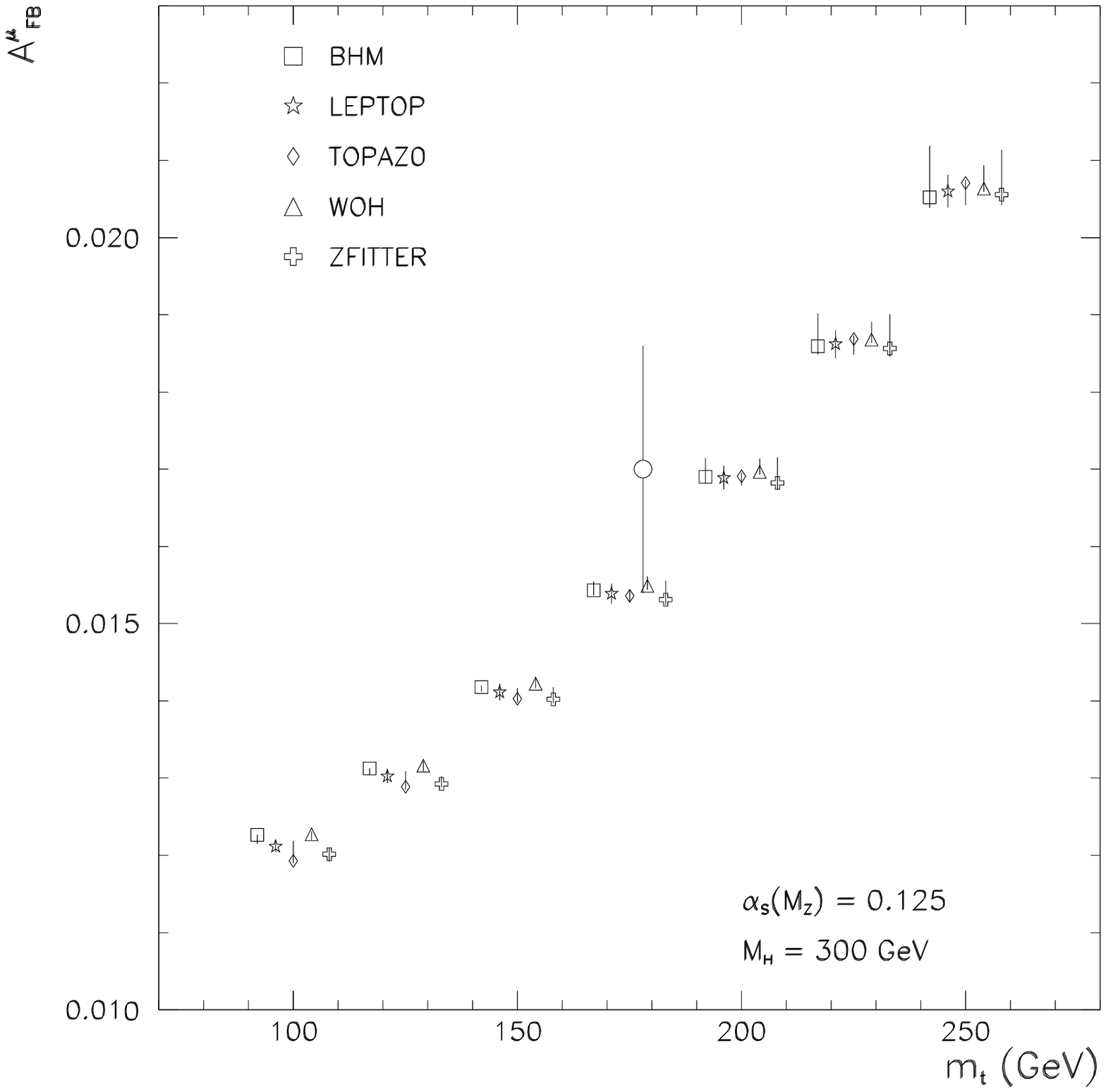,width=15cm,height=15cm}}
\end{center}
Figure 19: The {\tt BHM, LEPTOP, TOPAZ0, ZFITTER, WOH} predictions for
$\afb^{\mu}$, including an estimate of the theoretical error 
as a function of $m_t$,
for $\hm = 300\,$GeV and $\hat{\als} = 0.125$.
 
\newpage\vspace*{2cm}
\begin{center}
\mbox{\epsfig{file=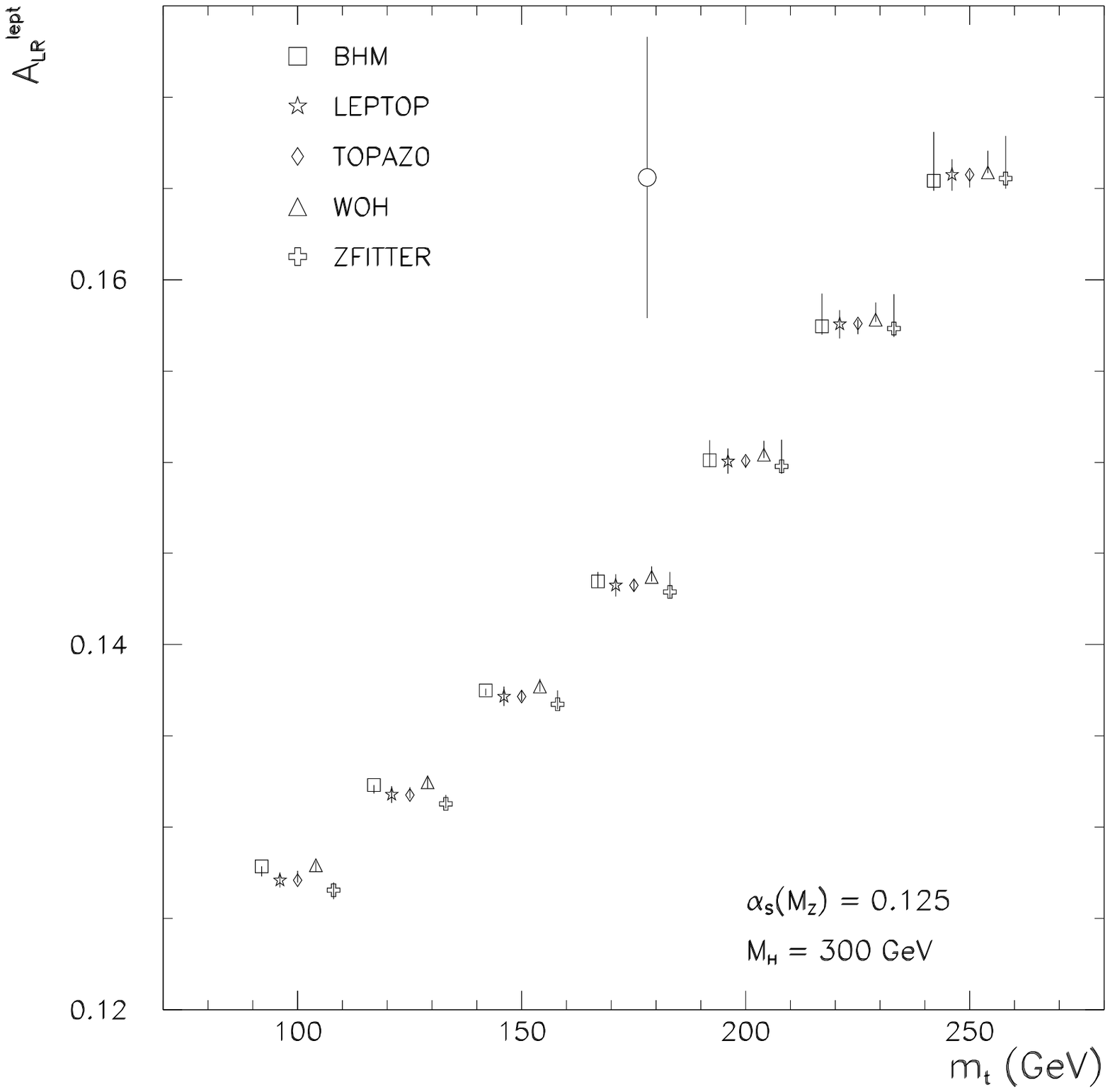,width=15cm,height=15cm}}
\end{center}
Figure 20: The {\tt BHM, LEPTOP, TOPAZ0, ZFITTER, WOH} predictions for
$\alr^{\rm {lept}}$, including an estimate of the theoretical error 
as a function of $m_t$,
for $\hm = 300\,$GeV and $\hat{\als} = 0.125$.
 
\newpage\vspace*{2cm}
\begin{center}
\mbox{\epsfig{file=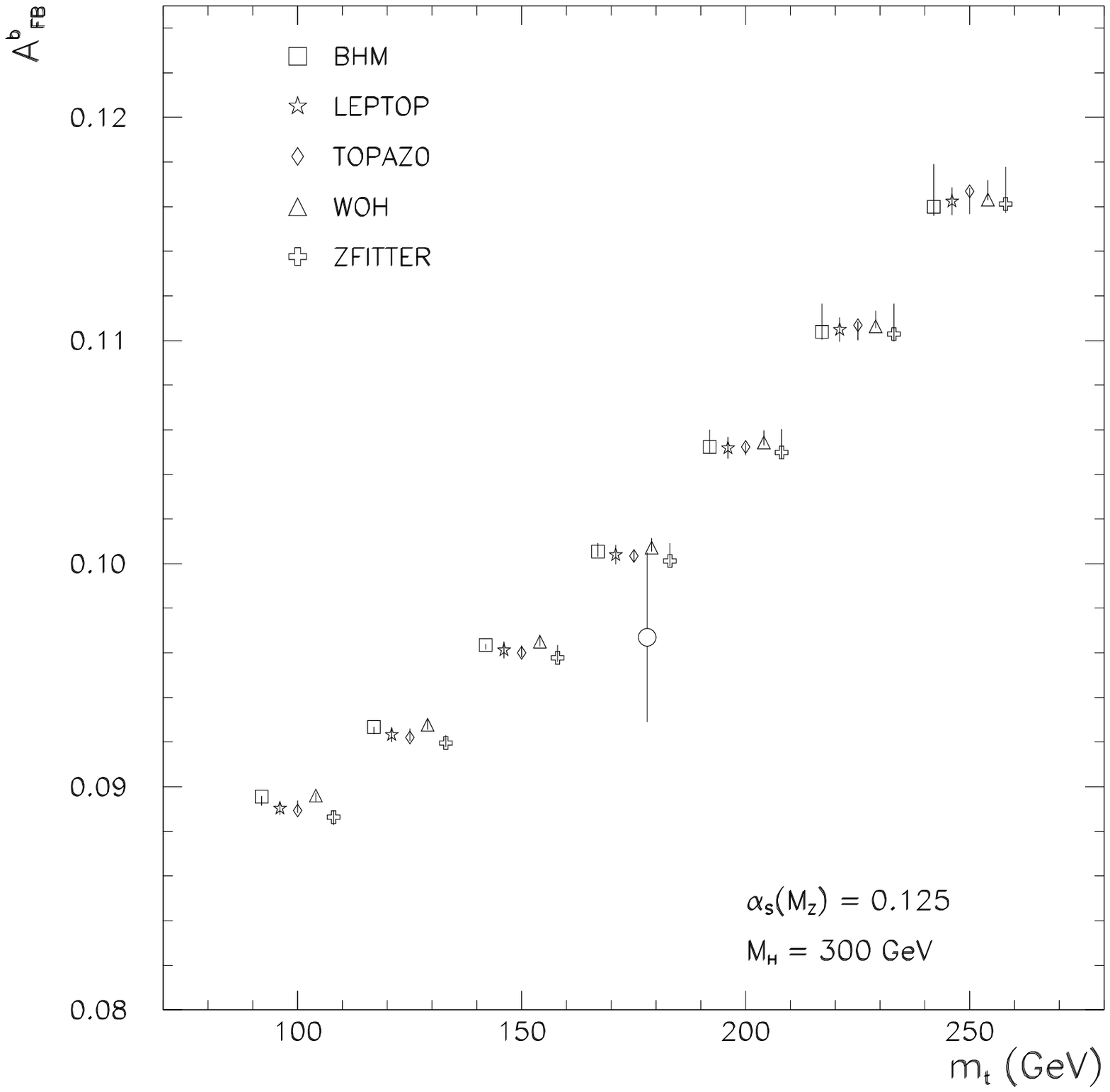,width=15cm,height=15cm}}
\end{center}
Figure 21: The {\tt BHM, LEPTOP, TOPAZ0, ZFITTER, WOH} predictions for
$\afb^b$, including an estimate of the theoretical error 
as a function of $m_t$,
for $\hm = 300\,$GeV and $\hat{\als} = 0.125$.
 
\newpage\vspace*{2cm}
\begin{center}
\mbox{\epsfig{file=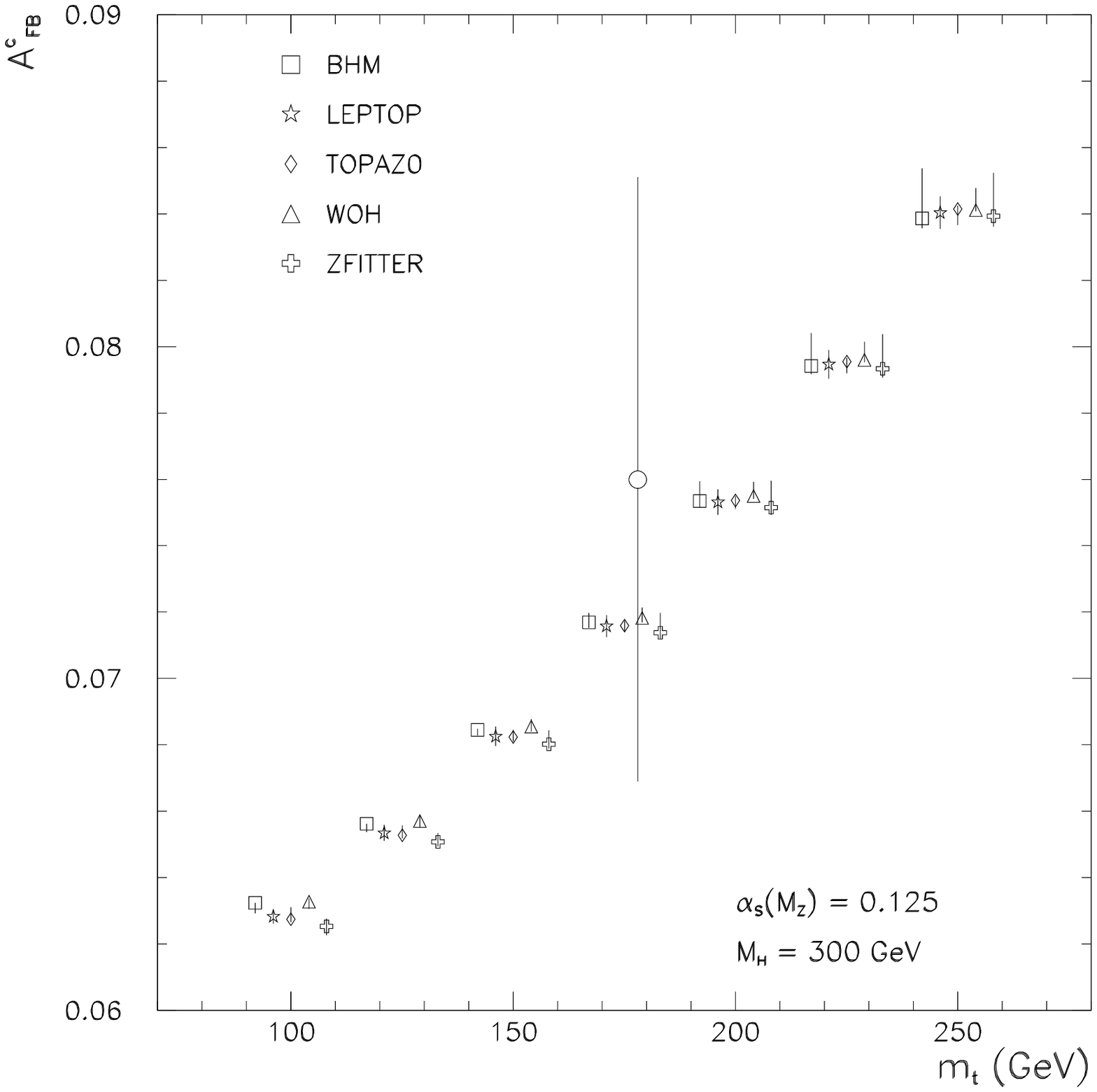,width=15cm,height=15cm}}
\end{center}
Figure 22: The {\tt BHM, LEPTOP, TOPAZ0, ZFITTER, WOH} predictions for
$\afb^c$, including an estimate of the theoretical error 
as a function of $m_t$,
for $\hm = 300\,$GeV and $\hat{\als} = 0.125$.
 
\newpage\vspace*{2cm}
\begin{center}
\mbox{\epsfig{file=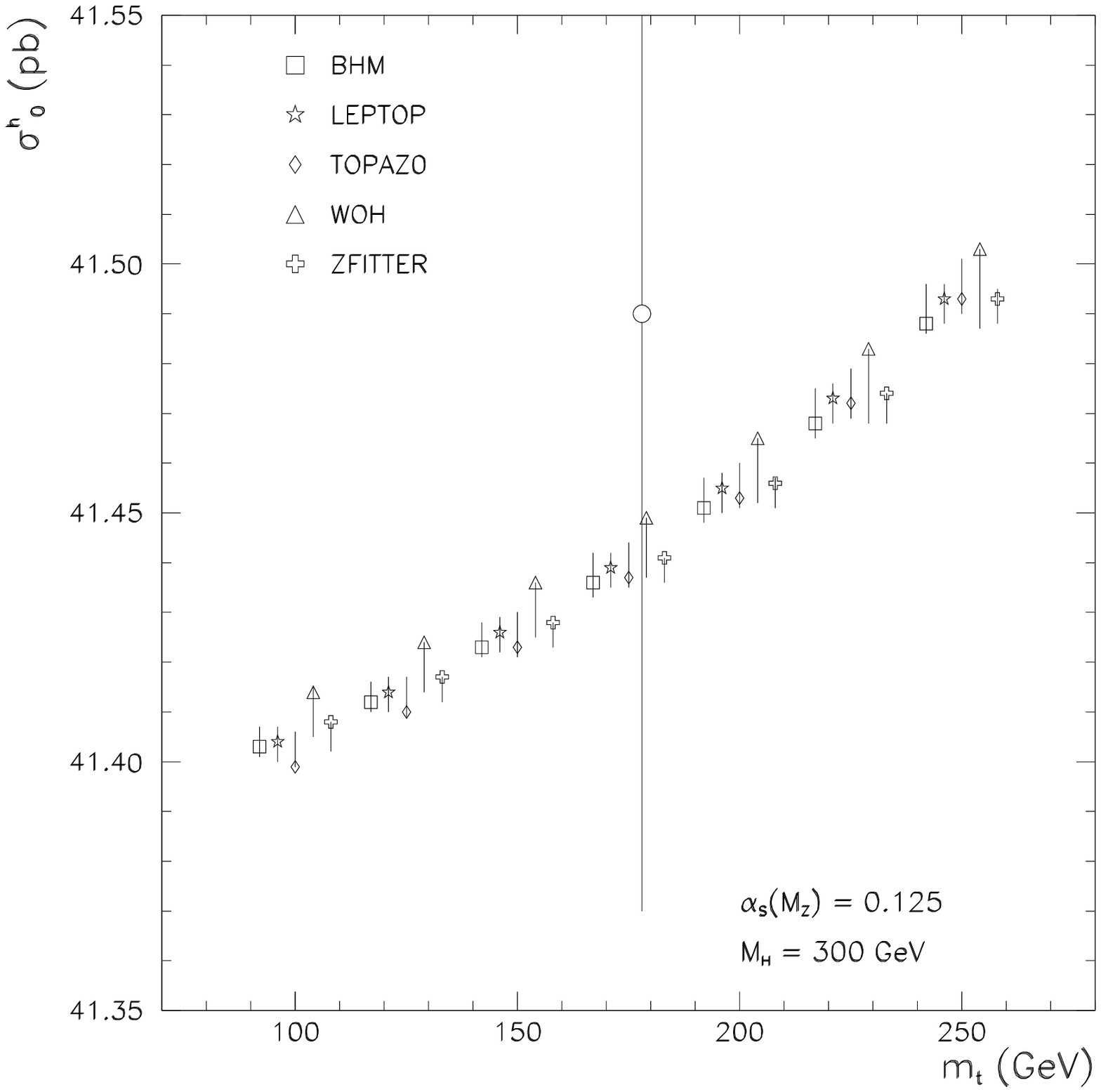,width=15cm,height=15cm}}
\end{center}
Figure 23: The {\tt BHM, LEPTOP, TOPAZ0, ZFITTER, WOH} predictions for
$\sigma^h_0$, including an estimate of the theoretical error 
as a function of $m_t$,
for $\hm = 300\,$GeV and $\hat{\als} = 0.125$.
 
\newpage
\subsection*{Realistic-observables}
\addcontentsline{toc}{subsection}{Realistic-observables}
\vspace*{1cm}
 
\begin{center}
\mbox{\epsfig{file=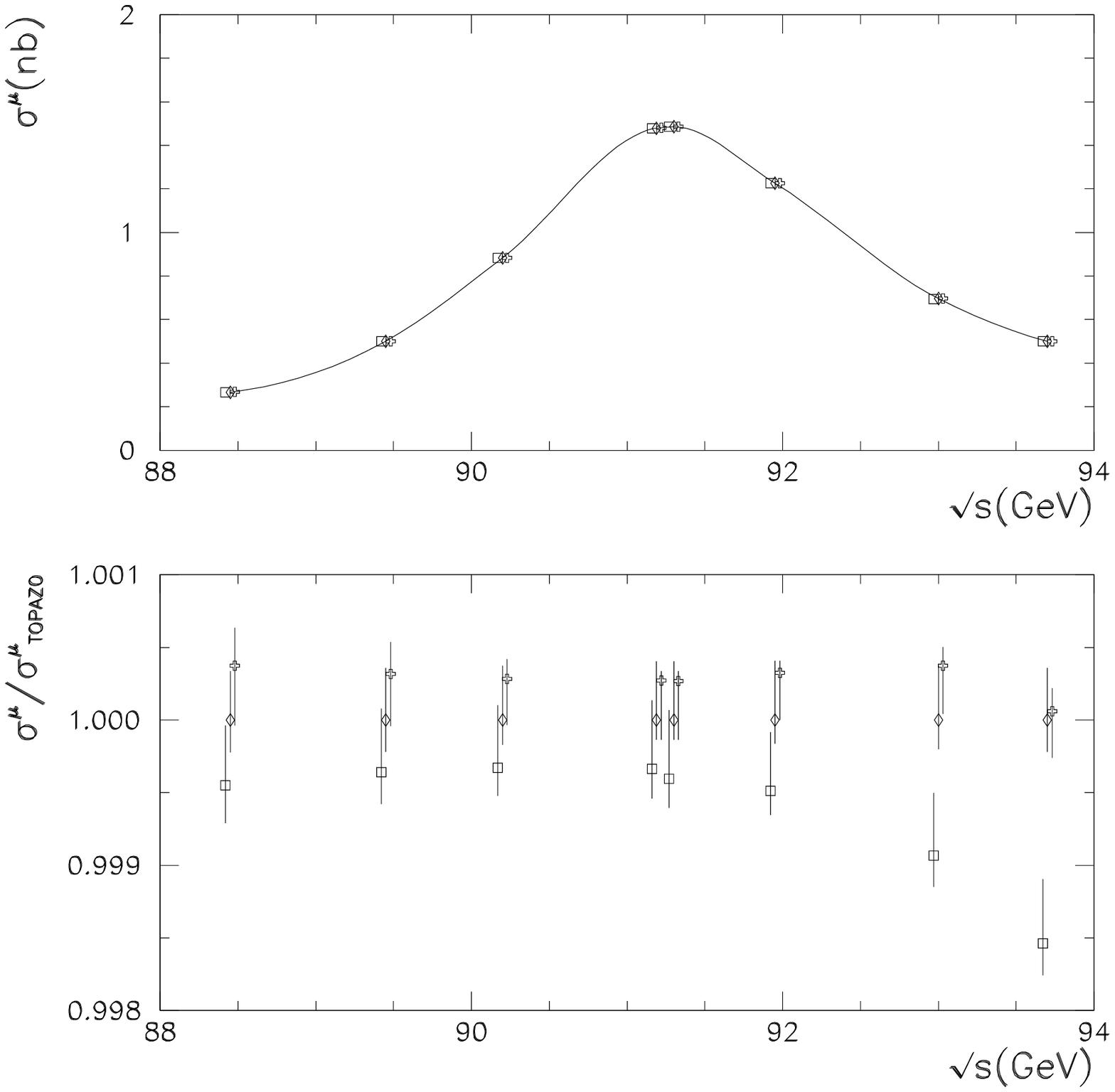,width=15cm,height=15cm}}
\end{center}
Figure 24: The {\tt BHM} (square), {\tt TOPAZ0} (diamond) and
{\tt ZFITTER} (cross) predictions, including an estimate of the theoretical
error, for $\sigma^{\mu}$ in a fully extrapolated set-up. Here $m_t = 175\,$GeV,
$\hm = 300\,$GeV and $\hat{\als} = 0.125$. In the lower part a comparison is
also shown with the relative deviation of {\tt BHM, ZFITTER} versus
{\tt TOPAZ0}.
 
\newpage\vspace*{2cm}
\begin{center}
\mbox{\epsfig{file=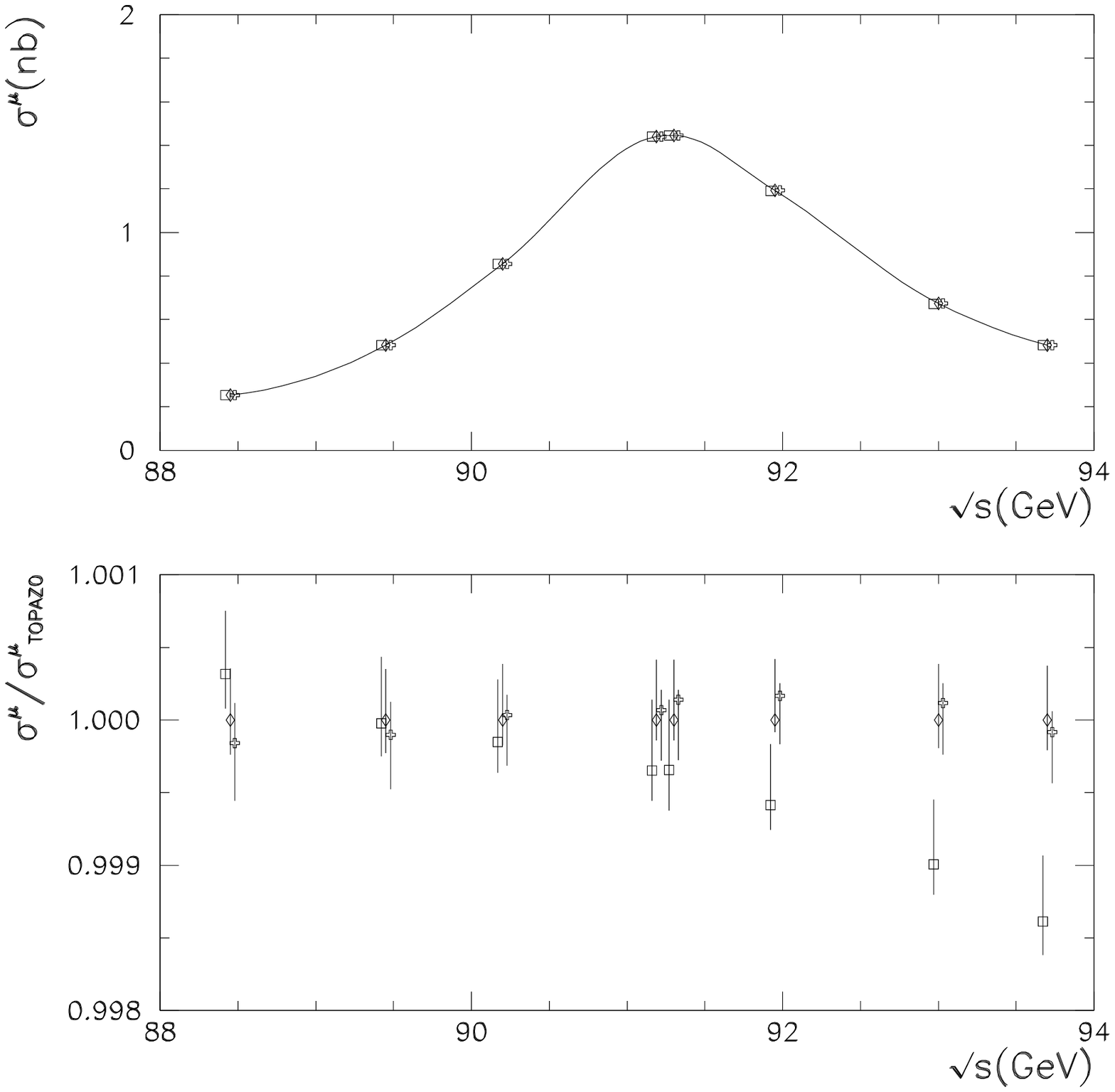,width=15cm,height=15cm}}
\end{center}
Figure 25: The same as in Fig. 24 with an $s'$ cut, $s' = 0.5 s$.
 
\newpage\vspace*{2cm}
\begin{center}
\mbox{\epsfig{file=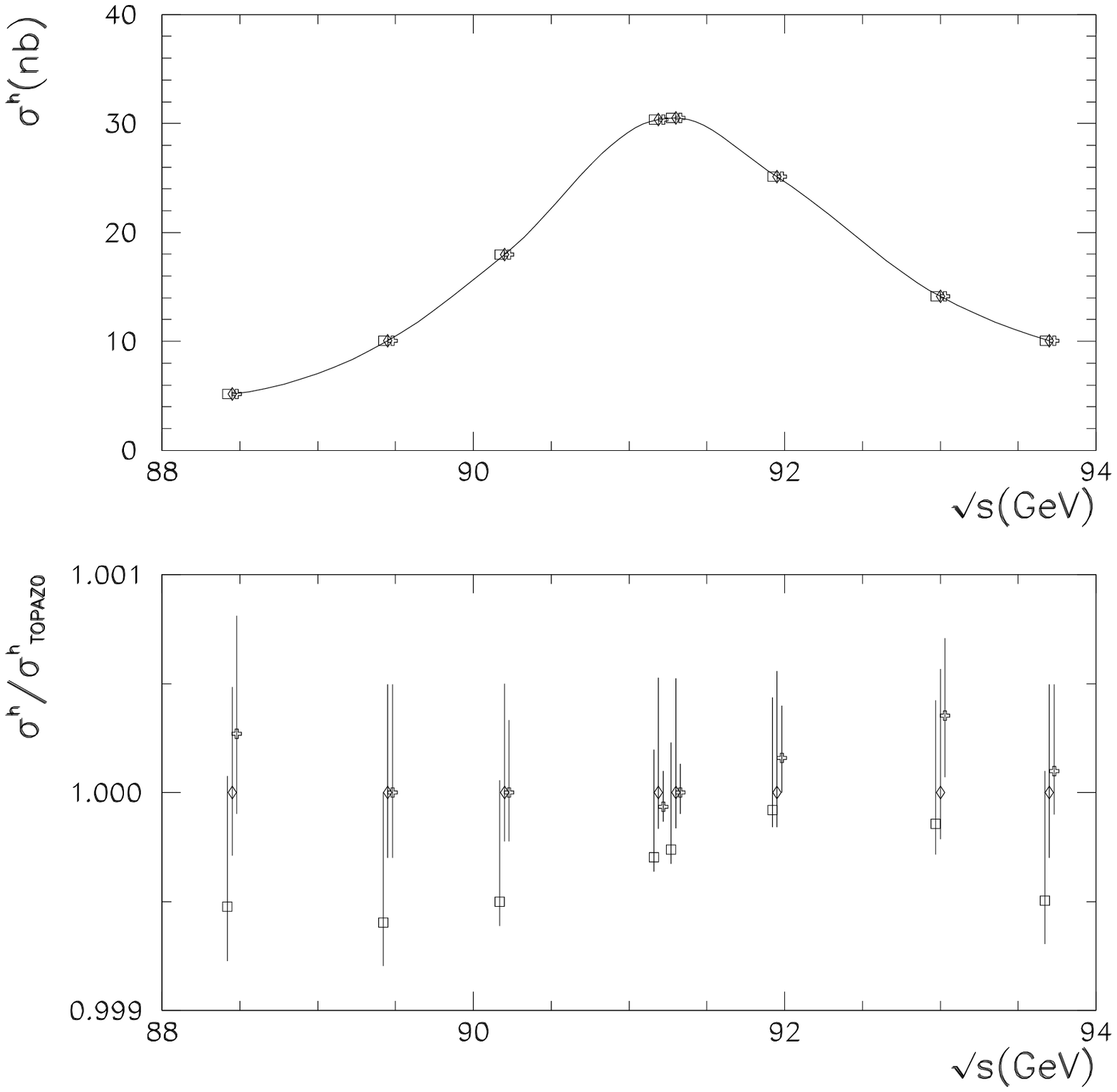,width=15cm,height=15cm}}
\end{center}
Figure 26: The same as in Fig. 24 for the hadronic cross-section,
$\sigma^h$.
 
\newpage\vspace*{2cm}
\begin{center}
\mbox{\epsfig{file=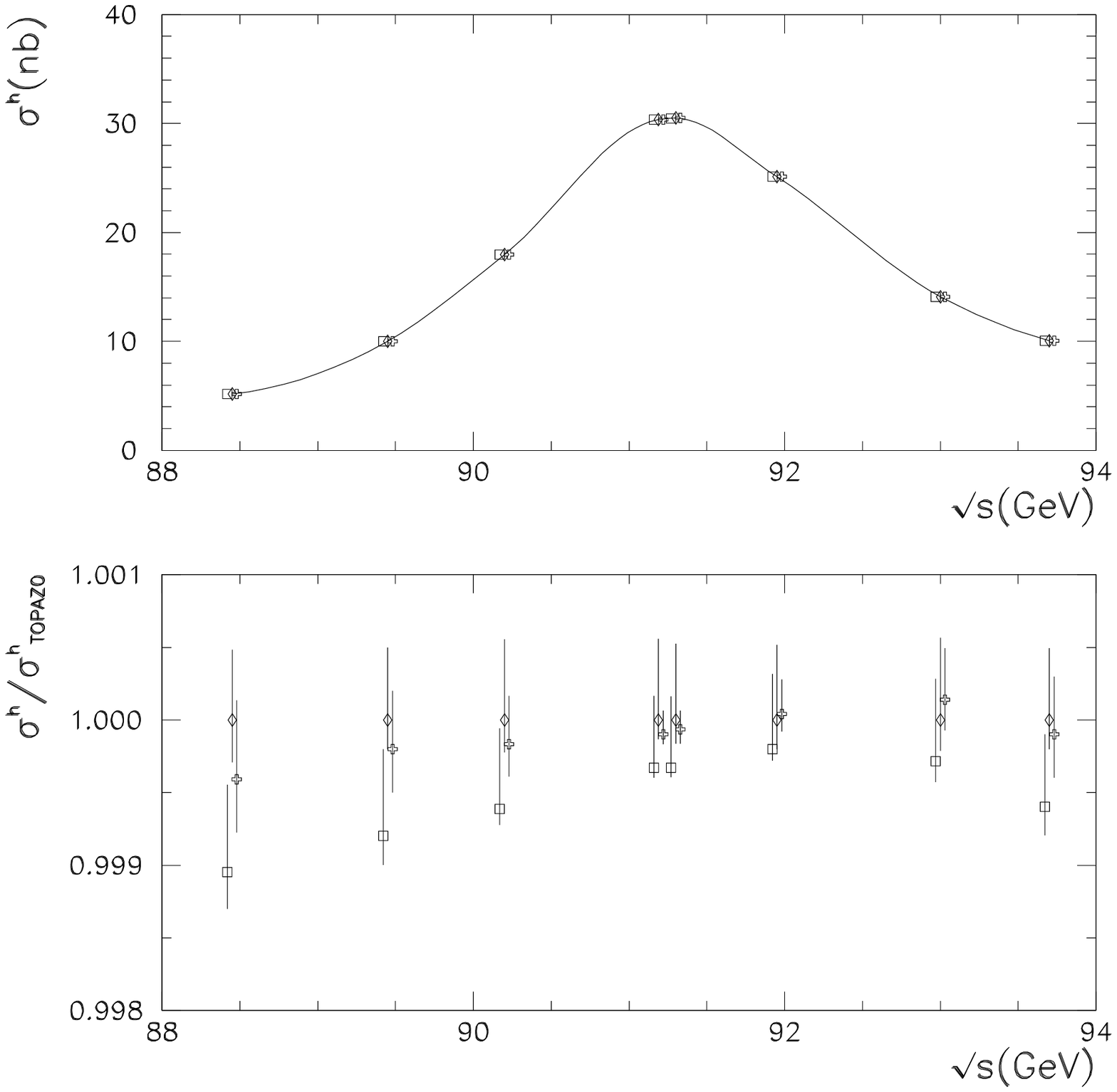,width=15cm,height=15cm}}
\end{center}
Figure 27: The same as in Fig. 26 with an $s'$ cut, $s' = 0.01 s$.
 
\newpage\vspace*{2cm}
\begin{center}
\mbox{\epsfig{file=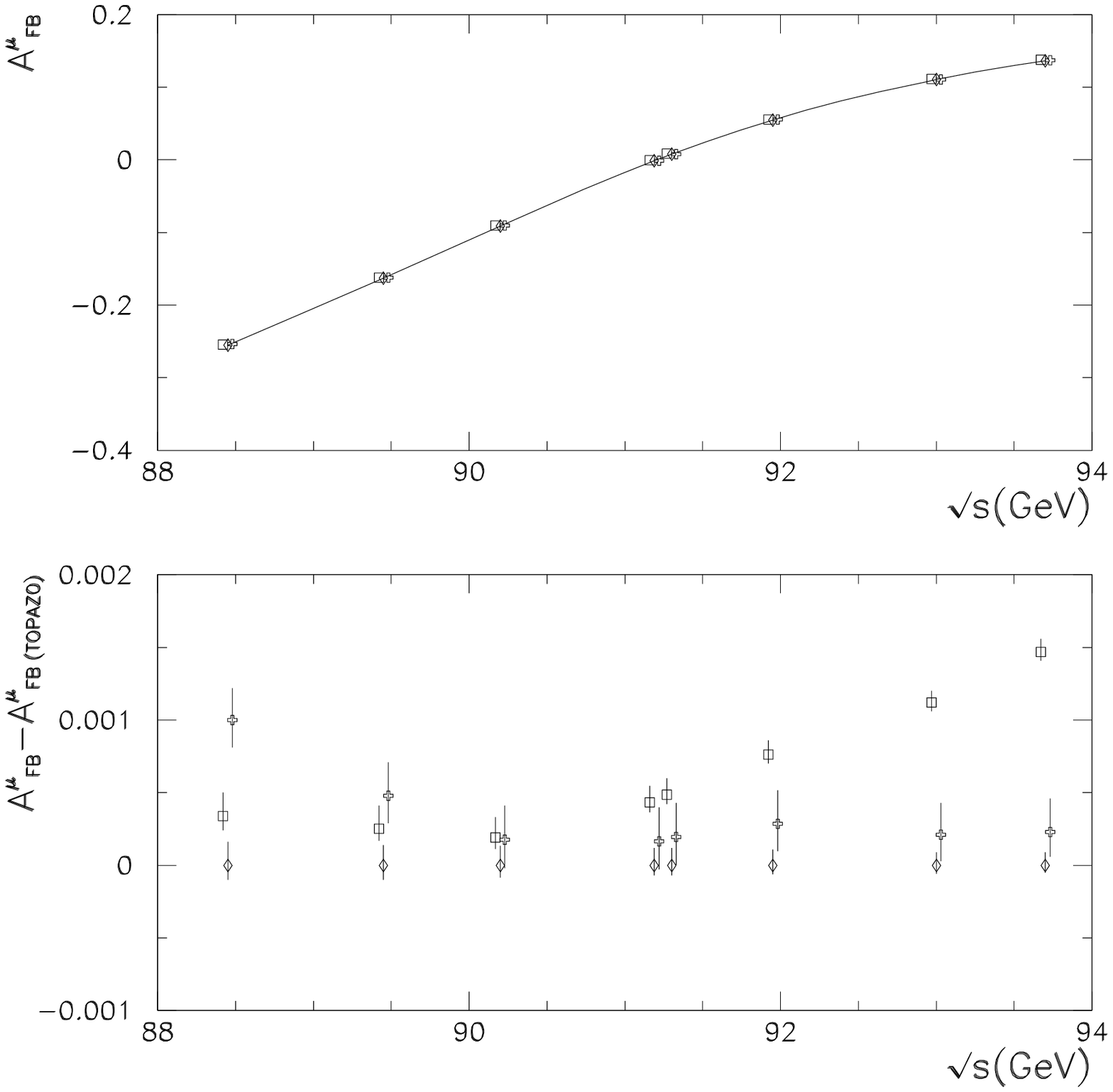,width=15cm,height=15cm}}
\end{center}
Figure 28: The same as in Fig. 24 for the leptonic forward--backward
asymmetry. In the lower part a comparison is also shown with the absolute
deviation of {\tt BHM, ZFITTER} versus {\tt TOPAZ0}.
 
\newpage\vspace*{2cm}
\begin{center}
\mbox{\epsfig{file=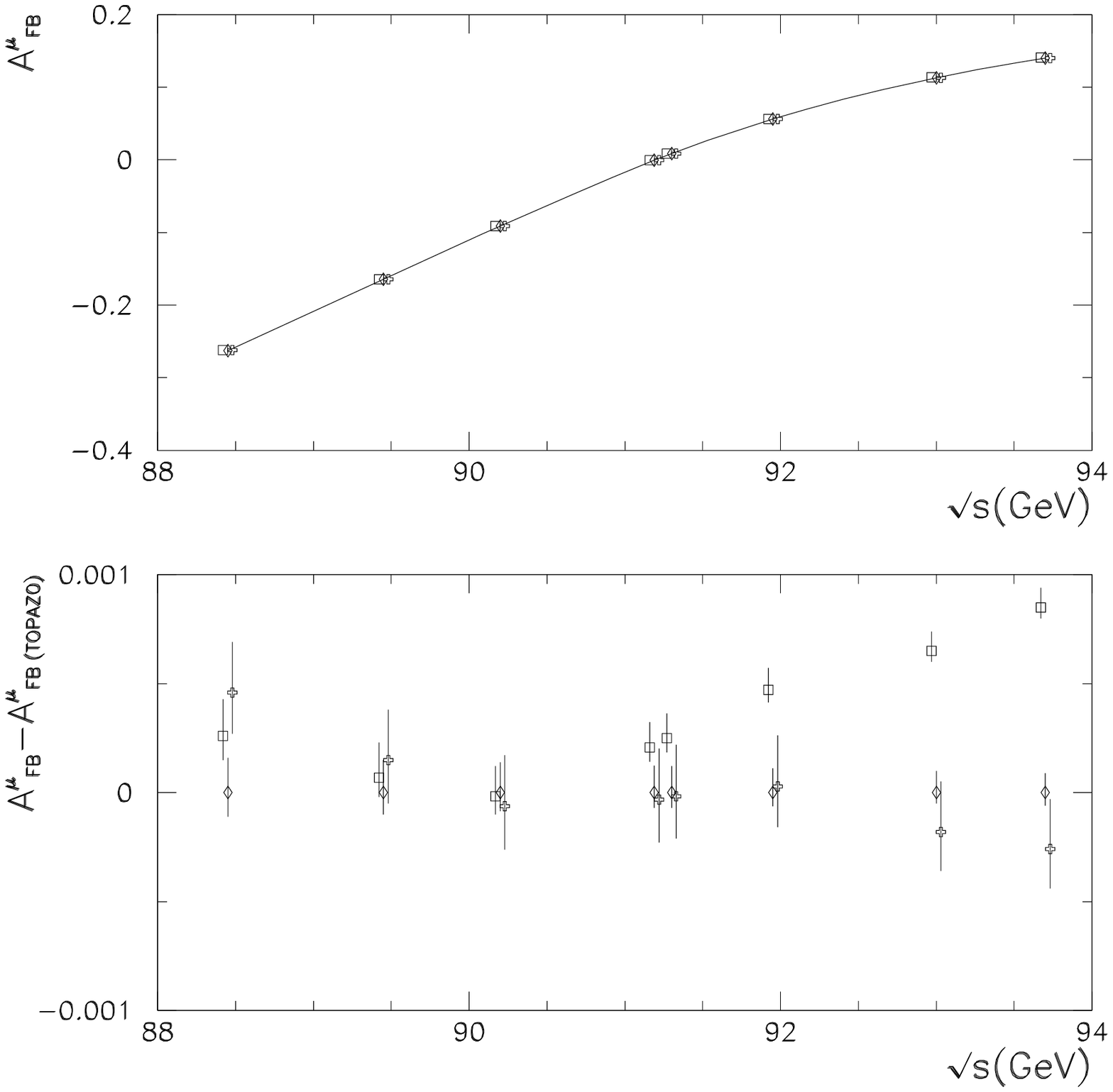,width=15cm,height=15cm}}
\end{center}
Figure 29: The same as in Fig. 28 with an $s'$ cut, $s' = 0.5 s$.
 
\newpage\vspace*{2cm}
\begin{center}
\mbox{\epsfig{file=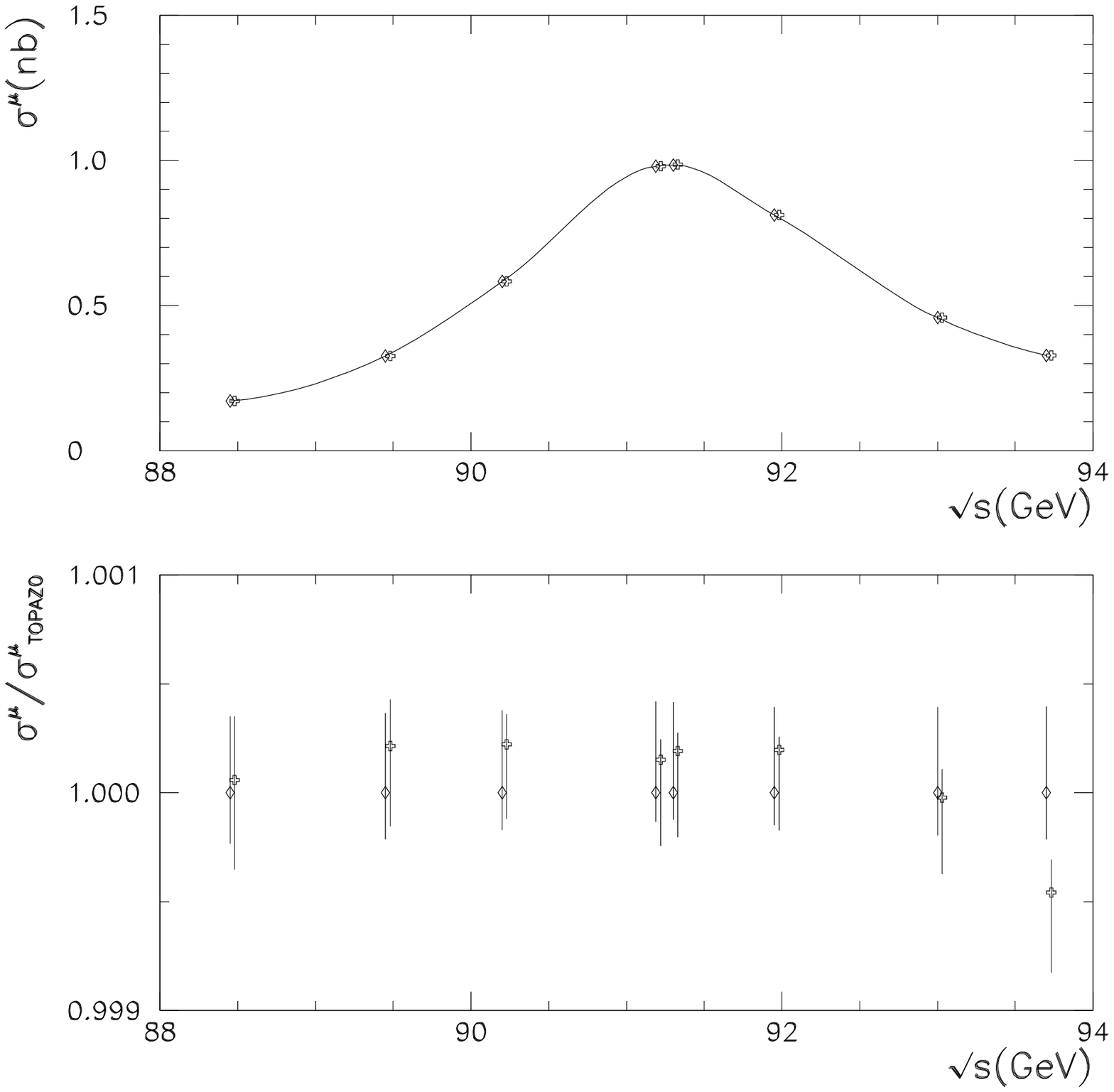,width=15cm,height=15cm}}
\end{center}
Figure 30: The {\tt TOPAZ0} (diamond) and {\tt ZFITTER} (cross) predictions
, including an estimate of the theoretical error, for $\sigma^{\mu}$ in the
following set-up: $40^{\circ} < \theta_-
< 140^{\circ}$, $\theta_{\rm {acoll}} < 10^{\circ}$ and
$E_{\rm {th}} = 20\,$GeV. Here $m_t = 175\,$GeV, $\hm = 300\,$GeV and 
$\hat{\als} = 0.125$.
In the lower part a comparison is also shown with the relative
deviation of {\tt ZFITTER} versus {\tt TOPAZ0}.
 
\newpage\vspace*{2cm}
\begin{center}
\mbox{\epsfig{file=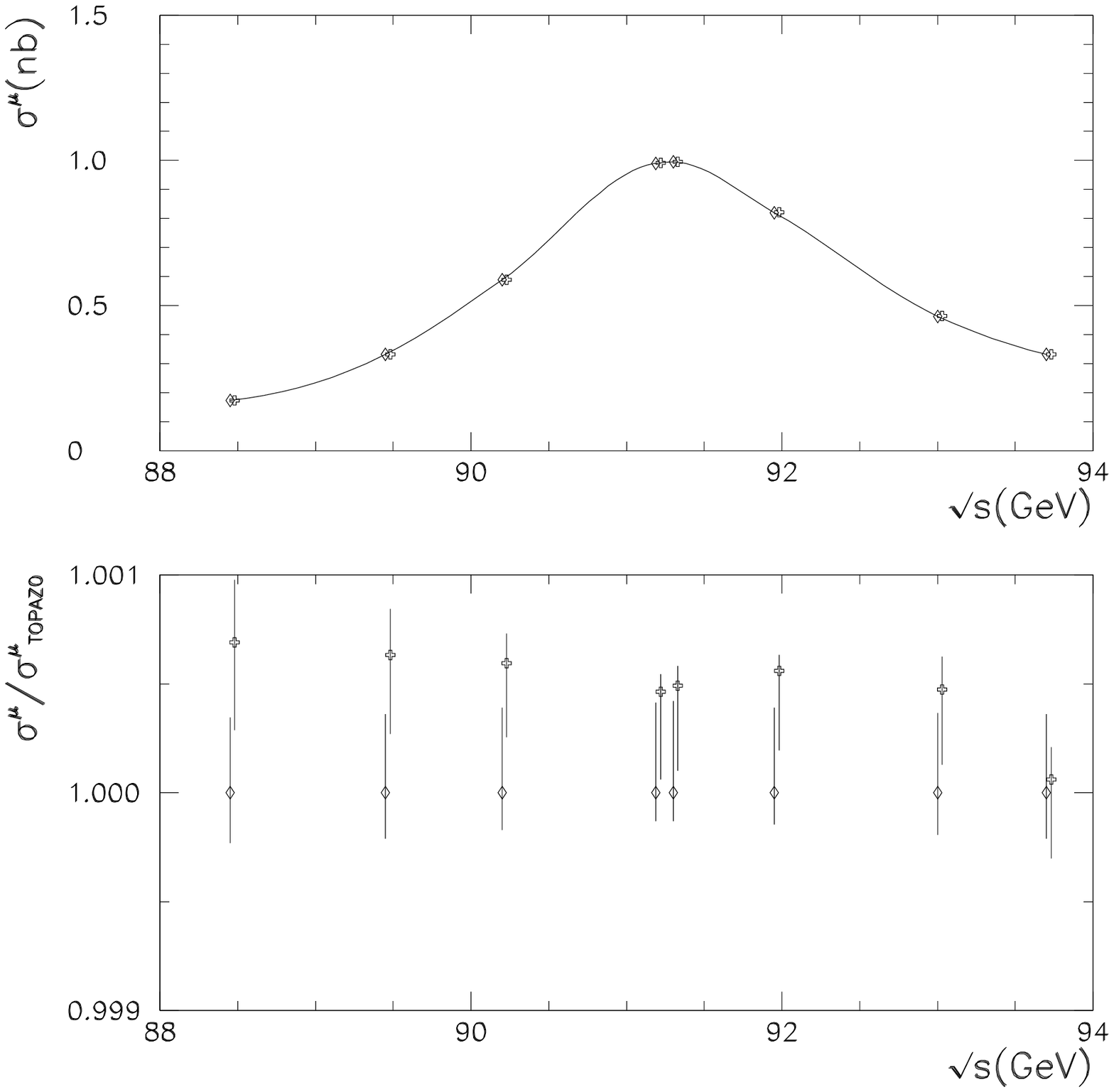,width=15cm,height=15cm}}
\end{center}
Figure 31: The same as in Fig. 30 for $\theta_{\rm {acoll}} < 25^{\circ}$.
 
\newpage\vspace*{2cm}
\begin{center}
\mbox{\epsfig{file=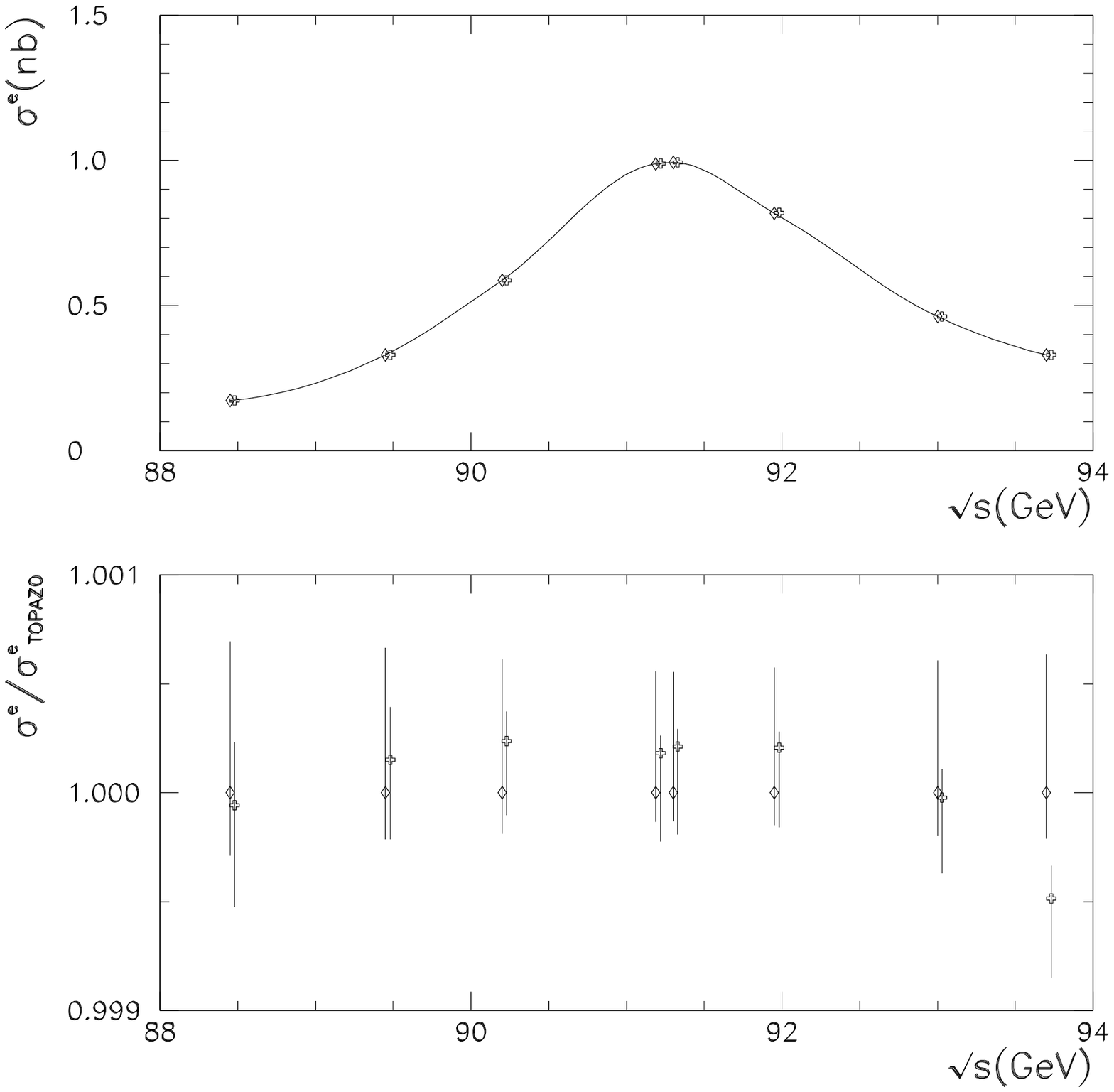,width=15cm,height=15cm}}
\end{center}
Figure 32: The {\tt TOPAZ0} (diamond) and {\tt ZFITTER} (cross) predictions,
including an estimate of the theoretical error, for $\sigma^e$ with
$s$-channel electrons, in the following set-up: $40^{\circ} < \theta_-
< 140^{\circ}$, $\theta_{\rm {acoll}} < 10^{\circ}$ and
$E_{\rm {th}} = 1\,$GeV. Here $m_t = 175\,$GeV,
$\hm = 300\,$GeV and $\hat{\als} = 0.125$. In the lower part a comparison is
also shown with the relative deviation of {\tt ZFITTER} versus {\tt TOPAZ0}.
 
\newpage\vspace*{2cm}
\begin{center}
\mbox{\epsfig{file=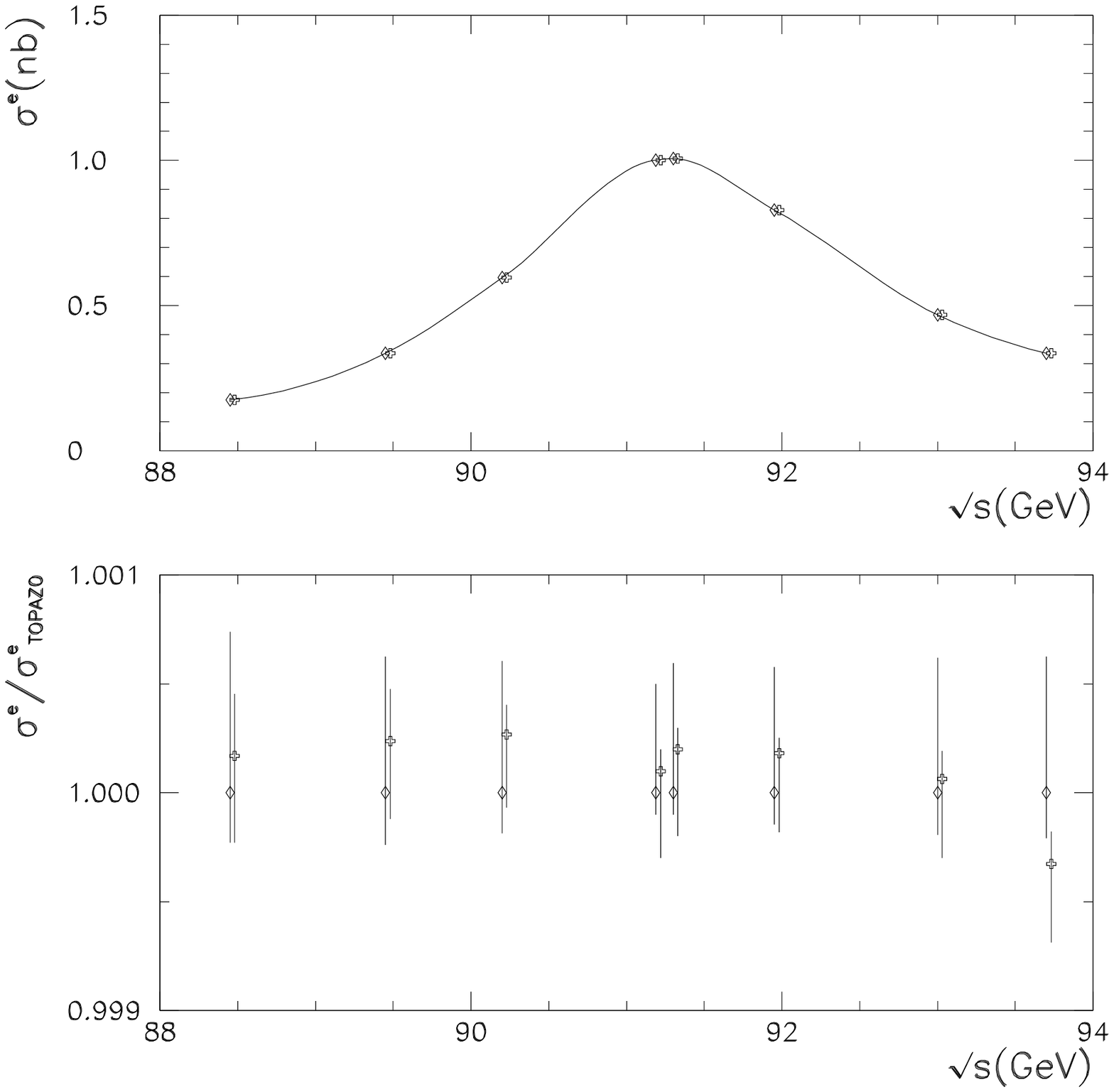,width=15cm,height=15cm}}
\end{center}
Figure 33: The same as in Fig. 32 for $\theta_{\rm {acoll}} < 25^{\circ}$.
 
\newpage\vspace*{2cm}
\begin{center}
\mbox{\epsfig{file=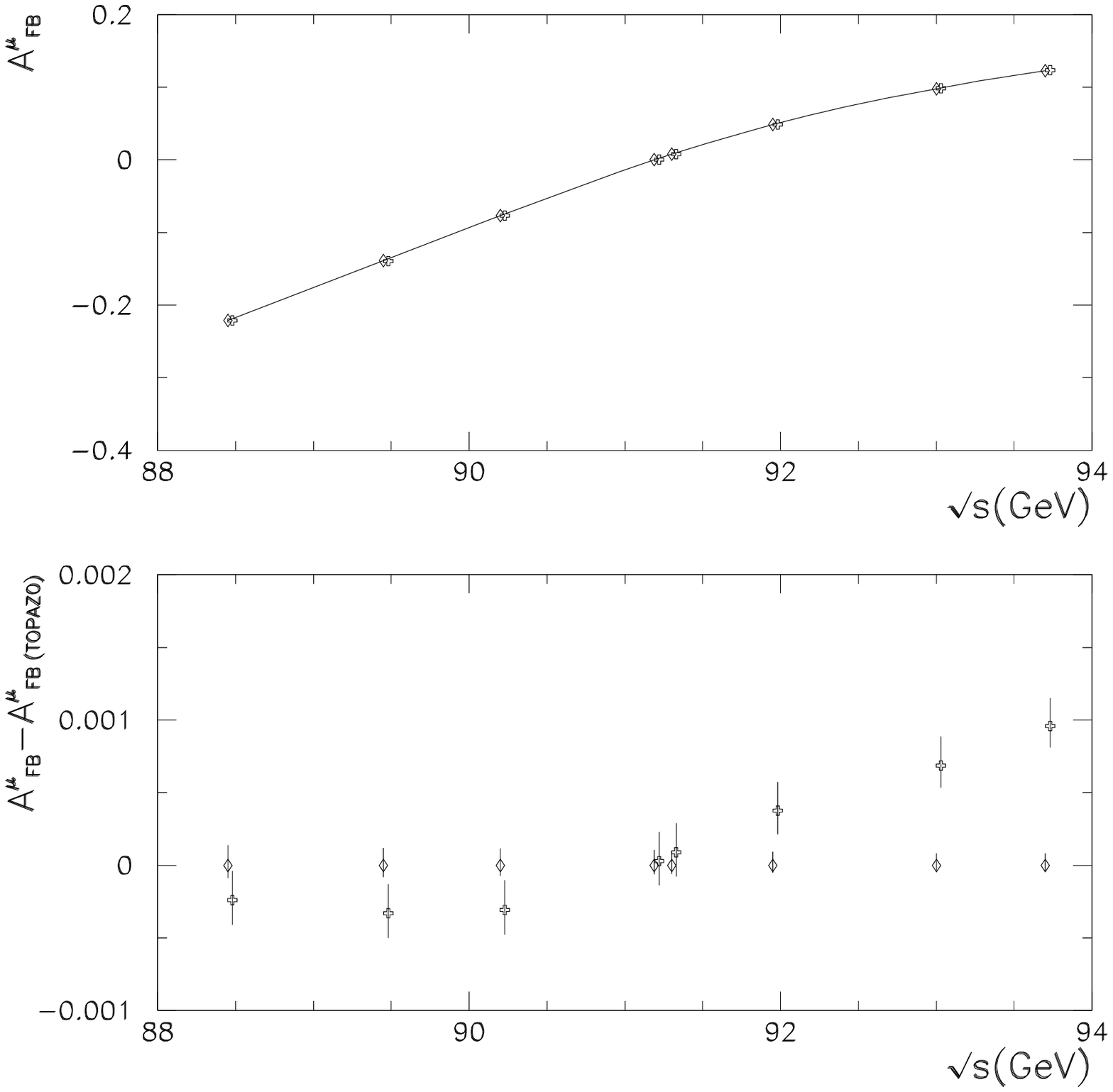,width=15cm,height=15cm}}
\end{center}
Figure 34: The {\tt TOPAZ0} (diamond) and {\tt ZFITTER} (cross) predictions,
including an estimate of the theoretical error, for $\afb^{\mu}$ in the
following set-up: $40^{\circ} < \theta_-
< 140^{\circ}$, $\theta_{\rm {acoll}} < 10^{\circ}$ and
$E_{\rm {th}} = 20\,$GeV. Here $m_t = 175\,$GeV, $\hm = 300\,$GeV and 
$\hat{\als} = 0.125$.
In the lower part a comparison is also shown with the absolute
deviation of {\tt ZFITTER} versus {\tt TOPAZ0}.
 
\newpage\vspace*{2cm}
\begin{center}
\mbox{\epsfig{file=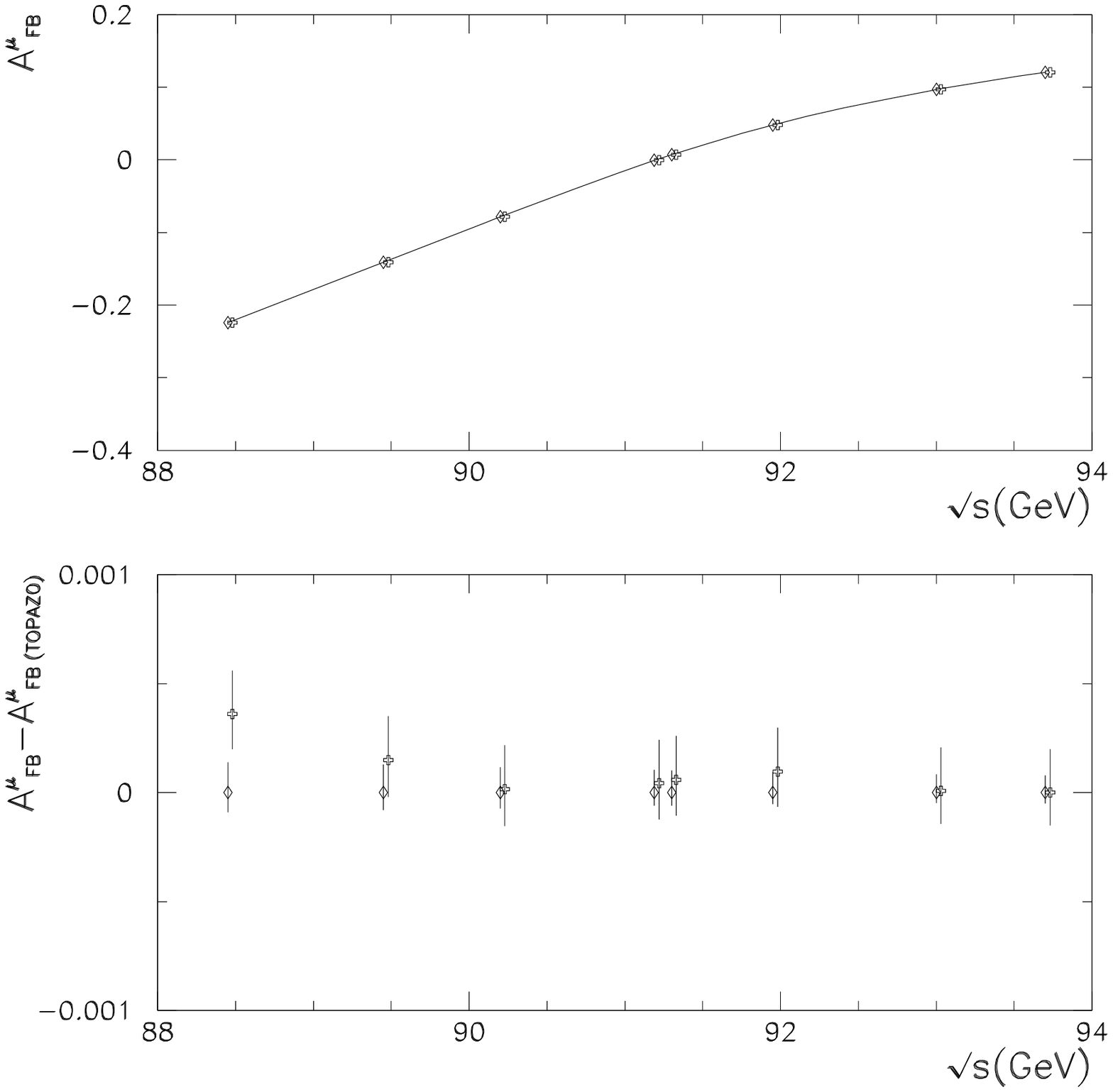,width=15cm,height=15cm}}
\end{center}
Figure 35: The same as in Fig. 34 for $\theta_{\rm {acoll}} < 25^{\circ}$.
 
\newpage\vspace*{2cm}
\begin{center}
\mbox{\epsfig{file=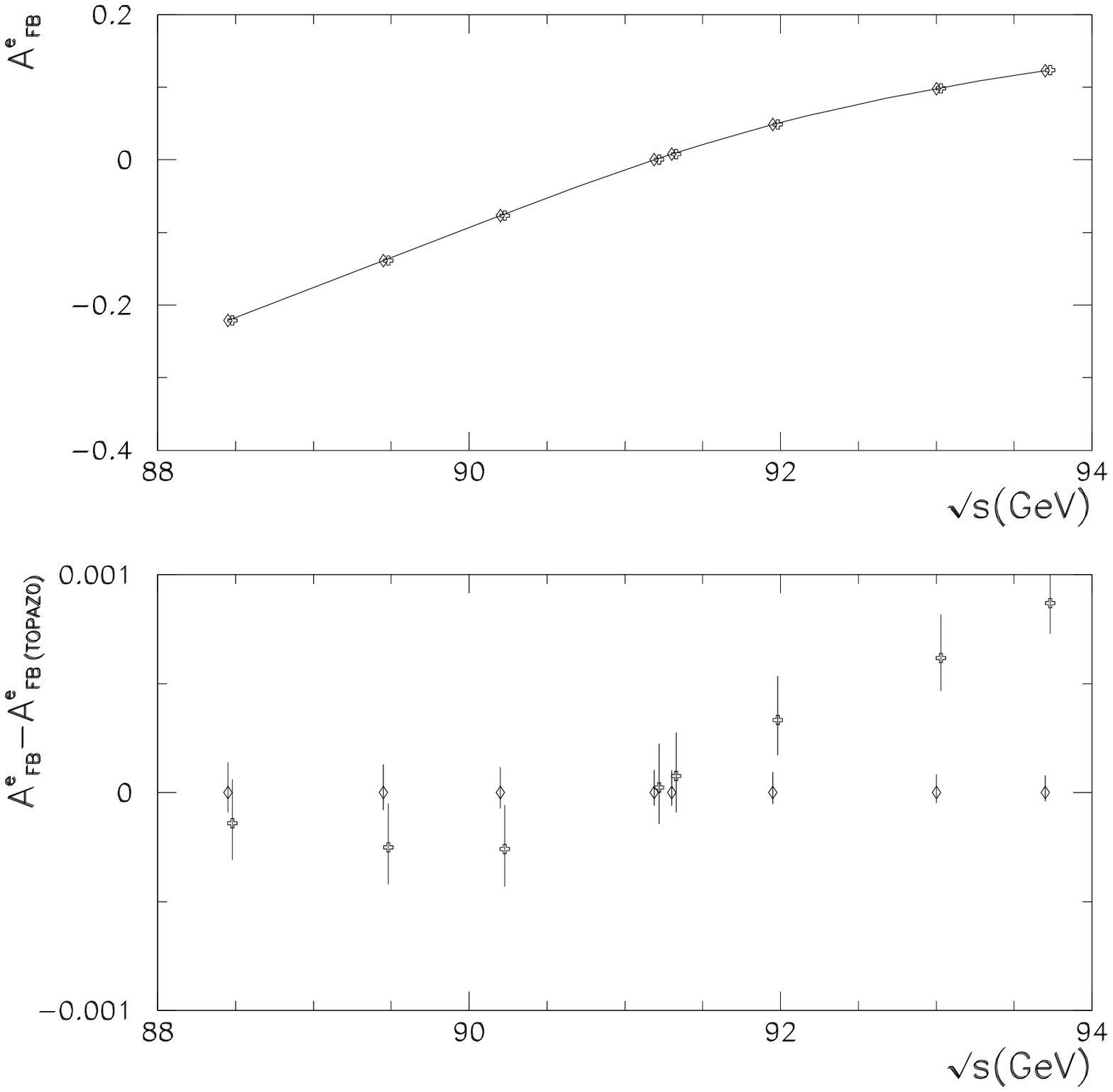,width=15cm,height=15cm}}
\end{center}
Figure 36: The {\tt TOPAZ0} (diamond) and {\tt ZFITTER} (cross) predictions,
including an estimate of the theoretical error, for $\afb^e$ with
$s$-channel electrons in the following set-up: $40^{\circ}
< \theta_- < 140^{\circ}$, $\theta_{\rm {acoll}} < 10^{\circ}$ and
$E_{\rm {th}} = 1\,$GeV.Here $m_t = 175\,$GeV,
$\hm = 300\,$GeV and $\hat{\als} = 0.125$.
In the lower part a comparison is
also shown with the absolute deviation of {\tt ZFITTER} versus {\tt TOPAZ0}.
 
\newpage\vspace*{2cm}
\begin{center}
\mbox{\epsfig{file=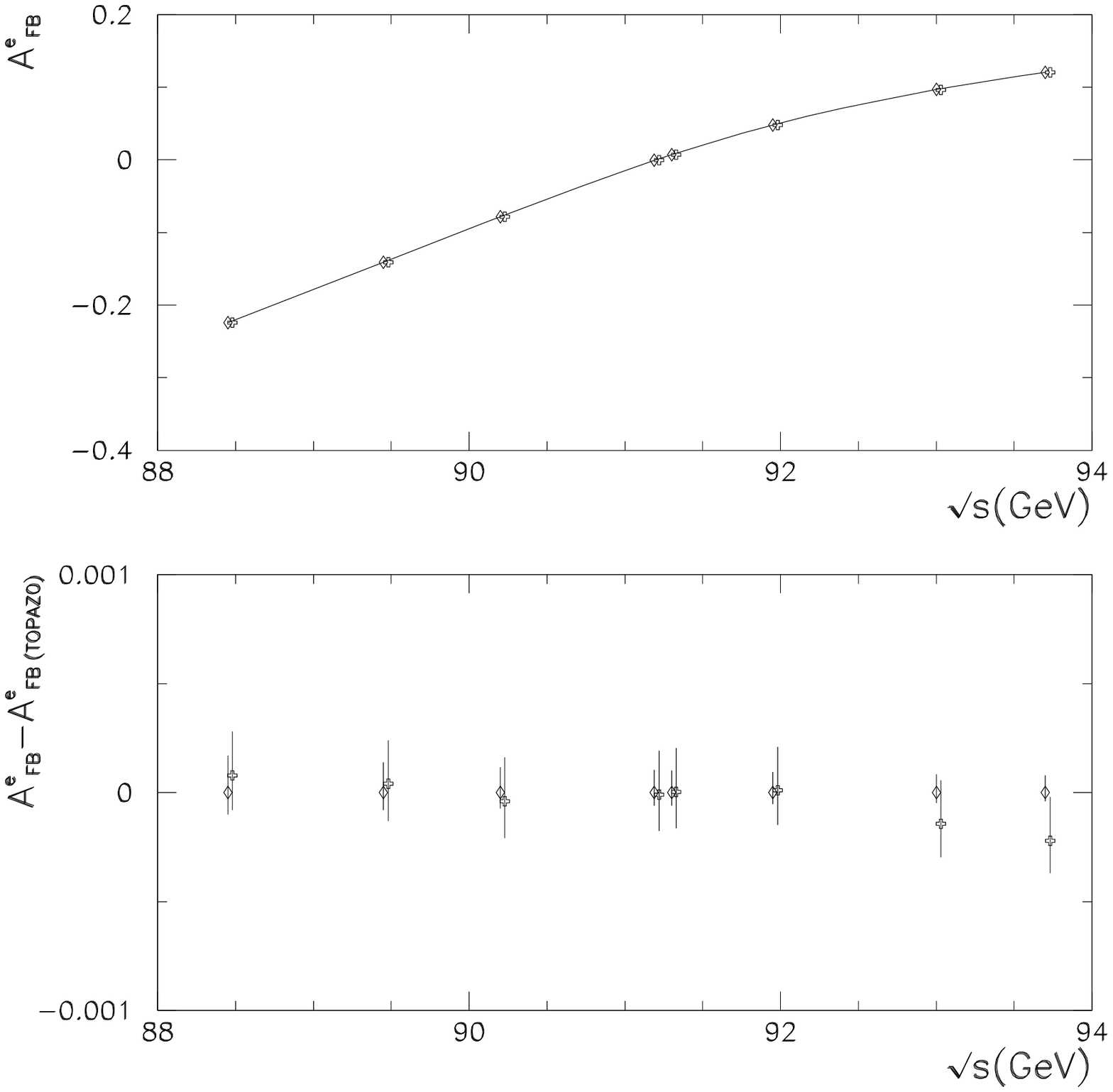,width=15cm,height=15cm}}
\end{center}
Figure 37: The same as in Fig. 36 for $\theta_{\rm {acoll}} < 25^{\circ}$.
 
\newpage
 
\section*{Tables}
\addcontentsline{toc}{section}{Tables}
 
\subsection*{Pseudo-observables}
\addcontentsline{toc}{subsection}{Pseudo-observables}
 
\begin{table}[htbp]\centering
\caption[]{Maximum Derivatives with respect to $\bar{\alpha}^{-1}$.}
\label{ta1}
\vspace{2.5mm}
\begin{tabular}{|c|c|c|c|c|}
\hline
Observables & {\tt BHM}  & {\tt LEPTOP} & {\tt TOPAZ0}  & {\tt ZFITTER} \\
\hline
         & & & & \\
$\wm\,$(GeV)
    &  0.13586       &  0.13537
    &  0.13541       &  0.13546  \\
$\gn\,$(MeV)
    &--0.66258$\tmt$ &--0.12495$\tmt$
    &--0.12805$\tmt$ &--0.18897$\tmt$\\
$\ge\,$(MeV)
    &  0.11994       &  0.12411
    &  0.12413       &  0.12174  \\
$\gmu\,$(MeV)
    &  0.11994       &  0.12411
      &  0.12413     &  0.12174  \\
$\gt\,$(MeV)
    &  0.11995       &  0.12411
      &  0.12413     &  0.12175  \\
$\gu\,$(MeV)
    &  1.3561        &  1.3656
    &  1.3664        &  1.3613   \\
$\gd\,$(MeV)
    &  1.2252        &  1.2371
    &  1.2374        &  1.2341   \\
$\gc\,$(MeV)
    &  1.3562        &  1.3656
    &  1.3665        &  1.3613   \\
$\gb\,$(MeV)
    &  1.2143        &  1.2282
    &  1.2330        &  1.2235   \\
$\stes^l$
    &--0.25824$\tmt$ &--0.25781$\tmt$
    &--0.25781$\tmt$ &--0.25798$\tmt$\\
$\stes^b$
    &--0.25968$\tmt$ &--0.25925$\tmt$
    &--0.25930$\tmt$ &--0.25947$\tmt$\\
$\afb^l$
    &  0.43992$\tmt$ &  0.43863$\tmt$
    &  0.43863$\tmt$ &  0.43781$\tmt$\\
$\alr$
    &  0.20344$\tmo$ &  0.20312$\tmo$
    &  0.20312$\tmo$ &  0.20327$\tmo$ \\
$\gz\,$(MeV)
    &  6.7115        &  6.7968
    &  6.8041        &  6.7685  \\
$R_l$
    &  0.46721$\tmo$ &  0.46367$\tmo$
    &  0.46446$\tmo$ &  0.46719$\tmo$ \\
$\s0h\,$(nb)
    &--0.12890$\tmo$ &--0.12305$\tmo$
    &--0.12363$\tmo$ &--0.13001$\tmo$ \\
$R_b$
    &--0.92933$\tmf$ &--0.91531$\tmf$
    &--0.89791$\tmf$ &--0.92040$\tmf$ \\
$\afb^b$
    &  0.14434$\tmo$ &  0.14411$\tmo$
    &  0.14411$\tmo$ &  0.14421$\tmo$ \\
$\gh\,$(MeV)
    &  6.3768        &  6.4335
    &  6.4408        &  6.4143     \\
$P^b$
    &  0.16748$\tmt$ &  0.16718$\tmt$
    &  0.16722$\tmt$ &  0.16738$\tmt$ \\
$\gi\,$(MeV)
    &--0.19877$\tmo$ &--0.37484$\tmt$
    &--0.38416$\tmt$ &--0.56691$\tmt$ \\
$\afb^c$
    &  0.11116$\tmo$ &  0.11095$\tmo$
    &  0.11098$\tmo$ &  0.11098$\tmo$ \\
$R_c$
    &  0.14772$\tmth$&  0.14744$\tmth$
    &  0.14730$\tmth$&  0.14698$\tmth$ \\
\hline
\end{tabular}
\end{table}
\normalsize
 
\begin{table}[htbp]\centering
\caption[]{Maximum Derivatives with respect to $m_b$.}
\label{ta2}
\vspace{2.5mm}
 \begin{tabular}{|c|c|c|c|c|}
 \hline
 Observables & {\tt BHM} &  {\tt LEPTOP} & {\tt TOPAZ0}  & {\tt ZFITTER} \\
 \hline
         &  & & & \\
$\gb\,$(MeV)
    &--0.77934       &--0.81333
    &--0.80247       &--0.79613  \\
$\gz\,$(MeV)
    &--0.81949       &--0.81344
    &--0.84278       &--0.79462  \\
$R_l$
    &--0.92763$\tmt$
    &--0.96913$\tmt$
    &--0.95275$\tmt$ &--0.94665$\tmt$  \\
$\s0h\,$(nb)
    &  0.73002$\tmt$
    &  0.76691$\tmt$
    &  0.74958$\tmt$ &0.74927$\tmt$  \\
$R_b$
    &--0.34685$\tmth$
    &--0.36581$\tmth$
    &--0.35739$\tmth$&--0.35819$\tmth$\\
$\gh\,$(MeV)
    &--0.80799
    &--0.81344
    &--0.83069       &--0.79462  \\
$R_c$
 &  0.76013$\tmf$
 &  0.80403$\tmf$
 &  0.78998$\tmf$ & 0.79395$\tmf$  \\
\hline
\end{tabular}
\end{table}
\normalsize
 
\clearpage
 
\vspace*{2cm}
\begin{table}[htbp]\centering
\caption[]{The experimental data.}
\label{ta3}
\vspace{2.5mm}
\begin{tabular}{|c|c|}
\hline
Observables & \multicolumn{1}{c|}
{Data} \\
\cline{2-2}
        &         \\
\hline
                 &                    \\
$\wm$(GeV)       &   $80.22 \pm 0.18$ \\
                 &                    \\
$\gl\,$(MeV)     &   $83.96 \pm 0.18$ \\
                 &                    \\
$\gz\,$(MeV)     &   $2497.4 \pm 3.8$ \\
                 &                    \\
$\sigma^h\,$(nb) &   $41.49 \pm 0.12$ \\
                 &                    \\
$R_l$            &   $20.795 \pm 0.040$  \\
                 &                    \\
$R_b$            &   $0.2202 \pm 0.0020$ \\
                 &                    \\
$R_c$            &   $0.1583 \pm 0.0098$ \\
                 &                    \\
$\stes^l$        &   $0.2321 \pm 0.0004$ \\
                 &                    \\
$\afb^l$         &   $0.0170 \pm 0.0016$ \\
                 &                    \\
$\afb^b$         &   $0.0967 \pm 0.0038$ \\
                 &                    \\
$\afb^c$         &   $0.0760 \pm 0.0091$  \\
                 &                    \\
$\alr$(SLD)      &   $0.1668 \pm 0.0077$ \\
                 &                    \\
\hline
\end{tabular}
\end{table}
\normalsize
 
\clearpage
 
\begin{table}[htbp]\centering
\caption[]
{The experimental data for $\wm, \gl=\ge, \gz$ and the theoretical
 predictions
corresponding to $m_t = 175\,$GeV, $\hm = 300\,$GeV and $\hat{\als}(\zm) =
0.125$ The first entry is {\tt BHM} then {\tt LEPTOP}, {\tt TOPAZ0}, {\tt WOH}
and {\tt ZFITTER}. The uncertainties quoted are obtained from a variation of
program options as described in Section 5.}
\label{ta4}
\vspace{2.5mm}
\begin{tabular}{|c|c|c|c|}
\hline
Observable & Exp. &  Theor. Predictions & Average  \\
\hline
                 &                    &                            & \\
                 &                    & $80.319^{+0.003}_{-0.007}$ & \\
                 &                    &                            & \\
                 &                    & $80.312^{+0.013}_{-0.013}$ & \\
                 &                    &                            & \\
$\wm\,$(GeV)     &   $80.22 \pm 0.18$ & $80.310^{+0.000}_{-0.007}$ & $80.315$\\
                 &                    &                            & \\
                 &                    & $80.319^{+0.004}_{-0.000}$ & \\
                 &                    &                            & \\
                 &                    & $80.317^{+0.007}_{-0.007}$ & \\
                 &                    &                            & \\
\hline
                 &                    &                            & \\
                 &                    & $83.919^{+0.020}_{-0.013}$ & \\
                 &                    &                            & \\
                 &                    & $83.930^{+0.023}_{-0.023}$ & \\
                 &                    &                            & \\
$\gl\,$(MeV)     &   $83.96 \pm 0.18$ & $83.931^{+0.015}_{-0.012}$ & $83.933$\\
                 &                    &                            & \\
                 &                    & $83.943^{+0.022}_{-0.022}$ & \\
                 &                    &                            & \\
                 &                    & $83.941^{+0.013}_{-0.021}$ & \\
                 &                    &                            & \\
\hline
                 &                    &                            & \\
                 &                    & $2497.4^{+0.9}_{-1.0}$     & \\
                 &                    &                            & \\
                 &                    & $2497.2^{+1.1}_{-1.1}$     & \\
                 &                    &                            & \\
$\gz\,$(MeV)     &   $2497.4 \pm 3.8$ & $2497.4^{+0.2}_{-0.5}$     & $2497.4$\\
                 &                    &                            & \\
                 &                    & $2497.4^{+1.5}_{-0.6}$     & \\
                 &                    &                            & \\
                 &                    & $2497.5^{+0.6}_{-0.5}$     & \\
                 &                    &                            & \\
\hline
\end{tabular}
\end{table}
\normalsize
 
\clearpage
 
\begin{table}[htbp]\centering
\caption[]{The same as in Table~\ref{ta4} for $R_l, R_b, R_c$.}
\label{ta5}
\vspace{2.5mm}
\begin{tabular}{|c|c|c|c|}
\hline
Observable & Exp. &  Theor. Predictions & Average  \\
\hline
                 &                    &                            & \\
                 &                    & $20.788^{+0.004}_{-0.008}$ & \\
                 &                    &                            & \\
                 &                    & $20.780^{+0.006}_{-0.005}$ & \\
                 &                    &                            & \\
$R_l$            & $20.795 \pm 0.040$ & $20.782^{+0.002}_{-0.005}$ & $20.782$\\
                 &                    &                            & \\
                 &                    & $20.780^{+0.013}_{-0.000}$ & \\
                 &                    &                            & \\
                 &                    & $20.781^{+0.006}_{-0.001}$ & \\
                 &                    &                            & \\
\hline
                 &                    &                            & \\
                 &                    & $0.21577^{+0.00010}_{-0.00011}$ & \\
                 &                    &                            & \\
                 &                    & $0.21564^{+0.00009}_{-0.00004}$ & \\
                 &                    &                            & \\
$R_b$            &$0.2202 \pm 0.0020$ & $0.21567^{+0.00003}_{-0.00012}$ &
$0.21569$ \\
                 &                    &                            & \\
                 &                    & $0.21567^{+0.00018}_{-0.00006}$ & \\
                 &                    &                            & \\
                 &                    & $0.21571^{+0.00001}_{-0.00002}$ & \\
                 &                    &                            & \\
\hline
                 &                    &                            & \\
                 &                    & $0.17236^{+0.00002}_{-0.00002}$ & \\
                 &                    &                            & \\
                 &                    & $0.17240^{+0.00002}_{-0.00003}$ & \\
                 &                    &                            & \\
$R_c$            &$0.1583 \pm 0.0098$ & $0.17237^{+0.00004}_{-0.00000}$ &
$0.17238$ \\
                 &                    &                            & \\
                 &                    & $0.17240^{+0.00001}_{-0.00003}$ & \\
                 &                    &                            & \\
                 &                    & $0.17236^{+0.00002}_{-0.00000}$ & \\
                 &                    &                            & \\
\hline
\end{tabular}
\end{table}
\normalsize
 
\clearpage
 
\begin{table}[htbp]\centering
\caption[]{The same as in Table~\ref{ta4} for $\stes^l, \stes^b, \afb^l$.}
\label{ta6}
\vspace{2.5mm}
\begin{tabular}{|c|c|c|c|}
\hline
Observable & Exp. &  Theor. Predictions & Average  \\
\hline
                 &                    &                            & \\
                 &                    & $0.23197^{+0.00004}_{-0.00007}$ & \\
                 &                    &                            & \\
                 &                    & $0.23200^{+0.00008}_{-0.00008}$ & \\
                 &                    &                            & \\
$\stes^l$        &$0.2321 \pm 0.0004$ & $0.23200^{+0.00004}_{-0.00004}$ &
$0.23199$ \\
                 &                    &                            & \\
                 &                    & $0.23194^{+0.00003}_{-0.00007}$ & \\
                 &                    &                            & \\
                 &                    & $0.23205^{+0.00004}_{-0.00014}$ & \\
                 &                    &                            & \\
\hline
                 &                    &                            & \\
                 &                    & $0.23331^{+0.00004}_{-0.00012}$ & \\
                 &                    &                            & \\
                 &                    & $0.23329^{+0.00008}_{-0.00010}$ & \\
                 &                    &                            & \\
$\stes^b$        &                    & $0.23330^{+0.00009}_{-0.00001}$ &
$0.23330$ \\
                 &                    &                            & \\
                 &                    & $0.23325^{+0.0004}_{-0.00007}$  & \\
                 &                    &                            & \\
                 &                    & $0.23335^{+0.00004}_{-0.00014}$ & \\
                 &                    &                            & \\
\hline
                 &                    &                            & \\
                 &                    & $0.01544^{+0.00011}_{-0.00007}$ & \\
                 &                    &                            & \\
                 &                    & $0.01539^{+0.00013}_{-0.00013}$ & \\
                 &                    &                            & \\
$\afb^l$         &$0.0170 \pm 0.0016$ & $0.01536^{+0.00008}_{-0.00007}$ &
$0.01540$ \\
                 &                    &                            & \\
                 &                    & $0.01549^{+0.00012}_{-0.00005}$ & \\
                 &                    &                            & \\
                 &                    & $0.01531^{+0.00024}_{-0.00007}$ & \\
                 &                    &                            & \\
\hline
\end{tabular}
\end{table}
\normalsize
 
\clearpage
 
\begin{table}[htbp]\centering
\caption[]{The same as in Table~\ref{ta4} for $\alr, \afb^b, \afb^c$.}
\label{ta7}
\vspace{2.5mm}
\begin{tabular}{|c|c|c|c|}
\hline
Observable & Exp. &  Theor. Predictions & Average  \\
 \hline
                 &                    &                            & \\
                 &                    & $0.14346^{+0.00052}_{-0.00031}$ & \\
                 &                    &                            & \\
                 &                    & $0.14326^{+0.00060}_{-0.00060}$ & \\
                 &                    &                            & \\
$\alr$           &$0.1668 \pm 0.0077$ & $0.14327^{+0.00028}_{-0.00031}$ &
$0.14332$ \\
                 &                    &                            & \\
                 &                    & $0.14372^{+0.00057}_{-0.00024}$ & \\
                 &                    &                            & \\
                 &                    & $0.14289^{+0.00110}_{-0.00032}$ & \\
                 &                    &                            & \\
\hline
                 &                    &                            & \\
                 &                    & $0.10053^{+0.00038}_{-0.00022}$ & \\
                 &                    &                            & \\
                 &                    & $0.10040^{+0.00043}_{-0.00042}$ & \\
                 &                    &                            & \\
$\afb^b$         &$0.0967 \pm 0.0038$ & $0.10033^{+0.00023}_{-0.00023}$ &
$0.10042$ \\
                 &                    &                            & \\
                 &                    & $0.10072^{+0.00041}_{-0.00006}$ & \\
                 &                    &                            & \\
                 &                    & $0.10013^{+0.00079}_{-0.00022}$ & \\
                 &                    &                            & \\
\hline
                 &                    &                            & \\
                 &                    & $0.07169^{+0.00029}_{-0.00017}$ & \\
                 &                    &                            & \\
                 &                    & $0.07158^{+0.00033}_{-0.00033}$ & \\
                 &                    &                            & \\
$\afb^c$         &$0.0760 \pm 0.0038$ & $0.07159^{+0.00016}_{-0.00017}$ &
$0.07161$ \\
                 &                    &                            & \\
                 &                    & $0.07183^{+0.00032}_{-0.00013}$& \\
                 &                    &                            & \\
                 &                    & $0.07138^{+0.00061}_{-0.00017}$ & \\
                 &                    &                            & \\
\hline
\end{tabular}
\end{table}
\normalsize
 
\clearpage
 
\begin{table}[htbp]\centering
\caption[]{Largest half-differences among central values ($d_c$) and among
maximal and minimal predictions ($d_g$) for $150\,\gev < m_t < 200\,\gev$,
$60\,\gev < \hm < 1\,\tev$ and $0.118 < \als(\zm) < 0.125$.}
\label{ta8}
\vspace{2.5mm}
\begin{tabular}{|c|c|c|}
\hline
Observable $O$ & $d_c O$  & $d_g O$ \\
\hline
            & & \\
$\wm\,$(GeV)          & $6.5\tmth$ & $1.9\tmt$\\
$\ge\,$(MeV)          & $1.7\tmt$  & $3.7\tmt$\\
$\gz\,$(MeV)          & $0.3$      & $1.6$\\
$\stes^l$             & $6.5\tmfv$ & $1.5\tmf$\\
$\stes^b$             & $6.0\tmfv$ & $1.6\tmf$\\
$R_l$                 & $4.0\tmth$ & $1.0\tmt$\\
$R_b$                 & $7.0\tmfv$ & $2.0\tmf$ \\
$R_c$                 & $3.0\tmfv$ & $5.0\tmfv$ \\
$\s0h\,$(nb)          & $7.5\tmth$ & $8.5\tmth$\\
$\afb^l$              & $1.2\tmf$  & $2.5\tmf$\\
$\afb^b$              & $3.5\tmf$  & $8.2\tmf$ \\
$\afb^c$              & $2.7\tmf$  & $6.3\tmf$ \\
$\alr$                & $5.0\tmf$  & $1.1\tmth$\\
\hline
\end{tabular}
\end{table}
\normalsize
 
\begin{table}[htbp]\centering
\caption[]{Largest half-differences among central values ($d_c$) and among
maximal and minimal predictions ($d_g$) for $m_t = 175\,\gev$,
$60\,\gev < \hm < 1\,\tev$ and $\hat{\als}(\zm) = 0.125$.}
\label{ta9}
\vspace{2.5mm}
\begin{tabular}{|c|c|c|}
\hline
Observable $O$ & $d_c O$  & $d_g O$ \\
\hline
            & & \\
$\wm\,$(GeV)          & $4.5\tmth$ & $1.6\tmt$\\
$\ge\,$(MeV)          & $1.3\tmt$  & $3.1\tmt$\\
$\gz\,$(MeV)          & $0.2$      & $1.4$\\
$\stes^l$             & $5.5\tmfv$ & $1.4\tmf$\\
$\stes^b$             & $5.0\tmfv$ & $1.5\tmf$\\
$R_l$                 & $4.0\tmth$ & $9.0\tmth$\\
$R_b$                 & $6.5\tmfv$ & $1.7\tmf$ \\
$R_c$                 & $2.0\tmfv$ & $4.5\tmfv$ \\
$\s0h\,$(nb)          & $7.0\tmth$ & $8.5\tmth$\\
$\afb^l$              & $9.3\tmfv$ & $2.2\tmf$\\
$\afb^b$              & $3.0\tmf$  & $7.4\tmf$ \\
$\afb^c$              & $2.3\tmf$  & $5.7\tmf$ \\
$\alr$                & $4.2\tmf$  & $8.7\tmf$\\
\hline
\end{tabular}
\end{table}
\normalsize
 
\clearpage
 
\vspace*{3cm}
\begin{table}[htbp]\centering
\caption[]{Largest half-differences among central values ($d_c$) and among
maximal and minimal predictions ($d_g$) for $m_t = 175\,\gev$,
$\hm = 300\,\gev$ and $0.118 < \hat{\als}(\zm) < 0.125$.}
\label{ta10}
\vspace{2.5mm}
\begin{tabular}{|c|c|c|}
\hline
Observable $O$ & $d_c O$  & $d_g O$ \\
\hline
            & & \\
$\wm\,$(GeV)          & $4.5\tmth$ & $1.3\tmt$\\
$\ge\,$(MeV)          & $1.2\tmt$  & $3.0\tmt$\\
$\gz\,$(MeV)          & $0.15$     & $1.4$\\
$\stes^l$             & $5.5\tmfv$ & $1.1\tmf$\\
$\stes^b$             & $5.0\tmfv$ & $1.1\tmf$\\
$R_l$                 & $4.0\tmth$ & $9.0\tmth$\\
$R_b$                 & $6.5\tmfv$ & $1.6\tmf$ \\
$R_c$                 & $2.5\tmfv$ & $4.5\tmfv$ \\
$\s0h\,$(nb)          & $6.5\tmth$ & $8.0\tmth$\\
$\afb^l$              & $8.9\tmfv$ & $1.9\tmf$\\
$\afb^b$              & $3.0\tmf$  & $6.1\tmf$ \\
$\afb^c$              & $2.3\tmf$  & $4.7\tmf$ \\
$\alr$                & $4.2\tmf$  & $8.6\tmf$\\
\hline
\end{tabular}
\end{table}
\normalsize

\clearpage
 
\vspace*{2cm}
\begin{table}[htbp]\centering
\caption[]
{Effects of additional options in {\tt TOPAZ0} for $m_t = 175\,$GeV, $\hm =
300\,$GeV and $\hat{\als}(\zm)= 0.125$.}
\label{ta11}
\vspace{2.5mm}
\begin{tabular}{|c|c|}
\hline
 Observables & \multicolumn{1}{c|}
{Predictions} \\
\cline{2-2}
        &         \\
\hline
                 &                    \\
$\wm$(GeV)       & $80.310^{+0.000}_{-0.007} \to 80.310^{+0.005}_{-0.007}$\\
                 &                    \\
$\gl\,$(MeV)     & $83.931^{+0.015}_{-0.012} \to 83.931^{+0.043}_{-0.031}$\\
                 &                    \\
$\gz\,$(MeV)     & $2497.4^{+0.2}_{-0.5} \to 2497.4^{+1.3}_{-0.6}$\\
                 &                    \\
$R_l$            & $20.782^{+0.002}_{-0.005} \to 20.782^{+0.010}_{-0.004}$\\
                 &                    \\
$R_b$            & $0.21567^{+0.00003}_{-0.00012} \to
                    0.21569^{+0.00000}_{-0.00016}$\\
                 &                    \\
$R_c$            & $0.17237^{+0.00004}_{-0.00000} \to
                    0.17237^{+0.00007}_{-0.00001}$\\
                 &                    \\
$\stes^l$        & $0.23200^{+0.00004}_{-0.00004} \to
                    0.23201^{+0.00004}_{-0.00026}$\\
                 &                    \\
$\afb^l$         & $0.01536^{+0.00008}_{-0.00007} \to
                    0.01538^{+0.00075}_{-0.00074}$\\
                 &                    \\
$\afb^b$         & $0.10033^{+0.00023}_{-0.00023} \to
                    0.10037^{+0.00137}_{-0.00067}$\\
                 &                    \\
$\afb^c$         & $0.07159^{+0.00016}_{-0.00017}  \to
                    0.07157^{+0.00115}_{-0.00053}$\\
                 &                    \\
$\alr$           & $0.14327^{+0.00028}_{-0.00031}  \to
                    0.14320^{+0.00193}_{-0.00039}$\\
                 &                    \\
\hline
\end{tabular}
\end{table}
\normalsize
 
\clearpage
 
\subsection*{Realistic-observables}
\addcontentsline{toc}{subsection}{Realistic-observables}
 
\vspace{2cm}
 
\begin{table}[htbp]\centering
\caption[]
{The hadronic cross-section (nb) in two different configurations, NN=NY/
YN=YY for inclusion of initial state pair production.}
\label{ta12}
\vspace{2.5mm}
\begin{tabular}{|c|c|c|c|}
\hline
$\sqrt{s} $~(GeV) & {\tt BHM} & {\tt TOPAZ0} & {\tt ZFITTER} \\
 \hline
           & &        & \\
   88.45   &  5.181 &  5.184 & 5.185\\
           &  5.168 &  5.171 & 5.172    \\
   89.45   & 10.062 & 10.068 &10.067\\
           & 10.036 & 10.042 &10.042     \\
   90.20   & 18.033 & 18.040 &18.039\\
           & 17.983 & 17.992 &17.992     \\
   91.1887 & 30.446 & 30.452 &30.451\\
           & 30.366 & 30.375 &30.373     \\
   91.30   & 30.585 & 30.590 &30.590\\
           & 30.506 & 30.514 &30.514     \\
   91.95   & 25.176 & 25.176 &25.181\\
           & 25.124 & 25.126 &25.130     \\
   93.00   & 14.131 & 14.127 &14.134\\
           & 14.117 & 14.119 &14.124     \\
   93.70   & 10.070 & 10.065 &10.072\\
           & 10.068 & 10.073 &10.074     \\
\hline
\end{tabular}
\end{table}
\normalsize
 
\clearpage
 
\begin{table}[htbp]\centering
\caption[]
{The $\mu$ forward-backward asymmetry in four different configurations,
NN/ YN/NY/YY for inclusion of initial state pair production and initial-final
QED interference.}
\label{ta13}
\vspace{2.5mm}
\begin{tabular}{|c|c|c|c|}
\hline
$\sqrt{s} $~(GeV) & {\tt BHM} & {\tt TOPAZ0} & {\tt ZFITTER} \\
\hline
           & & & \\
   88.45   &--0.2552 &--0.2551&--0.2546\\
           &--0.2552 &--0.2557&--0.2544\\
           &--0.2546 &--0.2543&--0.2541\\
           &--0.2546 &--0.2549&--0.2539\\
   89.45   &--0.1631 &--0.1632&--0.1630\\
           &--0.1631 &--0.1636&--0.1628\\
           &--0.1626 &--0.1624&--0.1625\\
           &--0.1626 &--0.1628&--0.1623\\
   90.20   &--0.0912 &--0.0914&--0.0913\\
           &--0.0912 &--0.0917&--0.0912 \\
           &--0.0907 &--0.0907&--0.0908\\
           &--0.0907 &--0.0909&--0.0907\\
   91.1887 &--0.0012 &--0.0015&--0.0014\\
           &--0.0012 &--0.0014&--0.0014\\
           &--0.0008 &--0.0011&--0.0011\\
           &--0.0008 &--0.0011&--0.0011\\
   91.30   &  0.0080 &  0.0076&  0.0077\\
           &  0.0080 &  0.0076&  0.0077\\
           &  0.0083 &  0.0078&  0.0080\\
           &  0.0083 &  0.0078&  0.0083\\
   91.95   &  0.0556 &  0.0550&  0.0552\\
           &  0.0556 &  0.0551&  0.0551\\
           &  0.0558 &  0.0549&  0.0554\\
           &  0.0558 &  0.0550&  0.0553\\
   93.00   &  0.1114 &  0.1105&  0.1108\\
           &  0.1114 &  0.1106&  0.1106\\
           &  0.1116 &  0.1104&  0.1109\\
           &  0.1116 &  0.1104&  0.1106\\
   93.70   &  0.1379 &  0.1368&  0.1372\\
           &  0.1379 &  0.1367&  0.1367\\
           &  0.1380 &  0.1367&  0.1372\\
           &  0.1380 &  0.1366&  0.1368\\
\hline
\end{tabular}
\end{table}
\normalsize
 
\clearpage
 
\begin{table}[htbp]\centering
\caption[]
{De-convoluted $\afb^{\mu}$ at $s = \zm^2$, first entry is the $\z0\z0$
part, then $\z0\z0+\gamma\gamma$, $\z0\z0 + \z0\gamma$ and total.}
\label{ta14}
\vspace{2.5mm}
\begin{tabular}{|c|c|c|c|}
\hline
Contribution & {\tt BHM}  & {\tt TOPAZ0}  & {\tt ZFITTER} \\
\hline
           & & & \\
   $\z0\z0$                & 0.015443 & 0.015358 & 0.015279 \\
   $\z0\z0+\gamma\gamma$   & 0.015351 & 0.015267 & 0.015189 \\
   $\z0\z0+\z0\gamma$      & 0.017026 & 0.016735 & 0.016725 \\
   Total                   & 0.016925 & 0.016636 & 0.016627 \\
\hline
\end{tabular}
\end{table}
\normalsize
 
\begin{table}[htbp]\centering
\caption[]{The de-convoluted $\sigma^{\mu}, \sigma^h$ and $\afb^{\mu}$.}
\label{ta15}
\vspace{2.5mm}
\begin{tabular}{|c|c|c|c|}
\hline
$\sqrt{s} $~(GeV) & {\tt BHM} & {\tt TOPAZ0} & {\tt ZFITTER} \\
\hline
           & & & \\
   88.45   &0.3491 &0.3492& 0.3492\\
           &7.020  &7.026&  7.022 \\
           &--0.2386 &--0.2388&--0.2384 \\
   89.45   &0.6884 &0.6887& 0.6887\\
           &14.073 &14.083& 14.077 \\
           &--0.1454 &--0.1456&--0.1453 \\
   90.20   &1.2453 &1.2457& 1.2458\\
           &25.650 &25.663& 25.655\\
           &--0.0749 &--0.0751&--0.0750 \\
   91.1887 &2.0015 &2.0019& 2.0022\\
           &41.400 &41.409& 41.402 \\
           &0.0169 &0.0166& 0.0166 \\
   91.30   &1.9812 &1.9816& 1.9820\\
           &40.983 &40.991& 40.987\\
           &0.0271 &0.0268& 0.0268 \\
   91.95   &1.4486 &1.4488& 1.4492\\
           &29.935 &29.936& 29.937\\
           &0.0855 &0.0853& 0.0851 \\
   93.00   &0.6534 &0.6533& 0.6536\\
           &13.411 &13.408& 13.411\\
           &0.1757 &0.1755& 0.1751 \\
   93.70   &0.4105 &0.4104& 0.4106\\
           &8.362  &8.358&  8.361\\
           &0.2323 &0.2320& 0.2316 \\
\hline
\end{tabular}
\end{table}
\normalsize
 
\clearpage
 
\vspace*{4cm}
\begin{table}[htbp]\centering
\caption[]
{$\delta_{\rm {conv}}$, as defined by Eq.(~\ref{deconv}) for the hadronic
cross-section.}
\label{ta16}
\vspace{2.5mm}
\begin{tabular}{|c|c|c|c|}
\hline
$\sqrt{s}\,$(GeV) & {\tt BHM} & {\tt TOPAZ0} & {\tt ZFITTER} \\
\hline
           & & & \\
   88.450           &--0.2638  &--0.2640  &--0.2634  \\
   89.450           &--0.2869  &--0.2869  &--0.2866  \\
   90.200           &--0.2989  &--0.2989  &--0.2987  \\
   91.189           &--0.2665  &--0.2665  &--0.2664  \\
   91.300           &--0.2556  &--0.2556  &--0.2555  \\
   91.950           &--0.1607  &--0.1607  &--0.1606  \\
   93.000           &+0.0526  &+0.0530  &+0.0532  \\
   93.700           &+0.2040  &+0.2052  &+0.2049  \\
\hline
\end{tabular}
\end{table}
\normalsize
 
\clearpage
 
\subsection*{Bhabha scattering}
\addcontentsline{toc}{subsection}{Bhabha scattering}
 
\vspace{1cm}
 
\begin{table}[htbp]\centering
\caption[]
{The {\tt ALIBABA}(A) - {\tt TOPAZ0}(T) comparison for the full Bhabha
cross-section (in pb) for the following set-up:
$40^o < \theta_- < 140^o$, $\theta_{\rm {acoll}} < 10^o$ and 
$E_{\rm {th}} = 1$~GeV.
I is the {\tt TOPAZ0} default for QED final state radiation, II is the
{\tt ZFITTER}-like default for QED final state radiation and III includes
initial state pair production.}
\label{ta21}
\vspace{2.5mm}
\begin{tabular}{|c|c|c|c|c|}
\hline
$\sqrt{s}\,$GeV & A &  T(I) & T(II) & T (III) \\
 \hline
& & & & \\
 88.45 & $ 457.52 \pm 0.27 $ & $ 457.30^{+0.17}_{-0.06}\pm 0.25$ &
           457.20 & 456.14 \\
& & & & \\
 89.45 & $ 643.95 \pm 0.31 $ & $ 644.37^{+0.26}_{-0.09}\pm 0.24$ &
           644.23 & 642.60 \\
& & & & \\
 90.20 & $ 908.99 \pm 0.39 $ & $ 910.46^{+0.38}_{-0.11}\pm 0.24$ &
           910.26 & 907.86 \\
& & & & \\
 91.19 & $1183.99 \pm 0.39 $ & $1184.03^{+0.48}_{-0.13}\pm 0.24$ &
          1183.78 &1180.79 \\
& & & & \\
 91.30 & $1163.56 \pm 0.45 $ & $1163.51^{+0.47}_{-0.13}\pm 0.24$ &
          1163.26 &1160.38 \\
& & & & \\
 91.95 & $ 876.90 \pm 0.28 $ & $ 874.02^{+0.35}_{-0.12}\pm 0.24$ &
           873.83 & 872.13 \\
& & & & \\
 93.00 & $ 481.35 \pm 0.14 $ & $ 477.51^{+0.17}_{-0.08}\pm 0.24$ &
           477.41 & 477.18 \\
& & & & \\
 93.70 & $ 355.57 \pm 0.13 $ & $ 352.52^{+0.14}_{-0.06}\pm 0.25$ &
           352.45 & 352.72 \\
& & & & \\
\hline
\end{tabular}
\end{table}
\normalsize
 
\clearpage
 
\begin{table}[htbp]\centering
\caption[]
{The same as in Table~\ref{ta21} for the forward backward asymmetry.}
\label{ta22}
\vspace{2.5mm}
\begin{tabular}{|c|c|c|c|}
\hline
$\sqrt{s}\,$GeV & A &  T(I) & T(III)  \\
\hline
& & & \\
 88.45 &$0.44611\pm 1.06\tmth$&$0.44534^{+0.02}_{-0.08}\pm 0.79\tmth$&0.44637\\
& & & \\
 89.45 &$0.34250\pm 0.90\tmth$&$0.34166^{+0.01}_{-0.07}\pm 0.51\tmth$&0.34252\\
& & & \\
 90.20 &$0.24956\pm 0.81\tmth$&$0.24977^{+0.02}_{-0.05}\pm 0.33\tmth$&0.25043\\
& & & \\
 91.19 &$0.13925\pm 0.66\tmth$&$0.13916^{+0.03}_{-0.08}\pm 0.23\tmth$&0.13951\\
& & & \\
 91.30 &$0.13050\pm 0.70\tmth$&$0.13035^{+0.03}_{-0.08}\pm 0.23\tmth$&0.13067\\
& & & \\
 91.95 &$0.10169\pm 0.63\tmth$&$0.10139^{+0.03}_{-0.09}\pm 0.30\tmth$&0.10158\\
& & & \\
 93.00 &$0.13110\pm 0.61\tmth$&$0.13055^{+0.03}_{-0.12}\pm 0.57\tmth$&0.13061\\
& & & \\
 93.70 &$0.18157\pm 0.68\tmth$&$0.17957^{+0.02}_{-0.10}\pm 0.83\tmth$&0.17944\\
& & & \\
\hline
\end{tabular}
\end{table}
\normalsize
 
\begin{table}[htbp]\centering
\caption[]{The same as in Table~\ref{ta21} for $\theta_{\rm {acoll}} < 25^o$.}
\label{ta23}
\vspace{2.5mm}
\begin{tabular}{|c|c|c|c|c|}
\hline
$\sqrt{s}\,$GeV & A &  T(I) & T(II) & T (III) \\
 \hline
& & & & \\
$88.45$&$     483.27 \pm 0.25$&$    484.97^{+0.07}_{-0.06}\pm 0.22$&$
              484.86$&$             483.74$ \\
& & & & \\
$89.45$&$     672.67 \pm 0.29$&$    673.99^{+0.13}_{-0.09}\pm 0.22$&$
              673.84$&$             672.13$ \\
& & & & \\
$90.20$&$     942.52 \pm 0.34$&$    942.96^{+0.25}_{-0.12}\pm 0.22$&$
              942.76$&$             940.27$ \\
& & & & \\
$91.19$&$    1218.66 \pm 0.40$&$   1219.24^{+0.36}_{-0.13}\pm 0.21$&$
             1218.98$&$            1215.90$ \\
& & & & \\
$91.30$&$    1198.23 \pm 0.36$&$   1198.42^{+0.35}_{-0.13}\pm 0.21$&$
             1198.17$&$            1195.20$ \\
& & & & \\
$91.95$&$     908.87 \pm 0.30$&$    905.35^{+0.24}_{-0.12}\pm 0.20$&$
              905.16$&$             903.39$ \\
& & & & \\
$93.00$&$     505.38 \pm 0.15$&$    504.24^{+0.10}_{-0.08}\pm 0.20$&$
              504.13$&$             503.88$ \\
& & & & \\
$93.70$&$     378.20 \pm 0.13$&$    377.94^{+0.06}_{-0.06}\pm 0.20$&$
              377.86$&$             378.14$ \\
& & & & \\
\hline
\end{tabular}
\end{table}
\normalsize
 
\clearpage
 
\begin{table}[htbp]\centering
\caption[]{The same as in Table~\ref{ta22} for $\theta_{\rm {acoll}} < 25^o$.}
\vspace{2.5mm}
\label{ta24}
\begin{tabular}{|c|c|c|c|}
\hline
$\sqrt{s}\,$GeV & A &  T(I) & T(III)  \\
 \hline
& & & \\
$88.45$&$ 0.45843 \pm 0.94\tmth$&$  0.46061^{+0.02}_{-0.19}\pm 0.65\tmth$&$
          0.46168$ \\
& & & \\
$89.45$&$ 0.35479 \pm 0.77\tmth$&$  0.35560^{+0.01}_{-0.17}\pm 0.43\tmth$&$
          0.67213$ \\
& & & \\
$90.20$&$ 0.26121 \pm 0.67\tmth$&$  0.26165^{+0.02}_{-0.14}\pm 0.29\tmth$&$
          0.26235$ \\
& & & \\
$91.19$&$ 0.15114 \pm 0.70\tmth$&$  0.15045^{+0.03}_{-0.12}\pm 0.20\tmth$&$
          0.15083$ \\
& & & \\
$91.30$&$ 0.14067 \pm 0.62\tmth$&$  0.14203^{+0.03}_{-0.12}\pm 0.20\tmth$&$
          0.14238$ \\
& & & \\
$91.95$&$ 0.11466 \pm 0.62\tmth$&$  0.11773^{+0.03}_{-0.17}\pm 0.25\tmth$&$
          0.11795$ \\
& & & \\
$93.00$&$ 0.15628 \pm 0.59\tmth$&$  0.15838^{+0.03}_{-0.29}\pm 0.45\tmth$&$
          0.15845$ \\
& & & \\
$93.70$&$ 0.21239 \pm 0.66\tmth$&$  0.21360^{+0.02}_{-0.36}\pm 0.65\tmth$&$
          0.21344$ \\
& & & \\
\hline
\end{tabular}
\end{table}
\normalsize
 
\begin{table}[htbp]\centering
\caption[]
{The same as in Table~\ref{ta21} with the percentage relative deviation.}
\label{ta25}
\vspace{2.5mm}
\begin{tabular}{|c|c|c|c|}
\hline
$\sqrt{s}\,$GeV & A &  T(II) & $\%$  \\
\hline
& & & \\
$88.45$&$   457.52$&$  457.20$&$ +0.07$ \\
$89.45$&$   643.95$&$  644.23$&$ -0.04$ \\
$90.20$&$   908.99$&$  910.26$&$ -0.14$ \\
$91.19$&$  1183.99$&$ 1183.78$&$ +0.02$ \\
$91.30$&$  1163.56$&$ 1163.26$&$ +0.03$ \\
$91.95$&$   876.90$&$  873.83$&$ +0.35$ \\
$93.00$&$   481.35$&$  477.41$&$ +0.82$ \\
$93.70$&$   355.57$&$  352.45$&$ +0.88$ \\
& & & \\
\hline
\end{tabular}
\end{table}
\normalsize
 
\clearpage
 
\begin{table}[htbp]\centering
\caption[]{The same as in Table~\ref{ta25} for $s$-channel alone.}
\label{ta26}
\vspace{2.5mm}
\begin{tabular}{|c|c|c|c|}
\hline
$\sqrt{s}\,$GeV & A &  T(II) & $\%$  \\
\hline
& & & \\
$88.45$&$       173.01$&$ 172.71$&$ +0.17$ \\
$89.45$&$       330.62$&$ 330.70$&$ -0.02$ \\
$90.20$&$       588.47$&$ 588.81$&$ -0.06$ \\
$91.19$&$       989.81$&$ 990.90$&$ -0.11$ \\
$91.30$&$       994.07$&$ 995.38$&$ -0.13$ \\
$91.95$&$       819.56$&$ 819.97$&$ -0.05$ \\
$93.00$&$       463.72$&$ 461.68$&$ +0.44$ \\
$93.70$&$       331.78$&$ 329.83$&$ +0.59$ \\
& & & \\
\hline
\end{tabular}
\end{table}
\normalsize
 
\vspace*{2cm}
\begin{table}[htbp]\centering
\caption[]
{{\tt ALIBABA-TOPAZ0} comparison for $s-t$ and $t-t$ contributions with the
following set-up: $40^o < \theta_- < 140^o$, $\theta_{\rm {acoll}} < 10^o$ and
$E_{\rm {th}} = 1$~GeV.}
\label{ta27}
\vspace{2.5mm}
\begin{tabular}{|c|c|c|c|c|c|}
\hline
$\sqrt{s}\,$GeV & $\sigma-\sigma(s)$ A &  $\sigma-\sigma(s)$ T &
$\delta$(A) & $\delta$(T) & $\%$\\
 \hline
& & & & & \\
$88.45$&$284.51$&$284.49$&$1.644$&$1.647$&$ -0.18$ \\
$89.45$&$313.33$&$313.53$&$0.948$&$0.948$&$ +0.00$ \\
$90.20$&$320.52$&$321.45$&$0.545$&$0.546$&$ -0.18$ \\
$91.19$&$194.18$&$192.88$&$0.196$&$0.195$&$ +0.51$ \\
$91.30$&$169.49$&$167.88$&$0.171$&$0.169$&$ +1.18$ \\
$91.95$&$ 57.34$&$ 53.86$&$0.070$&$0.066$&$ +5.88$ \\
$93.00$&$ 17.63$&$ 15.75$&$0.038$&$0.034$&$+11.11$ \\
$93.70$&$ 23.79$&$ 22.62$&$0.072$&$0.069$&$ +4.26$ \\
& & & & & \\
\hline
\end{tabular}
\end{table}
\normalsize
 
\begin{table}[htbp]\centering
\caption[]
{The same as in Table~\ref{ta27} for the forward-backward asymmetry.}
\label{ta28}
\vspace{2.5mm}
\begin{tabular}{|c|c|c|c|c|c|c|}
\hline
$\sqrt{s}\,$GeV & A s+t &  T s+t & A s & T s & A t & T t \\
\hline
& & & & & & \\
$88.45$&$ 0.44611$&$ 0.44534$&$ -0.22019$&$ -0.22060$&$0.66630$&$0.66594$\\
$89.45$&$ 0.34250$&$ 0.34166$&$ -0.13754$&$ -0.13847$&$0.48004$&$0.48013$\\
$90.20$&$ 0.24956$&$ 0.24977$&$ -0.07774$&$ -0.07649$&$0.32730$&$0.32626$\\
$91.19$&$ 0.13925$&$ 0.13916$&$ -0.00102$&$  0.00001$&$0.14027$&$0.13915$\\
$91.30$&$ 0.13050$&$ 0.13035$&$  0.00745$&$  0.00779$&$0.12305$&$0.12256$\\
$91.95$&$ 0.10169$&$ 0.10139$&$  0.05059$&$  0.04851$&$0.05110$&$0.05288$\\
$93.00$&$ 0.13110$&$ 0.13055$&$  0.09863$&$  0.09789$&$0.03247$&$0.03266$\\
$93.70$&$ 0.18157$&$ 0.17957$&$  0.12237$&$  0.12248$&$0.05920$&$0.05709$\\
& & & & & & \\
\hline
\end{tabular}
\end{table}
\normalsize
 
\clearpage
 
\begin{table}[htbp]\centering
\caption[]
{Comparison for the full Bhabha cross-section (in pb) for the following
set-up:
$40^o < \theta_- < 140^o$, $\theta_{\rm {acoll}} < 10^o$ and 
$E_{\rm {th}} = 1$~GeV.
First entry is {\tt ALIBABA}, second entry is the TOPAZ0 default for QED final
state radiation and no pair production, $\delta_{\rm {FSR}}$ is the uncertainty
on final-state QED radiation estimated by {\tt TOPAZ0}, while $\Delta_g$ is the
relative difference between the maximal and minimal predictions of
{\tt ALIBABA-TOPAZ0}.}
\label{ta29}
\vspace{2.5mm}
 \begin{tabular}{|c|c|c|c|c|}
 \hline
$\sqrt{s}\,$GeV & A &  T(I) & $\delta_{\rm {FSR}}(T) \%$ & $\Delta_g \%$ \\
 \hline
& & & & \\
$88.45$&$   457.52 \pm 0.27$&$  457.30^{+0.17}_{-0.06}\pm 0.25$&$
            0.02$&$   0.15$\\
& & & & \\
$89.45$&$   643.95 \pm 0.31$&$  644.37^{+0.26}_{-0.09}\pm 0.24$&$
            0.02$&$   0.19$\\
& & & & \\
$90.20$&$   908.99 \pm 0.39$&$  910.46^{+0.38}_{-0.11}\pm 0.24$&$
            0.02$&$   0.27$\\
& & & & \\
$91.19$&$  1183.99 \pm 0.39$&$ 1184.03^{+0.48}_{-0.13}\pm 0.24$&$
           0.02$&$  0.10$\\
& & & & \\
$91.30$&$  1163.56 \pm 0.45$&$ 1163.51^{+0.47}_{-0.13}\pm 0.24$&$
           0.02$&$  0.10$\\
& & & & \\
$91.95$&$   876.90 \pm 0.28$&$  874.02^{+0.35}_{-0.12}\pm 0.24$&$
            0.02$&$   0.37$\\
& & & & \\
$93.00$&$   481.35 \pm 0.14$&$  477.51^{+0.17}_{-0.08}\pm 0.24$&$
            0.02$&$   0.86$\\
& & & & \\
$93.70$&$   355.57 \pm 0.13$&$  352.52^{+0.14}_{-0.06}\pm 0.25$&$
            0.02$&$   0.95$\\
& & & & \\
\hline
 \end{tabular}
 \end{table}
 \normalsize
 
\clearpage
 
\subsection*{Effect of working options for different codes}
\addcontentsline{toc}{subsection}
{Effect of working options for different codes}
 
\begin{table}[htbp]\centering
\caption[]
{The effect of the working options of {\tt BHM} on theoretical errors.}
\label{ta17}
\vspace{2.5mm}
\begin{tabular}{|c|c|c|c|c|c|}
\hline
Observable & Default &  IRES err & +IQCD err & +IFAC err & +ITWO err \\
 \hline
$\wm\,$(GeV)  & 80.319 &  0.004  & 0.010 & 0.010 & 0.010 \\
$\ge\,$(MeV)  & 83.919 &  0.020 &  0.032 & 0.033 & 0.033 \\
$\gz\,$(MeV)  & 2497.4 &  0.68  &  1.07   & 1.71 & 1.92 \\
$R_l$         & 20.788 &  0.001 &  0.002 & 0.010 & 0.012 \\
$R_b$      & 0.21577 & 0.00000 & 0.00000 & 0.00011 & 0.00020 \\
$\afb^l$   & 0.015435 & 0.00011 & 0.00018 & 0.00018 & 0.00018 \\
$\afb^b$   & 0.10053 & 0.00037 & 0.00059 & 0.00059 & 0.00059 \\
$\stes^l$  & 0.23197 & 0.00007 & 0.00011 & 0.00011 & 0.00011 \\
$\stes^b$  & 0.23331 & 0.00007 & 0.00011 & 0.00011 & 0.00016 \\
\hline
\end{tabular}
\end{table}
\normalsize
 
\begin{table}[htbp]\centering
\caption[]
{The effect of the working options of {\tt LEPTOP}
 on theoretical errors.
The first two lines indicate parametric uncertainties caused by
 $\delta s^2 =
0.0003$ (which is equivalent to $\delta\bar{\alpha}^{-1} = 0.12$)
 and by
$\delta m_b = 0.3$GeV. The next six lines refer to the intrinsic
 theoretical
uncertainties. This table was calculated with
 $m_t$=175 GeV, $M_{_H}$=300 GeV
and $\hat{\alpha_s}$=0.125.}
\label{ta18}
\vspace{2.5mm}
\begin{tabular}{|c|r|r|r|r|r|r|r|r|r|r|}
\hline
   & $M_{_W}$ & $\Gamma_{l}$ & $\sin^2\theta^l_{\rm {eff}}$ & $\sigma^h_0(nb)$ &
  $ \Gamma_{_Z}$ & $\Gamma_h$ & $R_l$ & $\Gamma_b$ & $R_b\times$ &
  $R_c\times$ \\
  & (MeV) & (MeV) & & & (MeV) & (MeV) & &(MeV) & $10^5$ &
   $10^5$\\ \hline
 $\delta s^2$ &
 16    & .015    & .00031 & .0014 &  .8 &  .8 &  .0055 & .15 &
   1.1 & 1.8
 \\ \hline
  $\delta m_b$ &
  -    & -       & -      & .0023 &  .2 &  .2 &  .0029 & .24 &
   11.0 & 2.4
 \\ \hline \hline
  $\Delta V_i^{t^2}$ &
  9    & .015    & .00005 & .0002 &  .5 &  .4 &  .0009 & .08 &
     .2 &  .3
 \\ \hline
 $\Delta V_i^{\alpha^2_s}$ &
  5    & .008    & .00003 & .0001 &  .3 &  .2 &  .0005 & .04 &
     .1 &  .2
 \\  \hline
 $\Delta \Gamma_q$ &
  -    & -       & -      & .0029 &  .3 &  .3 &  .0037 & -   &
    3.8 & 1.5
 \\ \hline
 $\Delta \phi^{\alpha^2_s}$ &
  -    & -       & -      & .0010 &  .1 &  .1 &  .0012 & .10 &
    4.6 & 1.0
 \\ \hline
total &$+13$ &+.023 &+.00008 &+.0032 &+1.2 &+1.0 &+.0063 &+.23 &+8.7&
+1.9 \\
 &$-13$ &$-.023$ &$-.00008$ &$-.0042$ &$-1.1$ &$-0.9$
 &$-.0051$ &$-.13$ &$-4.1$&
$-2.9$ \\ \hline
\end{tabular}
\end{table}
\normalsize
 
\clearpage
 
\begin{table}[htbp]\centering
\caption[]
{The effect of the working options of {\tt TOPAZ0} on theoretical errors.}
\label{ta19}
\vspace{2.5mm}
\begin{tabular}{|c|c|c|c|c|c|c|}
\hline
Observable & Default &  OU1 err & + OU4 err & + OU5 err & + OU6 err &
+OU7 err \\
 \hline
$\wm\,$(GeV)  & 80.310 &  -  &  - &  0.001 & 0.001 & 0.007 \\
$\ge\,$(MeV)  & 83.931 &  0.014 &  0.014 & 0.015 & 0.015 & 0.027 \\
$\gz\,$(MeV)  & 2497.4 &  0.10  &  0.10 & 0.10 & 0.30 & 0.70 \\
$R_l$         & 20.782 &  0.004 &  0.004 & 0.004 & 0.006 & 0.007 \\
$R_b$      & 0.21567 & 0.00012 & 0.00012 & 0.00013 & 0.00014 & 0.00015 \\
$\afb^l$   & 0.015362 & 0.00006 & 0.00006 & 0.00008 & 0.00008 & 0.00015 \\
$\afb^b$   & 0.10033 & 0.00018 & 0.00019 & 0.00024 & 0.00024 & 0.00046 \\
$\stes^l$  & 0.23200 & 0.00004 & 0.00004 & 0.00004 & 0.00004 & 0.00008 \\
$\stes^b$  & 0.23330 & 0.00005 & 0.00005 & 0.00006 & 0.00006 & 0.00010 \\
\hline
\end{tabular}
\end{table}
\normalsize
 
\begin{table}[htbp]\centering
\caption[]
{The effect of the working options of {\tt ZFITTER} on theoretical errors.}
\label{ta20}
\vspace{2.5mm}
\begin{tabular}{|c|c|c|c|c|c|c|c|}
\hline
Observable & Default &  OZ1 err & +OZ2 err & +OZ3 err &
+OZ4 err & +OZ5 err & +OZ6 err\\
\hline
$\wm\,$(GeV)& 80.317   &  0.001  &  0.002  &  0.005  &  0.007  &  0.007
&  0.014  \\
$\ge\,$(MeV)& 83.941   &  0.004  &  0.006  &  0.014  &  0.014  &  0.021
&  0.034  \\
$\gz\,$(MeV)& 2497.5   &   0.1   &   0.2   &   0.2   &   0.2   &   0.6
&   1.1   \\
$R_l  $     & 20.781   &    -    &  0.001  & 0.002   &  0.003  &  0.005
&  0.007  \\
$R_b    $   & 0.21571  &    -    &    -    & 0.00001 & 0.00001 & 0.00002
& 0.00003 \\
$\afb^l $   & 0.01531  & 0.00003 & 0.00009 & 0.00015 & 0.00024 & 0.00024
& 0.00030 \\
$\afb^b $   & 0.10013  & 0.00011 & 0.00029 & 0.00050 & 0.00078 & 0.00079
& 0.00101 \\
$\stes^l$   & 0.23205  & 0.00002 & 0.00005 & 0.00009 & 0.00014 & 0.00014
& 0.00018 \\
$\stes^b$   & 0.23335  & 0.00002 & 0.00005 & 0.00009 & 0.00014 & 0.00014
& 0.00018 \\
\hline
\end{tabular}
\end{table}
\normalsize

\clearpage
       
\section*{Note added in proof}
\addcontentsline{toc}{section}{Note added in proof}

While making proof-reading of this contribution we have been informed on
some recent development concerning the AFMT term in $\Delta\rho$. A recent
calculation by K.G.~Chetyrkin, J.H.~K\"uhn and M.~Steinhauser~\cite{nafmt}
as well as a revised version of the AFMT calculation~\cite{afmt} have shown
that the correct coefficient of $\zeta(4)$ in a term proportional to $C^2_F$
of $\delta^{\rm {QCD}}_{(3)}$ is $4$ and not $188/5$.
This will correspond to some shift in our
predictions for the central values of the pseudo-observables, as a matter of
fact a shift common to all codes since the AFMT term is a common external
block. We have shown this shift in the following Table, where we compare our 
error bands with the shifted central values at the standard reference point.
The result can be simply summarized by saying that the updated central values
remain within our theoretical error bands. As for the theoretical bands
is concerned, we note first, that they are not dominated by the QCD-uncertainty
in the calculation of the $\rho$-parameter. Second, as it seen from the Table
added, the widths of the uncertainty bands is only marginally affected by the
shift of the AFMT correction.

\begin{table}[htb]\centering
\caption[]{Change of some of the observables due to the introduction of the
revised AFMT formulas. Here $m_t= 175\,$GeV, $\hm = 300\,$GeV and $\als= 0.125$.
In order to estimate the size of the
non-leading QCD effects, the $\delta^{^{\rm {QCD}}}$ correction factor has
been implemented according to the formulation of Ref.~\cite{sir}, with a scale
which gives the maximum variation with respect to the AFMT(revised) term ---
$\xi = 0.248(0.204)$ --- and the difference between this and the AFMT
(revised) calculation is used as an estimate of the corresponding uncertainty.
The numbers are calculated by
{\tt TOPAZ0}, first row, and {\tt ZFITTER}, second row.}
\label{ta40}
\vspace{2.5mm}
\begin{tabular}{|c|c|c|c|}
\hline
Observable & New AFMT &  AFMT & Central difference  \\
\hline
$\wm\,$(GeV)  & $80.307^{+0.000}_{-0.007}$ & $80.310^{+0.000}_{-0.007}$ & 
$-3\,\mev$\\
              & $80.314^{+0.007}_{-0.007}$ & $80.317^{+0.007}_{-0.007}$ &
$-3\,\mev$\\
\hline
$\gl\,$(MeV)  & $83.926^{+0.015}_{-0.012}$ & $83.931^{+0.015}_{-0.012}$ &
$-0.005\,\mev$ \\
              & $83.936^{+0.013}_{-0.021}$ & $83.941^{+0.013}_{-0.021}$ &
$-0.005\,\mev$ \\
\hline
$\gz\,$(MeV)  & $2497.2^{+0.2}_{-0.4}$     & $2497.4^{+0.2}_{-0.5}$     &
$-0.2\,\mev$ \\
              & $2497.3^{+0.7}_{-0.5}$     & $2497.5^{+0.6}_{-0.5}$     &
$-0.2\,\mev$ \\
\hline
$R_l$         & $20.782^{+0.002}_{-0.005}$ & $20.782^{+0.002}_{-0.005}$ & 
$0.000$\\
              & $20.780^{+0.006}_{-0.000}$ & $20.781^{+0.006}_{-0.001}$ &
$-0.001$\\
\hline
$R_b$      & $0.21568^{+0.00002}_{-0.00013}$ & $0.21567^{+0.00003}_{-0.00012}$ &
$+1\tmfv$\\
           & $0.21571^{+0.00001}_{-0.00002}$ & $0.21571^{+0.00001}_{-0.00002}$ &
$0\tmfv$\\
\hline
$\stes^l$  & $0.23201^{+0.00004}_{-0.00003}$ & $0.23200^{+0.00004}_{-0.00004}$ &
$+1\tmfv$ \\
           & $0.23206^{+0.00004}_{-0.00014}$ & $0.23205^{+0.00004}_{-0.00014}$ &
$+1\tmfv$ \\
\hline
$\stes^b$  & $0.23331^{+0.00009}_{-0.00000}$ & $0.23330^{+0.00009}_{-0.00001}$ &
$+1\tmfv$ \\
           & $0.23337^{+0.00004}_{-0.00015}$ & $0.23335^{+0.00004}_{-0.00014}$ &
$+2\tmfv$ \\
\hline
$\afb^l$   & $0.01533^{+0.00008}_{-0.00007}$ & $0.01536^{+0.00008}_{-0.00007}$ &
$-3\tmfv$ \\
           & $0.01528^{+0.00024}_{-0.00007}$ & $0.01531^{+0.00024}_{-0.00007}$ &
$-3\tmfv$ \\
\hline
$\alr$     & $0.14314^{+0.00028}_{-0.00030}$ & $0.14327^{+0.00028}_{-0.00031}$ &
$-1.3\tmf$ \\
           & $0.14275^{+0.00111}_{-0.00030}$ & $0.14289^{+0.00110}_{-0.00032}$ &
$-1.4\tmf$ \\
\hline
$\afb^b$   & $0.10024^{+0.00022}_{-0.00023}$ & $0.10033^{+0.00023}_{-0.00023}$ &
$-9\tmfv$ \\
           & $0.10004^{+0.00078}_{-0.00022}$ & $0.10013^{+0.00079}_{-0.00022}$ &
$-9\tmfv$ \\
\hline
\end{tabular}
\end{table}
\normalsize
 
\end{document}